# Synthesis, characterization and investigation of electrical transport in metal nanowires and nanotubes

A Thesis
Submitted for the Degree of
Doctor of Philosophy
In
Jadavpur University, Kolkata

*by*

### *M. Venkata Kamalakar*

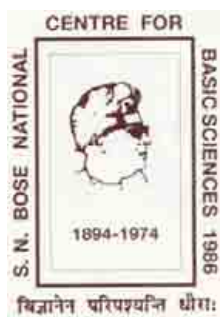

**DST Unit for Nanosciences**
Department of Material Sciences
S. N. Bose National Centre for Basic Sciences
Block-JD, Sector-III, Salt Lake
Kolkata – 700098, INDIA

June 2009

*To my family...*

# Contents

# List of Figures

x

# List of Tables

# CHAPTER 1

# Introduction

*One-dimensional nanostructures comprise an important category of nanoscience and technology. Their significance is not only limited to nano electronics but also to applications ranging from perpendicular high density magnetic recording to nanoelectromechanical systems (NEMS) comprising of sensors and actuators. In the last couple of decades, there has been a lot of development in the synthesis methods of these nanostructures and still the field has a large number of open questions. Metal nanowires and nanotubes represent a sub category of the one dimensional nanostructure and this thesis is an attempt to contribute in this direction to their synthesis, characterization and study of the intrinsic electrical transport in them due to size reduction. In this chapter, we start with the motivation of metal nanowires and nanotubes from the electrical transport point of view and present the basic mechanisms of electron transport, electrical transport in metals and a literature survey on such transport in metal nanowires and nanotubes (theoretical and experimental), followed by a short review of the synthesis of the metal nanowires and nanotubes synthesised earlier. We conclude the chapter with the outline of the thesis.*

This page is intentionally left blank

## 1.1 Motivation

The study of electrical transport properties of one dimensional nanostructure is important for their characterization, electronic device applications, and the investigation of unusual transport phenomena arising from one-dimensional quantum/finite size effects. Important factors that determine the transport properties of nanowires are the wire diameter, (important for both classical and quantum size effects), material composition, surface conditions, crystal quality, and the crystallographic orientation along the wire axis for materials with anisotropic material parameters, such as the effective mass tensor, the Fermi surface, or the carrier mobility.

## 1.2 Characteristic lengths

In order to understand the effect of size reduction on the various phenomena in one dimensional nanostructures, one needs to have an idea about certain characteristic lengths. These characteristic lengths determine the behaviour of the systems as the effects start to appear when the size of the system constrains them. In the case of electron transport properties, there are some characteristic length scales which determine whether the electrical transport in one-dimensional nanostructures is classical or quantum in nature. They are the Fermi Wavelength, the electron mean free path, the Phase relaxation length, the spin diffusion length etc.

### 1.2.1 Wavelength ($\lambda_F$)

The wavelength which is related to the kinetic energy of the electrons is called the de Broglie wavelength. At low temperatures the contribution to the current is mainly dominated by the electrons having energy close to the Fermi energy. Therefore, the Fermi wavelength is the relevant wavelength. The contribution from the electrons far from the Fermi level is usually neglected. The Fermi wavelength is given by

$$\lambda_F = \frac{2\pi}{k_F} = \sqrt{2\pi / n} \qquad (1.1)$$

where $k_F$ is the Fermi wave vector which is proportional to the square root of the electron density $n$ for 2-dimensional electron gas (2DEG). In 3DEG the corresponding relation is:

$$\lambda_F = \left( \frac{3n}{8\pi} \right)^{-1/3} \qquad (1.2)$$

In general, $k_F \sim (n)^{1/d}$, where $d$ represents the dimension of the system. For typical metallic systems the electron density is $n \sim 10^{29}/m^3$ so that the de Broglie wavelength of electrons in metal is of the order of $\sim 0.1 - 1$ nm.

## 1.2.2 Mean Free Path ($L_{mfp}$)

The mean free path is the average distance travelled by an electron (or a hole) before changing its momentum after a scattering process. The momentum change is related to the scattering of the electrons by static impurities, imperfections (lattice defects as well as internal or external surfaces and boundaries that contribute to the elastic mean free path. In addition lattice vibrations (phonons), spin wave modes (magnons) or other electrons inside the lattice also change the energy and contribute to the inelastic mean free path. For a pure metal sample, relatively free from defects, it can extend up to several microns at low temperature. The mean free path of a conduction electron $L_{mfp}$ is defined as

$$L_{mfp} = v_F \tau \tag{1.3}$$

Where $v_F$ is the velocity at Fermi surface and $\tau$ is the free time during which the electric field acts on the conduction electron.

## 1.2.3 Phase relaxation length ($L_\varphi$)

Electron wave functions can interfere among themselves forming standing waves leading to the substantial corrections to the Boltzmann conductivity, eventually leading to localization in extreme cases. For such interference to take place, the wave functions must be phase coherent. However, the phase coherence can be lost by dynamic scatters (such as electrons and phonons). The phase relaxation length ($L_\varphi$) is the length over which the conduction electrons lose their phase coherence. Hence, for a sample with length larger than the phase relaxation length, one cannot observe quantum interference of the electron wave functions. When the phase relaxation time $\tau_\varphi$ is defined as the time over which the phase fluctuations reach unity, the phase relaxation or coherence length $L_\varphi$ can be expressed as

$$L_\varphi = v_F \tau_\varphi \tag{1.4}$$

which is often the case of high mobility semiconductors [1]. But in the low mobility semiconductors or polycrystalline metal thin films, the momentum relaxation time $\tau_m$ can be considerably smaller than $\tau_\varphi$ leading to a diffusive regime. In such a case the electron trajectory over a time of $\tau_\varphi$ can be visualized as the sum of a number (=$\tau_\varphi / \tau_m$) of short trajectories each of length ~ $\upsilon_F \, \tau_m$. In this diffusive regime, the phase coherence length is expressed in terms of the diffusion constant D as [1]

$$L_\varphi^2 = D\tau_\varphi \qquad (1.5)$$

where $D = \upsilon_F^2 \, \tau_m / 2$. In transition metal heterostructures $L_\varphi$ ~ few tens of nanometers typically [1].

## 1.3 Electrical transport in 1-D nanostructures

As discussed above the dimensionality of nanostructures is determined by the characteristic length. Structurally if the a nanostructure is confined in two dimensions and free in the third dimension, then it can be called as one-dimensional provided its properties are affected by such formation. For example if the physical characteristic lengths get constrained in two dimensions, then the physical/chemical properties are bound to show changes due to finite size statistical mechanical effects/quantum mechanical effects and hence such systems can be called **one dimensional nanostructures**. Strictly going by the definition, unless there is some or the other forms of quantum mechanical confinement, such systems are termed as quasi one dimensional in nature.

Electronic transport phenomena in low-dimensional systems can be broadly categorized into mainly **ballistic transport** and **diffusive transport mechanisms.**

## 1.3.1 Ballistic transport

This kind of transport implies that electrons can travel across the sample length without undergoing any scattering inside the sample. Ballistic transport occurs when length over which the transport is measured which we call sample length $L$ is such that $L << L_{mfp}$, so that the electron suffers no scattering between the electrodes between which the voltage drop is measured. In this case the voltage drop occurs due to the contacts. This is the regime of classical Ballistic transport.

The Ballistic transport can also be quantum in nature. If the diameter of the wire is comparable to electron wavelength, the electronic energy levels become quantized and the electrical transport becomes Quantum in nature. For ballistic transport to take place, an electron must not overcome the energy difference ($\varepsilon_j - \varepsilon_{j-1}$) between subbands *j* and *j-1* by its thermal energy ($k_BT$). The conductance of such a wire connected to macroscopic reservoirs ("contact pads") is given by Landauer formula;

$$G = G_0 \sum_{i,j=1}^{N} T_{ij} \;\; ; \;\;\; G_0 = \frac{2e^2}{h} \tag{1.6}$$

where $e$ is the electronic charge, $h$ is the Planck's constant and $T_{ij}$ is the transmission probability from i$^{th}$ channel at one end of the wire to the j$^{th}$ mode at the other end. The summation is over all channels having a non-zero value of the transmission probability. In the ideal case with no backscattering at the contacts, $T_{ij}$ =1 for i=j and zero otherwise, $G = NG_0$ , the conductance is quantized into an integral number of universal conductance units $G_0$ [2,3], ($G_0^{-1} = h/2e^2 = 12.9\,k\Omega$) and $N$ is the number of channels available for conduction. The value of $N$ depends on the diameter of the wire and is given by $N \sim d/\lambda_F$. As stated before in the case of ballistic transport, the conduction is mainly determined by the contacts and is independent of the sample length provided the later is less than the mean free path. This together with the quantization of conductance gives rise to the discrete variation (unlike the continuous variation with diameter in case of Ohmic wire) of the conductance as a function of sample diameter.

Ballistic transport phenomena are usually observed in very short quantum wires, such as those produced using mechanically controlled break junctions (MCBJ) [4, 5] where the electron mean free path is much longer than the wire length and the conduction is a pure quantum phenomenon. As stated earlier, to observe ballistic transport, the thermal energy must also obey the relation $k_BT << \varepsilon_j - \varepsilon_{j-1}$, where $\varepsilon_j - \varepsilon_{j-1}$ is the energy separation between subband levels $j$ and $j-1$. The one-dimensional nanostructures, which exhibit such ballistic transport are said to be of **quantum nature.**

## Quantum conductance in 1-D metal systems

Since the discovery of quantized conductance in 1-D systems in 1988 [2, 3], the ballistic transport of 1-D systems has been extensively studied. The phenomena of conductance quantization occur when the diameter of the nanowire is comparable to the electron Fermi wavelength, which is on the order of 0.5nm for most metals [6]. Most conductance quantization experiments up to the date were performed by bringing together and separating two metal electrodes. As the two metal electrodes are slowly separated, a metal neck is formed due to strong metallic cohesive energy. This necks transforms into an atomically thin wire before it breaks completely (see Fig. 1.1(a)), and conductance in integral multiple values of $G_0$ is observed through these nano contacts. One such way of producing a nano contact is based on STM.

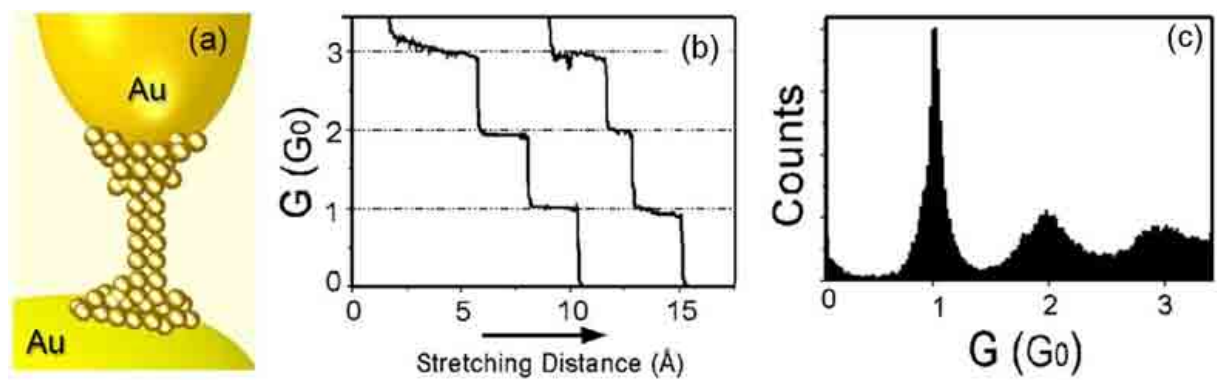

**Figure 1.1: (a)** Mechanical fabrication of a metal quantum wire with a STM setup. A STM tip is first pressed into a metal substrate and then pulled out of contact during which an atomically thin wire is formed before breaking. **(b)** Typical conductance versus stretching distance traces that show the quantized variation in the wire conductance. **(c)** Conductance histogram of Au wires in 0.1M NaClO4. The well defined peaks near integer multiples of $G_0$ = $2e^2/h$ have been attributed to conductance quantization. Each histogram was constructed from over 1000 individual conductance traces like the ones shown in (b) [7].

In this method the STM tip is brought in contact with the substrate and conductance is measured while retracing it back from the substrate. Gimzewski *et al*. first reported this kind of experiment to study the difference between electron transport through tunnelling and ballistic mechanisms [7]. Fig. 1.1(b) shows stepwise conductance traces during stretching a gold wire using STM. The steps of conductance are integral multiples of $G_0$ indicating quantization of conductance in atomic scale wires. The steps can deviate significantly from the exact values of $G_0$, $2G_0$, $3G_0$ etc. Thus in general a few thousands of times the experiment is repeated and a histogram is generated. Fig. 1.1(c) shows the histogram generated from 1000 individual conductance traces. The conductance quantization behaviour is found to be independent of the contact material, and has been observed in various metals, such as Au [8], Ag, Na, Cu [9], and Hg [10]. For semimetals such as Bi, conductance quantization has also been observed for electrode separations as long as 100 nm at 4K because of the long Fermi wavelength (≈ 26 nm) [6], indicating that the conductance quantization may be due to the existence of well-defined quantum states localized at a constriction instead of resulting from the atom rearrangement as the electrodes separation.

When the length $L$ of a wire is shorter than the electron mean-free path ($L \leq L_{mfp}$), the electron transport is ballistic, that is, it occurs without collisions along the wire. Thus, one would expect the resistance of a ballistic wire to be identically zero. Experimentally, however the resistance of a ballistic wire is measured to be have a finite value which equals $h/(2e^2N) = 12.9/N$ kΩ, where $N$ is the number of transverse modes in the wire. This was measured experimentally by Sharvin *et al.* [11] directly using point-contact to produce constrictions in a conductor smaller than the mean free path. This resistance is the "contact resistance" arising at the contact between the wire and the macroscopic pads connecting it to the measuring electronics. For metals the small value of $\lambda_F$ ensures that the number $N$ is very large and hence the contact resistance is very small.

Since conductance quantization is only observed in breaking contacts, or for very narrow and very short nanowires, most nanowires of practical interest (possessing lengths of several microns) lie in the diffusive transport regime, where the carrier scattering is significant and should be considered. .

## 1.3.2 Diffusive transport

For nanowires with lengths much larger than the carrier mean free path, the electrons (or holes) undergo numerous scattering events when they travel along the wire. In this case, the transport is in the diffusive regime, and the conduction is dominated by carrier scattering within the wires, due to phonons (lattice vibrations), boundary scattering, lattice and other structural defects, and impurity atoms. These wires are **classical** in nature. For wire diameters ($d$) comparable to or smaller than the carrier mean free path $L_{mfp}$, i.e., $d \approx L_{mfp}$ or $d < L_{mfp}$), but still much larger than the de Broglie wavelength of the electrons ($d >> \lambda_F$), the transport in nanowires falls into the **classical finite size regime**, where the band structure of the nanowire is still similar to that of bulk, while the scattering events at the wire boundary alter their transport behavior.

Thus, for most technological applications, the 1-D metal nanowires and nanotubes (diameter ~ 10nm-100nm) systems fall in the classical finite size regime. The length of such systems extend up to several microns and the finite diameter and enhanced surface to volume ratio play important roles in alteration of the transport behaviour. A quantitative analysis of electrons getting scattered from the finite boundary and quasi particles like phonons and magnons has never been done in such systems. Technologically it is a challenge to synthesise metal nanowires and nanotubes and their arrays and to understand the electron transport in them for various optimization issues related to different applications. This thesis as said earlier is thus an attempt to the synthesis, characterization and study of these effects of the finite size on the electron scattering phenomena in such systems.

Since, this thesis is about classical one-dimensional nanostructures and their electrical properties, it is essential introduce the general electrical transport in metals at this stage.

## 1.4 Electrical transport in Metals

The electrical transport studied by the electrical resistivity in metals has two main contributions, namely the residual resistivity ($\rho_0$) and the temperature dependent resistivity $\rho(T)$. The residual resistivity depends on the intrinsic defects, grain boundaries, impurities etc. In the case of nanowires, the finite diameter contributes to the residual resistivity. The temperature dependent part $\rho(T)$ can arise from various phenomena like electron-phonon interaction, electron-magnon interaction, electron localization (in case of disordered metals at low temperature). For a good non-magnetic metal, the temperature dependence originates mainly from electron-phonon interaction ($\rho_L$) described well by the Bloch-Grüneisen formula. However, in the case of magnetic metals, an additional temperature dependent magnetic contribution ($\rho_M$) arises from the electron-magnon interaction which has different temperature dependence in different temperature regimes. Thus for a metal the electrical resistivity is given by

$$\rho = \rho_0 + \rho(T)$$
$$\rho = \rho_0 + \rho_L + \rho_M$$

(1.7)

Quantitative estimation of these phenomena gives us a handle to tune the resistivity in case of metal nanowires and nanotubes.

As pointed out earlier, $\rho_M$ has different temperature dependence in different regimes. In case of a magnetic metal, the resistivity shows up a sudden change in behaviour near the ferromagnetic to paramagnetic phase transition at the Curie temperature ($T_C$). A theory has been proposed by Fisher and Langer in 1968 [12] describing the anomaly in resistance behaviour in terms of the divergence of specific heat near $T_C$. Since then, the resistance anomaly has been successfully used to describe the critical phenomena and derive the specific heat critical exponent in ferromagnets. However no such attempts were done to analyse the resistance data in case of magnetic nanowires. All these reasons, made us concentrate on the electrical transport in magnetic nanowires to study the above described phenomena. In the context of metal nanotubes, there are only a handful reports of synthesis of metal nanotubes because of the complexities involves which we will be discussing in Chapter 7. Thus, in this thesis we took challenge of inventing a method for the synthesis of pure metal nanotubes with little attention to the electrical transport in them. The other reason behind this was the electrical transport in metal nanotubes is similar to that in case metal nanowires as we observed.

## 1.5 Review of resistivity measurements in nanowires

Here we present a compilation of the literature on the resistivity measurements done on metal nanowires.

### 1.5.1 Theoretical work on resistivity of nanowires

Most of the classical theories of electrical resistivity of nanowires were developed long back, mainly to address surface scattering seen in thin films. However, many of these theories can be extrapolated to nanowires as well. The theoretical work on resistivity on nanowires is mainly done for the resistivity arising out of boundary wall scattering and grain boundary (in case the wire is polycrystalline) scattering. Dingle (1950) [13, 14] gave a relation for the resistivity of wires thin wires in terms of the bulk resistivity and the diameter of the thin wire. These complicated theoretical calculations are based on the Schondimer's (1932) work done on thin films of metals in terms of bulk resistivity. In accordance with Dingle's calculation, the resistivity ($\rho$) of a thin wire of diameter ($d$), follows as

$$\frac{\rho}{\rho_0} = \frac{\psi(\kappa)}{\kappa} \qquad (1.8)$$

Where $\kappa = \dfrac{d}{l}$; $l$ being the mean free path of electron in bulk.

$$\frac{1}{\psi(\kappa)} = \frac{1}{\kappa} - \frac{12}{\pi\kappa} \int_0^1 (1-t^2)^{1/2} S_4(\kappa t)\, dt \qquad (1.9)$$

$$S_4 = \int\limits_1^\infty e^{-ut}(1-t^2)^{1/2}t^{-n}\ dt \qquad (1.10)$$

The resistivity obtained from Eq. (1.8) does not involve specularity coefficient *P*, which determines the fraction of the elastically scattered (the other kind being diffusive scattering) electrons from the surface of the wire boundary. A relation for $p \neq 0$ can be obtained from $p = 0$ by means of the simple relation

$$\left(\frac{\rho_0}{\rho}\right)_{\kappa,p} = (1-p)^2 \sum_1^\infty np^{n-1}\left(\frac{\rho_0}{\rho}\right)_{n\kappa,\,p=0} \qquad (1.11)$$

For very thick and thin nanowires the formula takes the following forms

$$\frac{\rho}{\rho_0} = 1 + \frac{3}{4\kappa}(1-p) \qquad\qquad (\kappa \gg 1) \qquad (1.12)$$

$$\frac{\rho}{\rho_0} = \frac{1-p}{1+p}\ \frac{1}{\kappa} \qquad\qquad (\kappa \ll 1) \qquad (1.13)$$

The above theory does not consider the grain boundary scattering which can give rise to significant contribution to the residual resistivity. While in bulk material the mean grain size *Dg* is of the order of several micrometers, the *Dg* in thin polycrystalline wires/films is generally comparable to *l* . Thus, electron scattering from grain boundaries is expected to contribute to the electrical resistivity. The main theory of this effect on the resistivity of thin films was developed by Mayadas and Shatzkes [15] (MS model). Grain boundaries can be regarded as potential barriers which are randomly distributed. It is assumed that they exhibit partially reflecting surfaces perpendicular to the direction of the electric field in a conductor, providing additional scattering sites.

The fraction of electrons reflected from these potential barriers is described by the reflection coefficient *Rg*. From this theory, the contribution of grain-boundary scattering $\rho_g$ to the resistivity is given by

$$\frac{\rho_0}{\rho_g} = 3\left[\frac{1}{3} - \frac{\alpha}{2} + \alpha^2 + \alpha^3 \ln\left(1 + \frac{1}{\alpha}\right)\right] \qquad (1.14)$$

where

$$\alpha = \frac{l}{D_g}\frac{R_g}{1-R_g} \qquad (1.15)$$

In accordance with this theory, three factors, namely $Dg$, $l$, and $Rg$, determine the effect of grain-boundary scattering on resistivity.

An important drawback of the model of Mayadas and Shatzkes is that they assume the distribution of grain sizes to be Gaussian for mathematical simplicity while experimentally the distribution is almost always observed to be log-normal. Durkan *et al.* [16] incorporated this factor in their calculations and obtained almost the same result as Equation 1.9 with the average grain size $D_g$ replaced by an effective grain size $D_{eff}$ given by

$$D_{eff} = \frac{\frac{\pi}{4} \int\limits_{w}^{\infty} f(D) D \frac{D-w}{w} \, dD}{\int\limits_{w}^{\infty} f(D) \frac{D-w}{w} \, dD} \qquad (1.16)$$

where

$$f(D) = \frac{1}{\sigma D \sqrt{2\pi}} \exp\left\{ -\left[ \frac{1}{\sqrt{2}\sigma} \ln\left( \frac{D}{D_g} \right) \right]^2 \right\} \qquad (1.17)$$

$\sigma$ is the log-normal standard deviation of the grain diameters and $w$ the film width. These two theories together have been used quite successfully to explain the size dependence of resistivity in thin films or narrow wires.

Apart from these kind of analytical results, a recent theoretical study of the temperature dependence of resistivity in metallic nanowires has been carried out by Ratan Lal [17] using a tight-binding approach for the electronic structure, deformation potential approach for the electron-phonon interaction, and the Kubo formula for the conductivity. He estimated that at high temperatures, the resistivity should increase linearly with temperature for wires of diameter $d > 60M$ where $M$ is the width of the wire in units of the lattice constant. In the case of common metals like Ag and Cu this number turns out be around 25 nm. Below this diameter the theory predicts a super linear dependence of the resistivity on temperature. For example, for wires of diameter $20M$, the calculations predict that the resistivity should go as $T^{1.2}$. The slope of the curves also increases with a decrease in the wire diameter. The results also predict a diameter dependence of the resistivity in nanowires for diameters $d < 120M$ while for wires of larger diameter, the resistivity is expected to be independent of $d$.

## 1.5.2 Experimental Work:

The past experimental results can be mainly classified into two categories, namely the width/diameter dependence of resistivity and the temperature dependence of resistivity of the nanowires.

## 1.5.2.1 Width/Diameter dependence on the resistivity

The general trend seen in nanowires of various metals is that the resistivity increases with the decrease of the wire diameter due to the boundary scattering, grain boundary scattering limitations imposed by the finite wire diameter and is mainly dependent on the microstructure.

A detailed study of size dependence of the resistivity of nanowires at 300 K was carried out by Durkan *et al.* [16]. They fabricated polycrystalline Au nanowires of length 500 nm, thickness 20 nm by electron beam lithography and studied the width dependence of nanowire resistivity. The measurements were done before and after annealing.

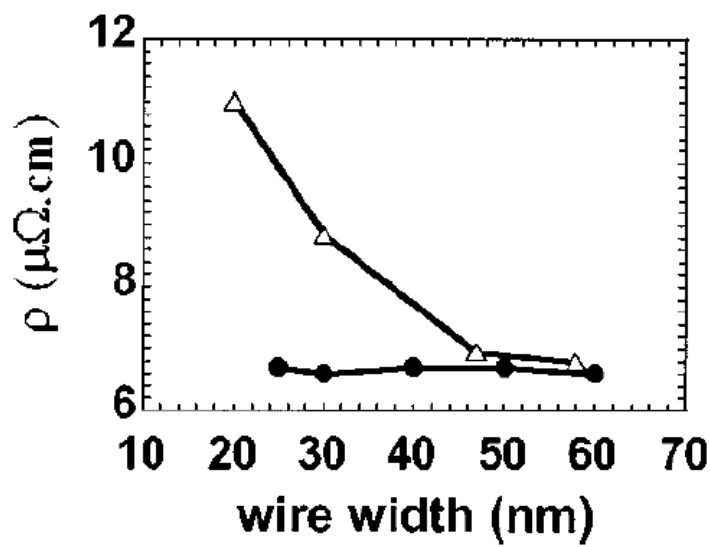

**Figure 1.2**: Resistivity of Au nanowires at 300 K as a function of the wire width as measured by Durkan *et al.*. The triangles are for samples with grain size 40 nm while the filled circles are samples with grain size 20 nm. Reprinted with permission from [16]. Copyright (2000) by the American Physical Society.

The plots shown in Fig. 1.2 reveals the width dependence of resistivity for the annealed (40nm mean grain size) and un-annealed (20 nm mean grain size). It is clear that the annealed wires with 40 nm grain size show width dependence. However no such width dependence is seen in case of measurements done before annealing where the grain size was 20 nm. They concluded that when the width of wire is comparable to the mean grain size of the film, grain boundary scattering is the dominant source of increased resistivity while, when wire width is below approximately half the mean grain size, surface scattering becomes important, approaching the same order of magnitude as grain-boundary scattering as the width decreases.

A similar study was carried out by Steinhöegl *et al.*[18, 19] on Cu nanowires of widths ranging from 40 nm to 800 nm on $SiO_2$ substrate by using e-beam lithography followed by anisotropic etching. From TEM analysis, it was found that the grain size increases linearly with the width until it saturates at a constant value of about 320-400 nm. The authors used a combined model of surface scattering (FS) and grain boundary scattering (MS) to explain their experimental data. They found that for wider structures the surface the surface scattering dominates over grain-boundary scattering. They argued that this is because these structures behave like thin films, where the limiting dimension for the effective mean free path is the thickness of the film (50-230 nm) and the average distance of the grain boundaries (320-400 nm).

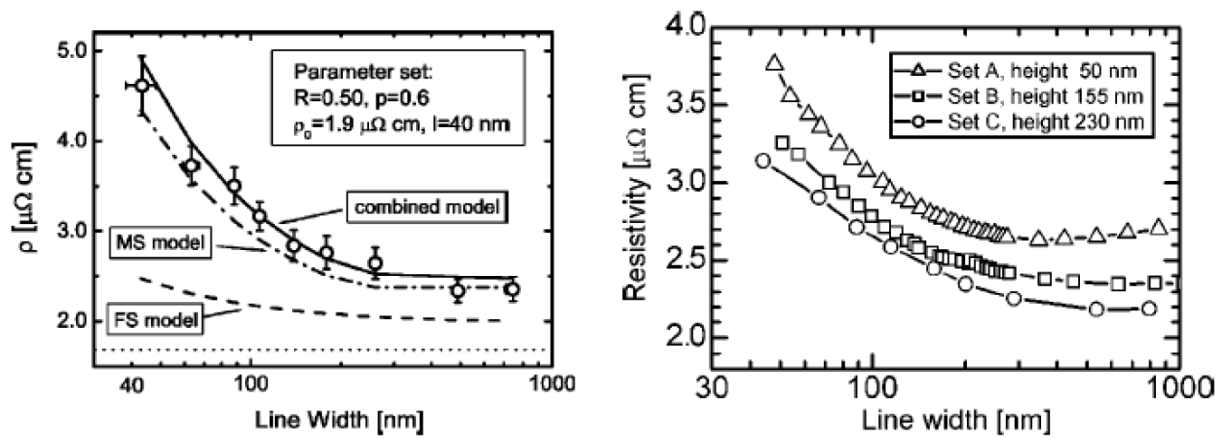

**Figure 1.3:** Resistivity of Cu nanowires at 300 K as a function of the wire width as measured by Steinhögl *et al.* (a) is reprinted with permission from [18] Copyright (2002) by the American Physical Society. (b) is reprinted with permission from [19]. Copyright [2004], American Institute of Physics.

For the narrow structures the contribution of surface scattering and grain boundary scattering is roughly the same. This is plausible because the average grain size is limited by the lateral geometrical dimensions. They also reported that the two parameters used to quantify the surface scattering and grain boundary scattering the conduction electrons (the specularity coefficient "*p*" and the reflectivity "*R*" respectively) could not be extracted independent of each other with high accuracy as both give a resistivity component which was inversely proportional to the film width.

Similar studies were carried out by Josell *et al.* [20] on Ag nanowires prepared by electrodeposition in substrates with prefabricated trenches of height 100-300 nm and width 50-840 nm. It was observed that the resistivity increased as the wire width and height were decreased.

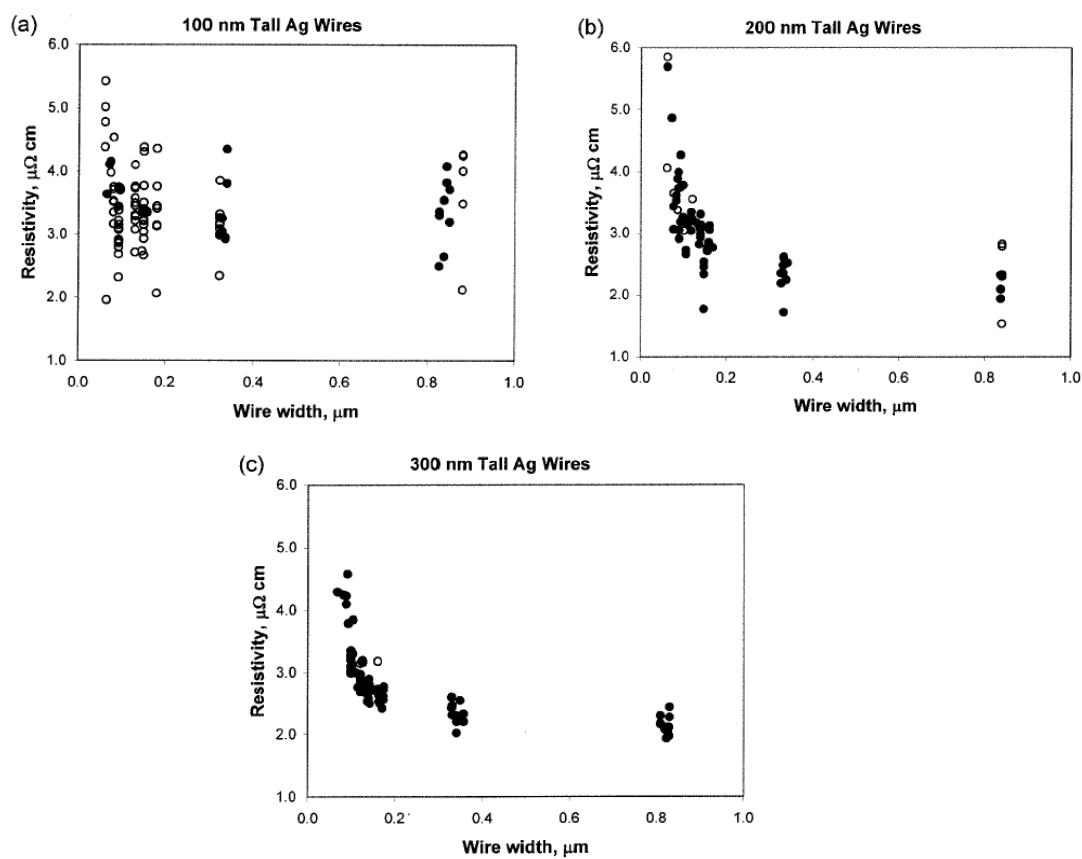

**Figure 1.4:** Resistivity of Ag nanowires at 300 K as a function of the wire width as measured by Josell *et al.* [20]. Reprinted with permission from [21]. Copyright [2004], American Institute of Physics.

Fig.1.4 (a) shows that the grain boundary contribution to the resistivity was found to be independent of width when the height was much less than the width because of constrained grain size. As the wire height is increased, the regular trend of grain boundary contribution is observed. However no attempts were made to distinguish between the grain boundary resistivity and the surface scattering contribution in the work.

Wu *et al.*[21] reported the measurements of resistivity on polycrystalline Cu nanowires of widths 90 nm to 300 nm by optical lithography & electroplating. The authors observed that the samples were polycrystalline with size of the grains being almost equal to the sample width. **Fig. 1.5** shows the plot of resistivities of nanowires of widths ranging from 90 nm to 300 nm at 273 K. The authors have attributed enhancement in the resistivity with decrease in diameter to the exclusive grain boundary scattering with negligible surface scattering. They note that the observed predominant grain boundary scattering may include scattering from the impurities as well, because both have the same temperature dependence. Since, the impurities are known to accumulate near the grain boundaries [22] due to purification of the grains during grain growth; it is more likely that their influence is manifested mainly through a change in the reflection coefficient $R$. Thus the grain size and the reflection coefficient $R$ mainly influence the enhanced resistivity at low dimension.

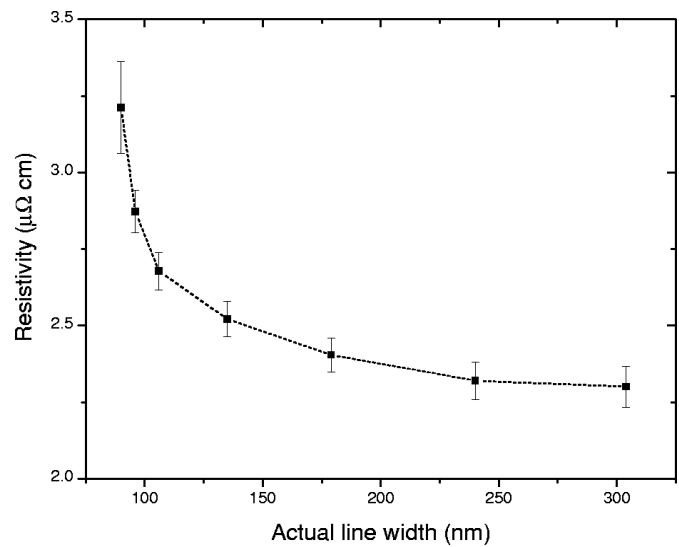

**Figure 1.5:** Resistivity of Cu nanowires at 300 K as a function of the wire width as measured by Wu *et al.* Reprinted with permission from [21]. Copyright [2004], American Institute of Physics.

The first main attempt to understand solely the surface scattering was done by A. Bid *et al.*[23] in single crystalline Ag nanowires. Here the authors studied the boundary wall scattering using single crystalline silver nanowires using Dingle and Chamber's surface scattering model for thin wires. The authors noticed no grain boundaries in TEM and obtained the specularity coefficient $p = 0.5$ which matches with that in thin films. The Dingle's model fit to the data obtained by Aveek Bid *et al.* is shown in Fig. 1.6.

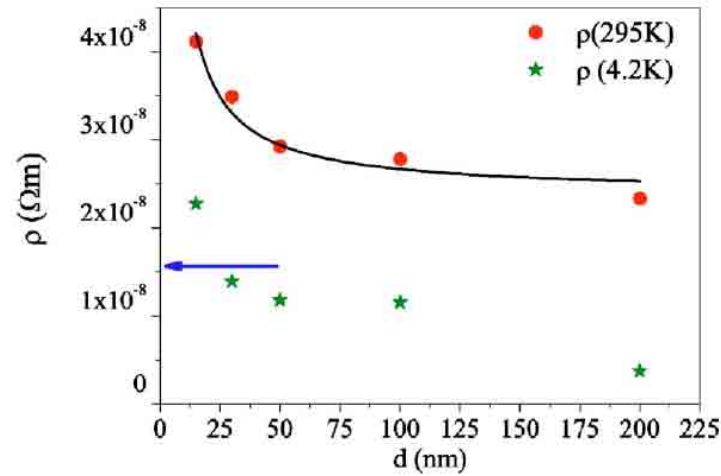

**Figure 1.6:** Resistivity of Silver nanowires at 295 K and 4.2 K as a function of the wire diameter as measured by A. Bid *et al.* [23].

## 1.5.2.2 Temperature dependence of resistivity

The temperature dependence of the resistivity reflects the main mechanism electron transport involving the scattering by the lattice and spin wave modes and electron localization effects. Giordano *et al.*[24] studied the temperature and size dependence of the resistivity of Au nanowires with diameters as small as 8 grown in nuclear track etched porous mica by electroplating of Au.

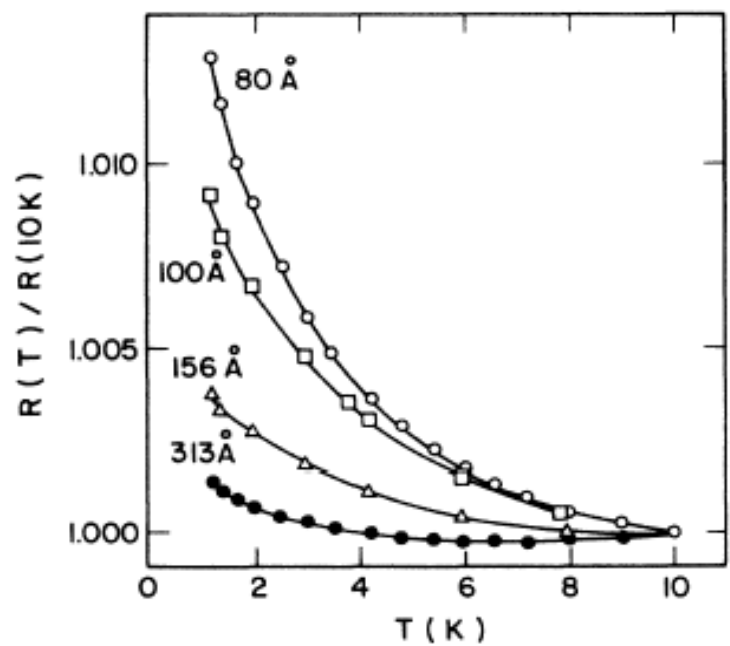

**Figure 1.7:** Fractional change in the resistance of disordered Au nanowires of various diameters as the samples are cooled down from 10 K to 1.2 K. Reprinted with permission from Giordano *et al.* [24]. Copyright (1986) by the American Physical Society.

The upturn in resistance shown in Fig. 1.7 reveals that the wires were disordered. For lower diameter nanowires the relative increase in resistance with decreasing temperature was found to be more. Beyond 10 K the wires showed an increase in resistance due to normal electron-phonon scattering as explained by the authors. No quantitative analysis of the temperature dependence of the resistance in the temperature range above 10 K was carried out by the authors in this case. The origin of upturn of resistance can have two sources; (a) electron localization (b) electron-electron interaction. It has been predicted theoretically that for a wire having a length larger than $L_{loc}$ (where $L_{loc}$ is the length of the wire at which its residual resistance becomes $\approx h/e^2$ 25.8 KΩ) all the electronic states will be localized and hence electrically the wire will behave as an insulator with a negative value of $dR/dT$ [25, 26, 27]. The relative change in the resistance with decreasing temperature in this case will go as $\Delta R/R \sim ( D\tau_i^{1/2} ) / L_{loc}$ where $\tau_i$ is the inelastic scattering time which depends on temperature as $\tau_i \sim T^{-p}$ with $p = 1 - 2$. The other mechanism that can cause the upturn in resistance at low temperatures is electron-electron interactions in which case the relative change in the resistance with decreasing temperature will go as $\Delta R/R \sim T^{-1/2}$ [25, 26, 27]. The authors find that at low temperatures ($T < 3$ K), their data followed a $T^{-1/2}$ behaviour consistent with electron-electron scattering theory. At higher temperatures (3 K$< T <$10 K) the plots deviate from $T^{-1/2}$ behaviour with the value of $\Delta R/R$ falling below what it should have been for a $T^{-1/2}$ behaviour. This, as claimed by the authors, could be due to the effect of localization, as for a system with weak spin scattering, localization decreases the value of resistance [25, 26].

The $T^{-1/2}$ dependence of the resistance of disordered metallic nanowires at low temperatures have been repeatedly observed experimentally [28-31] suggesting that the low temperature electrical resistance of disordered metallic nanowires is dominated by electron-electron interactions.

The upturn at low temperature due to the localization due to constrained phase coherence length in Zn nanowires fabricated by vapour deposition in porous silica or alumina is reported by J. P. Heremans *et al.* [32]. For nanowires with diameters comparable to the phase-breaking length, their transport properties may be further influenced by localization effects. It has been predicted that in disordered systems, the extended electronic wave functions become localized near defect sites, resulting in the trapping of carriers and giving rise to different transport behaviour. Localization effects are also expected to be more pronounced as the dimensionality and sample size are reduced.

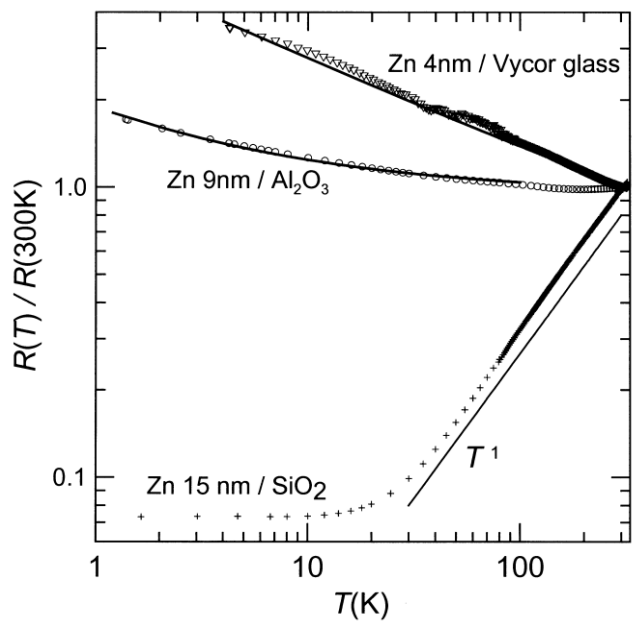

**Figure 1.8:** Temperature dependence of the resistance of Zn nanowires synthesized by vapour deposition in various porous material templates (Heremans et al., 2002). The data are given as points, the full line are fits to a $T^1$ law for 15nm diameter Zn nanowires in $SiO_2$ template, denoted by $Zn/SiO_2$. Fits to a combined $T^1$ and $T^{-1/2}$ law were made for the smaller nanowire diameter composite 9nm $Zn/Al_2O_3$ and 4nm Zn/Vycor glass samples.

While 15 nm Zn nanowires exhibit an $R(T)$ behavior with a $T^1$ dependence as expected for a metallic wire, the $R(T)$ of 9 nm and 4 nm Zn nanowires exhibits a temperature dependence of $T^{-1/2}$ at low temperatures, consistent with 1-D localization theory. Thus, due to this localization effect, the use of nanowires with very small diameters for transport applications may be limited.

A lot of work has been done by Zhang *et al.* [33] on the resistivity of Bi and Sb nanowires because of low carrier concentrations ( and hence the large Fermi wave vector which enables to see quantum effects at even larger diameters ) than normal metals like Ag, Au, Cu with high conductivity. For pure annealed Bi nanowires of diameter 90 nm, the TCR changed sign from negative below 70 K while for the wire of smaller diameter 65 nm, the TCR remained negative over the entire range of temperature (Fig. 1.9). The resistance of Bi nanowires depends on two competing factors: the carrier density that decreases with T, and the carrier mobility that decreases with T. The interplay between them explains the transition from semimetal to a semiconducting behaviour in 90 nm nanowire. The increase in carrier concentration with increasing T for T > 70 K outweighs the decrease of carrier mobility for the nanowire sample and therefore their resistivities decrease with increasing T over this temperature range. However at low temperature the dominant scattering mechanism for carriers is the wire boundary scattering, making the carrier mean free path and carrier mobility relatively insensitive to T. It was seen for T < 70 K, the carrier mobility of the 90 nm sample increases faster with decreasing temperature than that of the 65 nm sample, consistent with the fact that the wire boundary scattering for carrier is dominant as wire diameter is reduced.

For the 90 nm sample at T<50 K, the increase in carrier mobility with decreasing T can therefore outweigh the small decrease in carrier concentration, so that the total resistivity decreases with decreasing T. However, for the 65 nm the T dependence of the carrier mobility not only is very small, but actually indicates a slightly decreasing carrier mobility with decreasing T at T<10 K. Consequently, the overall resistivity of the 65 nm sample continues to increase with decreasing T at low temperatures. For the un-annealed samples there was not much of a diameter dependence of resistivity, which the authors attribute to the higher impurity and defect levels in the as prepared samples. The 60 nm-diameter (0.1 at.% Te-doped) Bi nanowire shows a much more metallic R(T) behaviour than pure Bi samples. The authors do not attribute any reason for this difference of behaviour between pure Bi and Te-doped Bi nanowires.

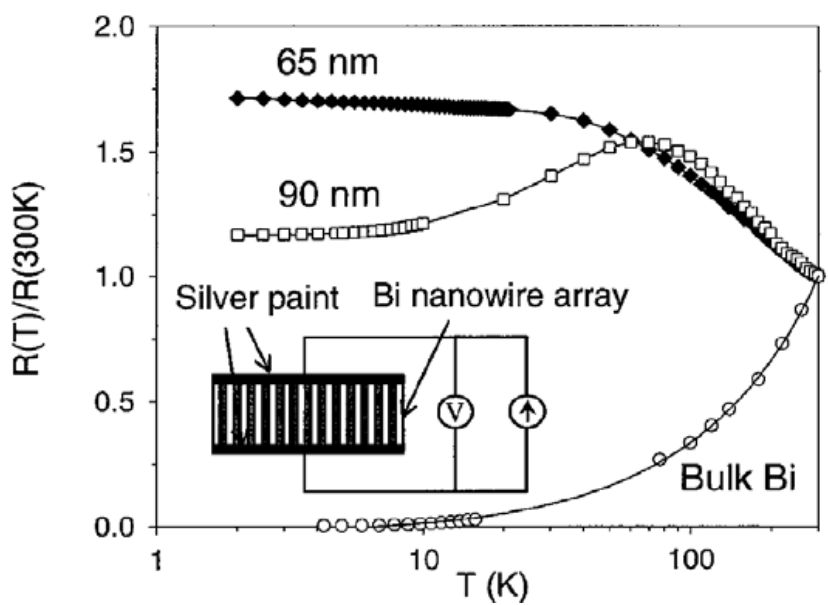

**Figure 1.9:** Resistivity of Bi nanowires of different wire widths as a function of temperature as measured by Zhang *et al.*. Reprinted with permission from [33]. Copyright (2000) by the American Physical Society.

Similar observations have been reported by Wang *et al.* [34] on Bi nanowires of diameter 20 nm, 50 nm and 70 nm electrochemically grown in anodic alumina template using a three-electrode electrochemical cell and by Chiu *et al.* [35] on rectangular cross-section Bi nanowires of thickness 50 nm and width 70-200 nm fabricated by electron beam lithography.

The clear signature of the semimetal-to-semiconductor transition observed in Bi nanowires was absent in Sb nanowires (prepared in anodic alumina template using the vapour phase technique) for the measured diameter range [36]. For Bi nanowires with diameters ranging from 10 to 200 nm, the measured resistance changes from a monotonic temperature dependence to a non-monotonic dependence as the width of the wire increases, while resistance of Sb nanowire arrays in this range of width exhibits less variation with temperature. Qualitatively variation of resistance of Sb wires with temperature can be explained in terms of the relative importance of electron-phonon scattering, which increases with temperature, and electron-boundary scattering, which is relatively independent of temperature. Unlike Bi, for the Sb nanowires, in the measured diameter range, the carrier density is a weak function of T, and the resistance is mainly determined by the temperature dependence of the various scattering mechanisms.

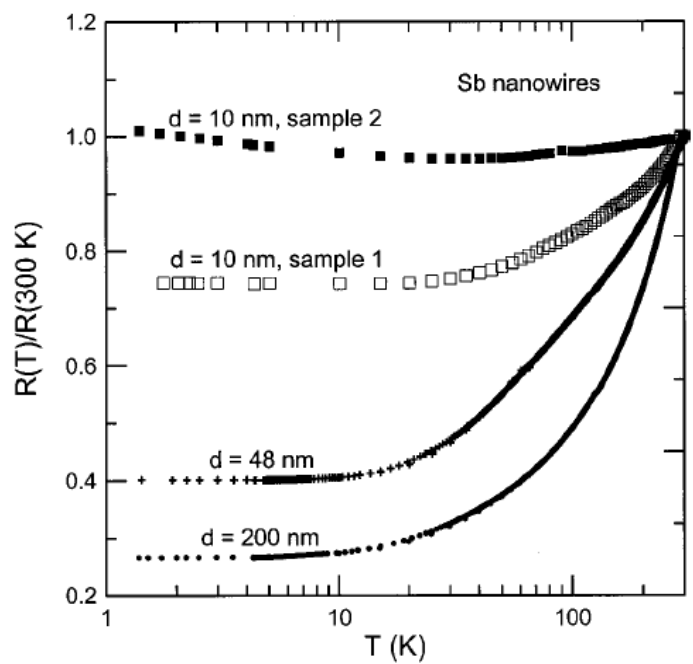

**Figure 1.10:** Experimental temperature dependence of the resistance of Sb nanowires of various diameters, normalized to the resistance at 300 K. Reprinted with permission from [36]. Copyright (2000) by the American Physical Society.

At high temperatures (T>100 K), the phonon scattering dominates and the resistance rises with increasing T. As the temperature is lowered, the boundary scattering becomes dominant, giving rise to a less T-dependent resistance for T< 100 K, which explains the general features of the measure R(T) for the various Sb nanowire arrays. As the wire diameter decreases, boundary scattering becomes relatively more important in them than it is for wires of larger diameter and so, smaller-diameter wires exhibit less temperature variation than larger-diameter wires.

More recently Aveek Bid *et al.* [23] carried out first detailed study temperature dependence of electrical resistivity of the nanowires of Ag and Cu with diameters ranging from 15-200 nm synthesized by electrodeposition of the respective metals in nanoporous polycarbonate membranes. It was observed that the resistivity could be well described by Bloch-Grüneisen formula with a characteristic Debye temperature successfully over the whole range of temperature which implies that the basic electron-phonon interaction remains the same. The Debye temperature was found to decrease as the wire diameter is decreased. The authors have attributed the increase in residual resistivity of the nanowires with the decrease in diameter to the enhanced surface scattering.

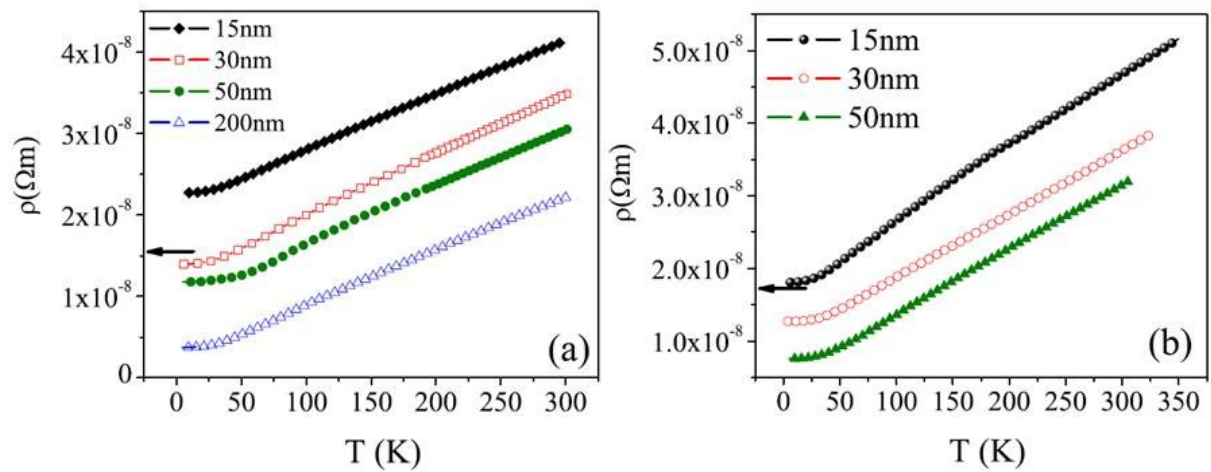

**Figure 1.11:** The resistivity of (a) Ag and (b) Cu nanowires as a function of temperature. The arrows show the value of the resistivity of high purity Ag and Cu at 300 K.

From this review of the resistivity of metal nanowires we understand that the resistivity increases as the width/diameter of the nanowire is reduced. This is because the electron mean free path becomes comparable and constrained by either the wire diameter or grain width or both. We notice that for disordered metal nanowires of low diameter at low temperature, the resistivity take upturn as it is being dominated by electron localization and electron-electron interaction. For pure metal nanowires, the temperature dependent resistivity is mainly dominated by the electron-phonon scattering. There are no reports of systematic studies of electrical transport in the case of magnetic nanowires. There are only a handful of experiments performed on the temperature dependence of metal nanowires. There are no measurements performed for temperatures beyond 300 K. Hence, there are plenty of questions left unanswered on the electrical transport on the magnetic nanowires starting from low temperatures to their phase transition temperatures.

### 1.5.3 Electrical transport measurements on metal nanotubes

Because of a handful number of reports of synthesis of metal nanotubes, there are not many reports of electrical transport in them. This is because metals cannot be reduced to particular shapes in absence of templates. Nevertheless in the last few years, there have been some efforts to synthesize metal nanotubes, particularly arrays of metal nanotubes in anodic alumina templates. Still, the synthesis of pure metal nanotubes remains a challenge as a general method of controlled synthesis is needed. The metal nanotubes so far produced are mainly classical if we try to classify them in terms of quantum and classical transport mechanisms discussed earlier in this chapter. The electrical properties of classical metal nanotubes are supposed to be similar to that of classical nanowires and thin films of similar dimensions because of low Fermi wavelength in case of metals.

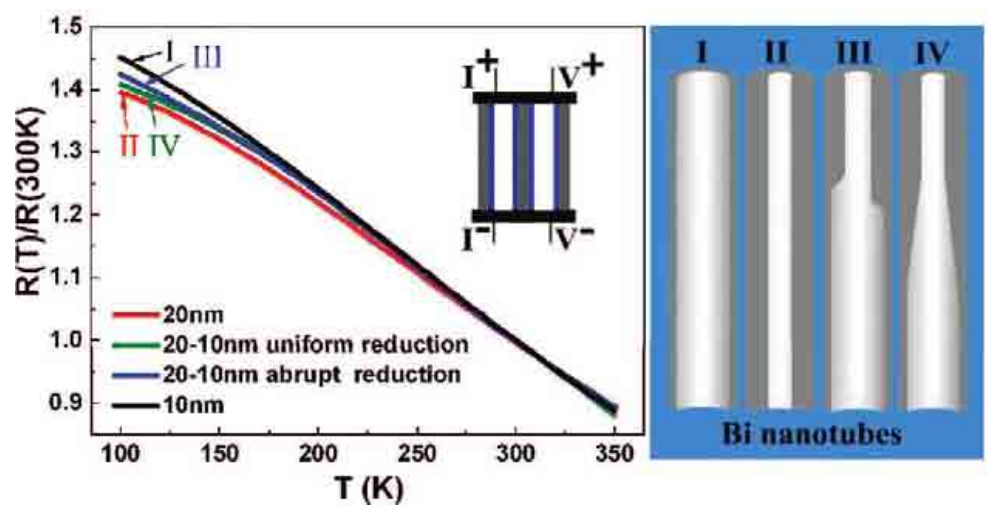

**Figure 1.12:** Temperature dependence of the resistance of BiNT arrays with a designed wall thickness (normalized to resistance at 300 K). Curves correspond to BiNTs with wall thicknesses of 10 nm (I) and 20 nm (II) and wall thickness reductions from 20 to 10 nm abruptly (III) and uniformly (IV), respectively. D. Yang *et al.* [38].

Li *et al.* [37] reported the measurement of electrical resistivity of Bi nanotube arrays fabricated by electrodeposition. The authors could substantiate the observed results of metal to semiconductor transition using the same arguments (as in case of nanowires) of competing factors; the carrier density and the carrier mobility. They gave the same qualitative description for the observed temperature dependent resisitivity with size as usually given in case of Bi nanowires. More recently D. Yang *et al.* [38] reported the semiconducting behaviour of thin Bi nanotubes of thickness 20 nm and diameter 100 nm prepared by electrodeposition in anodic alumina templates. The authors observed the semiconductor behaviour of these nanotubes as observed in case of nanowires of similar dimension.

## 1.6 Review of Synthesis of metal nanowires and nanotubes

In the last decade, there are an extensive number of reports of metal nanowire synthesis, but there are only a few reports of metal nanotube synthesis due to the complexities involved in synthesis. Here we give a short review on the synthesis of metal nanowires and nanotubes synthesized. The main method of synthesis which is followed in case of metal nanowires and nanotubes is the template based synthesis. A template in this case is a material having nanoporous channels. Template based synthesis generally involves the filling of the pores of the template with the material required. Among various kinds of templates available for the growth of metal nanowires, Anodic aluminium Oxide (AAO) and polycarbonate are widely employed for the synthesis. There are also other kinds of nanoporous materials available like porous silica, zeolites, porous mica etc.

### 1.6.1 Polycarbonate (PC) Membranes

This kind of membranes are usually prepared by track-etch method in which, a nonporous sheet of the desired material and thickness (standard thickness range from 6 to 20 mm) is bombarded with nuclear fission fragments to create damage tracks in the material. The tracks are then chemically etched into pores. The resulting porous materials contain randomly distributed nanochannels of uniform diameter (as small as 10 nm) with pore densities as high as $10^9$ pores/cm2. Such membranes are also often called nuclear track filters or screen membranes. The commercially available (e.g. Nuclepore from Whatman [39]) filtration membranes are usually prepared from polycarbonate or polyester. The structure of track-etched membranes can be modified by high frequency discharge plasma in air. By varying the etching conditions, pores of different size and geometry can be produced.

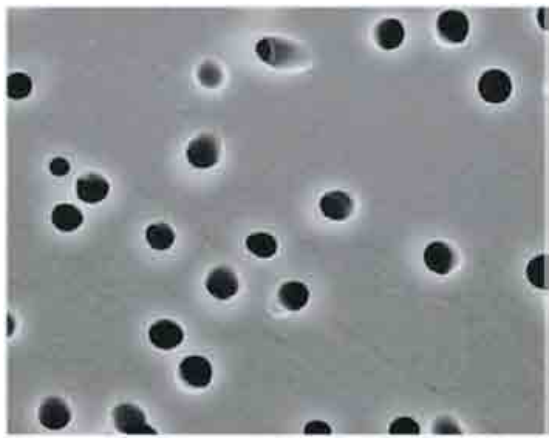

**Figure 1.13:** Typical SEM image of a polycarbonate membrane.

Fig. 1.13 shows the typical SEM image of a polycarbonate membrane. These membranes are very suitable as templates for electrochemical deposition of nanowires. Since a wire acquires the exact pore shape, this template technique enables the fabrication of wires with different shapes, e.g. conical or cylindrical, with dimensions down to few nm. These templates suffer from the drawback that the channels are neither always formed parallel to each other nor are regularly spaced. Being made of polymer, these templates cannot stand high temperatures (450 K is the maximum working temperature of polycarbonate membranes).

## 1.6.2 Anodic Aluminium Oxide (AAO)

The anodic aluminium oxide membranes are widely used for the synthesis of metal nanowires as they have regularly arranged arrays of pores. The pores can have diameter < 10 nm and their lengths can extend up to 100 μm. Thus wires having high aspect ratios can be obtained by depositing metals inside the pores of anodic alumina. These templates are prepared by a two step anodization process in which high purity (99.99%) aluminium foil is anodized in acidic medium ($H_2SO_4$ or $(COOH)_2$ or $H_3PO_4$) in two steps. The anodized aluminium foil is etched after 1$^{st}$ anodization to obtain an array of regular pits which then develop into pores during the second step of anodization under the same conditions. The pores form with uniform diameter throughout their length because of a fine balance between electric-field-enhanced diffusion which determines the growth rate of the alumina, and dissolution of the alumina into the acidic electrolyte. The pores self order because of mechanical stress at the aluminum-alumina interface due to expansion during the anodization. This stress produces a repulsive force between the pores, causing them to arrange in a hexagonal lattice. Depending upon the anodization conditions the pore density of the nanowires can be varied from $10^9$-$10^{11}$pores/cm$^2$.

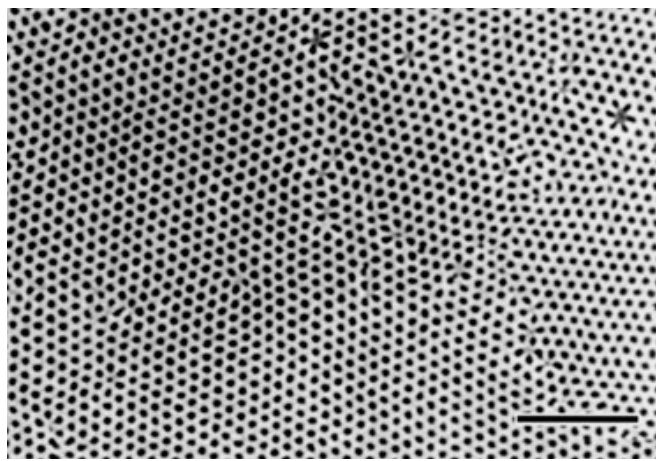

**Figure 1.14:** Low magnification SEM view of the long-range ordered anodic porous alumina formed in sulphuric acid at 25 V [40]. The scale bar is 200 nm.

The lowest pore diameter being reported was 7nm [41]. The large arrays of hexagonally arranged parallel pores of high aspect ratio makes the AAO templates ideal for growing arrays of nanostructures. Apart from these features, the AAO templates have high working temperatures up to 700 K. Nanowires and nanotubes of a number of metals have been made in the past as tabulated below.

## 1.6.3 Template synthesis of nanowires and nanotubes

As stated earlier, though till date there are plenty of reports of nanowire synthesis using electrochemical deposition, there are fewer number reports for the synthesis of metal nanotubes and the field is still challenging. In fact electrochemical deposition in nanoporous templates is the only widely used techniques for the synthesis of metal nanowires. As listed in Table 1.1, nanowires of various metals like Ni, Co, Fe, Cu, Ag, Au etc., have been synthesized till date. However, except for a few noble metals, it is still a challenge to prepare single crystalline oriented nanowires of metals like Nickel which we have succeeded in this thesis. The process consists of electrodeposition of metals inside the cylindrical pores from solutions of their respective salts. A conducting layer of a noble metal like Ag or Au is first evaporated onto the one side of the template which acts as a cathode/working electrode. A platinum electrode is used as anode/counter electrode with saturated calomel electrode (SCE) acting as the reference electrode. Deposition is carried out at a constant optimum deposition potential with respect to the SCE with the uncoated surface pores exposed to the solution.

**Table 1.1:** Nanowires and nanotubes of metals synthesised by electrodeposition

| Metal | Electrodeposition Technique | Quality/Reference |
|-------|------------------------------|-------------------|
| **Nanowires** | | |
| **Au** | Pulsed Electrochemical deposition | Gold nanoparticles inside AAO [42], Single crystalline [43] |
| **Ag** | DC Electrochemical deposition | Single crystalline highly ordered arrays in AAO[44] |
| **Cu** | DC Electrochemical deposition | Single crystalline highly ordered arrays in AAO [45], Polycarbonate in PC [46] |
| **Fe** | DC Electrochemical deposition | Oriented single crystalline [47] |
| **Zn** | DC Electrochemical deposition | Polycrystalline nanowire arrays in AAO [48] |
| **Ni** | DC Electrochemical deposition, Pulsed Deposition | Polycrystalline nanowires in PC membrane[49], Highly ordered with equal spacing between the nanowires in AAO templates [50] |
| **Co** | DC Electrochemical deposition, Pulsed Deposition | [51], Oriented single crystalline[47] |
| **Bi** | DC Electrochemical deposition | Polycrystalline nanowires in PC membrane [52–53], Polycrystalline in Anodic alumina[54] |
| **Pd** | Step decoration, DC Electrochemical deposition | Polycrystalline nanowires on HOPG steps[55] |
| **Nanotubes** | | |
| **Ni** | Template replication assisted DC electrochemical deposition, Pore modification assisted DC electrochemical deposition, Current density controlled growth DC electrochemical deposition | Polycrystalline [56, 57, 58] |
| **Co** | Current density controlled growth DC electrochemical deposition | Polycrystalline [58] |
| **Fe** | Current density controlled growth DC electrochemical deposition | Polycrystalline [58] |
| **Bi** | partial coating of working electrode technique DC electrochemical deposition | Highly oriented (202) and single crystalline[37], Single crystalline[38] |

The kind of deposition described above is known as potentiostatic deposition. However, in literature, there are other kinds of electrode configurations are also used. For instance, the platinum electrode is replaced by a desired metal electrode, when the metal ions in the solution remain constant and solution does not need to be replenished from time to time. Usually in this kind of deposition only two electrodes (anode and cathode) are used and the reference electrode is not used. This kind of deposition usually carried out at constant voltage or constant current (also called as galvanostatic mode).

Several other techniques have been also tried for the growth of nanowires in templates, although with much less success. Au and Ag nanowires have been grown by a gas-phase reaction inside the templates [59] and by electroless deposition [60]. Nanowires have also been grown by physical vapour deposition (PVD) [61], chemical vapour deposition (CVD) [62], metal-organic chemical vapour deposition (MOCVD) [63] inside template.

There are a fewer number of reports of synthesis of nanotubes of metals. Ni nanotube arrays have been done by using surfactant assisted electrodeposition [56]. Ni nanotubes have been synthesised by template replication method [57]. In these cases, the complexities involved in the synthesis challenge the purity of the nanowires and the time of synthesis. H. Cao *et al.* [58] reported the synthesis of nanotubes of Ni, Co, Fe by controlling the current density and tried to substantiate the preferential deposition on walls mechanism by using low current densities. With these fewer number of works, the synthesis of large arrays of oriented metal nanotubes of high purity still remain a major technological challenge.

## 1.7 Structure of the thesis:

The thesis has been presented in such a way that it starts with the significance and the physics of the field, necessary literature of the earlier work done in the field of metal nanowires and nanotubes.

The **2$^{nd}$ Chapter** gives a description of the synthesis and characterization of the samples.

The **3$^{rd}$ Chapter** describes the experimental setups developed for the work. It also describes various advanced techniques like, Optical -lithography and ion-beam lithography used in this work for making contacts to single nanowires.

The **4$^{th}$ Chapter** deals with the high temperature resistivity measurements made on arrays of nickel nanowires and the detailed study of the effect of size reduction on the electrical resistivity of nanowires near the ferromagnetic Curie temperature.

The **5$^{th}$ Chapter** is devoted to the low temperature electrical study of single crystalline nickel nanowire arrays. The effect of size reduction on the Debye temperature, the electron-phonon and the electron-magnon interaction in nanowires where their diameter puts a limit to the electron mean free path are described in this chapter.

The **6$^{th}$ Chapter** deals with the electrical measurement made on a single nickel nanowire. The results are compared with those measurements made on arrays of nanowires.

The **7$^{th}$ Chapter** is dedicated to synthesis of novel tubular structures. It describes a novel method of synthesis of metal nanotubes with the help of a rotating electric field. The computer simulations of the idea are presented and as a generic example, synthesis of copper nanotubes is demonstrated. The electrical resistance of the nanotube array is also measured and compared with nanowire arrays.

The **last Chapter** is the summary and conclusions of this thesis. It involves a discussion of the achievements of the thesis and the future challenges.

# CHAPTER 2

## Synthesis and characterization

*In this chapter we present the synthesis of the samples by electrodeposition and their characterization techniques using techniques like X-ray diffraction (XRD), Transmission Electron Microscopy (TEM), Scanning Electron Microscopy (SEM), Vibrating sample Magnetometer (VSM). Superconducting quantum interference device (SQUID). Here, for the sake of completeness, we describe the techniques in brief followed by the description of the data we obtained on our samples.*

This page is intentionally left blank

## 2.1 Electrodeposition

Electrodeposition, better called as an 'art', is as old as the electricity. Electrodeposition of metals and alloys involves the reduction of metal ions from aqueous, organic, and fused-salt electrolytes by the passage of an electric current. The success of electrodeposition lies in its application in the industries ranging from electronics, micro-macro, and opto-electronics to automobile industry. The first quantitative formulation of electrodeposition was done by Michael Faraday, popularly known as Faraday's laws of electrodeposition. The laws can be found in any high school chemistry text book and give an estimate of the amount of material *(W)* deposited is given by the product of electrochemical equivalent $\left(\dfrac{E}{F}\right)$ and the charge passing through the electrolyte

$$W = \frac{E}{F} ct \tag{2.1}$$

where *E* is equivalent weight and *c, t* are current and time respectively with F= 96485 Coulomb/mol. This formulation tacitly assumes the complete reduction of ions by the charge passed between electrodes. However in practise the problem is much complicated and factors like sticking of electrodeposited atoms, Joule heating play affect the efficiency of the process. Hence, in practise one needs a deeper knowledge of electrochemistry to achieve a high efficient electrodeposition process and it is proved to be a practically viable technique for large scale regular synthesis of materials (even ternary alloys). It is widely used in industry for coating metals in a process called as electroplating [1]. It is also used in electroforming [2], i.e., to give shapes to electrodeposits. The electrodeposits general follow the electrode shape as the electrolyte being a liquid assumes the shape of the electrode. The feature that the electrodeposits can take shapes of objects in controlled way makes electrodeposition a unique technique for synthesis of materials at nanoscale. When the deposition is confined to the pores of template membranes, nanowires/nanotubes arrays are produced. Before we discuss the synthesis technique in details, it is important to discuss some basics of electrodeposition. When a solid is immersed in a polar solvent or an electrolyte solution, surface charge will develop and the electrode attains a potential given by the Nernst equation [3],

$$E = E_0 + \frac{R\,T}{n\,F} \ln \left( a_{ion} \right) \tag{2.2}$$

In equilibrium we have metal ions get discharge to metal atoms and vice-versa.

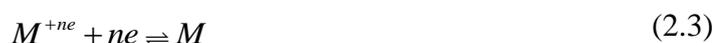

$$M^{+ne} + ne \rightleftharpoons M \tag{2.3}$$

where $E_0$ is the standard electrode potential (or the potential difference between the electrode and the solution) when the activity $a_{ion}$ of the ions is unity, $F$ is Faraday's constant, $R$ is the gas constant, and $T$ is the temperature. The equilibrium potential is always expressed with respect to a standard electrode.

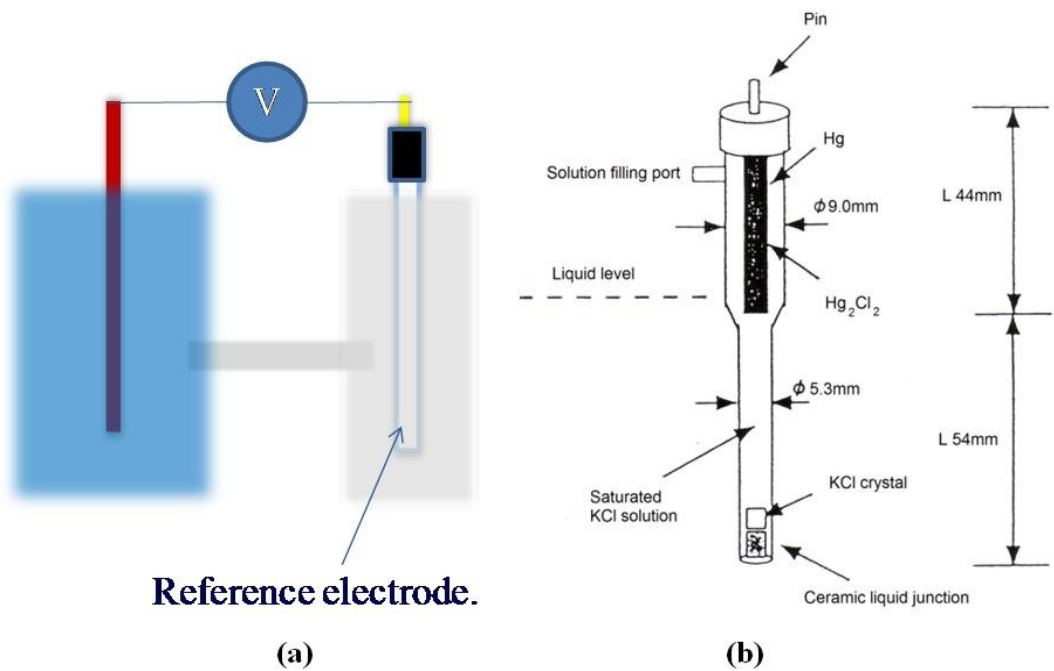

**Figure 2.1: (a)** A Schematic of electrode potential of a half cell being measured with respect to a standard reference electrode. **(b)** The construction of a saturated calomel electrode (SCE).

A standard electrode is such an electrode whose electrode potential remains always constant because of the constant activity of the ions [4], Ag/AgCl, $Hg/Hg_2Cl_2$ (saturated Calomel Electrode shown in Fig. 2.1(b)), $Cu/CuSO_4$ are some examples of standard electrodes used in laboratory. A detailed literature regarding various kinds of reference electrodes can be found in standard text books dedicated to electrochemistry [5, 6].

When the electrode potential is higher than the energy level of a vacant molecular orbital in the electrolyte, electrons will transfer from the electrode to the solution and the electrolyte will be reduced. On the other hand, if the electrode potential is lower than the energy level of an occupied molecular orbital in the electrolyte, the electrons will transfer from the electrolyte to the electrode, resulting in electrolyte oxidation. These reactions stop when equilibrium is achieved. Thus an external potential difference is needed to take the system away from equilibrium and one can achieve deposition or dissolution of the metal and this process is termed as electrodeposition.

Depending upon the number of electrodes and current, voltage configuration, electrodeposition is classified into mainly two ways

## 2.1.1 Galvanostatic Electrodeposition

The deposition is carried out by passing a constant current between two electrodes immersed in an electrolyte. In this kind of deposition, the electrode potential of the electrodes is not given much importance and deposition is standardized at optimum current.

## 2.1.2 Potentiostatic Electrodeposition

This kind of deposition is carried out in an electrolytic cell containing an extra reference electrode [7] in addition to the regular working electrode (cathode) and counter or auxiliary electrode. The deposition is carried out at a constant potential of the working electrode with respect to the reference electrode [7].

The quality of electrodeposits mainly depends upon the conditions of electrodeposition and hence it becomes important to determine a particular potential where the rate of deposition is optimum with good quality. A further increase or decrease in applied potential will not increase the efficiency and the current. Since the quality and forms of deposits are the main concerns in nanoscale deposition, potentiostatic deposition is widely preferred in case of synthesis of nanowires and nanotubes. Fig. 2.2(a) shows the scheme of potentiostatic deposition where effectively current between C and W is monitored by maintaining W at a constant potential w.r.t R[7] with required electronics.

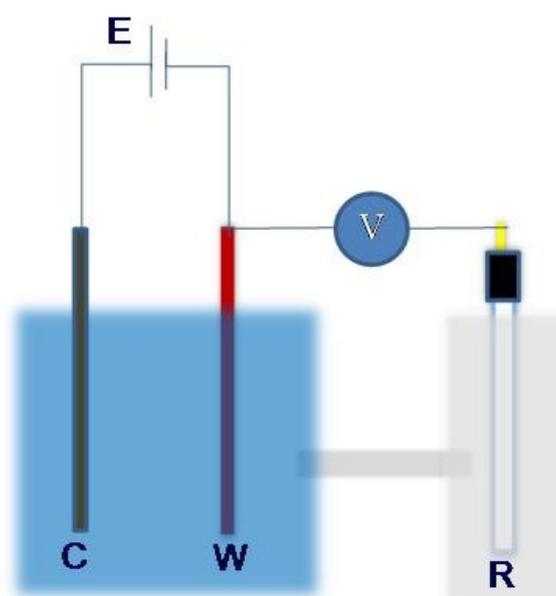

**Figure 2.2(a)** The scheme of three electrode potentiostatic configuration.

In general E can be varied (while taking a feedback from the reference electrode) to achieve a particular value of V (potential of the working electrode w.r.t Standard reference electrode) and V can be maintained. Thus a potentiostat can vary E to determine the values of V needed for optimum deposition.

## Electronic circuit representation of a potentiostat

A potentiostat is used for most voltammetric measurements. The potentiostat uses an electrical feedback loop to control the potential of the working electrode (the electrode at which the reaction of interest occurs) with respect to a reference electrode, even in the presence of ohmic drop. A third electrode, the auxiliary/counter, is used to supply the current flowing at the working electrode. The use of a third electrode eliminates current flow through the reference electrode, permitting smaller reference electrodes and more accurate potential control. A simplified diagram of a potentiostatic circuit is given in Figure 2.2(b) (5).

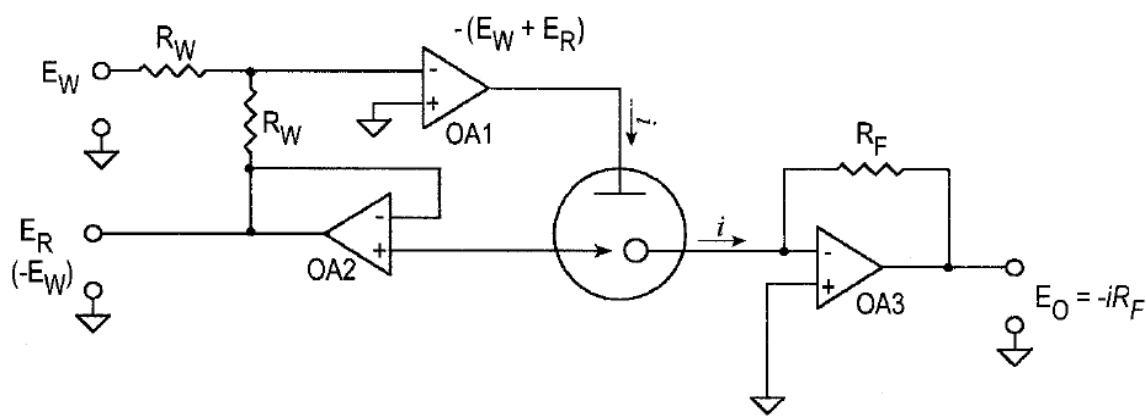

**Figure 2.2(b):** Electronic circuit representation of a potentiostat [5, 8].

Most modern electrochemical equipment employ versatile high-input impedance differential amplifiers known as operational amplifiers (OAs) as circuit elements. In the potentiostat circuit, OA1 provides the auxiliary electrode with the required voltage and current to maintain the desired potential difference between the reference and working electrodes. OA2 is a buffer amplifier (voltage follower) that prevents significant current draw through the reference electrode and outputs a low-impedance measure of the reference electrode potential. In operation, the amplifier OA1 adjusts the current and voltage at the auxiliary electrode to whatever is necessary to minimize the voltage difference at the inputs. The negative feedback loop that exists between the output and input of OA1 includes OA2, the reference and auxiliary electrode, and the solution resistance between reference and auxiliary.

Thus, the voltage at the auxiliary electrode will be the inverted sum of the reference electrode voltage ($E_R$), any externally applied voltage ($E_W$), and the ohmic voltage arising from current flow through the solution resistance. This means that, if the reference electrode is close to the working electrode, the effect of the ohmic drop is minimized and the potential difference between the working electrode and reference electrode is maintained or ''clamped'' close to $E_W$. Clamping the voltage at $E_W$ is the basis for the name ''voltage clamp''; a term used outside the electrochemical literature as a more descriptive name for the potentiostat circuit. The working electrode is maintained at the circuit common potential (which may not be earth ground) in the potentiostat circuit of Fig. 2.2(b). The current-transducer amplifier, OA3, maintains the electrode potential at common potential and provides a voltage output proportional to the current input. As before, the amplifier uses negative feedback to minimize the potential difference between the inputs by adjusting the output. In the process, the inverting input (-) is set to be virtually the same as the potential of the non-inverting input (+), and thus the inverting input is at ''virtual ground.'' Note that the output voltage across the feedback resistor, $R_F$, produces a current to oppose the current flowing into the input and changing $R_F$ changes the signal gain. Although the circuit in Fig. 2.2(b) could be used in voltammetric experiments, most potentiostat circuits are considerably more sophisticated. Additional circuitry is added to provide damage protection from excessive signal inputs or outputs, additional current or voltage range (i.e., compliance) at the auxiliary electrode, and for variable gain and filtering of the input current.

To determine the optimum deposition potential (**V**) corresponding to the maximum current, a Cyclic Voltammetry (CV) [5] is performed. In CV, the potential of the working electrode with respect to the reference electrode is scanned at a constant rate linearly to obtain the optimum deposition potential. Fig. 2.3 shows a typical Cyclic voltammogram taken for 1M $NiCl_2.7H_2O$ solution. The optimum deposition potential is related to the potential corresponding to the maximum cathodic current. A deposition potential is the optimum potential in a sense that the ions near the cathode get conveniently discharged rather than creating any charge cloud which can hinder further deposition process. Once the deposition potential is determined, the potential of the working electrode with respect to the reference electrode is maintained at the deposition potential and the electrodeposition is carried out. In our experiments, we used **a CH-instruments 600B electrochemical analyzer** [9] for CV and electrodeposition. The deposition is generally carried out at a potential close to the potential corresponding to maximum current and is better determined by analysing the quality of the electrodeposits obtained.

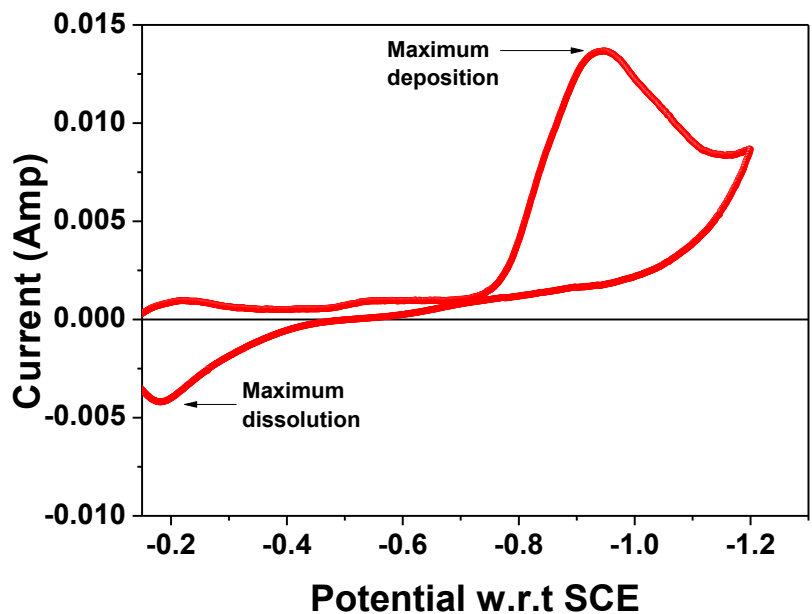

**Figure 2.3:** Cyclic voltammogram taken for 1M $NiCl_2.7H_2O$ solution taken at a sweep rate of 20mV/Sec.

## 2.2 Template assisted synthesis

A review of the template assisted synthesis of nanowires and nanotubes is presented in the previous chapter. In this section, we describe the details of each step involved in the synthesis.

### 2.2.1 DC electrodeposition

In this kind of deposition, the working electrode is maintained at a constant potential with respect to the reference electrode and the deposition is carried out with the current flowing between working electrode and the counter electrode noted as a function of time.

### 2.2.2 Pulsed Electrodeposition

In the case of Pulsed electrodepositon, the potential of working electrode is made to vary as a pulse between two potentials ($V_1$ and $V_2$) with a particular duty cycle and time period depending upon the requirement. It is seen that pulsed electrodeposition often improves the quality of the deposits. It is particularly very effective in depositing multilayer of metals where one metal gets deposited at $V_1$ and the other preferentially gets deposited at $V_2$. Using this scheme of deposition multilayered nanowires are also deposited in templates [10].

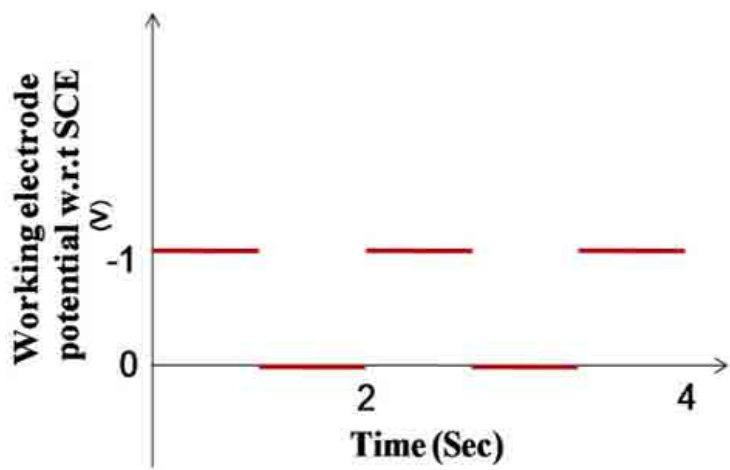

**Figure 2.4:** A typical pulse for deposition of voltage $V_1$=-1 and $V_2$=0 for the deposition Nickel.

Pulsed deposition is useful in situation where DC electrodeposition cannot achieve the required results. It ensures better quality of electrodeposits by relaxing the overcrowded charge accumulation, other electrochemical reactions (reductions involving the gases) near the working electrode. The pulse deposition gives a way to remove the end products of redundant electrochemical reductions (which often hinder the deposition) improving the quality of electrodeposits.

## 2.2.3 The Steps involved in the template assisted synthesis of nanowires

**Step 1:** The porous anodic alumina (AAO) [11, 12] /polycarbonate [13] templates are taken and are cleaned with acetone to remove any greasy substance in contact. Presence of such a substance prevents wetting of the template.

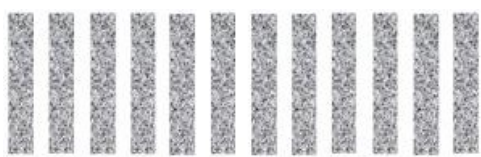

**Figure 2.5.1:** A schematic representation of an uncoated template.

The detailed images of the topography and the cross section of an anodic alumina template can be seen in Fig 2.5.2, where we show the SEM images of a 35 nm pore diameter template.

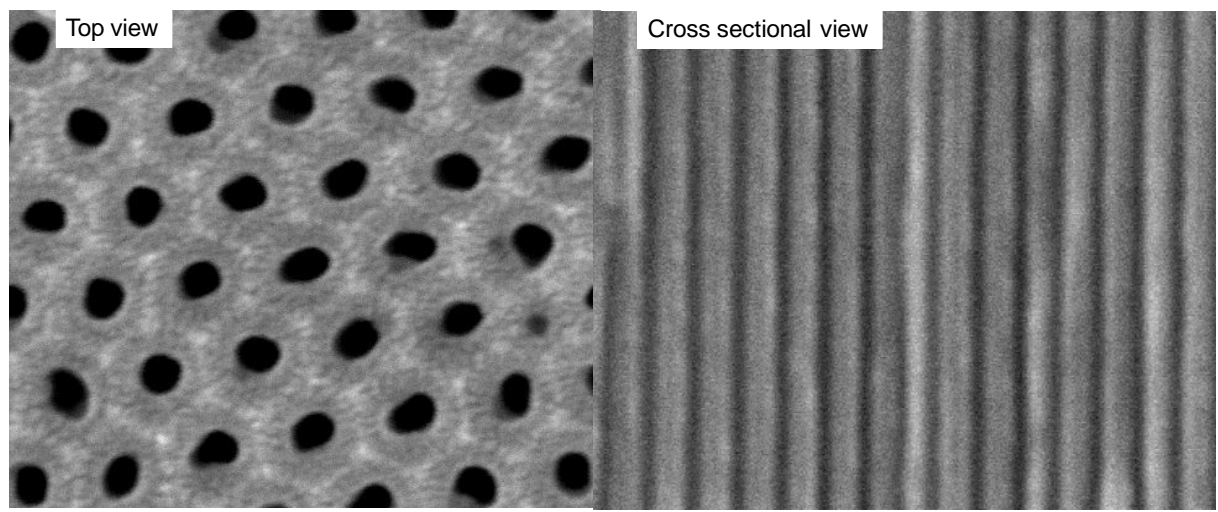

**Figure 2.5.2:** The top view and cross sectional SEM images of a 35 nm porous anodic alumina template.

**Step 2:** One side of the template is coated with a considerably thick layer (~200nm) of gold/silver metal by thermal evaporation/sputtering. The exposed portion of this coating through the pores acts as the **working electrode (WE)** in the three electrode potentiostatic electrodeposition [7] configuration as shown in the Fig. 2.4.2.

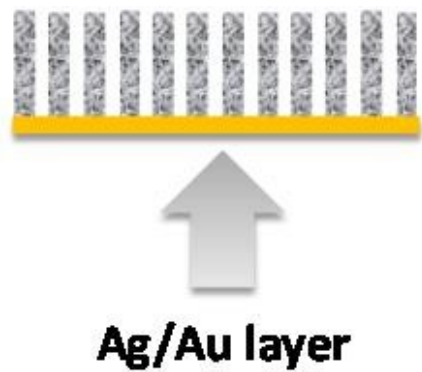

**Figure 2.5.3:** A schematic representation of a template coated on onside.

**Step 3:** The potential of the working electrode with respect to a standard reference electrode **RE** (a **saturated calomel electrode**[6] in our case) is varied to determine the optimum deposition potential using Cyclic Voltammetry technique. A typical voltammogram is shown in Fig. 2.3.

**Step 4:** The deposition is carried at the optimum potential determined as shown in the Fig. 2..3.

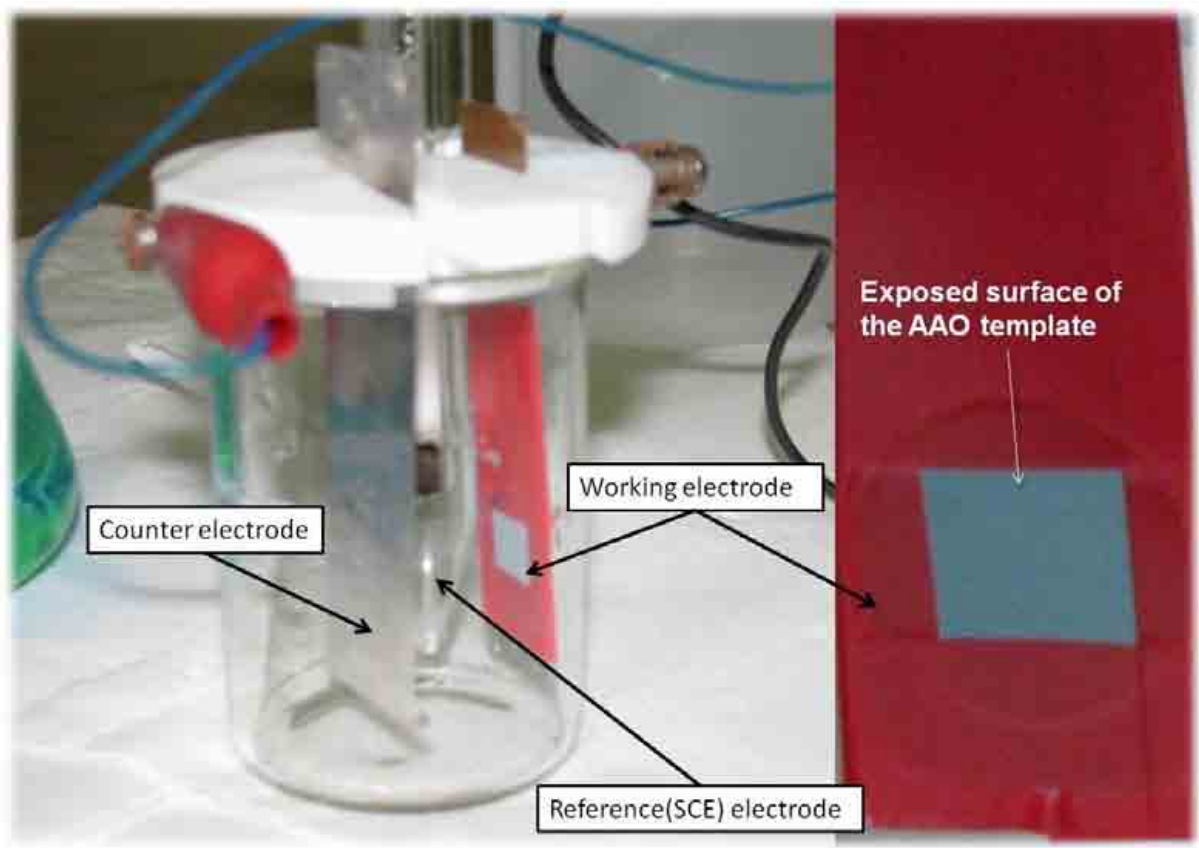

**Figure 2.5.4:** The electrodeposition Cell used for deposition.

The coated surface of tempate is attached to copper base plate as shown in the above Fig. 2.5.4 with the help of insulating tape. Electrolyte enters into the pores  wet only the exposed part of the AAO template.  A thick platinum foil is used as the counter electrode. The electrodeposition cell shown in Fig. 2.5.4 consists of a galss beaker capped with a Teflon holder containing slots for inserting the working electrode, counter electrode and a saturated calomel reference electrode. During depositiong, the ions inside/entering the pores get deposited on the 200 nm thick coating of silver and the eventually the deposits fill template pores with the electroforms resulting in nanowire arrays. The deposition can also be acheived in the configuration shown in Fig. 7.5(b) in chapter 7 later, which we used for the synthesis of nanotubes. During the deposition the current is monitered by a computer to which the potentiostat is interface via RS 232 port. The electrodepositon setup during deposition is shown in Fig. 2.5.4. The current between the working electrode and the counter electrode is recorded with respect to time during deposition. The deposition is monitored by a current versus time plot as shown in Fig. 2.5.6 for the deposition of nickel in a 200 nm pore diameter AAO template.

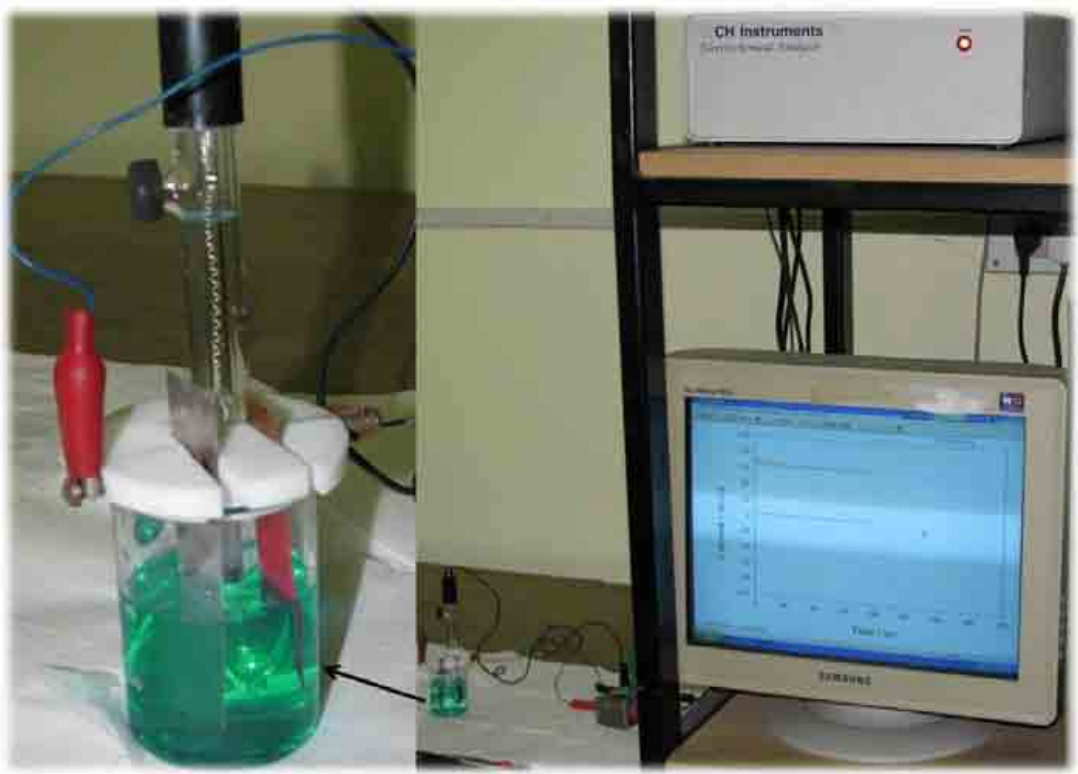

**Figure 2.5.5:** The electrodeposition Setup during deposition.

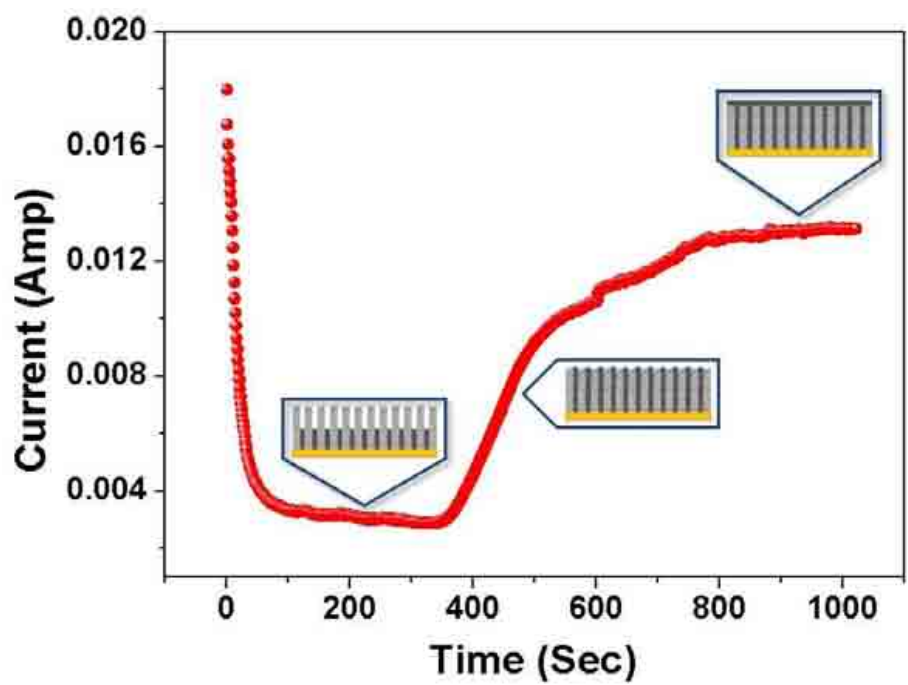

**Figure 2.5.6:** Typical Current vs. Time plot during potentiostatic electrodeposition taken during the synthesis of Ni nanowires of 200 nm diameter. The various stage of pore filling during deposition are shown as insets in the figure at the respective current-time positions.

To completely fill the pores, the deposition is carried out to the stage of current saturation observed after a continuous increase in current.

In the case of **pulsed deposition**, where the deposition is intermittent a current versus time is shown in Fig. 2.5.7. The deposition is carried out to sufficiently long time to completely fill the pores with the sufficient amount of charge which is used in DC electrodeposition.

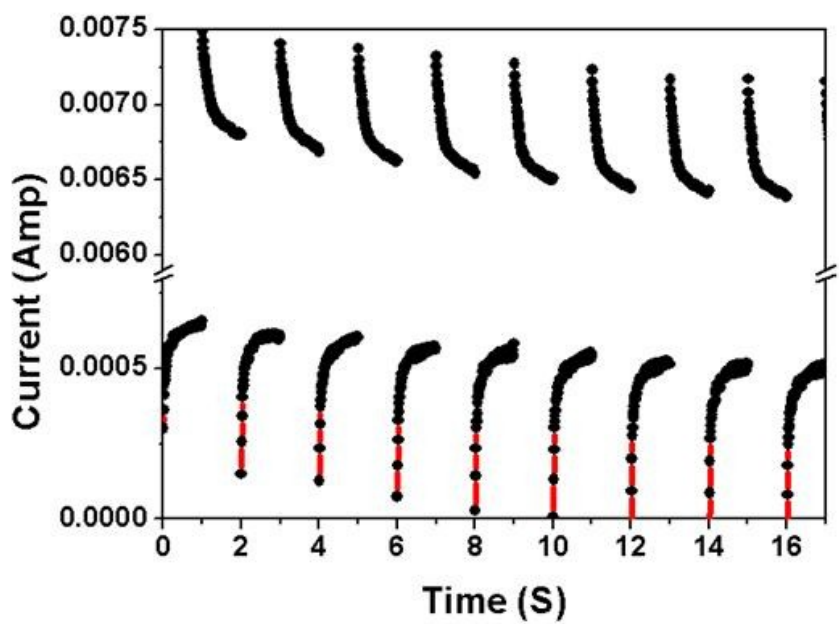

**Figure 2.5.7:** Typical Current vs. Time plot during pulsed potentiostatic electrodeposition corresponding to pulsed shown in Fig.2.4.

**Step 5:** Once the templates are completely filled, the overgrowth can be removed by mechanical polishing. Fig. 2.5.9 shows the polished surface of a template of 55 nm diameter pores filled with nickel.

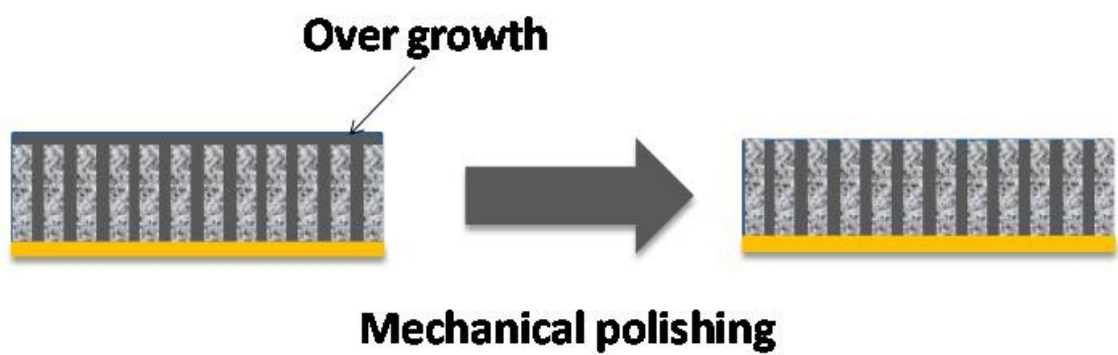

**Figure 2.5.8:** Removal of over growth by polishing.

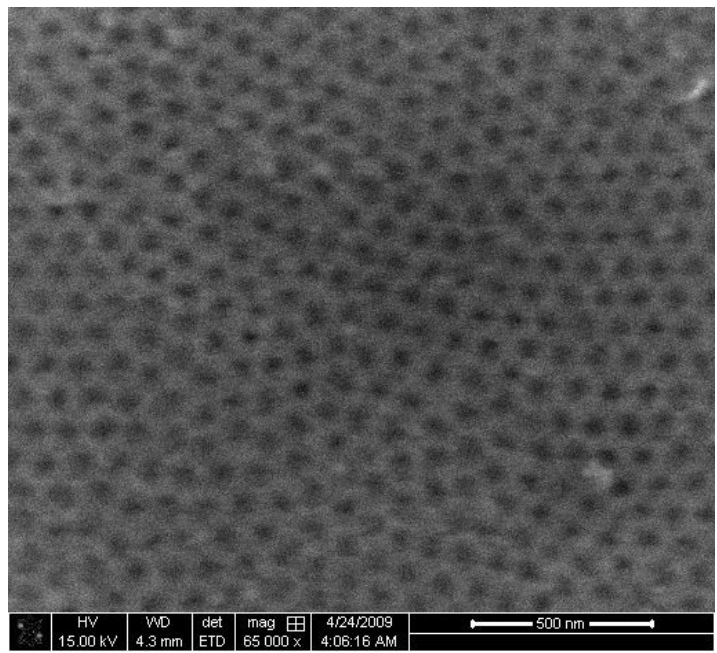

**Figure 2.5.9:** Polished surface of a template of 55 nm diameter pores filled with nickel.

## 2.2.4 Characterization

**X-ray diffraction/ Magnetic Measurements/ Electrical Measurements:** The as prepared sample can be used for X-ray diffraction, magnetic measurement (parallel and perpendicular configurations) and electrical measurements.

**Microscopy:** For **scanning electron microscopy(SEM)**, the template containing the nanowires/nanotubes is partially etched with 3M NaOH in case of anodic alumina and with chloroform in case of polycarbonate template. The etched sample is washed several times with millipore water. For **transmission electron microscopy(TEM)**, the sample is completely etched with the respective etchants and then washed with millipore water. The resulting solution containing nanowires/nanotubes is dropped (1-2 drops) on carbon coated 300 mesh copper grid.

## 2.3 CHARACTERIZATION TECHNIQUES

The samples as discussed above are mainly characterized using X-ray diffraction (XRD), Scanning Electron Microscopy (SEM), Transmission electron microscopy (TEM) and Energy dispersive spectroscopy (EDS). Here we discuss these various characterization techniques and their basic principle and the results we obtained on our samples.

## 2.3.1 X-ray Diffraction (XRD)

X-ray Diffraction [14, 15] characterization is the first step in determining the phase of any material synthesized. In this section, we give a brief overview of the technique and the diffractometer, followed by the data of the nanowires synthesized by DC electrodeposition as well as pulsed deposition.

**Principle:** Interaction between matter and x − rays with suitable wavelengths results in inelastic and elastic scattering of the radiation. It is the elastic scattering from the electrons inside the matter which leads to the observed diffraction of the incident x − ray beam by the sample. By recording and analyzing the diffraction pattern, information on the crystal structure of the material is obtained. Although the scattering takes place from individual atoms, the observed patterns can also be explained by assuming that planes of atoms are sources for the scattered radiation wave. This picture leads to Bragg's law [14], which states that large diffracted intensity takes place only when the phases of the scattered waves differ by an integral multiple of wavelengths ($n\lambda$). When this is the case, the waves sum up leading to constructive interference and a peak is seen in the diffraction pattern. Otherwise, the waves cancel each other causing destructive interference and a lower intensity is observed. Bragg's law is illustrated in Fig. 2.6, where n = 1, 2, 3,…..; $\lambda$ = wavelength, $d_{hkl}$ = distance between adjacent planes of atom indexed as hkl. The condition $n\lambda = 2d_{hkl}\sin\theta$, derived from simple geometry (as can be seen from the figure), has to be valid for the waves 1 and 2 to interfere constructively.

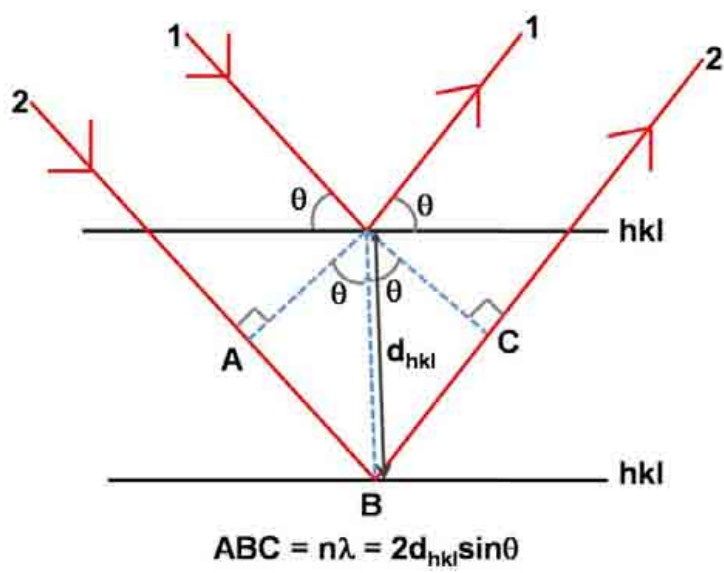

**Figure 2.6:** Bragg's law.

The X-Ray diffraction measurements were done with X'Pert PRO Diffractometer [16]. This diffractometer works in the Bragg – Brentano geometry with a θ:θ arrangement i.e. the sample remains fixed while the tube rotates at a rate of $-θ^0$/min and the detector rotates at a rate of $+θ^0$/min. A schematic representation is shown in Fig.2.7.

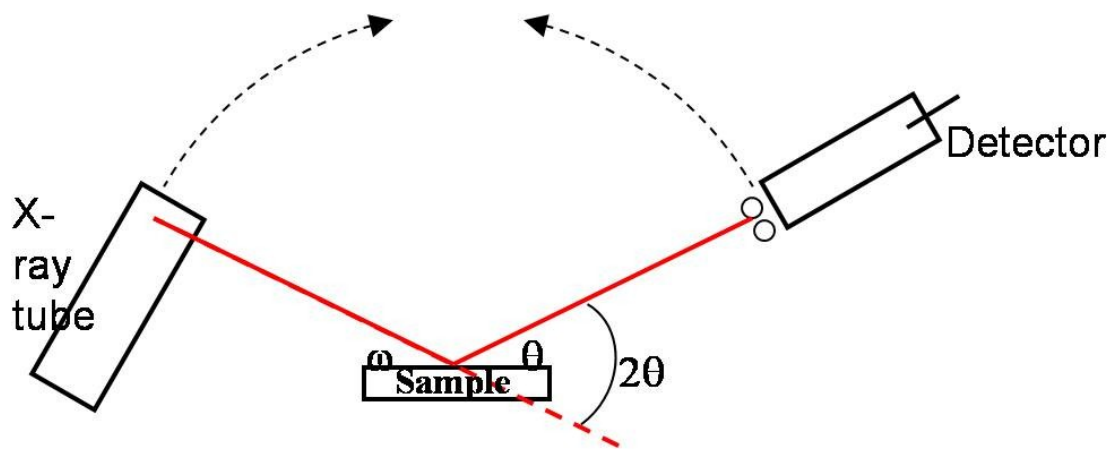

**Figure 2.7:** Schematic representation of the Bragg – Brentano geometry.

The x – ray tube (PW3373/10 Cu LFF DK 184115) used is a ceramic tube with a maximum tube power of 1.8 kW. The x – ray radiation is generated in the tube which consists of a source of electrons and a metallic anode (made of Cu) kept under high vacuum. The source of electrons is a filament of tungsten heated by an electric current, which expels electrons by the thermionic effect. A high voltage of 45 kV is applied between the source of electrons (cathode) and the metallic anode which accelerates the electrons. The radiation produced has a discrete number of wavelengths and a broad background. For conventional purposes, we have used Cu Kα radiation (λ = 1.5418 Å). LFF stands for Long Fine Focus and has a focusing dimension of 12 mm X 0.4 mm. This focusing dimension gives optimum resolution combined with high intensity and is useful for phase analysis. The detector used is a gas proportional detector (PW3011/20). It has a typical energy resolution (for Cu Kα radiation) of 20%. (The energy resolution of a detector is a measure of its ability to resolve two X – ray photons of different energy). The resolution of this instrument in terms of scanning angle is $0.001^0$.

In Fig. 2.8, we show the data taken on the nickel nanowires grown in anodic aluminum oxide templates with various pore diameters (18nm-200nm) by DC elctrodeposition. The structure of the nanowires is FCC and the wires are polycrystalline as we obtained the peaks in case of a polycrystalline sample with peaks of various intesities. A close examination of the peaks reveal broadening of the peaks in Fig. 2.8 as the diameter of the nanowire decreases.

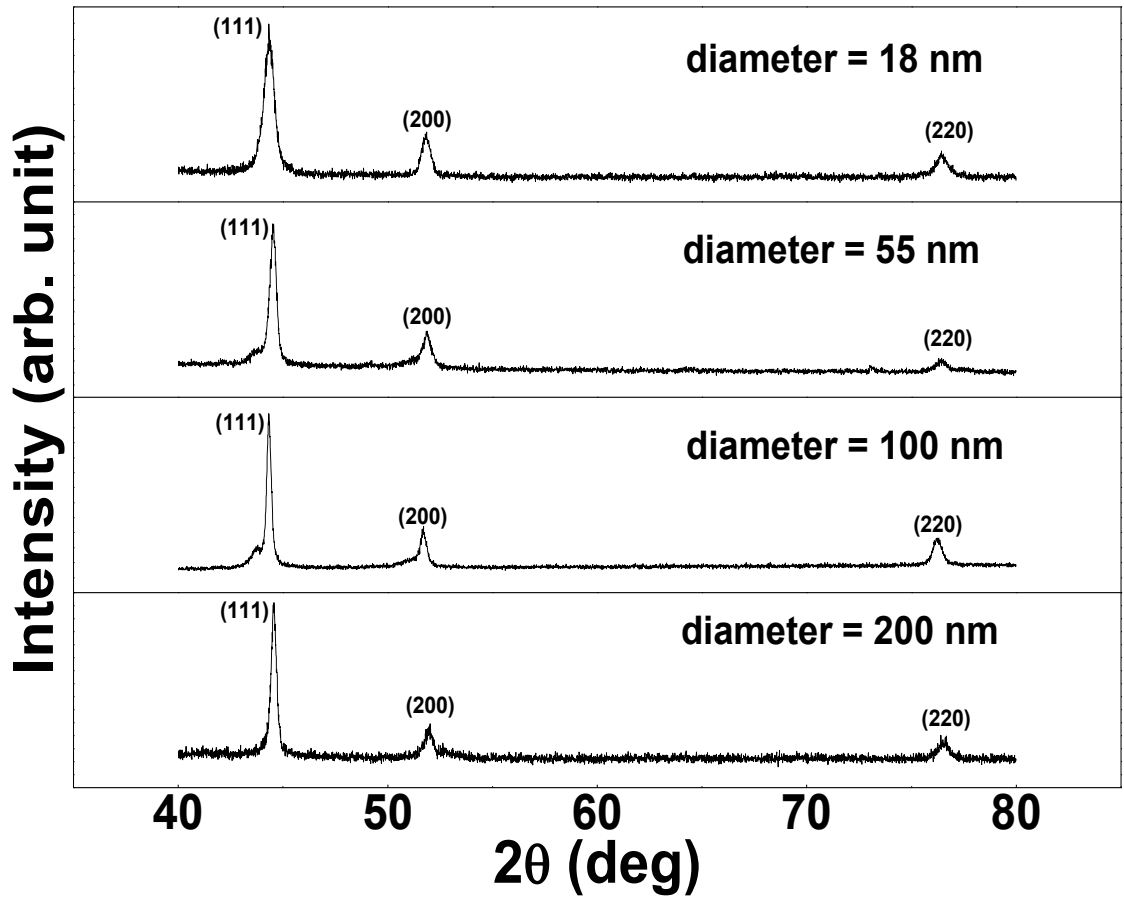

**Figure 2.8:** X-ray diffraction patterns of nickel nanowires of various diameters grown in anodic alumina templates by DC potentiostatic deposition.

The particle size can be calculated from x-ray line broadening using the Scherrer formula : $d = \dfrac{\lambda}{\beta_{size}\cos\theta}$ [17]. The strain is given by $\varepsilon = \dfrac{\beta_{strain}}{2\tan\theta}$. Size and strain broadening show different $\theta$ dependence. This provides a way to separate the two effects. **Williamson Hall plot** [18] is a method of deconvoluting size and strain broadening by looking at the peak width as a function of $2\theta$.

$$\beta_{obs} = \beta_{size} + \beta_{strain}$$
$$\beta_{obs} = \frac{\lambda}{d\cos\theta} + 2\varepsilon\tan\theta \qquad (2.4)$$
$$\frac{\beta_{obs}\cos\theta}{\lambda} = \frac{1}{d} + \varepsilon\left(\frac{2\sin\theta}{\lambda}\right)$$

Thus, plotting $\beta_{obs}\cos\theta/\lambda$ on the y-axis and $2\sin\theta/\lambda$ on the x-axis, we can estimate the particle size from the inverse of the intercept on the y-axis, and the strain from the slope. An example, analysis done on a 55 nm XRD data is shown in Fig. 2.9.

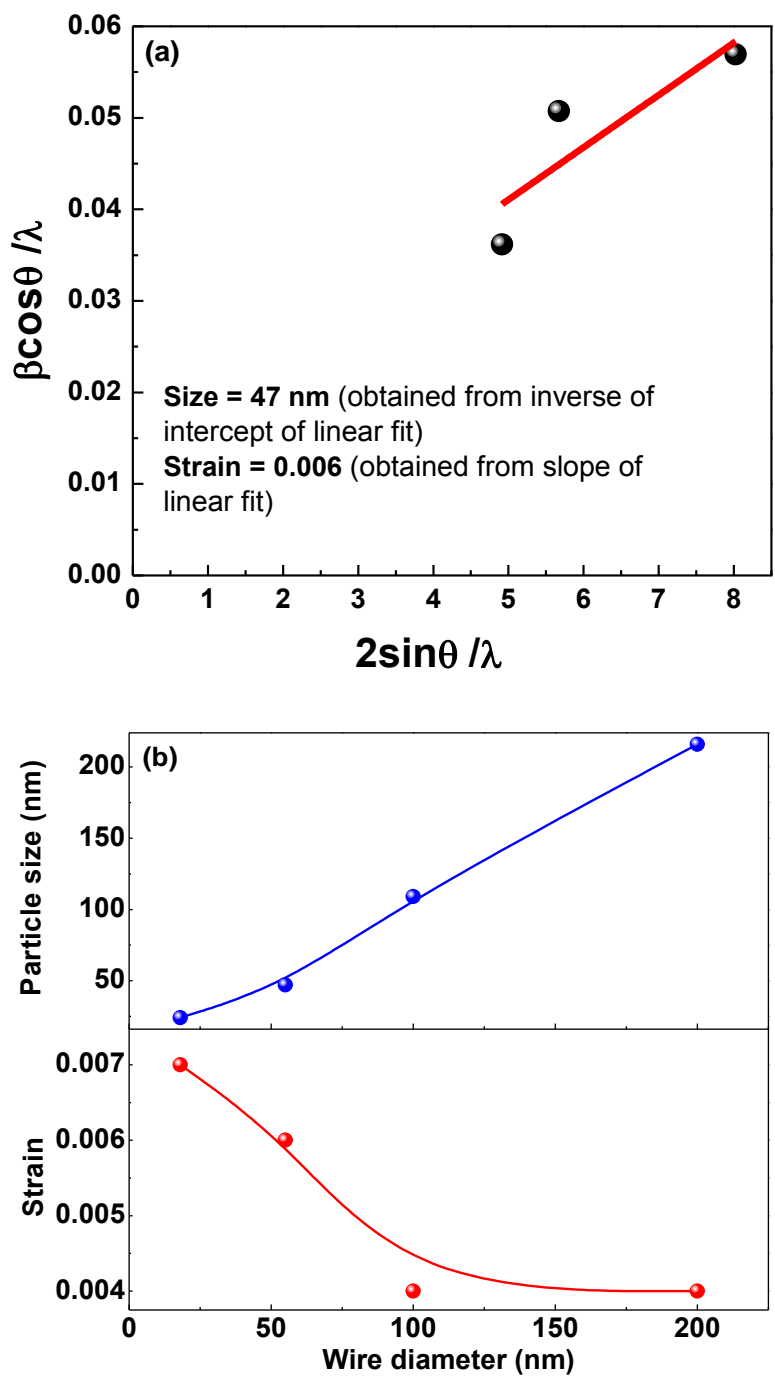

**Figure 2.9:** (a) Particle size and strain estimated from Williamson-Hall plot [18] for 55 nm diameter nanowire array data. (b) Variation of particle size and strain for nanowires grown in AAO templates of various pore diameters plotted together.

The particle sizes in polycrystalline nanowires of nickel estimated from Williamson-Hall plot [18] match closely with the pore diameters of the AAO templates which we used for synthesis of the nanowires of respective diameters. Also we observed an increase in microstrain (as shown in Fig. 2.9(b)) with the decrease in wire diameter. We also show the Full Width at Half Maximum (FWHM) of various peaks in the polycrystalline nanowires in Fig. 2.10.

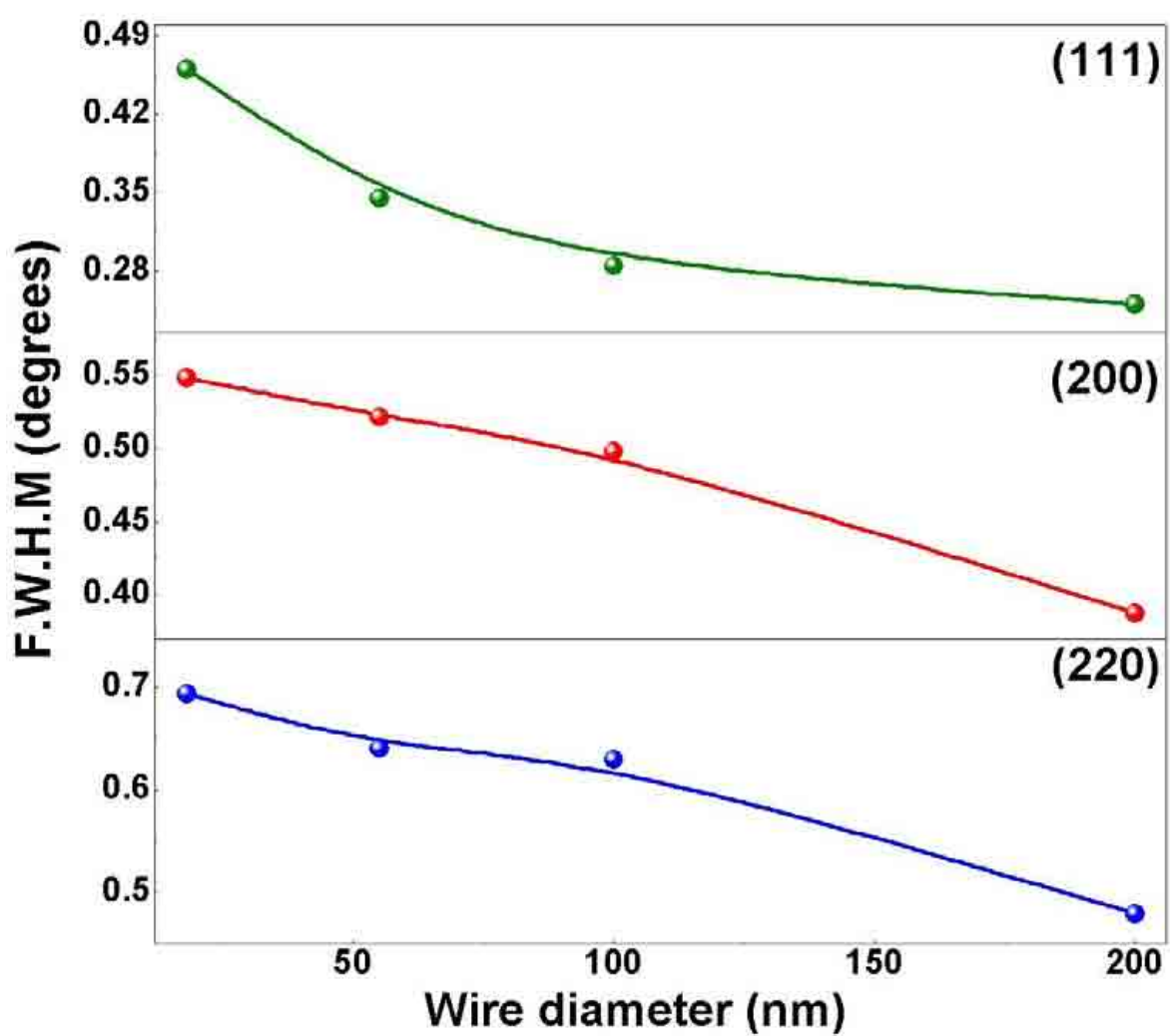

**Figure 2.10:** Full Width at Half Maximum (FWHM) of various peaks in the polycrystalline nanowires.

In Fig. 2.11, the X-ray diffraction data of the nanowires grown by pulsed electrodeposition in anodic alumina templates with pore diameters ranging from 55nm-13nm are shown. The pulsed potentiostatic deposition was carried out in AAO templates at a potential of -1 V with a duty cycle of 80% and pulse period of 1 second. One can clearly see that the nanowires grown are oriented along the specific plane (220). For the single crystalline samples, particle size and strain could not be evaluated using this method because firstly, it has only one peak. So Williamson Hall plot is not possible. Secondly, because the wires are single crystalline, the x-ray is seeing 60 μm (width of the template) of sample instead of nanometer dimension. So the peaks are very narrow (peak width arising only because of instrumental broadening). For example, for the 13 nm diameter sample, peak width ~ 0.1$^{\circ}$, which is much smaller than the peak width due to size broadening (see plot of F.W.H.M. versus wire diameter for the polycrystalline samples). So, particle size calculation using Scherrer formula is also not possible.

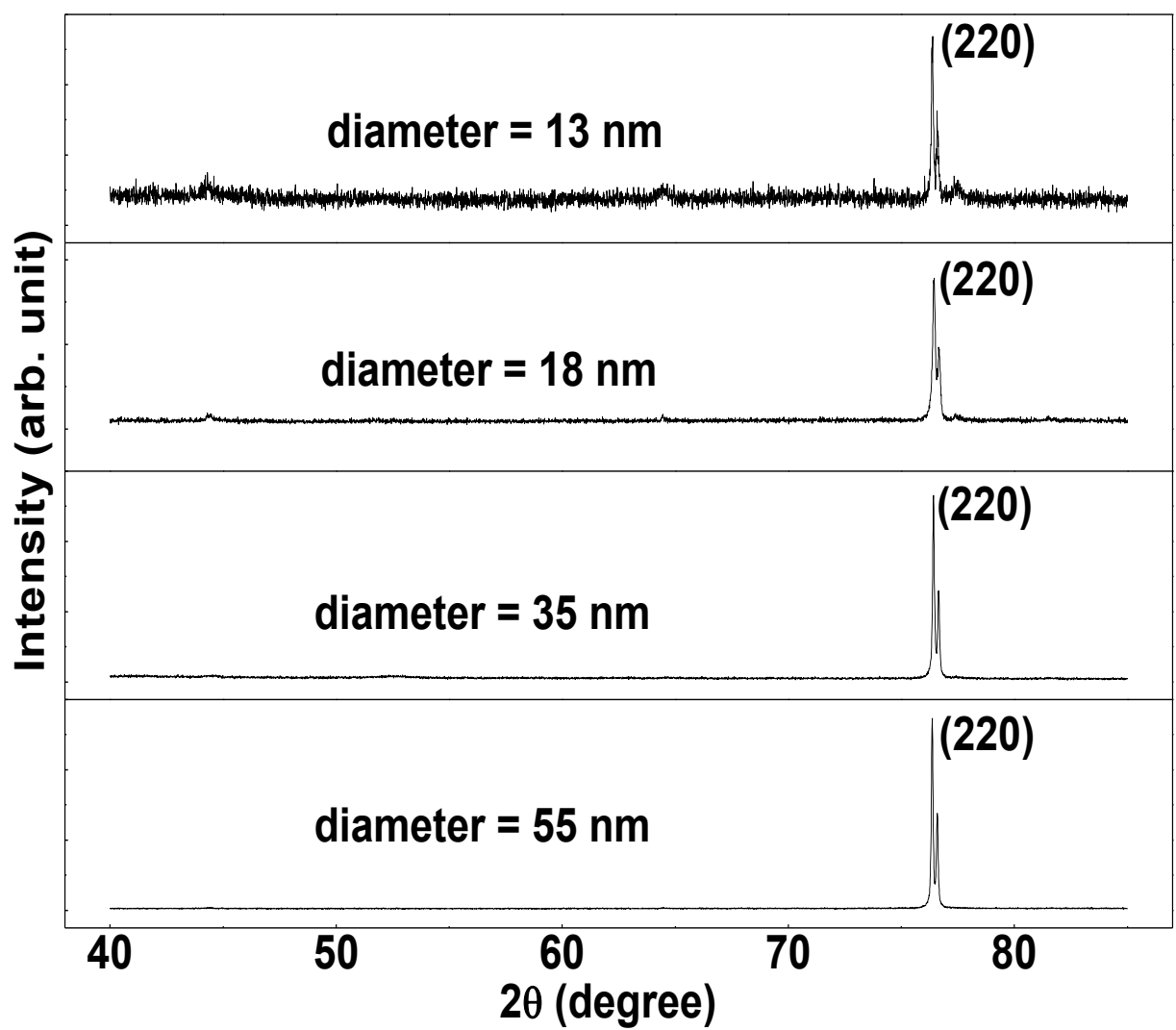

**Figure 2.11:** X-ray diffraction patterns of nickel nanowires of various diameters grown in anodic alumina templates by pulsed potentiostatic deposition. The nanowires are single crystalline with specific orientation of (220) plane. (The split in the line arises from the Cu K$_\alpha$ lines).

In Table 2.1 we present the cumulative data of the XRD data for both polycrystalline and single crystalline nanowires as compared with bulk Ni standard data. In case of single crystalline nanowires we did not calculate the d values for other planes because of negligible or low intensities.

**Table 2.1:** Compiled data of the XRD data for both polycrystalline and single crystalline nanowires as compared with bulk Ni standard data.

| Sample | (111) | | (200) | | (220) | |
|---|---|---|---|---|---|---|
| | 2Θ (degree) | d (Å) | 2Θ (degree) | d (Å) | 2Θ (degree) | d (Å) |
| Ni (Standard) | 44.37 | 2.04159 | 51.596 | 1.77 | 76.084 | 1.25098 |
| Polycrystalline Ni Nanowire samples | | | | | | |
| 18 nm | 44.32 | 2.04377 | 51.855 | 1.76313 | 76.425 | 1.24624 |
| 55 nm | 44.555 | 2.03354 | 51.885 | 1.76218 | 76.229 | 1.24894 |
| 100 nm | 44.325 | 2.04356 | 51.705 | 1.76789 | 76.225 | 1.24901 |
| 200 nm | 44.535 | 2.03441 | 51.95 | 1.76013 | 76.504 | 1.24515 |
| Single crystalline Ni Nanowire samples | | | | | | |
| 13 nm | - | - | - | - | 76.385 | 1.24679 |
| 18 nm | - | - | - | - | 76.455 | 1.24583 |
| 35 nm | - | - | - | - | 76.425 | 1.24624 |
| 55 nm | - | - | - | - | 76.365 | 1.24707 |

# Imaging tools

We have used the regular imaging tools Scanning Electron Microscope (SEM) (both FESEM and ESEM) and Transmission Electron Microscope (TEM) for imaging of the nanowires and nanotubes. In this section the details of the microscopy techniques are described.

## 2.3.2 Scanning Electron Microscopy (SEM)

SEM [19, 20] is most widely used form of electron microscope in the field of material science. It uniquely combines the simplicity and ease of sample preparation with much of the performance capability and flexibility of the more expensive and complex Transmission Electron Microscope (TEM) [21, 22]. In a scanning electron microscope, electrons emanating from a hot tungsten filament (thermionic emission)/sharp tungsten tip (field emission type) are accelerated by a high potential difference and collimated to a sharp beam. The beam traversing through a series of electromagnetic lenses is made to raster the surface of the sample under observation. The high energetic electrons after during interaction with the sample give rise to various signals collected by the respective detectors as a function of beam spot position on the sample giving rise to information of the topography of the sample.

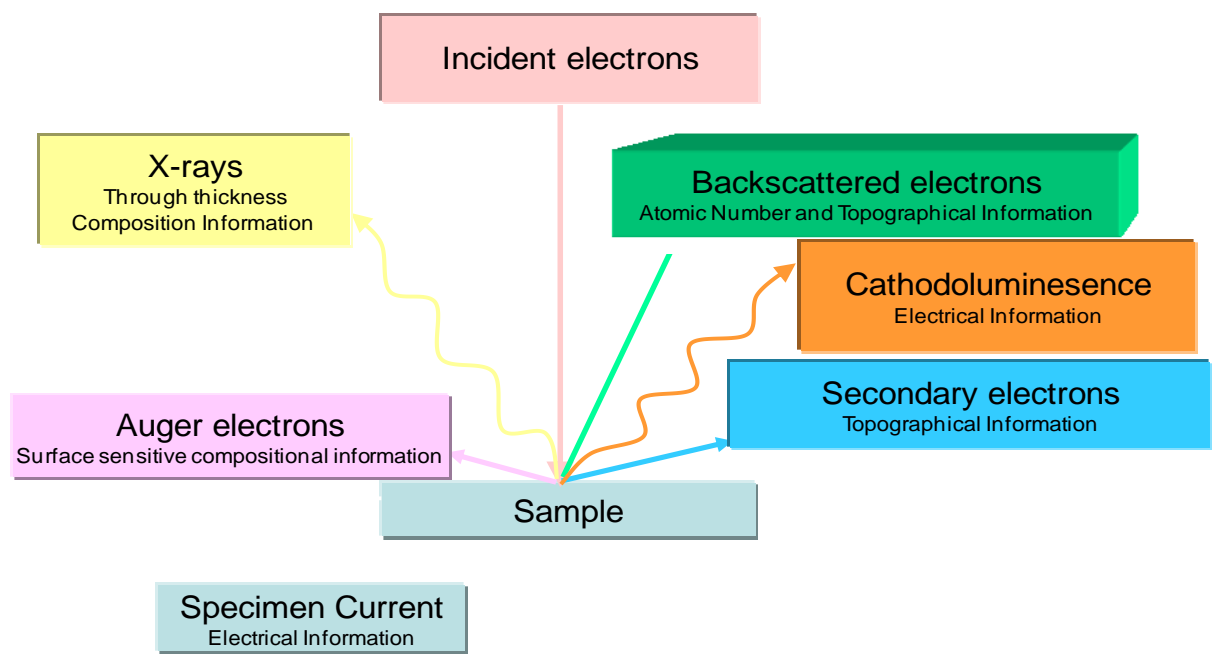

**Figure 2.12:** Electron interaction with matter

When the electron beam strikes the sample, both photon and electron signals are emitted. The various signals are depicted in Fig. 2.12. The interaction of the electrons can be mainly classified into elastic (interaction with the nucleus) and inelastic (interaction with the nucleus). Except when the electrons are backscattered, all other interaction lead to the loss of energy of the incident electrons.

The **backscattered electrons (E~KeV)** constitute a fraction of the incident electrons and their quantity depends on the atomic number of the atomic nucleus scattering them i.e. heavier the nucleus, greater the number of backscattered electrons. This property is often used as Z-contrast in alloy images. Apart from the Z-contrast, the backscattered electrons also play a role in revealing the topography of the sample. BSE detectors are usually either of scintillator or semiconductor types. However, the main topography information is obtained from the **secondary electrons (E~ eV)**, which are ejected from the valence shells because of the incident electrons. The secondary electrons are detected by an Everhart-Thornley detector [23] which is a type of scintillator-photomultiplier system. The secondary electrons are first collected by attracting them towards an electrically-biased grid at about +400V, and then further accelerated towards a phosphor or scintillator positively biased to about +2000V. The accelerated secondary electrons are now sufficiently energetic to cause the scintillator to emit flashes of light (cathodoluminescence) which are conducted to a photomultiplier outside the SEM column via a light pipe and a window in the wall of the specimen chamber. The amplified electrical signal output by the photomultiplier is displayed as a two-dimensional intensity distribution that can be viewed and photographed on an analogue video display, or subjected to analog-to-digital conversion and displayed and saved as a digital image.

For scanning electron microscopy (SEM) of nanowires, the template containing the nanotubes is etched partially with 3M NaOH solution to dissolve most of the aluminium oxide layer. The remaining template after washing several times with Millipore water was used for the imaging. SEM images were taken mostly with the Quanta 200 ESEM (FEI Co.), Quanta 200 FEG SEM (FEI Co.)[24] and JEOL FEG SEM [25]. Fig. 2.13 shows some of the SEM images of the nanowires grown in Anodic Aluminium Oxide membrane with various pores sizes by DC and Pulsed electrodeposition. It is observed that for taking better scanning electron micrographs to is necessary that no over growth (i.e., the metal should not fill the template completely) takes place during electrodeposition.

Completely filled templates with over growth can hamper the etching and hence the dissolution of AAO template. This is because in such a case the etchant solution is stopped to reach the inner AAO by the top layer of metal layer which is created by over growth. One can polish the layer and then do the etching to get the required results. But for preliminary images, it is always important to avoid overgrowth to get images of nanowire arrays. The images presented here in Fig. 2.13 are that of metal nanowires grown by potentiostatic deposition (both DC and pulsed).

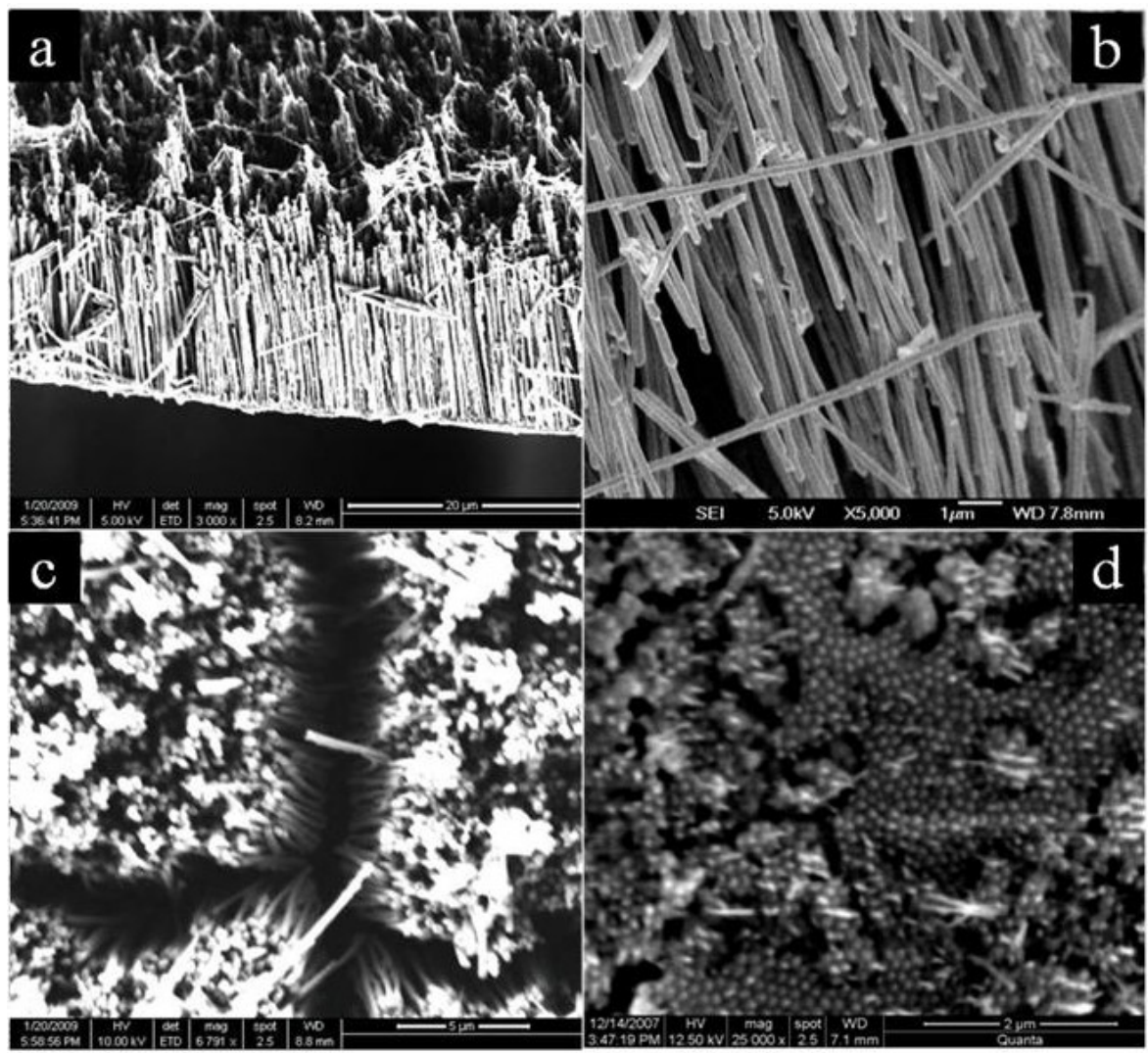

**Figure 2.13:** SEM images of nickel nanowires (a) A partially etched AAO template showing 200 nm diameter nanowires. (b) 100 nm diameter nickel nanowires grown by DC electrodeposition. (c) Nanowires of copper grown by DC electrodeposition at -0.5 Volts of the working electrode w.r.t SCE. (d) Top surface of polished template with pore ~55 nm showing nanowires grown by pulsed electrodeposition.

### 2.3.3 Transmission Electron Microscopy

The Transmission Electron Microscope (TEM) is a versatile instrument, capable of characterizing the internal structure of materials with a wide range of imaging and analytical methods. It uses diffraction of electrons to image materials. In transmission electron microscopy, a beam of electrons (with a typical wavelength of less than 1 Å) is sent through a thin specimen (to image powder samples, a very dilute solution of the powder in ethanol is dispersed on a TEM grid). A fraction of the electrons in the beam get either elastically or inelastically scattered. The elastically scattered (diffracted) electrons contribute to the image contrast. Thus, while passing through the specimen, the electron beam undergoes diffraction according to Bragg's law. The non – diffracted as well as the diffracted beams are focused on the back focal plane of the objective lens (which is the first lens below the specimen), resulting in an electron diffraction pattern. The diffraction pattern of the back focal plane undergoes an inverse Fourier transformation and a highly magnified image is formed. In amplitude or diffraction contrast imaging, an objective aperture (a small hole in a metallic plate) is introduced into the beam in the back focal plane. Thus, a significant amount of electrons are removed from the beam before further magnification. This results in an intensity contrast in the final image.

For Transmission Electron Microscopy (TEM), the template was completely etched with 6M NaOH solution and washed 10 times with Millipore water before spraying on a carbon coated copper grid. Images were obtained by a JOEL, JEM-2010 having $LaB_6$ electron gun for operation between 80-200kV [25]. Template etching is a crucial step for TEM sample preparation of nanowires. An under etched template leaves debris surrounding the wires giving a completely different idea of the diameter of the nanowire. Thus it is important to remove the debris after etching over a considerably long time (1-2 hrs) and washing at least 10 times with Millipore water. The nanowires deposited at constant potential of the working electrode (the metal coating facing electrolyte inside the porous membrane) with respect to the saturated calomel electrode (SCE), are mainly polycrystalline. In case of the nanowires grown in polycarbonate membranes, the nanowires have poor crystallinity with similar growth conditions. Thus, we used anodic alumina templates for the growth of nanowires. In Fig. 2.14, we show the TEM images of nanowires of nickel of various diameters grown in anodic alumina and polycarbonate templates grown by DC and pulsed electrodeposition.

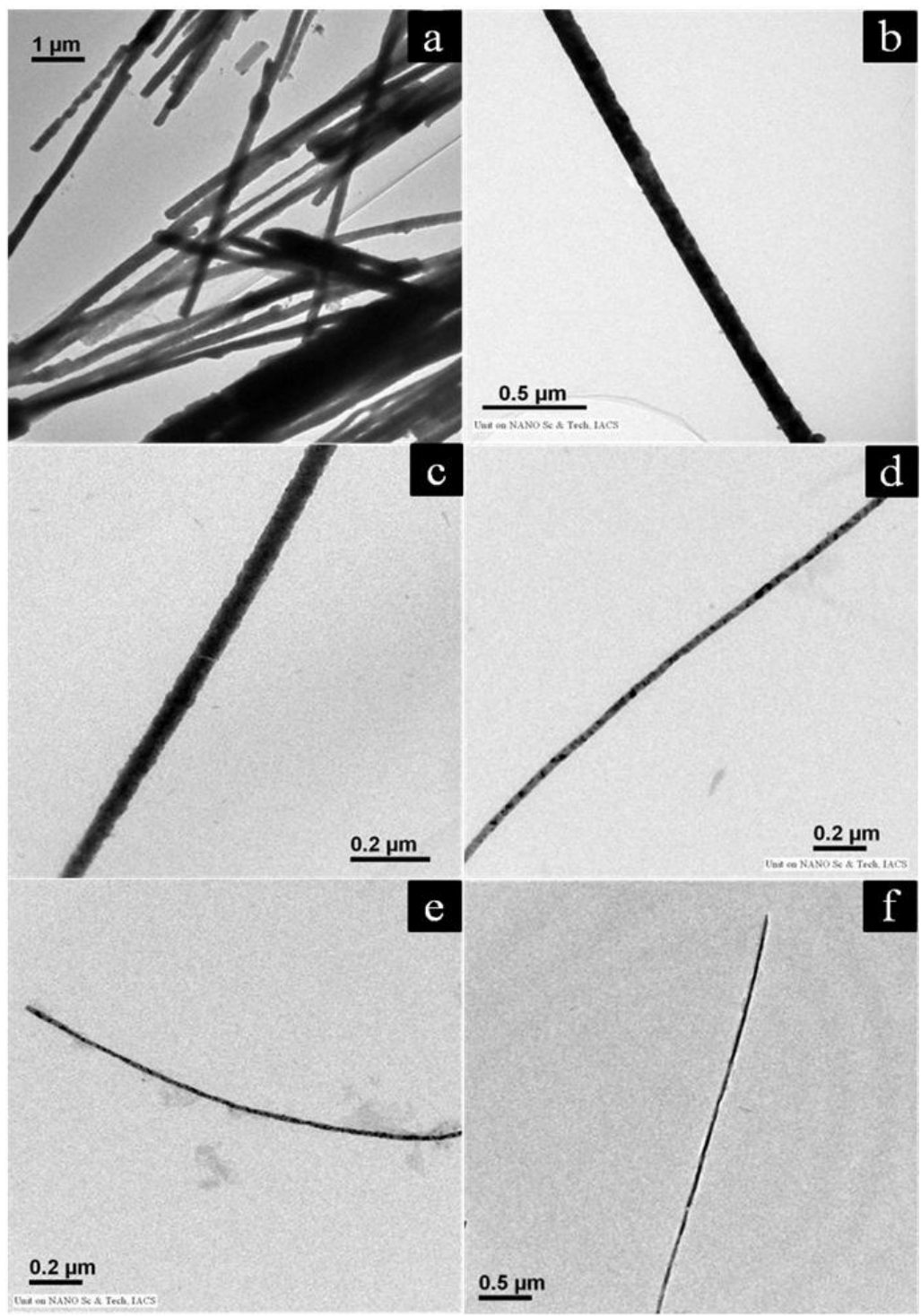

**Figure 2.14:** TEM images of nanowires (a) A 200 nm diameter nickel nanowires (b) 100 nm diameter nickel nanowires grown by DC electrodeposition (c) Nanowires of nickel grown by DC electrodeposition in polycarbonate membrane. (d) 35 nm diameter nanowire (e) 18nm diameter nickel nanowire (f) 55 nm diameter nickel nanowire; nanowires (d), (e), (f) are grown by pulsed deposition.

**2.3.3.1 HRTEM:** The other basic imaging mode is *phase contrast imaging or high resolution TEM (HRTEM) imaging* mode. It can show structures at the atomic scale, lattice fringes, and hence can be used to look at the arrangement of atoms in a crystal. Here, the aperture in the back focal plane is larger (sometimes the aperture is dispensed with altogether) and multiple beams are allowed to contribute to the final image. One or more diffracted beams interfere with the directly transmitted beam to form the image, and the image contrast depends on the relative phases of the various beams. HRTEM, thus, requires interference of the unscattered and the scattered beams.

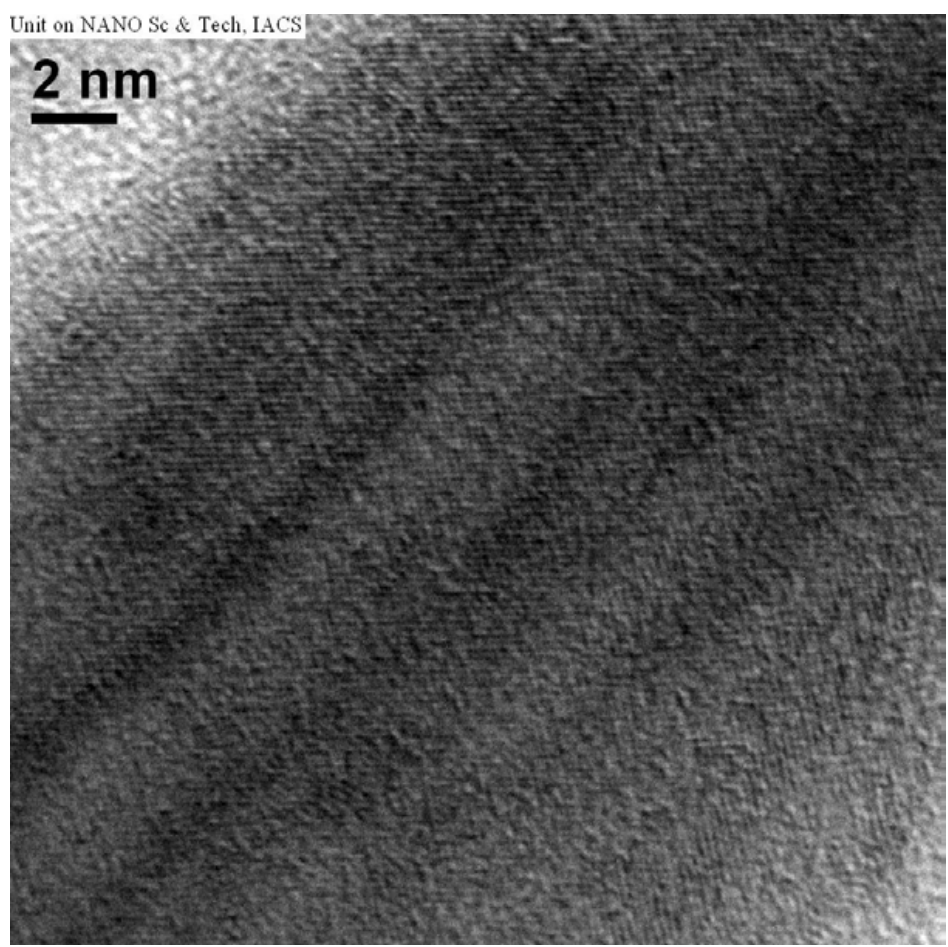

**Figure 2.15:** A high resolution image taken on a nanowire grown in an anodic alumina template with average pore diameter of 18 nm.

In Fig. 2.15, we show the high resolution image of nickel nanowire grown by pulsed deposition in anodic alumina template with average pore diameter of 18nm. The lattice planes can be really seen with no grain boundaries over the width of the nanowire. One can also see a variation in the contrast along the diameter which can be attributed to the uneven thickness or slight presense of amorphous unetched alumina matrix in which the nanowires are grown.

Fig. 2.16 shows the high resolution image of a nickel nanowire grown by the same pulsed electrodeposition condition as the earlier 18 nm diameter one. The arrows show the interplaner spacing for (111) plane.

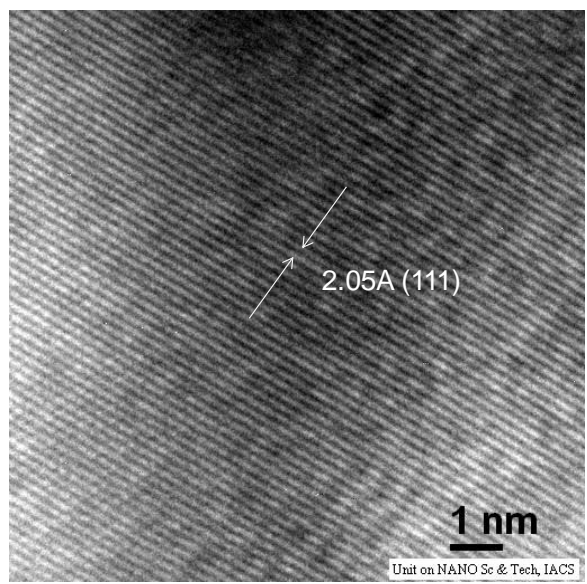

**Figure 2.16:** A high resolution image taken on a nanowire grown in an anodic alumina template with average pore diameter of 55 nm.

### 2.3.3.2 Indexing of the electron diffraction patterns:
For small angle of diffraction $\theta$, $\mathrm{Sin}\theta$ in the Bragg's law can be replaced by $\theta$. Using this reduced Bragg condition, we get $\lambda = 2d_{hkl}\theta$. Now, if $L$ is the distance between the screen and the sample and $R$ is the distance of the diffracted spot of interest from the central spot, then $\tan 2\theta = \dfrac{R}{L}$. This can be seen in Fig.2.17. Again for small $\theta$, $\tan 2\theta \approx 2\theta$. Thus, $\lambda L = Rd_{hkl}$.

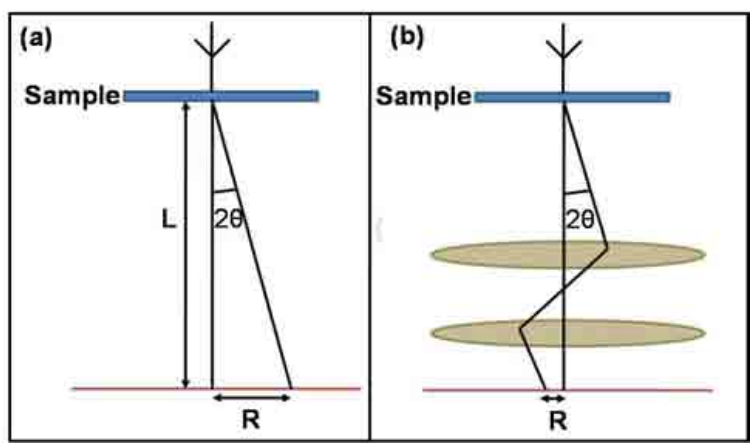

**Figure 2.17:** Camera length L in the presence and absence of imaging lenses.

The effective length $L$ in the presence and absence of imaging lenses are shown in Fig.2.17. For ring patterns, the radius of each ring can be converted into interplaner spacing by the above equation. For spot patterns (as in our case), the distances of the diffracted spots from the central spot is measured. These distances correspond to the interplaner spacing of the diffracting planes. Thus, by measuring $R$ and calculating the corresponding plane spacing ($d_{hkl}$), the diffraction pattern can be indexed. With advent of modern microscopes, a scale of inverse of d is introduced into the pattern image itself taking into account the above calculations. Once the scale is given, it becomes easier to index the spots as the inverse of the distance of adjacent spot from the centre spot gives the $d_{hkl}$. Thus once it is known, a set of spots can be assigned (h k l) and the rest of the spots are indexed by simple laws of vector addition.

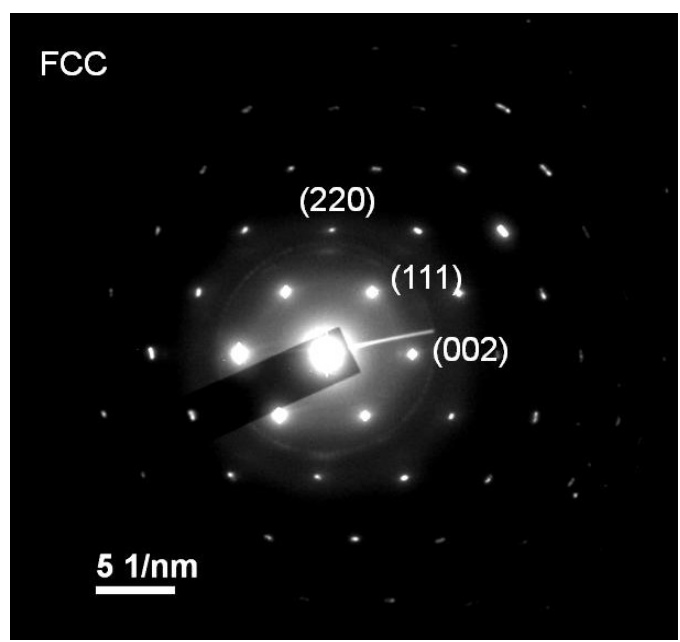

**Figure 2.18:** Electron diffraction pattern of a single crystalline nickel nanowire of 55 nm diameter.

The electron diffraction pattern of a single crystalline nickel nanowire of diameter 55nm, being grown by pulsed electrodeposition is shown with indexing the main planes in Fig. 2.18. To understand the single crystalline of a nanowire, it is important to obtain the electron diffraction patterns at various points along the length of the nanowire. Fig. 2.19 show a nanowire grown 55 nm pore diameter AAO template by pulsed electrodeposition. The various portion of the nanowire along with the electron diffraction patterns are shown as insets closer to the respective places of the nanowire. One can see throughout the nanowire the same kind of electron diffraction pattern is obtained confirming the single crystalline and oriented nature of the nanowire.

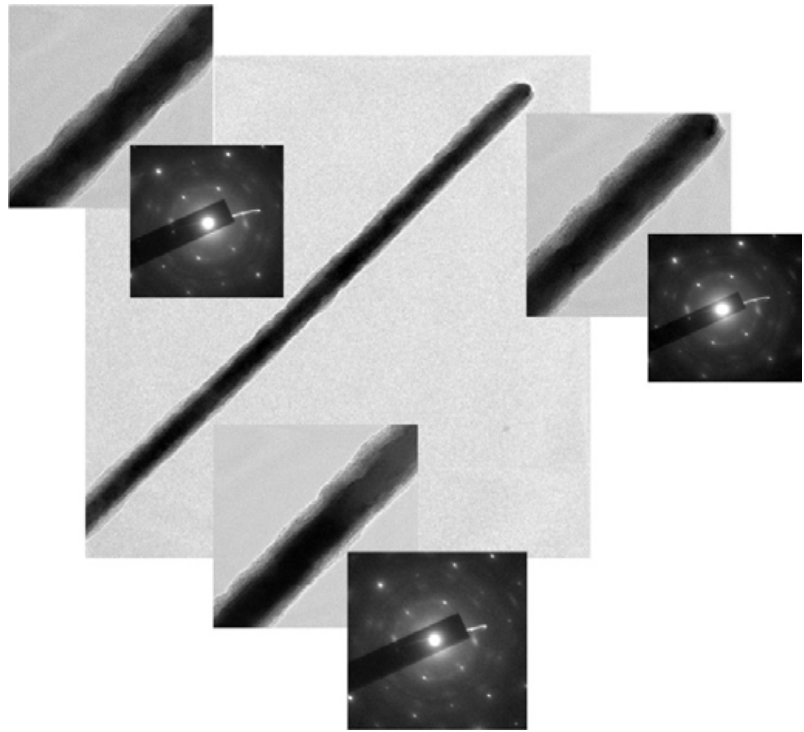

**Figure 2.19:** A single nanowire grown in 55 nm pore diameter AAO template. Insets show electron diffraction pattern of the various regions of the nanowire closer to the respective points in the nanowire.

Unlike the single crystalline nanowire electron diffraction pattern, the electro diffraction patterns have rings instead of regular spots correspoding to the respective lattice planes. Fig. 2.20 shows the electron diffraction pattern taken on DC electrodeposited 200 nm diameter nanowire.

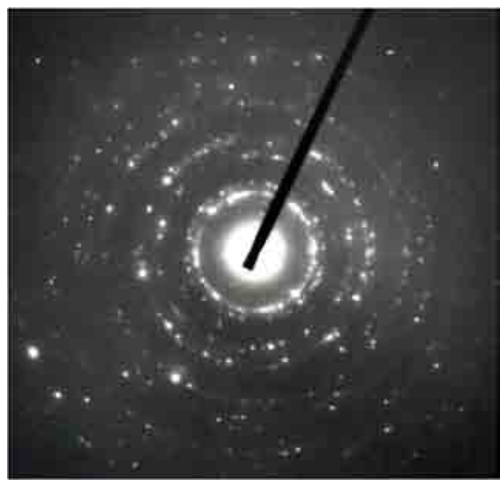

**Figure 2.20:** Electron diffraction pattern of a typical polycrystalline nanowire.

### 2.3.4 Energy Dispersive X-ray Spectroscopy (EDXS)

The technique of energy dispersive x-ray spectroscopy (EDXS) utilizes the characteristic spectrum of x − rays that are emitted by a sample following initial excitation of an inner shell electron to a vacant higher energy level by the high energy electron beam. Highly localized information about the elemental composition of the sample can be obtained, with the spatial resolution determined primarily by the probe size. EDXS is a relatively straightforward technique experimentally, but it requires pre − calibration of the specific EDXS analysis system using standards of known composition.

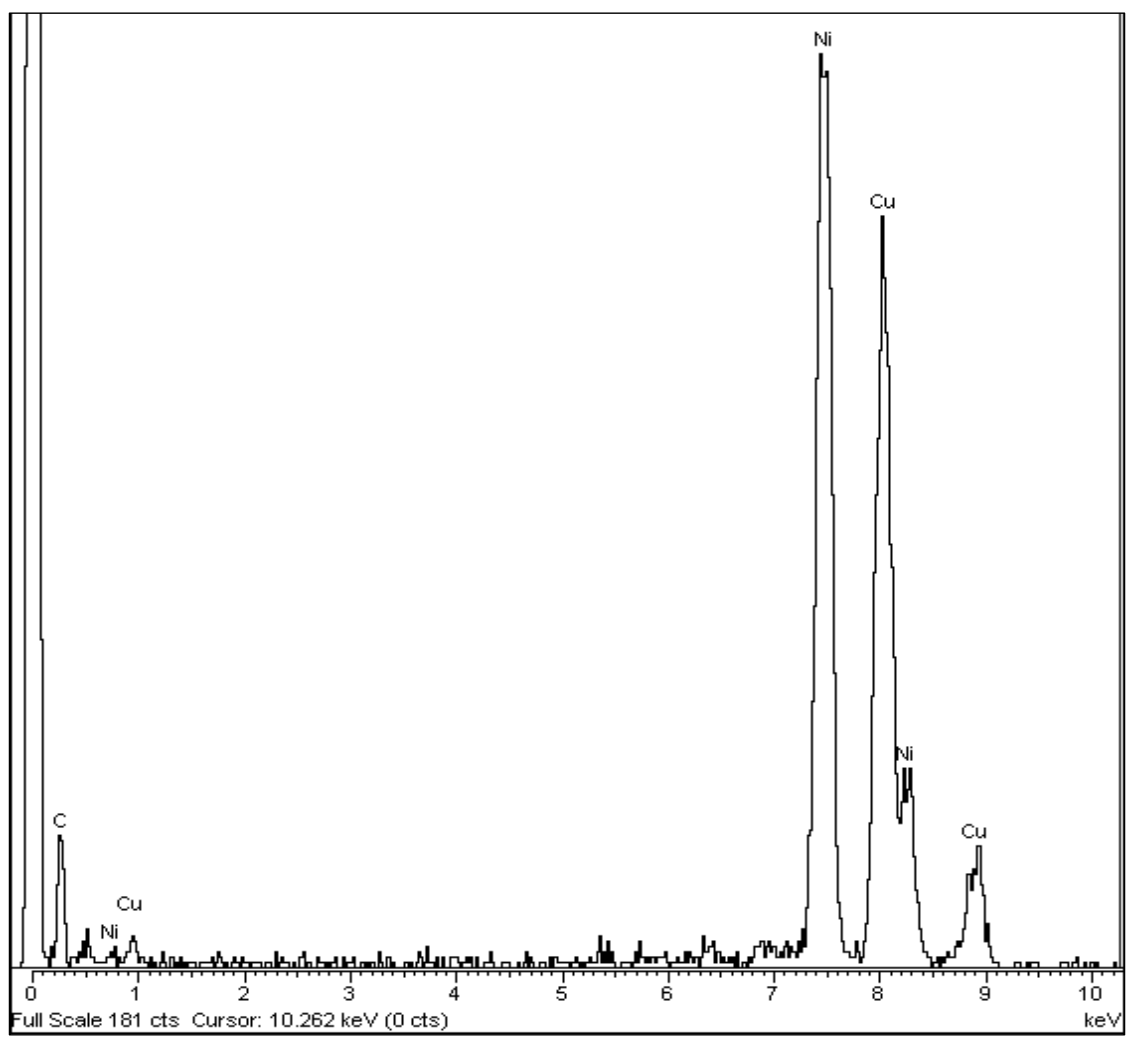

**Figure 2.21:** Energy dispersive spectrum taken on a nickel 100 nm nickel nanowire showing nickel as the main element present.

In addition to the Ni peaks in the EDXS of 100 nm diameter nanowire (as shown in Fig 2.21), there are also some copper peaks. The copper signal is obtained because of the background reflections from copper grid.

## 2.3.5 Magnetization measurements

The magnetic measurements were carried out using a commercial vibrating sample magnetometer (VSM)[26] and SQUID magnetometer[27]. Since we worked with Nickel nanowires, it becomes important to know how magnetic the wires are. Since the electrical transport properties are dependent are on the magnetic nature of the sample in such itinerant systems, it is important to characterize them magnetically.

### 2.3.5.1 Vibrating Sample Magnetometer (VSM)

A vibrating sample magnetometer (VSM)[26] operates on Faraday's Law of Induction which tells us that a changing magnetic field will produce an electric field. This electric field can be measured and can give us information about the changing magnetic field. Thus, the sample is placed in a constant magnetic field and oscillated with a fixed frequency near a set of detection (pickup) coils. If the sample is magnetic, the constant magnetic field induces a magnetic dipole moment in the sample. This magnetic dipole moment creates a magnetic field around the sample. As the sample is oscillated, this magnetic flux changes as a function of time and the voltage induced (proportional to the rate of change of flux) in the pickup coils is synchronously detected. The voltage will also be proportional to the magnetic moment of the sample. In Fig.2.22 we show (a) schematic of VSM and (b) detailed configuration near the pick-up coils.

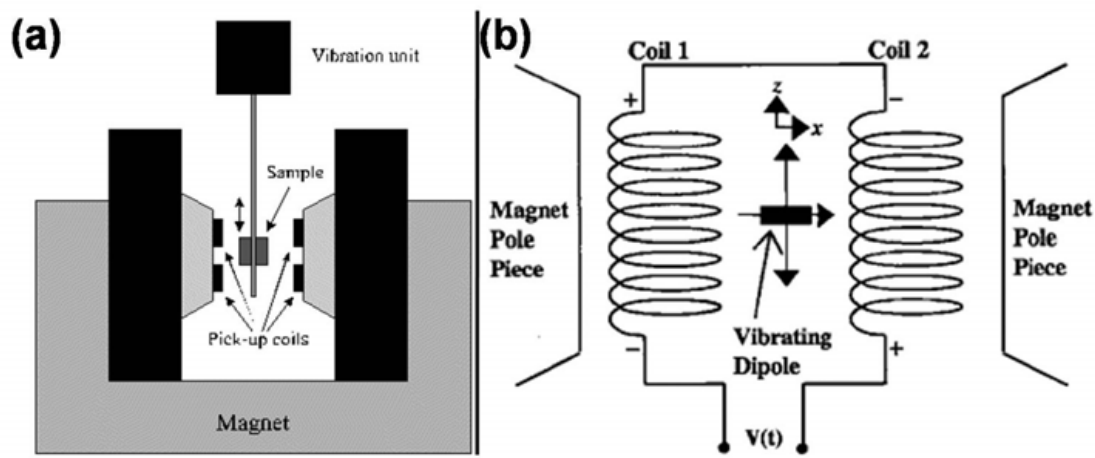

**Figure 2.22:** Schematic illustration of (a) VSM and (b) details near the pickup coils.

The system detection capability can be maximized by optimizing the geometry of the pickup coils (Fig.2.23), and by having oscillation amplitude that is relatively large (1 − 3 mm peak). The time dependent voltage induced in the pickup coils is given by $V_{induced} = (d\phi/dt) = (d\phi/dz)(dz/dt)$, where, $\phi$ represents the magnetic flux, the axis of oscillation of the sample is conventionally chosen to be the $z-$axis, and z, therefore, represents the position of the sample along this axis and t is the time. If the sample is made to oscillate sinusoidally, then the induced voltage in the pickup coils will have the form $V_{induced} = cmA\omega \sin \omega t$, where, c is a coupling constant, m is the DC magnetic moment of the sample, A is the amplitude of oscillation, and $\omega = 2\pi f$, where, f is the frequency of oscillation of the sample. The detection of the magnetic moment of the sample, thus, amounts to measuring the coefficient of sinusoidal voltage response induced in the detection coil.

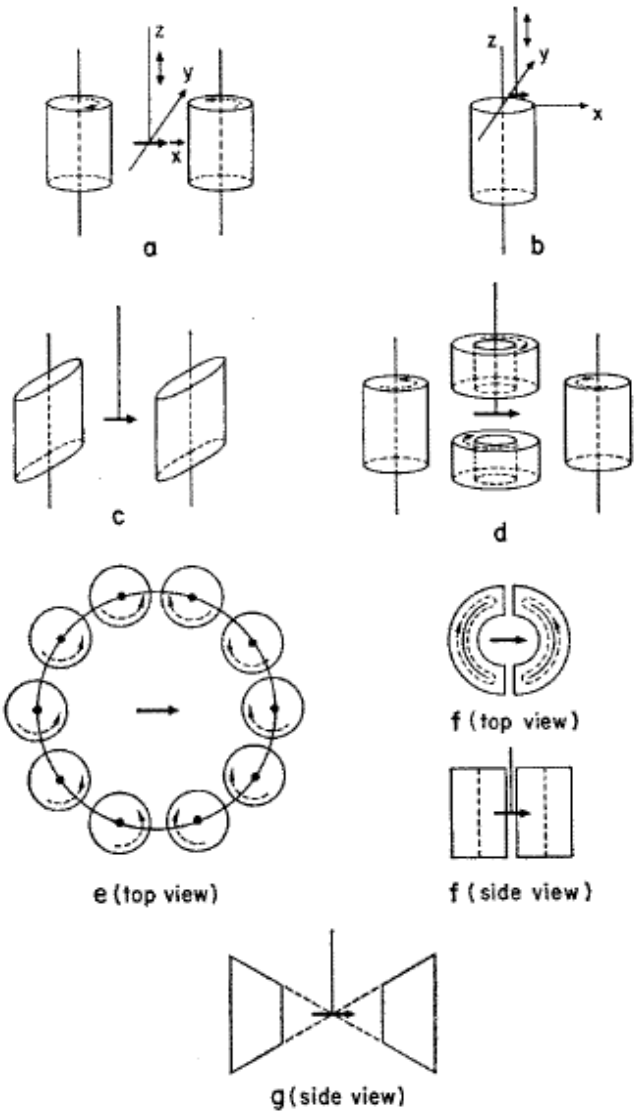

**Figure 2.23:** Examples of different pickup coil arrangements. The sample, indicated by the heavy arrow, is vibrated along the z–direction. (Reproduced from [28]). The VSM that we have used [29] has the coil arrangement shown in (a).

Fig. 2.24 shows the compiled magnetization curves (M-H) at room temperature of nanowires of different diameters. This study establishes that the nanowires are ferromagnetic in nature with characteristic hysteresis loops. As said earlier, this magnetic characterization is important before we measure the electrical transport properties and do the analysis.

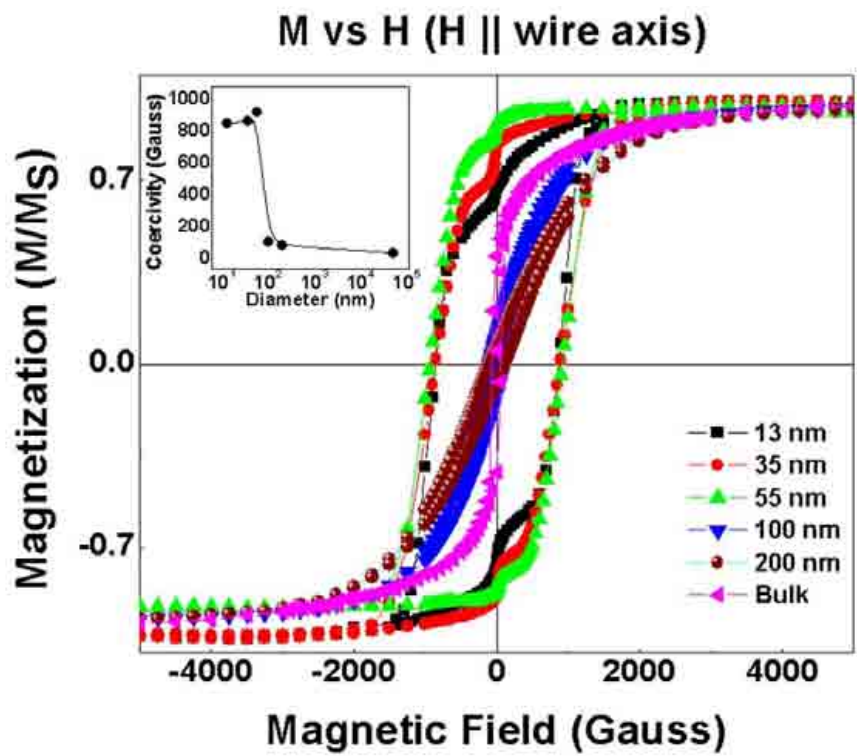

**Figure 2.24:** The room temperature M-H curves of nickel nanowires of various diameters, each having a characteristic hysteresis loop showing that the nanowires are ferromagnetic up to the lowest diameter. The inset shows the coercivity as a function of wire diameter.

Few things are noted from the magnetization curve that becomes important from the view point of magnetism in them. First of all down to 100nm the M-H curves are rather similar to the bulk with very low coercive field. For diameter d ≤ 55nm, the hysteresis loop is qualitatively different. The M-H curve tends to become nearly rectangular when the size is below 55nm with $H_c$ rather large (~850 Gauss). This is expected of a one dimensional magnetic wire. In Ni the domain wall width $\lambda_{DW}$ is 35-40nm. If d < $\sqrt{\pi}$ $\lambda_{DW}$ and length >> $\lambda_{DW}$, the wire has one dimensional magnetic behavior [30] In this case, for wires are with diameter d ≤ 55nm, and thus the wires are expected to show one dimensional behaviour with magnetization rotating in unison at the coercive field. We find, however, that there are nucleation of reverse magnetization within small volumes for H << $H_c$ that lowers the remnant magnetization. Such nucleation can occur at the defect sites as well as the ends.

### 2.3.5.2 SQUID Magnetometer

SQUID magnetometers have a higher sensitivity than VSMs (the resolution obtained in a SQUID magnetometer is of the order of $10^{-8}$ emu as compared to a resolution of only $10^{-6}$ emu for VSM). As such, we used SQUID magnetometer for measuring those samples which have a lower magnetic moment (especially those samples showing an antiferromagnetic and charge ordering transition). The working principle of a SQUID is based on the quantum interference of wave functions that describe the state of the superconducting charge carriers (the Cooper pairs). A SQUID is based on an interferometer loop in which two weak links (Josephson contacts) are established. A weak link is established by interrupting a superconductor by a very thin insulating barrier. The function of the SQUID is to link the quantum mechanical phase difference of the Cooper pairs wave functions over a weak link with the magnetic flux penetrating the interferometer loop.

The components of a SQUID magnetometer typically consist of the following: a detection coil which senses changes in the external magnetic field and transforms them into an electrical current, an input coil which transforms the resulting current into a magnetic flux in the SQUID sensor, electronics which transform the applied flux into a room temperature voltage output and acquisition hardware and software for acquiring, storing and analyzing data. Both the SQUID amplifier and detection coils are superconducting devices.

**Superconducting detection coil:** The detection coil used in a SQUID is a single piece of superconducting wire in a set of three coils configured as a second order gradiometer. The coil arrangement is shown in Fig.2.25. The coil sits outside the sample space within the liquid helium bath. The upper coil is a single turn would clockwise, the centre coil comprises two turns wound counter-clockwise and the bottom coil is a single turn wound clockwise. The coils are positioned at the centre of the superconducting magnet outside the sample chamber such that the magnetic field from the sample couples inductively to the coils as the sample is moved through them. The gradiometer configuration is used to reduce noise in the detection circuit caused by fluctuations in the large magnetic field of the superconducting magnet. It also minimizes background drifts in the SQUID detection system caused by relaxation in the magnetic field of the superconducting magnet. Ideally, if the magnetic field is relaxing uniformly, the flux change in the two turn centre coil will be exactly cancelled by the flux change in the single turn top and bottom coils. On the other hand, the magnetic moment of the sample can still be measured by moving the sample through the detection coils because only the counter wound coil set measures the local changes in magnetic flux density produced by the dipole field of the sample.

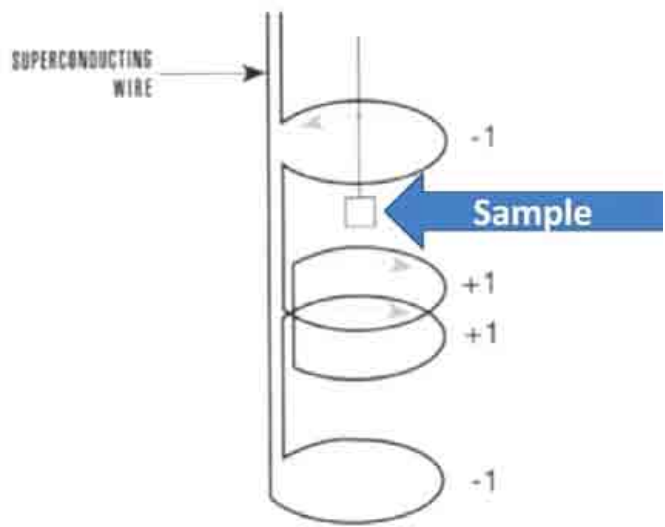

**Figure.2.25:** Second order gradiometer coil configuration (reproduced from [28]).

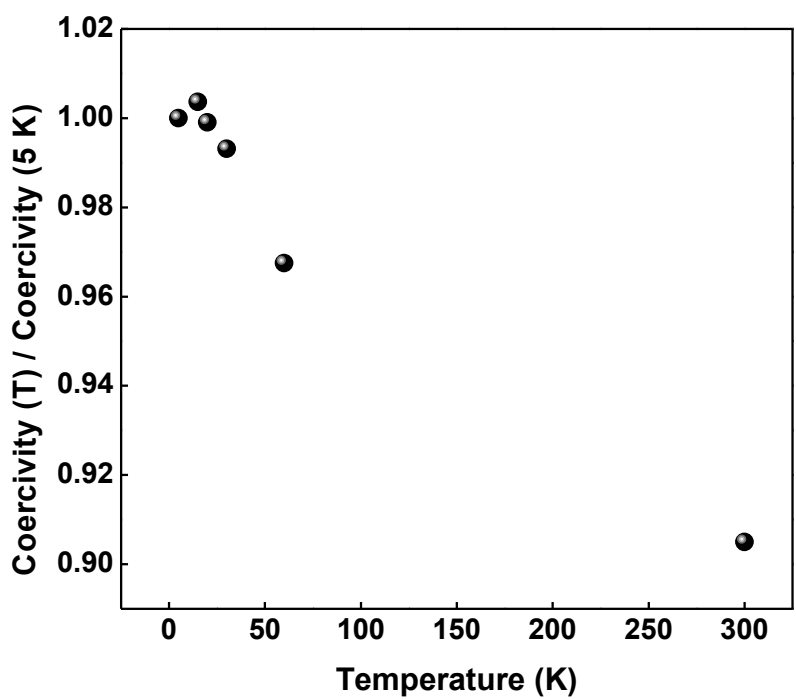

**Figure.2.26:** The variation of normalized coercivities of 55 nm Ni nanowire array obtained by Squid magnetometer in the temperature range 5-300 K.

Fig. 2.26 shows the normalized coercive field as a function of temperature measured by squid magnetometer on a 55 nm Ni nanowire. It can be clearly seen that the Coercive field has shallow temperature dependence (changes by less than 10 % in the temperature range 5-300 K). This would mean that the thermal assisted magnetization reversal rate is low and the wires have rather high barrier to thermally assisted reversal [30].

To see the temperature dependence on magnetization and the functional behavior, we have done the Zero Field Cooled (ZFC) and Field Cooled (FC) Magnetization-Temperature measurements on the nanowires of 55 nm and 18 nm diameter. The results are shown in Fig. 2.27 (a) & (b).

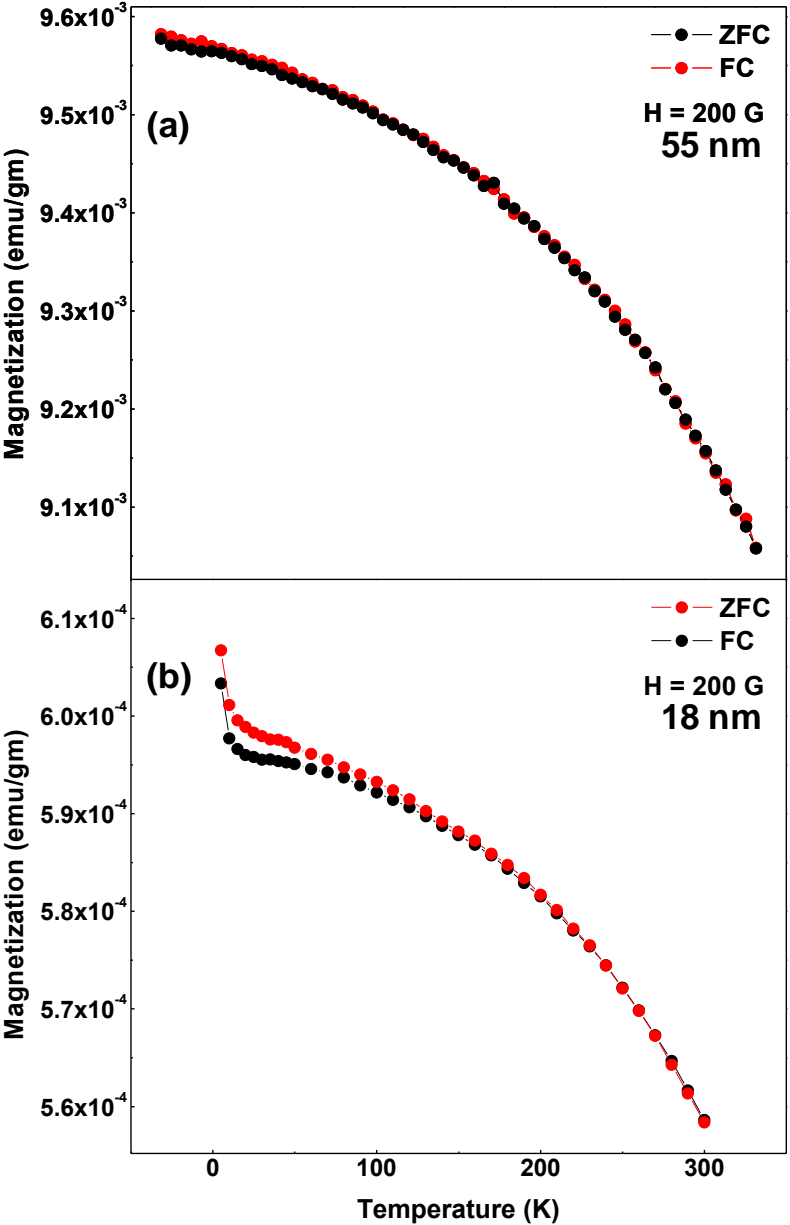

**Figure.2.27:** Zero Field Cooled (ZFC) and Field Cooled (FC) Magnetization-Temperature measurements on the nanowires of diameter(a) 55 nm and (b) 18 nm.

While for the 55nm wire the data clearly show that for a field H (=200 Gauss) the FC and ZFC curve do not deviate (since $H_c$~850 Gauss) as expected and M follows the temperature dependence of a bulk ferromagnet (see below). However, for the 18 nm wire there is a small deviation between the ZFC and FC curves below 25K. There is also small fall in M at T< 10 K as T is raised. The fall at low T can be due to thermal assisted magnetization reversal at local regions around surfaces where the barrier to such reversal may be low. On the diameter is less, the surface area increases and such a phenomena can occur.

To understand the power law behavior of the magnetization M(T), we plot the M(T) as a function of $T^{3/2}$ to see whether the magnetization follows similar behavior as bulk with temperature. Fig. 2.28 shows that the magnetization in case of 55 nm nanowire follows a straight line when plotted as a function of $T^{3/2}$ verifying the validity of Bloch's $T^{3/2}$ law in case of 55 nm diameter nanowires. This establishes the ferromagnetic character of these nanowires.

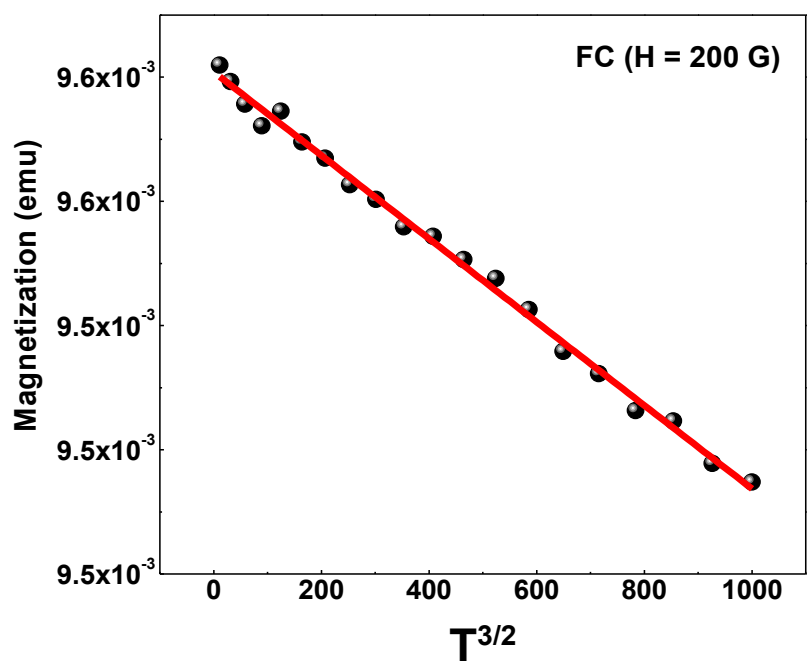

**Figure.2.28:** Fit of Bloch's equation to the M-T data for 55 nm Ni diameter nanowire array.

We note that the magnetic measurements in this thesis have been used to mainly characterize the magnetic state of the samples and show that the sample's are magnetic. This is to make sure that we are measuring the electrical transport properties of ferromagnetic nanowires. The quantitative analysis of the magnetization data and other important measurements are not part of the thesis.

# CHAPTER 3

## Experimental Techniques

*In this chapter we discuss the various experimental techniques which have been setup and used in measurement of resistance of the samples. For studying the electrical transport, two setups: a high temperature (300 K – 700 K) and a low temperature (3 K - 300 K) resistance measurement setup have been developed. The measurement setups were automated by GPIB interfacing using LabVIEW 8. In the following section we describe the setups, followed by the most crucial part of sample mounting. We also discuss here some experimental techniques like photolithography and Focused Ion Beam lithography which are used in this thesis to make contacts to single nanowires for electrical measurement purpose.*

This page is intentionally left blank

## 3.1 Instrumentation for Resistance measurements

Mainly we used A.C resistance measurement technique instead of D.C to avoid measurement using high current and the problem of thermo e.m.f arising out of temperature gradient across the sample. The A.C. technique one can use very low signal to excite the sample and the problem of thermo emf [1] generation due to Seebeck effect can be avoided. Also the problems of electromigration caused by high current density can be avoided in this method. In case of A.C resistance measurement technique, a low frequency A.C current is passed through a sample and the voltage across the sample is measured by a phase sensitive detection [2] technique to detect the signal corresponding to the excitation frequency. The signal so measured has a very high signal to noise ratio making the measurement reliable at very low current amplitudes. Low current amplitude also ensures safety of the sample against electro-migration. A schematic view of the measurement is shown Fig. 3.1.

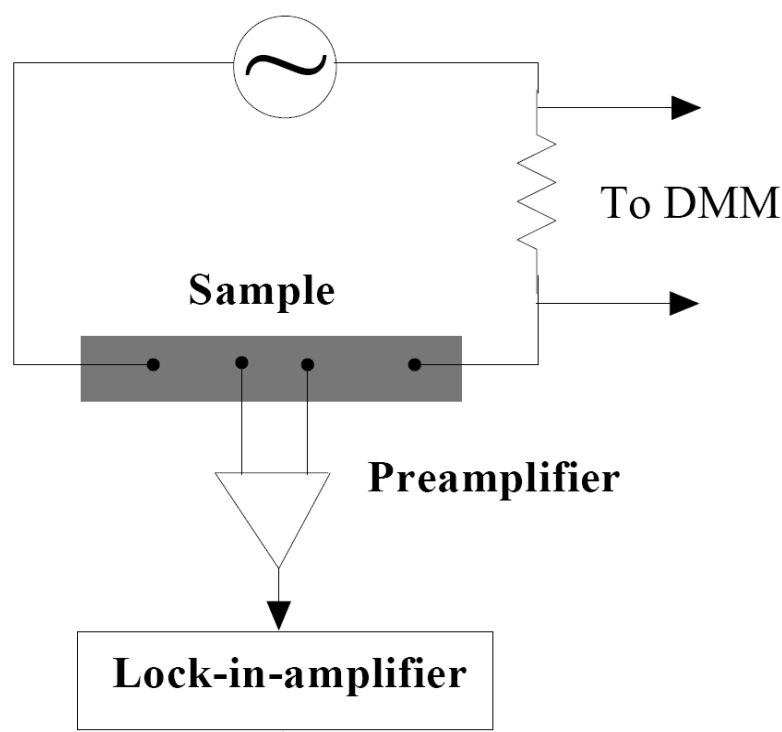

**Figure 3.1:** Basic principle of A.C. resistance measurement.

For metals, the resistivity and thus the resistance are very low as in our case. A.C resistance measurement technique plays a vital role in measurement of resistance of metallic samples with resistance below a few ohms. In the actual experiment, as seen from the Fig. 3.1, the A.C. signal with the series constant resistance 'R' constitutes the A.C current source used to excite the sample. The resistance of the sample is measured by the four probe method [1] as shown in Fig. 3.1 and is given by the ratio of the voltage amplitude drop across the voltage terminal to the current amplitude of the A.C passing through the circuit. To avoid any error due to drift at any moment, the voltage drop across R is measured simultaneously by a digital multimeter (DMM) and the exact value of current amplitude is evaluated. A more detailed scheme of temperature dependent resistance (R-T) measurement is shown in Fig. 3.2.

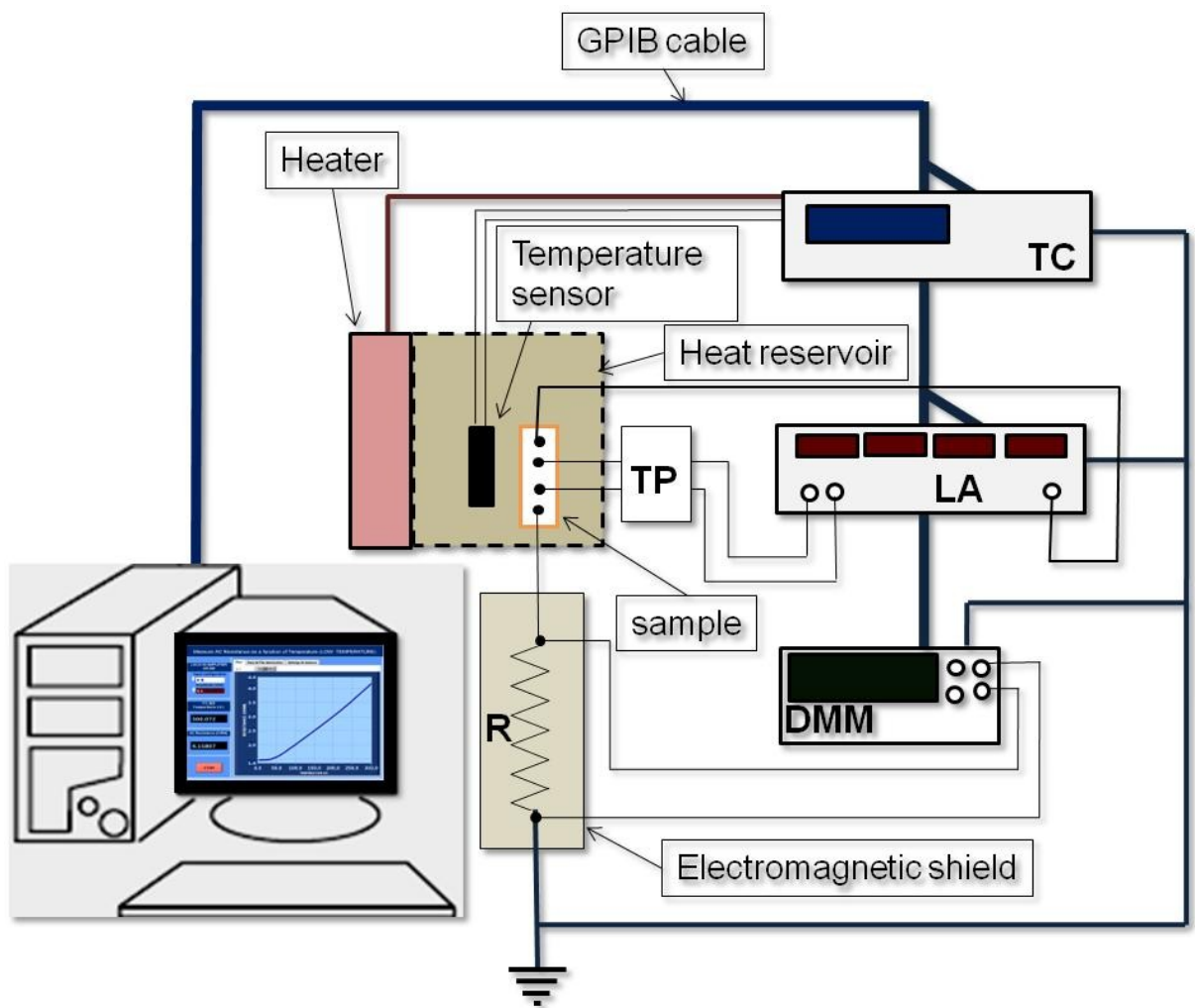

**Figure 3.2:** Detailed A.C. resistance measurement technique used for the measurement of sample resistance ranging between 1mΩ-10 Ω. TC=Temperature Controller, LA=Lock-in Amplifier and TP= Transformer Preamplifier.

In case of temperature dependent measurement, the temperature of the sample is also measured and controlled simultaneoulsy by a temperature controller (TC). In our measurement measurement we used a Lakeshore 340[2] temperature controller for high temperature measurements and a ITC 503[3] temperature controller for low temperature measurements. The heat reserviour ensures a constant temperature equilibrium at any given instant for both the sample and the sensor. The heater in contact with the reserviour is used to rise the temperature of the sample and is controlled by the temperature controller to acheive a particular temperature or a constant ramp of the sample temperature. The four probe measurement eliminates the error due to the contact resistace. The signal source of the Lock-in Amplifier LA (SR 830) used to excite the sample. The voltage drop across the sample's voltage terminals is amplified by a Trasfomer preamplifier (TP) before being fed to the Lockin amplifier (LA) (SR 830 from standford instruments [4,5]) for phase sensitive detection. The trasformer preamplifier (TP) is a simple step up trasformer with a constant gain in the range 10 Hz-1000 Hz (SR 554 from standford instruments [5,6]). To avoid any drift in the standard metal film resistor (1 k$\Omega$-10 k$\Omega$), it was enclosed inside a electromagnetic shielded box. The A.C signal across R was measured by a digital mutimeter DMM (Keithley 2000) [7]. With the resolution of measurement of the Lock-in Amplifier (SR 830) and the Digital Multimeter being 1 ppm, the acheived resolution is nearly ~ 2 ppm. The temperature variation was acheived by the heater in case of high temperature setup. For low temperature measurements, the sample stage of a closed cycle refrigerator acted as the reservior. Below we describe both the measurement setups in detalis. Finally, the system is computer interfaced for automation of the experiment.

## 3.1.1 High Temperature Electrical Transport setup

The high temperature measurement setup is used for basic electrical measurements such as electrical resistance (Using AC or DC), capacitance, I-V characteristics etc. It consists of three main parts comprising of hardware and software namely the high temperature chamber for producing high temperature, external instrumentation for measurement, computer program for automation.

**3.1.1.1 The high temperature Sample chamber:** The sample chamber consisted of a copper block onto which the sample and PT100 sensor were mounted, each sealed within thin sheets of mica screwed to the copper block for proper mechanical and thermal contact. A heater wire (made of Nichrome) of 25 ohms was kept inside the copper block. This arrangement was inserted in a thick copper cylinder surrounded by a thick layer of insulation for thermal equilibrium. The whole chamber was kept in a vacuum of $10^{-6}$ mbar and the measurements were done at such high vacuum. The temperature was controlled by a Lakeshore 340 temperature controller with millikelvin resolution. Much attention has been paid to the fact that both the sample and the sensor for temperature measurement are under the same caloriemetric conditions.

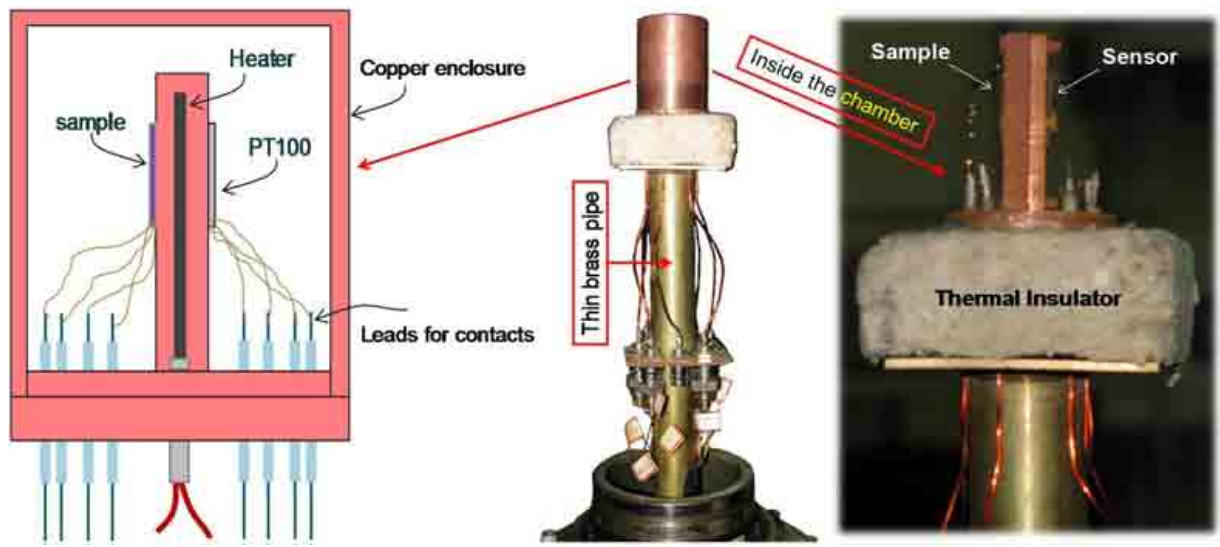

**Figure 3.3:** The high temperature resistance measurement setup along with the schematic and image of the sample chamber (being shown on either side of the setup).

The design has been standardized using a different PT100 in the sample's place and also exchanging the positions of the PT100 and the sample. The measurement setup is shown in Fig. 3.3 with sufficient details. The leads for contacts and the contacts to the heater coil are well insulated from the surrounding metal base and are introduced in the chamber by enclosing them in ceramic tubes.

**3.1.1.2 Instrumentation:** The measurement part instrumentation consisted of measurement of a low frequency AC signal using a phase sensitive detection technique with a Lock-in amplifier (SR 830, Stanford Instruments). The current through the sample is measured by measuring the voltage drop across a standard resistance using a Digital Multimeter (Keithley DMM 2000). A standard calibrated PT 100 sensor was used as the sensor for temperature measurement with a temperature controller with millikelvin resolution (Lakeshore 340). Fig 3.2 shows the schematic view of the experiment. The complete experimental setup for measuring the A.C resistance is shown later in Fig. 3.8. In case of High temperature measurement, the Lakeshore 340 temperature controller (shown in the setup) is used for measuring temperature of the sample with the rest of the instruments remaining the same for the low temperature resistance measurements.

**3.1.1.3 Automation of the experiment:** The computer program for automation has been done by GPIB interface using mainly programs like C++ (initially) and later using LabVIEW 8 [8]. A screen shot taken during measurement of the front panel is shown in Fig. 3.4. The screenshot of the program was taken during the measurement of resistance of a 200 nm diameter nickel nanowire array by pseudo four probe method (discussed later in section 3.2). The corresponding block diagram (program) is shown Fig. 3.5. KPCI-488 [9] was used for the interfacing.

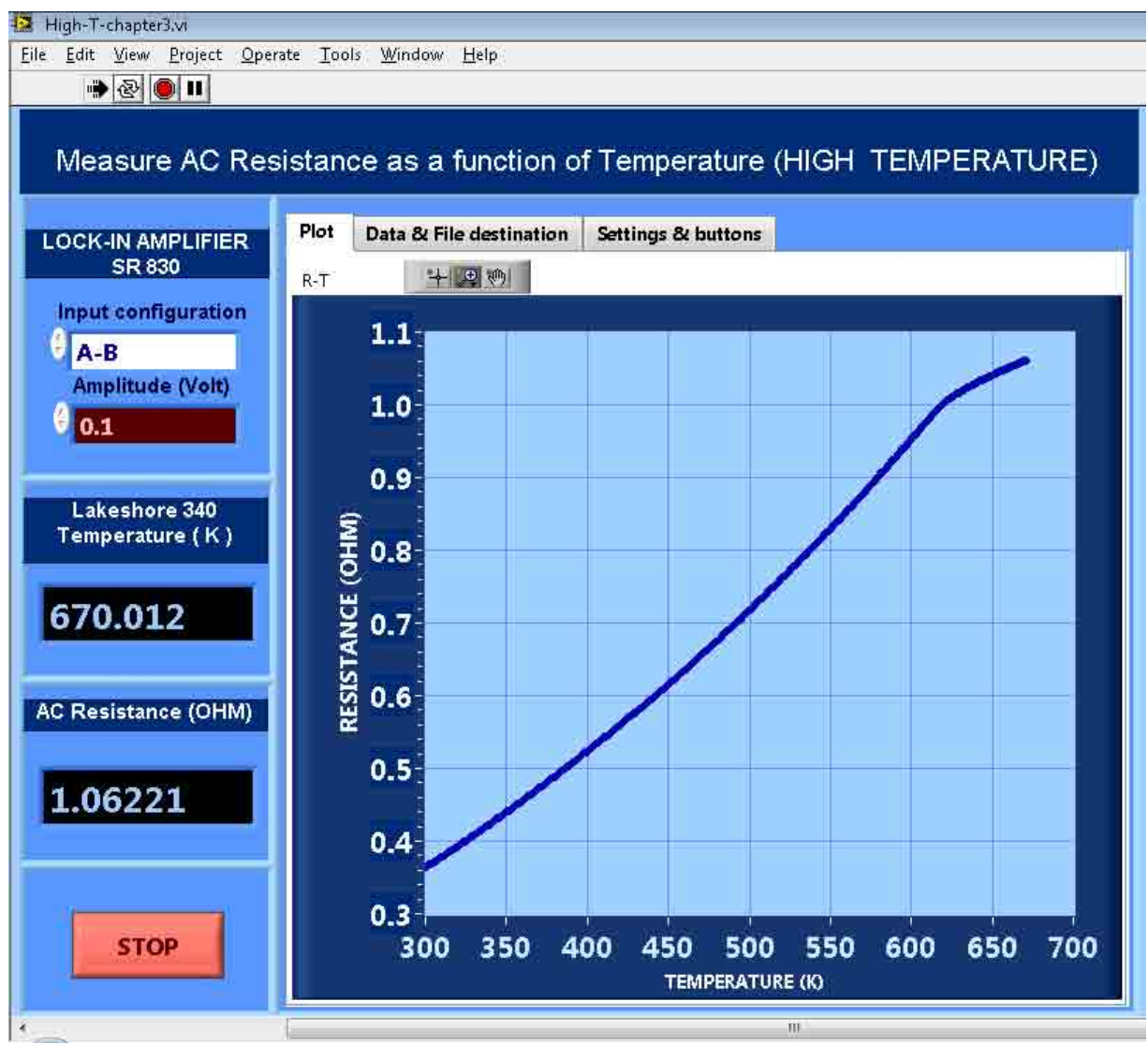

**Figure 3.4:** The screen shot of the resistance measurement program in LabVIEW 8 during the measurement of the resistance of a 200 nm diameter nickel nanowire in the temperature range 300 K to 675 K.

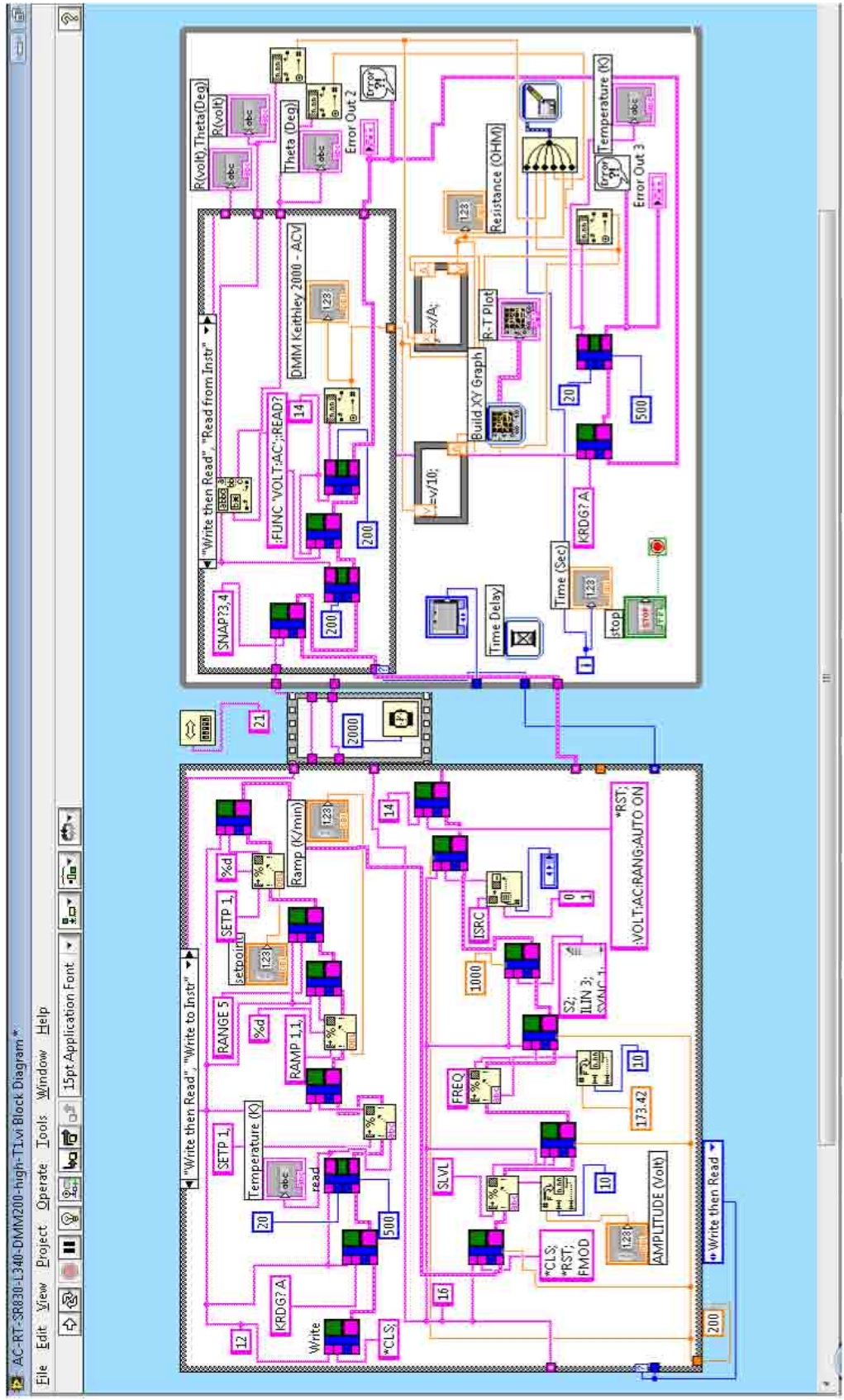

**Figure 3.5:** The screen shot of the block diagram of a high temperature resistance measurement program.

### 3.1.2 Low Temperature Electrical Transport setup

The low temperature measurement setup is used for basic electrical measurements such as electrical resistance (Using AC or DC), capacitance, I-V characteristics etc. It consists of three main parts comprising of hardware and software namely the crycooler for producing low temperature, external instrumentation for measurement, computer program for automation.

**3.1.2.1 Pulsed tube cryocooler:** In general any cryocooler consists of three parts namely the compressor, the displacer (a free floating piston) and a regenerator (a tube with heat exchangers at both ends for free expansion of the working fluid). Such cryocoolers connected in series give rise to multistage cooler giving rise to cryogenic temperatures and enhancing the efficiency of the system. The basic description of such cryocoolers can be found in standard text books [10]. In our experiments we have used the pulse tube refrigerator for low temperature measurements. Below we give a brief description of the said system. The basic concept of pulse tube cooler was first described by Gifford and Longsworth in 1963[11]. The pulse tube cryocooler differs from other cryocoolers in the fact that it does not any moving part making it vibration free and does not need maintenance. This enhances the life time of the pulse tube systems. Fig.3.6 shows a scheme of a single stage pulse tube cryocooler.

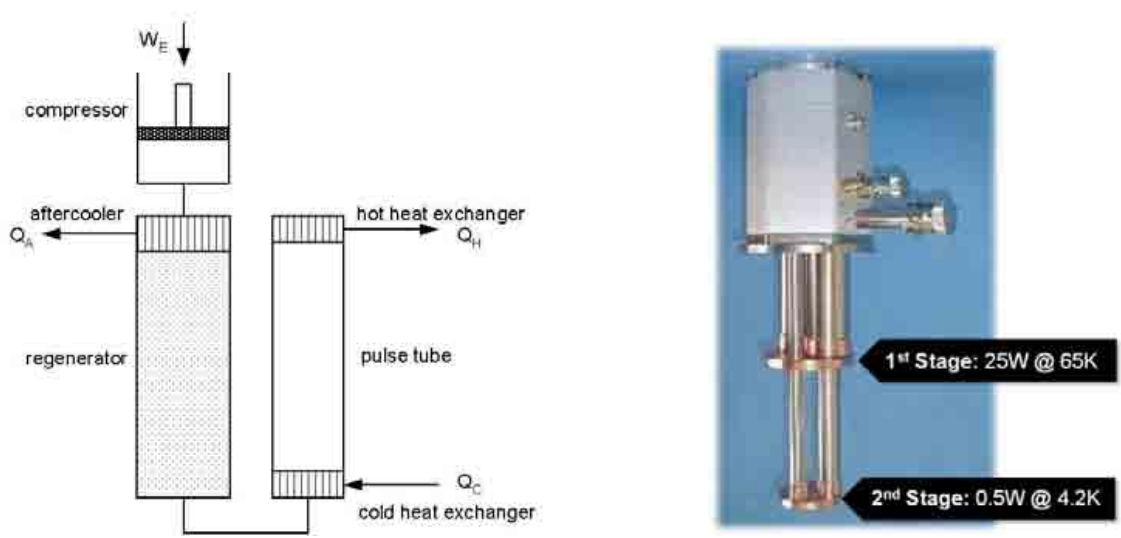

**Figure 3.6:** The scheme of a single step pulse tube (left image). The pulse tube cryocooler from Cryomech model PT405 (right image).

The main moving part as said earlier in a cryocooler is the displacer which is a free floating piston. In a pulse tube cryocooler, the displacer is replaced by a pulse tube as shown in the Fig. 3.6 scheme. In conventional cryocoolers, the gas compression is achieved by the displacer. The basic cooling effect relies on a periodic pressure variation and a displacement of the working gas in the pulse tube. The pressure variation is produced by the compressor. In response to the pressure changes, mass flows through the regenerator and exchanges heat with it. The regenerator serves as a heat reservoir. The gas within the pulse tube acts as a compressible displacer. The gas leaving the cold end of the regenerator does work on the gas inside the pulse tube, producing cooling. The heat of compression is taken out at the hot-end heat exchanger. In our experiments we used a two stage pulse tube crycooler [12] as shown in Fig. 3.6 (b) with lowest attainable temperature of ~ 2.7 K. The sample is mounted at the cold head of the second stage.

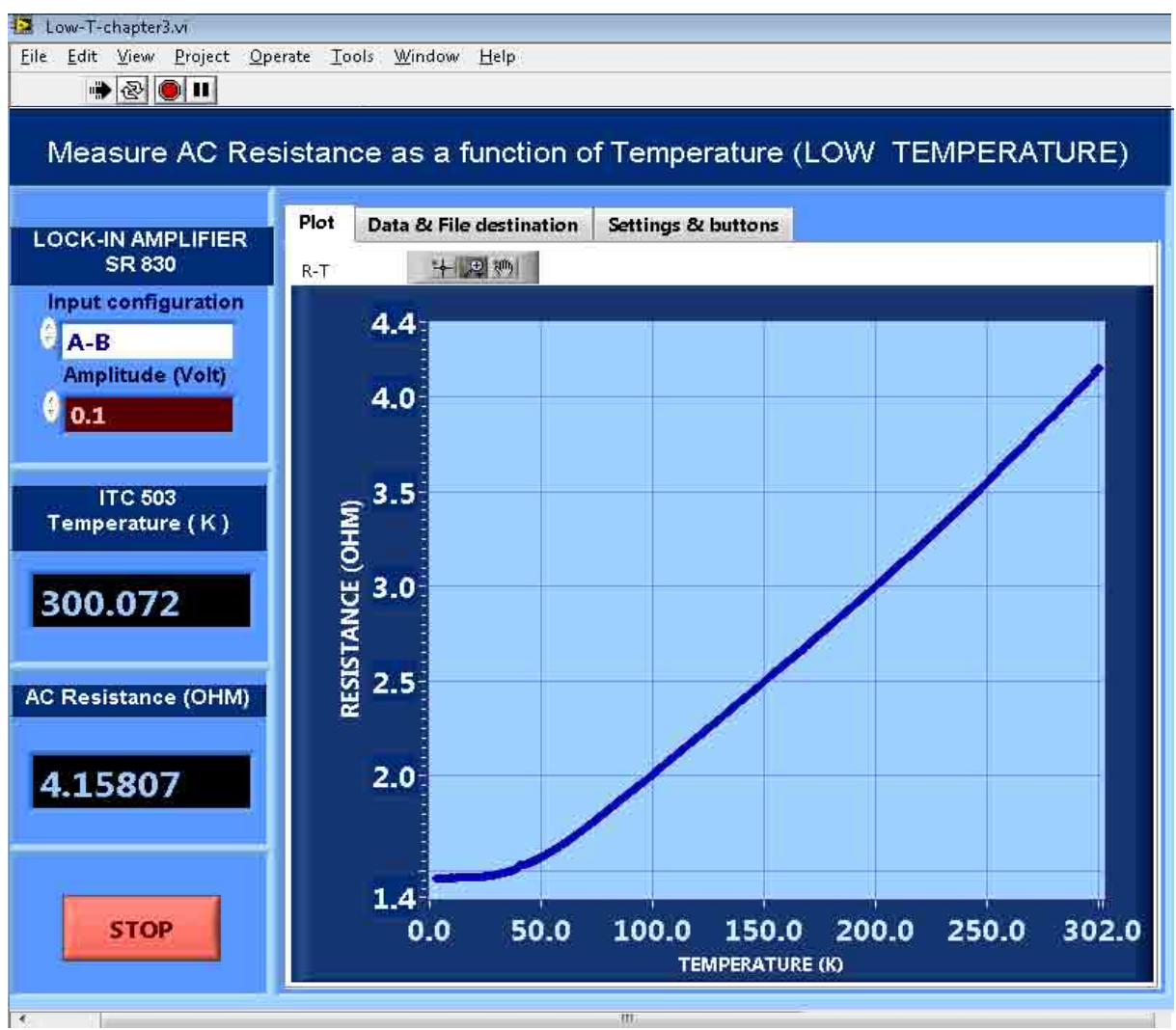

**Figure 3.7:** A screen shot of the resistance measurement program in LabVIEW 8 for low temperature during the measurement of the resistance of a 55 nm diameter nickel nanowire.

### 3.1.2.2 Instrumentation & Automation

The instrumentation is the same in this case also except the temperature controller is ITC 503. As in case of high temperature setup, the programming and automation of the experiment was done using LabVIEW 8. The complete setup for automated measurement for resistance measurement is shown in Fig. 3.8

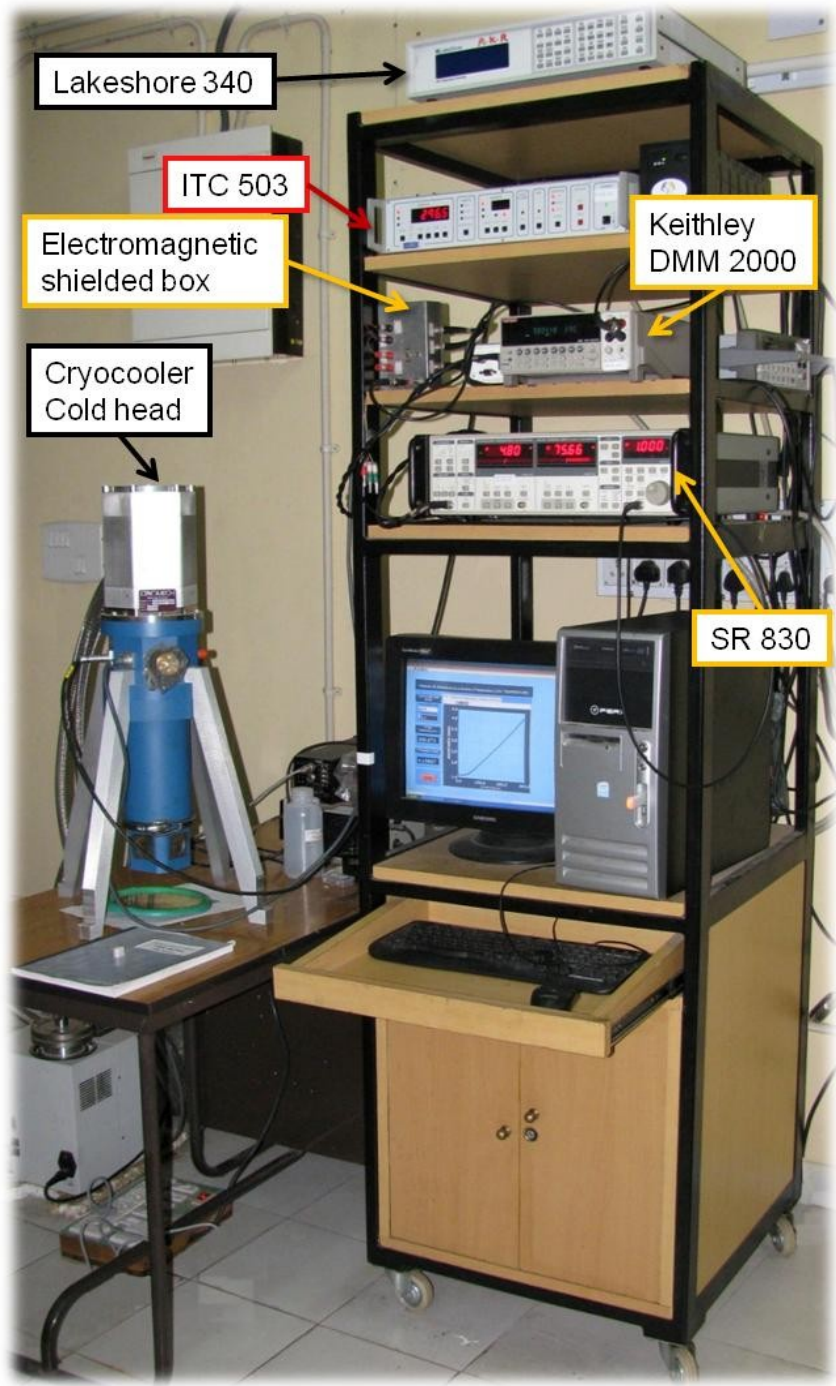

**Figure 3.8:** Setup for measuring the A.C resistance.

## 3.2 Sample mounting

The sample mounting is the most important part of measurement as any mistake in mounting can lead to spurious resistance data. The measurements done on our samples are mainly four probe and pseudo four probe [13] in nature. The four probe and pseudo four probe configurations are shown in Fig. 3.9

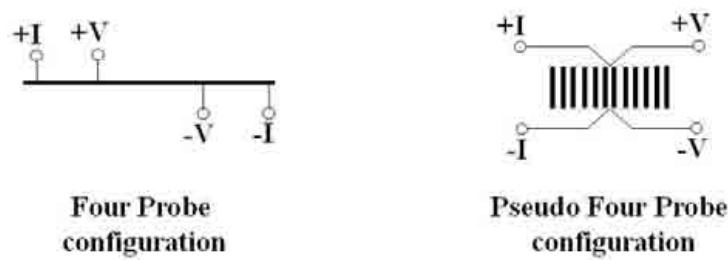

**Figure 3.9:** The four probe and pseudo four probe configurations used in measurements.

The four probe method is used for measurement of resistance of single wire. For measurement on arrays of nanowires/nanotubes grown in the template, the pseudo four probe configuration is the only way to measure the resistance with certain degree of accuracy [13]. Two thin copper wires are attached (as current and voltage probe) to each side of a template containing the sample with silver epoxy. The resulting template containing the nanowire/nanotube arrays is enclosed inside thin sheets of high purity mica to avoid any kind of electrical contact with the heat reservoir. The resulting sample is attached to the cold head of a closed cycle refrigerator (for low temperature measurements) using low temperature silicon grease. For high temperature measurements the sample prepared in similar way is mechanically screwed to the sample stage shown in Fig. 3.3. To avoid any error in temperature measurement, the sensor is also mounted in the similar way.

## 3.2.1 Sample mounting for single nanowire measurements

The sample mounting process for electrical measurements for a single nanowire is a complicated process consisting of various steps. In our experiment we used photolithography [14-16] and Focused Ion Beam (FIB) [17-20] based platinum deposition for making contacts to single nanowires. A flow chart is shown in Fig. 3.10 describing the sequential steps involved in the whole process. A template containing nanowires is etched using 6M NaOH solution followed by washing several times with Millipore water to remove the debris. The resulting solution is sonicated well and diluted and a drop is of this solution is sprayed at the centre of lithographically made gold contact pads as shown in Fig. 3.10

**Flow chart:** Making contacts to a single nanowire

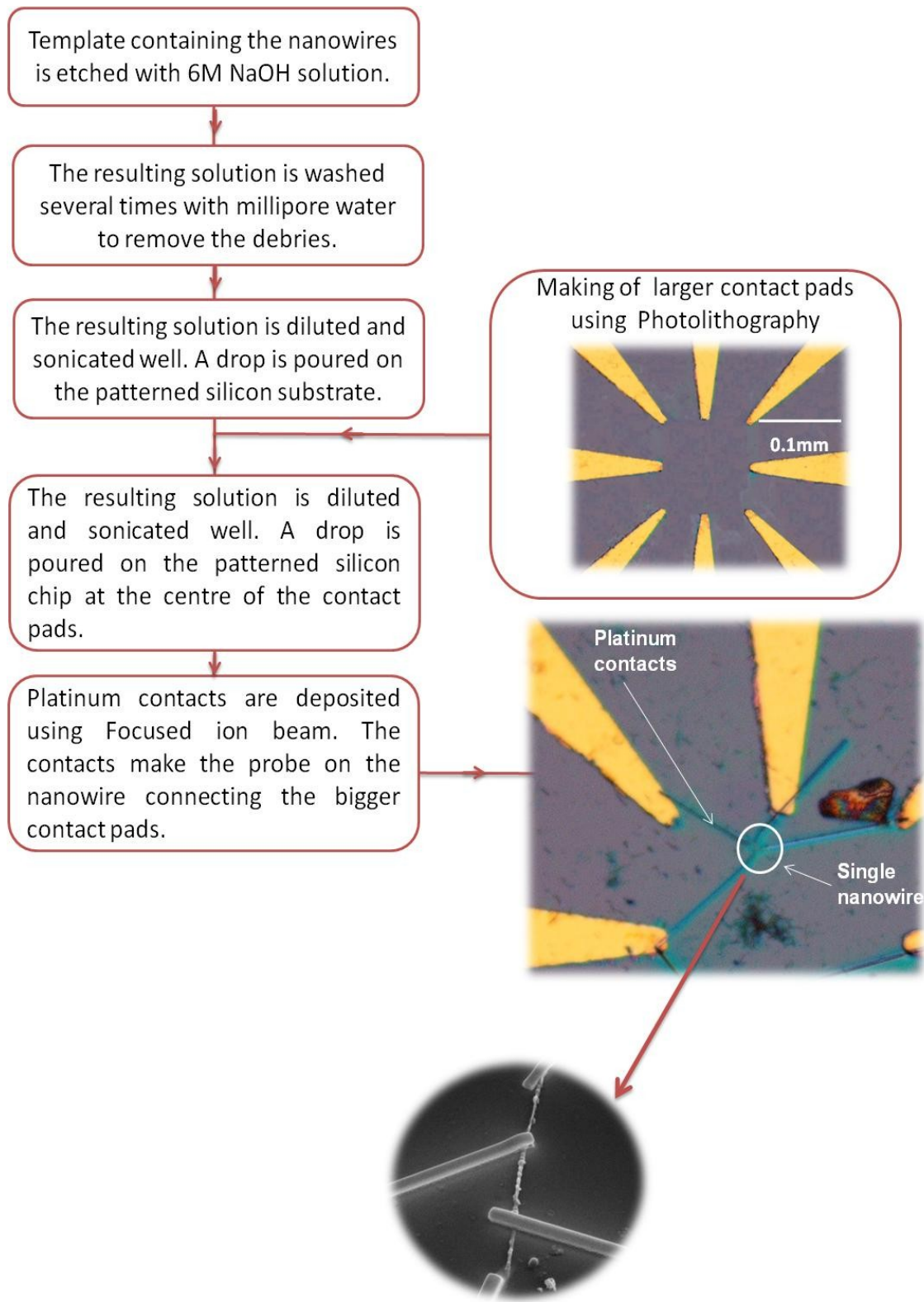

**Figure 3.10:** Making sample for single nanowire electrical measurements; the four probes made on a 55 nm nickel nanowire.

In the whole process, finding the nanowires is the most crucial step. For finding the nanowire it is important to image it using SEM mode of the dual beam FIB system. One should not use ion-beam for imaging because the gallium can induce impurities in the sample and damage it severely. However, a nanowire imaged in the SEM mode, not necessarily stay at focussed when the stage is tilted by 52° to the ion beam mode. The best recipe is to choose a nanowire/nanostructure on the substrate closer to the nanowire on which the metal contacts are to be made. This nanostructure is then used as a reference and imaged to best possible magnification with best focus in both the e-beam mode and the i-beam mode. After standardizing by lot of trials, the deposition of contact on the nanowire of interest is made. The sample is then attached to the cold head of a pulsed tube using a double sided copper adhesive tape for temperature dependent measurement. In the following sections we describe Photolithography and focused ion beam (FIB) techniques which have been used in this thesis to make contacts on a single nanowire.

## 3.2.2 Photolithography

Photolithography has been widely used in the last five decades in the semiconductor industry for microfabrication [14-16]. We have used photolithography to fabricate gold contact pads for making electrical transport measurements in single nanowire. Making contacts to nanowires for electrical measurements as described in the earlier section has two steps. Neither photolithography, nor FIB alone can be used for this purpose because both have limitations and effective in different length scales with photolithography being efficient in the range of several millimeters to a micron) and FIB being efficient in the range of several microns to a few nanometers. The various steps involved in the process are depicted in Fig. 3.11. The substrate at first is cleaned by RCA cleaning [21]. The silicon substrate is then coated with a considerably thick layer (few microns, 7 microns in our case) of photoresist (positive resist SPR220-7 in our case) by spin coating (rpm ~ few thousand rpm, 3000 rpm for 45 seconds in our case). The kind of resists determines the various parameters like the thickness and rate of spin coating and layers of coating required. The silicon substrate with the layer of resist is baked for some time (1-2 minutes at a temperature of 115°C for 90 sec in our case). In the next step the resist is exposed to UV radiation via a mask which contains the pattern to be transferred. The whole process is carried in an instrument called mask aligner. The masks are usually made of chrome glass. For an inexpensive alternative, a Mylar transparency print from a high-resolution photo-quality copier (also called "film plotting") can be used as a contact mask in place of a standard photolithography mask.

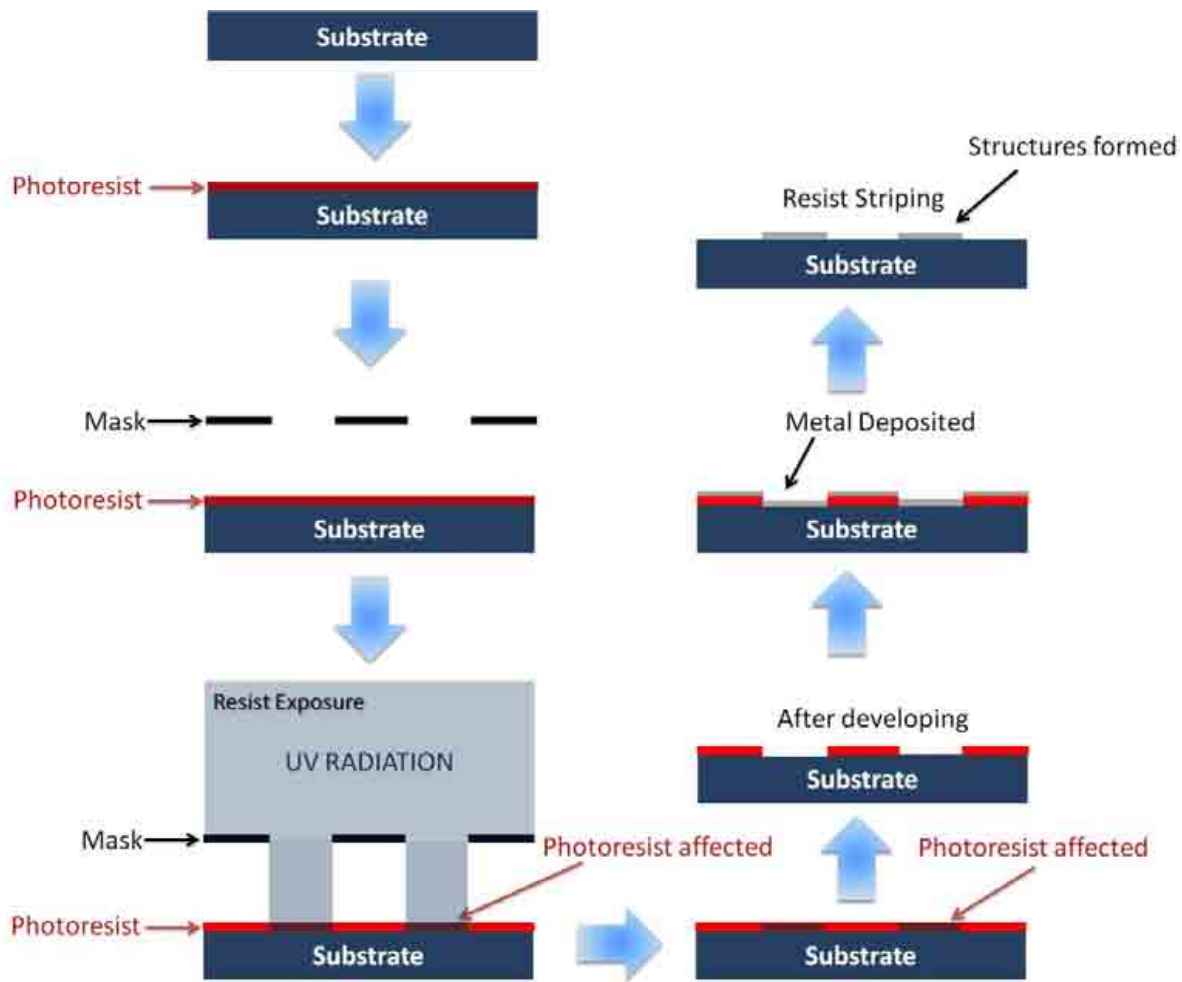

**Figure 3.11:** The sequential steps involved in photolithography. The scheme depicted here is for positive resist.

Transparency films don't wear well and have limited feature size resolution (depending on the print quality), but they are cost-effective for limited patterning of larger structures. A caveat: conventional laser printers do not print densely enough for to sufficiently block UV illumination and only film plotters with print density of at least 3800 dpi work well for this purpose. We used such mask for our work. The exposure is carried out in the scheme shown in Fig. 3.11 with the exposure times depending upon the kind of resist used. The exposure time has to be optimum as any less time of exposure results in the underdevelopment of the resist and more exposure time results in over development of the resist both of which affect the pattern features and the resolution. In case of positive resist where the exposed regions of the resist get monomerised and on subsequent developing are removed, the exposure time is of few seconds (12 seconds in our case). However, in the case of negative resists where the exposed regions only remain after development, the exposure time extends up to few minutes. In our experiment we used a mask aligner Canon PLA – 501FA with a broadband wavelength 300nm – 500nm.

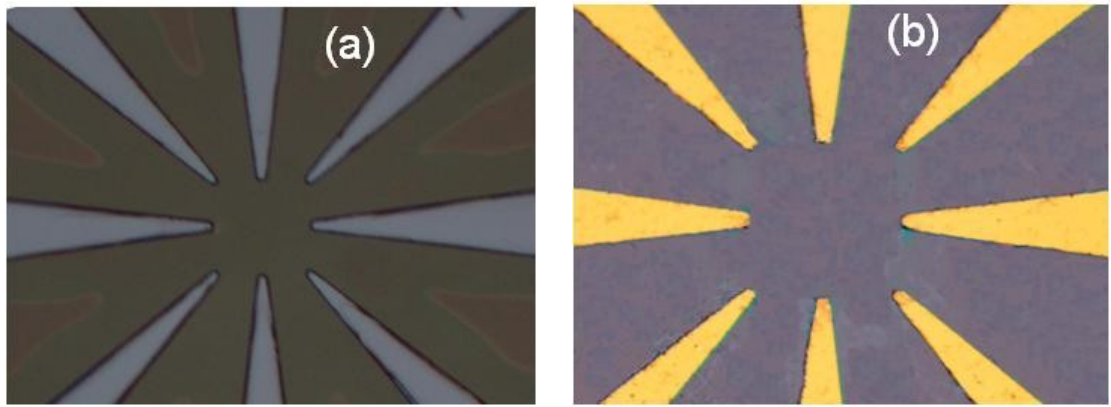

**Figure 3.12:** Patterns transferred by lithography (a) after developing stage (b) after liftoff.

After exposure and developing with developer solution, gold is evaporation on to the patterned surface. This is followed by stripping the resist with acetone. The gold that is directly on the silicon substrate (where from the resist is removed after developing) remains stuck to the surface with deposits in other places coming out when washed with acetone. This process is called lift off. Fig 3.12(a) shows a pattern transferred after developing stage. Fig. 3.12(b) shows gold deposited in the patterns. These gold deposited pads are used as used the secondary contacts for making measurements on single nanowire.

### 3.2.3 Focused Ion Beam Lithography

Focused ion beam (FIB) uses a liquid metal ion source (LMIS) [18] of gallium and the extracted ions are accelerated, collimated and focused by a series of apertures and electrostatic lenses on a surface to achieve various kinds of structures at much lower scale than conventional photolithography. It is extensively used in semiconducting industry for failure analysis. FIB has three important roles to play (a) Material removal (Sputtering material/polishing/ milling at micro-nanoscale). (b) Material deposition (C, Au, W, Pt etc.) (c) Microscopy. The basic FIB as shown in Fig. 3.13(a) consists of a the ion beam column, a detector and several gas injecting needles to perform functions like enhanced gas etching, deposition of various materials. The ion beam column consisting of various electrostatic lenses is depicted in Fig. 3.13(b). The beam spot size can be varied from 8 nm to 500 nm, which makes it suitable for nanofabrication.

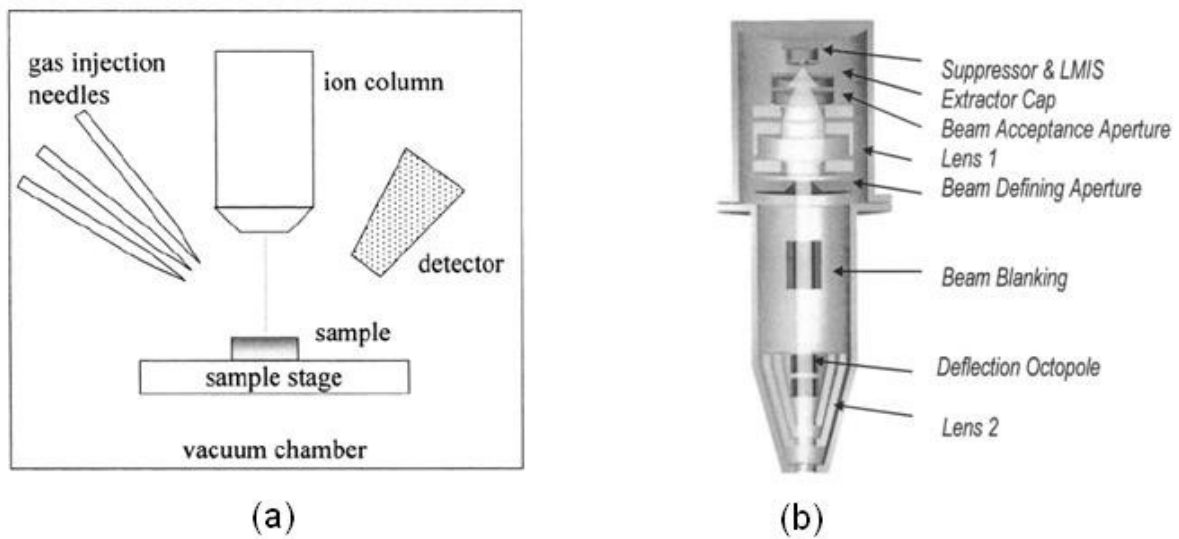

**Figure 3.13:** The focused ion beam (a) The basic instrument scheme (b) The ion column showing various parts.

The Fig. 3.14 shows the interaction mechanism of gallium ions with matter. The Focused Ion Beam (FIB) tool can cut away (mill) material from a defined area with dimensions typically in square microns or deposit material onto it. Milling is achieved by accelerating concentrated gallium ions to a specific site, which etches off any exposed material, leaving a very clean hole or surface.

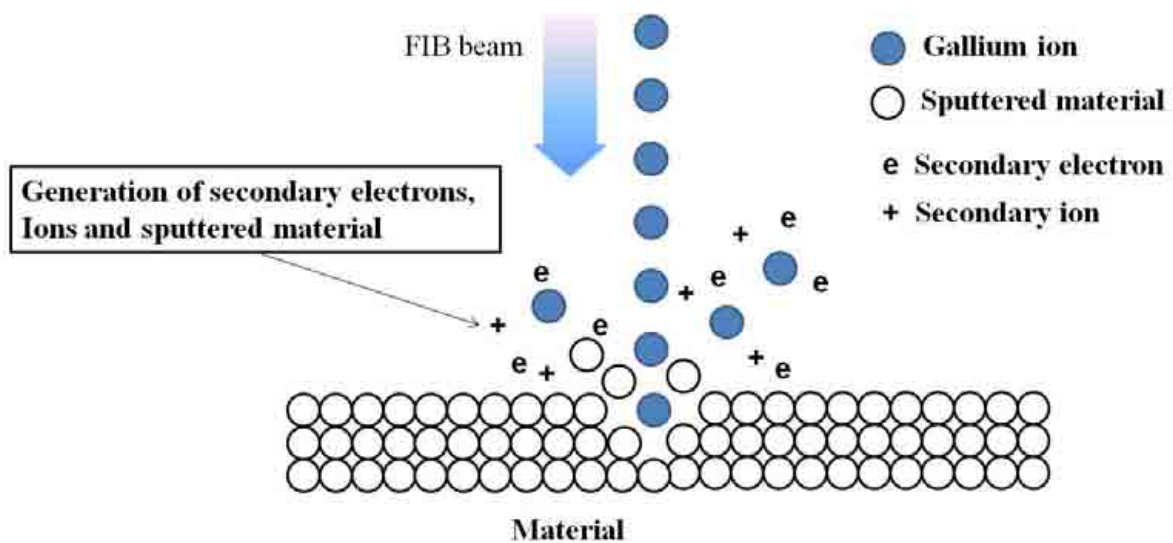

**Figure 3.14:** Ga+ ions upon striking the surface of the material generate electrons, ions and sputtered material.

The FIB is used for such tasks as site-specific cross-sectioning for interfacial microstructure studies, preferential removal of certain metals or oxides, semiconductor device editing or modifications, site-specific TEM sample preparation, and grain imaging. By introducing gases or an organic gas compound, the FIB can selectively etch one material much faster than surrounding materials, or deposit a metal or oxide. In case of metal deposition the precursor gas molecules contain the metal atoms to be deposited are injected through the gas injecting needles shown in Fig. 3.13(a).

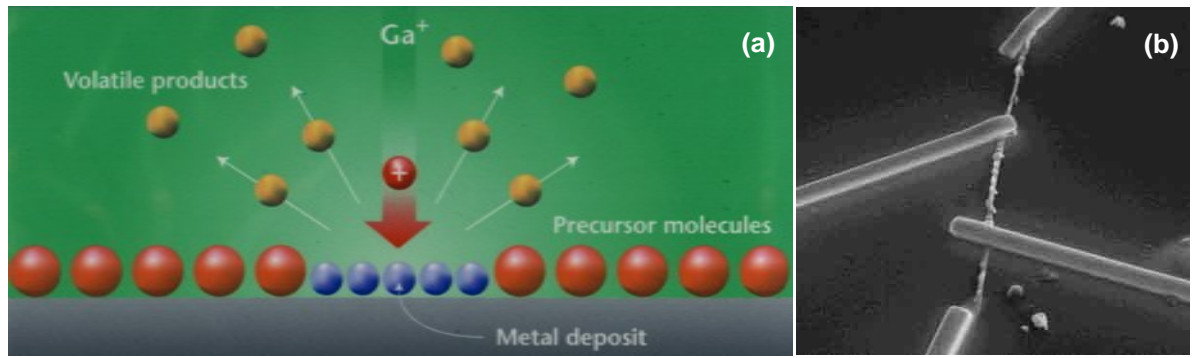

**Figure 3.15:** (a) Metal deposition scheme using ion beam. (b) Platinum contacts being made to a single nanowire in a dual beam FIB.

The deposition mechanism is well depicted in Fig. 3.15(a), where the accelerated highly energetic gallium ions decompose the precursor gas molecules to metal deposits and volatile products leaving the metal deposits on the surface of the substrate. Fig. 3.15(b) shows four platinum contacts being made on a 55 nm diameter nickel nanowire using FIB.

In our work we have used FIB (**FEI Strata DB235** FIB/SEM dual beam system) mainly for making contacts to single nanowires for measuring their electrical transport properties. A more complete description of the various aspects of focused ion beam and its applications can be found in standard text books dedicated to the subject [18, 19]. For platinum deposition we have used the precursor gas methylcyclopentadienyl platinum trimethyl $(CH_3)_3(CH_3C_5H_4)$ Pt. The details of the contacts to the single nanowire are described in details in chapter 6, which is dedicated to the electrical measurements on a single ferromagnetic nanowire.

# CHAPTER 4

# Electrical resistance of Ni nanowires near the Curie Temperatures

*In this chapter, we describe the electrical transport measurements on nickel nanowires of diameters down to 20 nm in the region close to its paramagnetic-ferromagnetic transition temperature $T_C$ (reduced temperature $|t| \leq 10^{-2}$). The size reduction leads to reduction in $T_C$ following the finite size scaling laws although with a different value of exponent. The data analysis done in the frame work of critical behavior of resistance near $T_C$ shows that the critical behavior persists even down to the lowest diameter wire of 20 nm. However, there is a suppression of the amplitude of the critical behavior of the resistivity as measured by the critical exponents and the parameters quantifying the resistivity anomaly. The spin system shows an approach to a quasi one-dimensional spin system.*

This page is intentionally left blank

## 4.1 Introduction

Resistivity ($\rho$) of a metallic nanowire is a topic of considerable current interest. In addition to effects such as Quantum transport, even in the regime of classical transport there are a number of effects that have received scant attention. One such phenomenon is the issue of electrical resistance near the Curie temperature ($T_C$) for paramagnetic-ferromagnetic transition. In this range, the issues associated with Critical point phenomena become important [1]. While in bulk ferromagnets and to some extent in thin films of ferromagnetic materials like Ni, Co and Fe, the anomaly in resistivity near the transition temperature $T_C$ has been well studied and analyzed in terms of critical phenomena[2, 3], there exists no such study in ferromagnetic nanowires whose diameters can become smaller than the spin correlation length. One important issue that becomes crucial when the size is reduced is finite size scaling, which can describe the shift of the ferromagnetic $T_C$ as a function of size [4]. In the context of ferromagnetic nanowires, this particular point (shift of $T_C$ with size) has been looked into in only one study through determination of $T_C$ from magnetization measurements [5]. Though ferromagnetic nanowires of reasonable quality have been grown over the years, issues like critical exponents have not been investigated. While the resistivity anomaly has been investigated in bulk ferromagnets near the critical region, there are no reports of investigation of critical exponents of resistivity in ferromagnetic nanowires. The finite size scaling in ferromagnetic systems (2D films) have been investigated mainly by magnetization measurements [6-8]. However, to our knowledge high precision resistance measurements have not been used to study critical phenomena even in films. This part of thesis is thus the first attempt to investigate an unexplored yet seemingly interesting region.

In this chapter we report an extensive investigation of the anomaly associated with the resistivity near the $T_C$, through an electrical transport measurement (Resistance ($R$) vs. Temperature($T$)) in the temperature range 300 K < T < 675 K in Ni nanowires of diameter ranging from 200 nm to 20 nm. We note that, to the best of our knowledge, such resistance measurements up to this temperature range have not been done in metal nanowires before. This measurement, made close to the critical temperature $T_C$ ($|t| \leq 10^{-2}$, where, $t = \dfrac{T - T_C}{T_C}$), allows us to investigate whether the critical anomaly in $\rho$ persists even in nanowires with diameters less than the correlation length. This investigation is of significance in view of the recent observation in Zn nanowires with sizes much less than the superconducting coherence length, the specific heat anomaly near the transition temperature is essentially unaltered and there is not much "rounding-off" of the critical region [9]. Investigation of the critical region in systems with reduced size is not an easy task because the data analysis invariably involves effects due to rounding-off and related issues. In this thesis we did make an attempt to quantitatively separate out the region marked by rounding- off errors.

### 4.1.1 Resistance Anomaly and evaluation of critical exponent α

The anomaly in the electrical resistivity near the ferromagnetic transition has been probed extensively over years in the bulk. It has been established that at least in the bulk material the information on the critical part of the specific heat can be obtained from the critical behavior of the resistivity derivative [10]. Resistivity anomaly near a continuous transition has been associated with the specific heat anomaly($C$) such that $\frac{1}{\rho}\frac{d\rho}{dT} \propto C$.

$$\frac{1}{\rho}\frac{d\rho}{dT} \propto C \propto |t|^{-\alpha}$$
$$\Rightarrow \frac{d\rho}{dT} \propto |t|^{-\alpha} \qquad (4.1)$$
$$\Rightarrow \frac{d\rho}{dT} = A\,|t|^{-\alpha}$$

Where α is the critical exponent associated with the specific heat near critical point and $t = \frac{T-T_C}{T_C}$ is the reduced temperature. Thus α can also be evaluated from the resistance data. Since a higher precision can be achieved in the resistance measurement, the resistance can give higher accuracy in the evaluation of critical exponent as justified from the earlier reports[2, 3, 11]. However, published resistance measurements on nickel [12] ($\alpha = 0.1\pm0.1(t<0); -0.3\pm0.1\,(t>0)$) and iron [13-15] ($\alpha = 0.12\pm0.01$) have given rather differing results for the critical parameters. In contrast, modern theoretical methods give very accurate predictions of the critical indices [16]. One source of error in previous analysis of experimental data is the use of the simple function Eq. (4.1). To obtain good results one must not only have good data, but also have a functional form of the singular temperature dependence that can describe the data to a corresponding degree of accuracy [17]. This implies that in general it is necessary to provide for higher-order correction, leading to function with confluent singularities, of the form

$$\frac{dR}{dT} = A\,|t|^{-\alpha}\,(1 + D\,|t|^{-z})$$ (4.2)

in order to obtain correct results even for the leading exponent and amplitude ratio. A non-linear least square method of applied directly to the expression (4.2) was first applied by Ahlers [18, 19] to specific heat measurements. The first applications to resistance data seem to have been done by Balberg *et al.* [20] and by Malmström and Geldart [21] to the high precision measurement on the antiferromagnet Dysprosium by Rao *et al*. [22]. Since numerical differentiation of the resistance data can involve errors, it is always convenient to use the resistance data directly and use the following expression of resistance [2]

$$R(t) = A\,|t|^{1-\alpha}\,(1 + D\,|t|^{-z})$$ (4.3)

It has been demonstrated [2,3] in case of bulk Ni, that an expression which takes into account the residual resistance and linear temperature dependence of resistance added to the above expression gives an accurate determination of the exponent α.

$$R(t) = R_0 + R^{'}(t) + A|t|^{(1-\alpha)}\left(1 + D|t|^{z}\right)$$ (4.4)

Accounting for a spread in the value of $T_C$ (described later on) further enhances the precision of evaluation of the critical exponent and amplitude ratios.

## 4.1.2 Resistance Anomaly in nanowires

The critical exponent α determined from both specific heat as well as resistivity anomaly were found to be similar in bulk Ni. There is no theory or experiment that has been done to prove the validity of the Fischer-Langer theory [10] in the region of nanowires where the effects of finite size should be visible. However, we note that the premises of the theory depends on short range spin-correlation which we believe may be sufficiently unaltered in these wires and thus there may be reason enough to assume the validity of the above theory. However, it is noted that our analysis of the results, which we present in this chapter in no way depend on the validity of the above relation.

In this work, we report the measurements of resistance ($R$) of nickel nanowires of various diameters (20 nm, 55 nm, 100 nm and 200 nm) in the temperature region 300 K-675 K which encompasses the region T ~ T$_C$, where the critical behavior shows up. The $R$ of a 50 µm diameter nickel wire (99.999% purity) is also measured as the reference bulk wire. The synthesis of nickel nanowires is done by electrodeposition of nickel from a 1M NiCl$_2$.7H$_2$0 solution into the pores of nanoporous Anodic Alumina Membranes (AAM) having thickness in the range 50 µm-60 µm. The diameters of pores inside the membranes have a very narrow size distribution (rms value < ± 5%) and the wires grown in them have nearly the same diameter as the pore.

## 4.2 Synthesis and characterization

The details of the method of preparation are well described in details in chapter 2. X-ray diffraction (XRD) of the nickel filled membranes revealed FCC structure of the nickel nanowire arrays with lattice constant of 3.54 Å. Energy dispersive spectroscopy further confirmed the purity of the nanowires. A Transmission electron micrograph and HRTEM of 20 nm diameter nanowires are shown in Fig. 4.1(a) and Fig. 4.1(b) respectively.

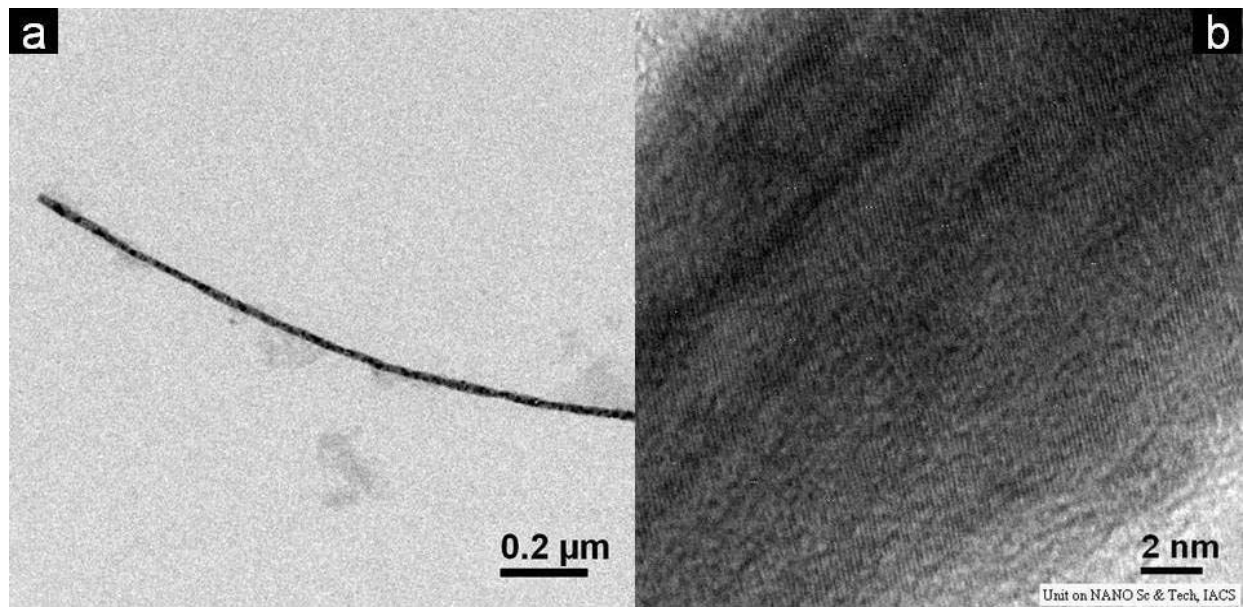

**Figure 4.1:** (a) TEM image of a 20 nm diameter nickel nanowire, (b) HRTEM image of an edge of a 20 nm diameter nickel nanowire.

To improve the homogeneity and to reduce the resistance due to grain boundary contribution, the nanowires were annealed at 670 K for 24 hrs in a vacuum of $10^{-6}$ mbar. XRD pattern of the Ni nanowires after vacuum annealing did not show any peaks of NiO ruling out any oxidation of the samples. The annealing increased the average grain size to a size approximately similar to their diameters and also made the wire stable and the resistance ($R$) decreased to a low value. (Note: The average grain size and the micro strain were estimated from the XRD line widths using the Williamson Hall plot [24] and the values are mentioned in chapter 2).

## 4.3 Resistance Measurement and Analysis

In this section we will discuss the resistance measurement and describe the basic resistance data followed by the determination of the Curie temperature from the resistance anomaly and the scaling behavior of the Curie temperatures and finally we analyze the resistance data near the critical region to extract the parameter discussed in the previous sections.

### 4.3.1 Basic resistance data

The resistance of the bulk nickel wire (50 μm) was measured by four probe method while for the nanowire arrays a pseudo-four probe technique [25] (described in Chapter 3) was used by retaining the wires inside the templates as shown in the inset of Fig. 4.2. The leads were attached using high temperature silver paste. It has been shown by us in chapter 6 that measurements made by the pseudo-four probe technique gives the same resistivity as that obtained from a proper measurements done on a single Ni nanowire of the same diameter. Such a single Ni nanowire measurements, however, cannot be extended to high temperatures because the FIB fabricated contacts are not stable up to that temperature range. For measurement, a phase-sensitive detection technique using a low frequency AC signal was used and the samples were kept in a high vacuum ($10^{-6}$ mbar) in a specially designed high temperature resistivity measurement setup (detailed description is given in Chapter-3). The temperature was measured and controlled by a temperature controller (with PT100 sensor) that allows    resolution and control to a mK. Close to $T_C$, the resistance of the samples was measured at a slow ramp of 0.1 K/min. The measurement of high temperature resistance is a very challenging task, especially in case of nanowire arrays. The problems faced frequently in these kinds of measurements on pseudo four probe measurements are the drift in resistance due to change in the number of wires in the array exactly if the contact is not robust enough to overcome any variation in the physical parameters. Electro-migration though can create change in resistance, is not a problem because we use low current density during measurements. So, it is important to train the samples to high temperatures and make the contacts robust enough to tackle the problem of drift. This we achieved by annealing the wires with the silver contacts on it to the maximum possible temperature of measurement (675 K in the present case). Such annealing helps to achieve stability in measurement.

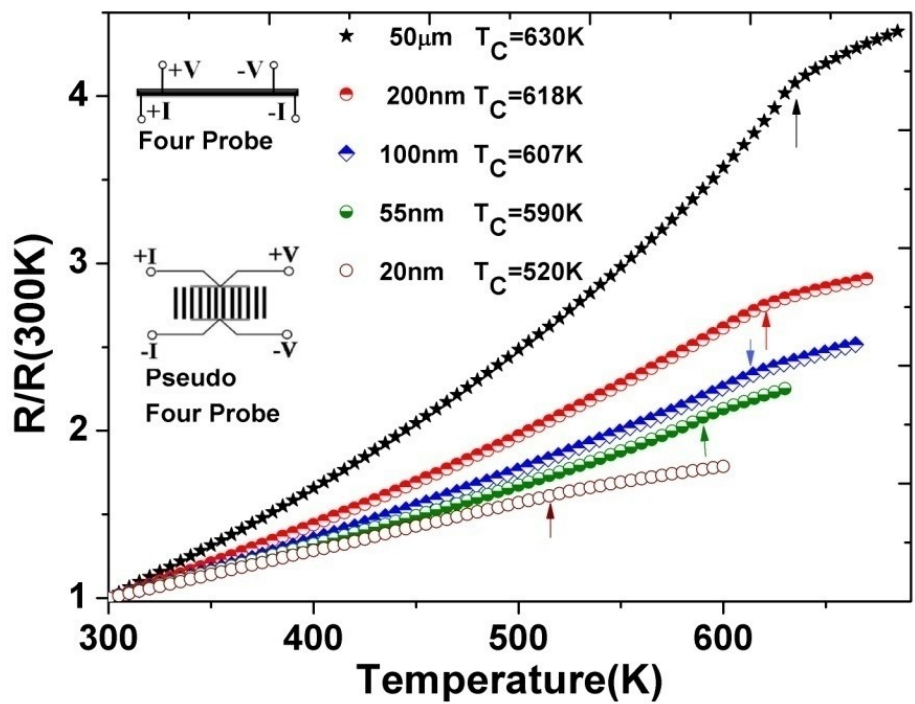

**Figure 4.2:** Normalized plot of *R* vs. *T* for nickel nanowires of varying diameters. The Curie temperatures determined from the resistance anomaly are shown by arrows.

Fig. 4.2 shows the *R* data of the nanowire arrays (normalized with the value of *R* at 300 K) as compared with *R* of the bulk. The Curie temperatures ($T_C$) as determined by quantitative analysis, described later on, are marked by arrows. The values of $T_C$ can be more clearly seen in the normalized plots of derivatives shown in Fig. 4.3. The resistance anomaly in the resistance data can be seen in the even in the nanowire with lowest diameter.

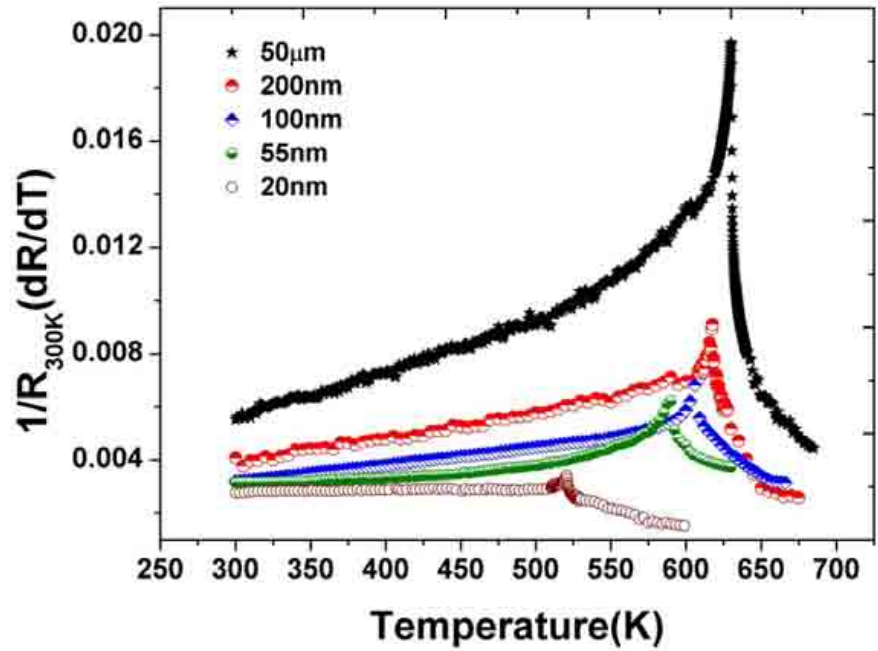

**Figure 4.3:** Plot of $\dfrac{1}{R_{300K}}\dfrac{dR}{dT}$ vs. $T$ for nickel nanowires.

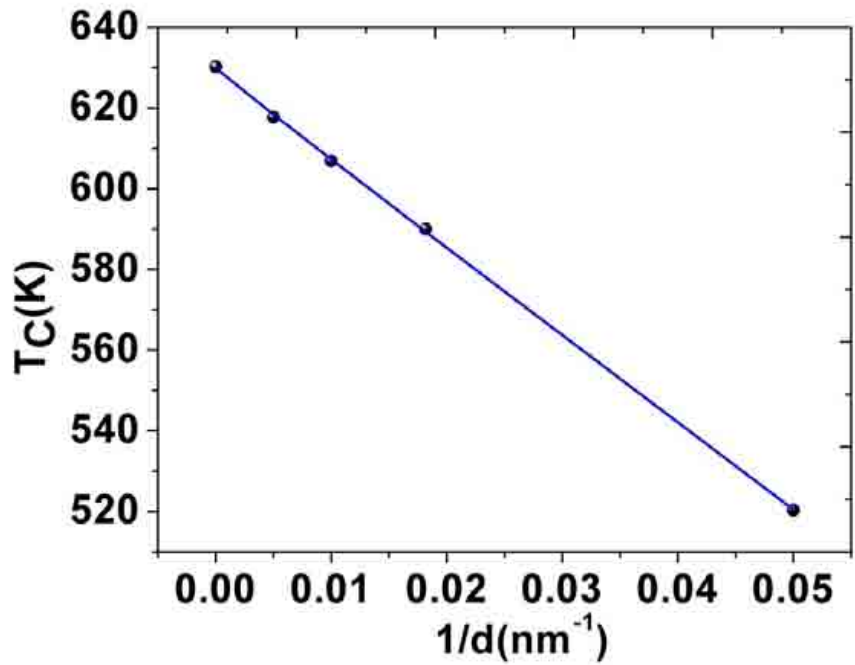

**Figure 4.4:** The variation of $T_C$ with $1/d$.

## 4.3.2 Determination of $T_C$ from resistance anomaly and its dependence on size

First we present some of the basic observations. In Fig. 4.3, it can be seen that the anomaly in the resistance reduces as the diameter is reduced (along with a distinct reduction in $T_C$). However, the anomaly, though reduced is present even in the wire with the smallest diameter. To accentuate the anomaly in $R$ we show, in Fig. 4.3, the derivative of the resistance $\left( \dfrac{1}{R_{300K}} \dfrac{dR}{dT} \right)$ as a function of $T$. The presence of the distinct feature in the derivative near $T_C$ is present in all the wires, though much reduced in the wires of smaller diameter. The dependence of $T_C$ on wire diameter is shown in Fig. 4.4 plotted as $T_C$ vs. $1/d$. The reduction in $T_C$ (as determined from the resistance data) with diameter $d$ is very similar to that obtained from magnetization measurements [5] (Fig. 4.5). The dependence of $T_C$ can be described by the finite-size scaling laws [4]. The finite size effect can be observed as a change of $T_C$ when the correlation length $\xi(T)$ is comparable to $d$. The asymptotic behavior of the correlation length $\xi(T)$ of a magnetic system close to the bulk transition temperature is described by [4]

$$\xi(T) = \xi_0 |t|^{-\nu} \tag{4.5}$$

where $\xi_0$ is the extrapolated value of $\xi(T)$ at $T = 0$ K and $t = \dfrac{T - T_C}{T_C}$ is the reduced temperature. The critical temperature $T_C(d)$ scales with the diameter of the nanowires as

$$\frac{T_C(\infty) - T_C(d)}{T_C(\infty)} = \left( \frac{\xi_0}{d} \right)^{\lambda} \tag{4.6}$$

where $\lambda = \dfrac{1}{\nu}$. We obtain $\lambda = 0.98$. The value obtained from the results of resistivity anomaly agrees very well with those obtained from magnetization measurements ($\lambda = 0.94$) for nanowires[5]. However it is lower than the values predicted by 3D Heisenberg Model ($\lambda = 1.4$) and 3D Ising Model ($\lambda = 1.58$) [1, 26]. The value of $\xi_0$ (= 3.35 nm), is larger than but comparable to that obtained from the magnetization measurements $\xi_0$ (= 2.2 nm) [5].

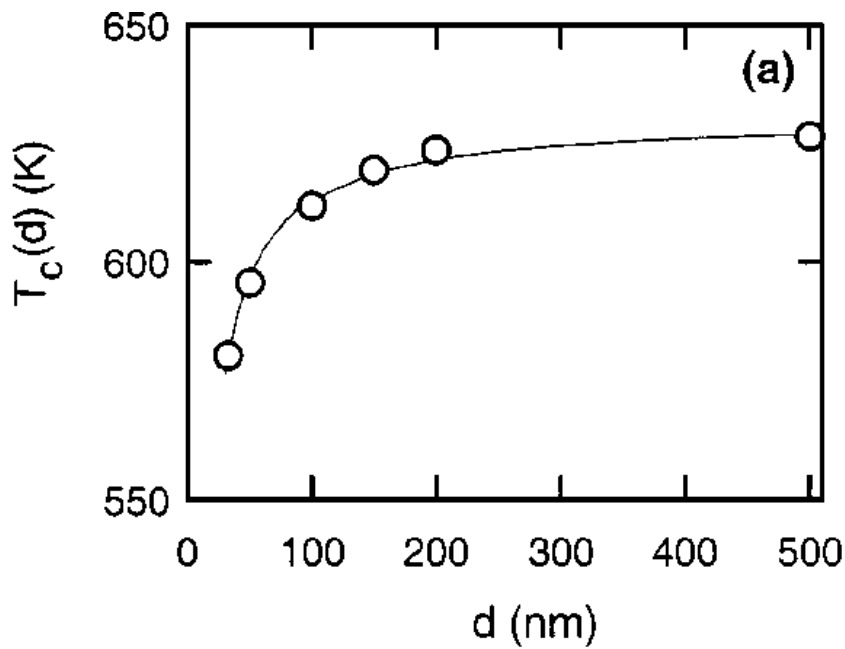

**Figure 4.5:** Curie temperature $T_C(d)$ of nickel nanowire arrays vs. wire diameter $d$. Adopted from [5].

A lower value of $\lambda$ (~1) has been also observed in nickel and other ferromagnetic thin films [27, 28, and 29]. As shown by the analytical calculation on films done by Fisher and Barber [4], the finite size and the 'constant density' due to the constraint imposed on the system have been attributed to such a value of $\lambda$ ~1 in case of thin films in these reports. Similar argument applies also in case of nanowires embedded inside porous alumina matrix imposing a 'constant density' of the nickel. The pressure acting on the nanowires because of the thermally generated strain can alter the Curie temperatures significantly leading to a change in $\lambda$ when such strain correction is taken into account. However, such a process involves a careful measurement of the elastic modulus, coefficient of expansion and the Curie temperature dependence on pressure. Another way is to measure the resistance of freely suspended single nanowires in ultra high vacuum to see the temperature of resistance anomaly and exactly calculate $\lambda$.

### 4.3.3 Determination of critical exponent α

We now investigate the crucial issue of whether the resistance anomaly near $T_C$ can indeed be analyzed using the frame work of critical phenomena. In nanowires such issues like strain as well as other effects can lead to a rounding-off of the transition and we would have to address this issue when analyzing the data in the critical region. The resistivity data near $T_C$ in bulk Ni wires have been adequately analyzed by precision experiments in two earlier reports and there are well accepted values for the critical exponents [2, 3]. This has been summarized in Table 4.1. To standardize our analysis we use the data on the 50 μm sample and analyze it using the protocol as described earlier [2, 3] in the region around $T_C$ for $|t| \leq 10^{-2}$. We use the equations

$$R(t) = R_0 + R'(t) + A_-|t|^{(1-\alpha)}\left(1 + D_-|t|^z\right) \qquad \text{for } t < 0 \qquad (4.7)$$

$$R(t) = R_0 + R'(t) + A_+t^{(1-\alpha)}\left(1 + D_+t^z\right) \qquad \text{for } t \geq 0 \qquad (4.8)$$

The terms $R_0$ and $R'(t)$ mainly consist of the non-critical part of the resistance of the sample. The third term is the critical part of the resistance responsible for the power law behavior. $\alpha$ is the critical exponent with $A_-$ and $A_+$ being the critical amplitudes of the leading term. We note that in the frame work of the Fisher-Langer theory this $\alpha$ is the same as the specific heat exponent. $Z$ is the exponent and $D_-$ and $D_+$ are the amplitudes of the second order correction to scaling or the confluent singularity. $Z = 0.55$ as predicted by the renormalization group theory [30]. Though the above equation for $R(t)$ represents the behavior of resistance near $T_C$ for a homogeneous material with no spread in $T_C$, one can correct for the inhomogenetes by using a convolution of the form[2,3]

$$R^*(T, T_C, \sigma) = \int R(T, T_C - x)\, g_\sigma(x)\, dx \qquad (4.9)$$

where $g_\sigma(x)$ is a Gaussian in $x$ of width $\sigma$ and $x$ is the temperature variance from an average $T_C$. $R(T, T_C - x)$ is obtained by replacing $T_C$ in the expression for $R(t)$ by $T_C - x$.

$$R^*(T, T_C, \sigma) = \int R(T, T_C - x)\frac{1}{\sigma\sqrt{2\pi}}e^{-\frac{x^2}{2\sigma^2}}dx \qquad (4.10)$$

$$R^*(T, T_C, \sigma) = \frac{1}{\sqrt{\pi}}\int R\left(T, T_C - \sqrt{2}\,\sigma X\right)e^{-X^2}dX \qquad (4.11)$$

The integration was numerically solved using a 32 point Gauss-Hermite quadrature [see Appendix-II] and a least square fitting was done with variable parameters $R_0$, $R'$, $\alpha$, $A_-$, $A_+$, $D_-$, $D_+$, $T_C$, $\sigma$ with a maximum fit error of 0.04%. Fitting results of the nanowires of various diameters along with the bulk wire are shown in Fig.4.6. In the region very close to $T_C$ there is a deviation due to rounding-off which is discussed later on in the section on discussion. As mentioned earlier the fitting in each case is done on resistance data for $T_C \pm 10$ K.

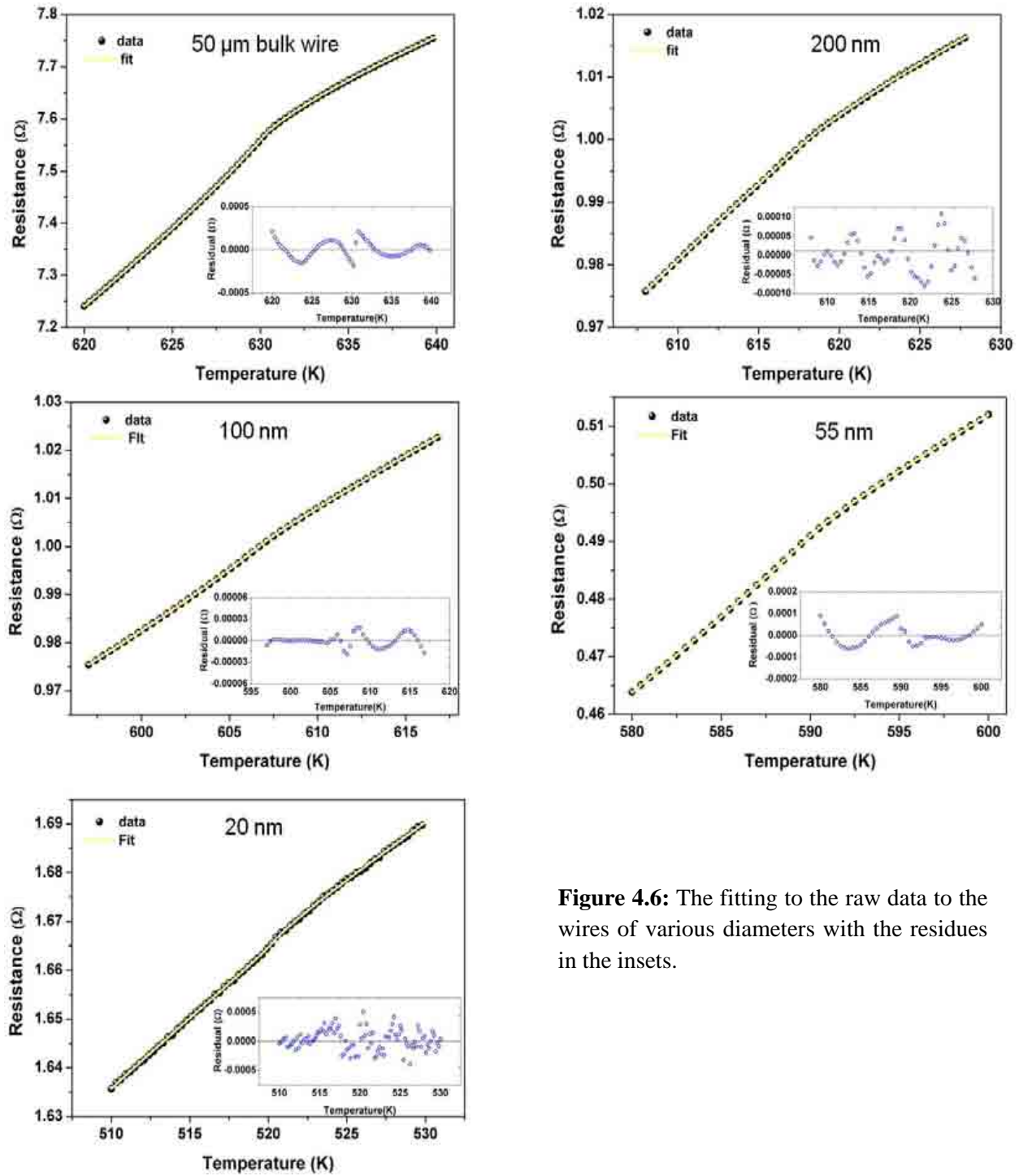

**Figure 4.6:** The fitting to the raw data to the wires of various diameters with the residues in the insets.

The extracted parameters from fitting are tabulated in Table 4.1 along with the exponents obtained from the previous work on bulk samples of Ni. Interestingly we can compare the important numbers that we obtain with those reported in the literature for the bulk sample. $T_C$ determined by us and Ref 2 have comparable accuracy and these agree to within .034K or ~53ppm. This is a very good agreement in absolute value because the materials are obtained from different sources. In comparison the bulk $T_C$ obtained from Ref 3 disagree to about .05% and this is reasonable noting that the $T_C$ determination has about 2 orders less accuracy. This is also reflected in the values of α (agreement to within 4.5 and amplitude ratio $A_+/A_-$ (agreement within 1.5%). The strain broadening is also a within these two samples. The comparison shows that the analysis used by us is standardized and thus can be used to analyze the data for the nanowires.

**Table 4.1:** The exponents obtained for the bulk and the nanowires by fit as compared with previous data.

| Reference/ Diameter (nm) | $T_C(K)$ | $α_r$ | $A_+/A_-$ | $D_+/D_-$ | $σ$ |
|---|---|---|---|---|---|
| Bulk Ref. 2 | 630.284±0.003 | -0.095±0.002 | -1.53±0.02 | -0.8±0.1 | 0.13±0.12 |
| Bulk Ref. 3 | 630.6±0.4 | -0.089±0.003 | -1.48±0.03 | -1.2±0.2 | 1.28±0.03 |
| 50000 | 630.248±0.001 | -0.0996±0.0001 | -1.506±0.001 | -0.98±0.03 | 0.141±0.006 |
| 200 | 617.752±0.003 | -0.0864±0.0006 | -1.42±0.01 | -0.84±0.03 | 0.43±0.02 |
| 100 | 606.90±0.02 | -0.081±0.001 | -1.38±0.02 | -0.74±0.15 | 0.321±0.001 |
| 55 | 590.004±0.008 | -0.0638±0.0005 | -1.29±0.01 | -0.10±0.02 | 0.44±0.01 |
| 20 | 520.33±0.01 | -0.0309±0.0001 | -1.003±0.002 | 0.47±0.05 | 0.52±0.02 |

For comparison of the change in critical behavior, a plot of resistance (normalized by the resistance value at $T_C$) near $T_C±10$ K for the bulk wire and the smallest diameter nanowire of 20 nm is shown in Fig. 4.7. Fit of the equation for $R*(T,T_C,σ)$ to the data for 20 nm diameter nanowire is also shown in Fig. 4.7. The data can be fitted to the given equation with reasonable accuracy which is comparable to that obtained for the bulk data. The exponents and other parameters obtained show a very clear trend as the diameter is progressively reduced (see Table-4.1).

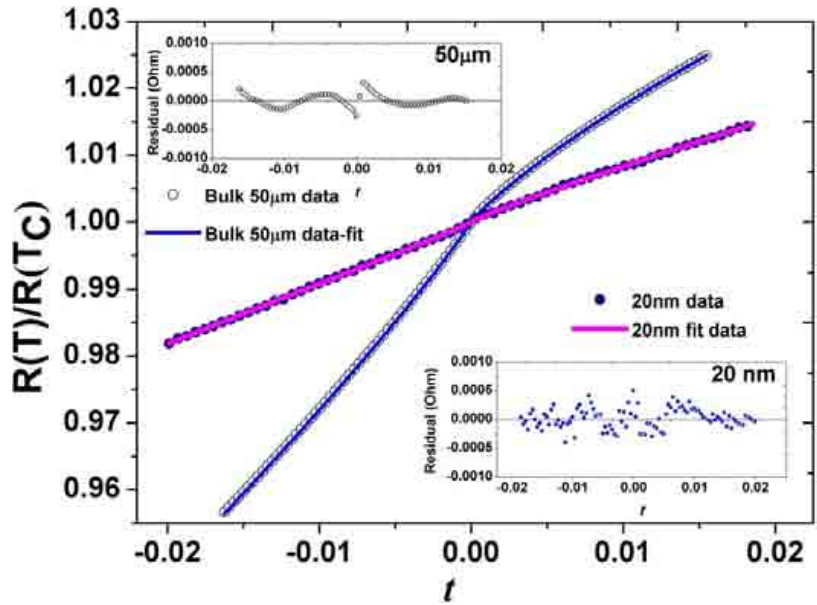

**Figure 4.7:** $R(T)/R(T_C)$ vs. $t$ (along with the fits) for the 20 nm and bulk wire about T$_C$ ($t = 0$). The insets show the residuals for bulk and 20 nm diameter nanowire data fits.

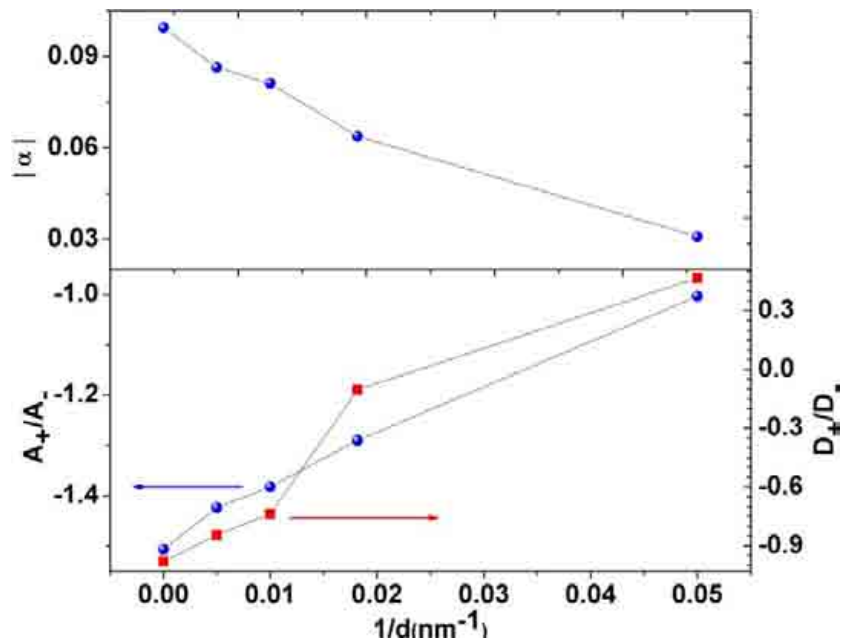

**Figure 4.8:** Variation $\alpha$ and amplitude ratios $\dfrac{A_+}{A_-}$ and $\dfrac{D_+}{D_-}$ for the wires studied as determined from the scaling relation.

In case of the nanowires, there is a considerable decrease in the magnitude of $\alpha_r$ showing a slowing down of the growth of the $dR/dT$ near the critical point. In addition, changes in $A_+/A_-$ and $D_+/D_-$ are also prominent. The variations in the values of the $\alpha_r$, amplitude ratios and the correction terms are plotted in Fig. 4.8. The above analysis establishes two points: (1) The critical behavior as manifested in the resistance anomaly is present even in the smallest diameter nanowire (although suppressed) and it is describable within the mathematical frame work given in the equation for $R*(T, T_C, \sigma)$ (Eqns. 4.7-4.9) and (2) the amplitude of the critical behavior and also the correction to scaling are severely suppressed as the diameter is reduced.

While analyzing the data near the critical point the rounding off becomes an important issue. This is present even in bulk systems like Ni and in nanowires, in particular, when we do the measurements in an array of nanowires, these can be severe. To avoid the effect of rounding-off in the estimation of the exponents often the small regions in temperatures ($|t| \approx 10^{-3}$) around the $T_C$ is left out in data analysis. We have estimated the rounding-off region below and find that it is limited to a region ($\approx \pm 2$ K) even in the smallest diameter wire.

## 4.4 Discussion

The observed spread in $T_C$ as measured by the width of $\sigma$ of the Gaussian in the equation for $R*(T, T_C, \sigma)$ is a good empirical measure of the minimum rounding off that we may expect. In the bulk sample we used $\sigma \approx 0.14$ K which is comparable to that seen in most bulk samples before (see Table 4.1). In the nanowires, irrespective of diameter, the typical $\sigma \approx 0.4$-$0.5$ K. The spread in $T_C$ and the resulting rounding-off in the resistance anomaly in our measurements arise from the following two principal sources: (1) Spread due to strain inhomogeneity, (2) Spread in the wire diameters as the measurement is done in an array. We can estimate them and compare that with experiments. From the X-ray micro-strain measurements which give the strain inhomogeneity in the wires we can estimate the expected spread in $T_C$ due to strain inhomogeneity using the bulk modulus of Ni and the dependence of $T_C$ on pressure $(dT_C/dP)$[3]. We find that for all the nanowires the expected maximum spread in $T_C$ is in the range of 0.4 - 0.7 K which is of the same magnitude as $\sigma$ that one obtains from the fit to the resistance data. The expected spread in $T_C$ from the distribution in size can be calculated from the measured distribution and the finite size scaling relation as follows

$$\Delta T_C(d) = \frac{\lambda T_C(\infty)(\xi_0)^\lambda \Delta d}{d^{\lambda+1}} \qquad (4.12)$$

The estimated spread $\Delta T_C(d)$ is $\leq 0.9$ K for the larger diameter wires and $\leq 2$ K for the smallest diameter wire. Thus it appears that most of the contribution causing spread in $T_C$ comes from the wire size distribution.

The above analysis establishes that the resistivity anomaly exists in ferromagnetic nanowires at their transition temperature $T_C$ even in nanowires. The anomaly and the associated critical behavior is progressively but systematically suppressed (reduction in absolute values of the ratio $A_+ / A_-$ as well as $\alpha$ as the diameter of the wire is reduced). The correction to scaling as quantified by the ratio $D_+ / D_-$ shows a systematic change from negative to positive sign.

The effect of finite size on the critical exponents should be visible in the temperature and size regime we are working. As the transition temperature is approached, the correlation length gets constrained by the wire diameter. Hence the truncation of the divergence of the correlation length begins as the temperature approaches $T_C$. Using $\xi_0 = 3.35$ nm and Eq. 4.5, we can see that this happens at $|t| \leq 0.17$ in the case of 20 nm wire and $|t| \leq 0.036$ in the case of 100 nm wire. This is the range of $|t|$ where we observe critical behavior. In case of strictly one dimensional nanowires, there should not be a second order transition if one considers the 1D Ising model [31]. Mathematically speaking, the critical part of $R$ has to vanish or become a constant. This implies the ratio $A_+ / A_- \rightarrow -1$ and $\alpha \rightarrow 0$ with $D_-$ and $D_+$ becoming insignificant. We observed $\alpha_r \rightarrow 0$ and also the observed ratio $A_+ / A_- \rightarrow -1$. This would imply that the spin system is tending towards 1-D behavior. We note again, as we have pointed out before, that we are not sure whether in such systems as ours we will have $\alpha$ (from heat capacity) same as $\alpha_r$. However, we also note that in one class of systems with restricted diameter where the value of $\alpha$ from specific heat experiment has been evaluated rigorously is in superfluid transition of He$^4$ in pores with diameter in the same range. It has been seen that in these 1D constrained systems the value of $\alpha$ is -0.02[32] which is comparable with the value we obtained ($\approx$ -0.03 in 20 nm wires).

## 4.5 Conclusion

In summary, we have presented precision measurements of the resistance of nickel nanowires of various diameters in the critical region (T$_C$ ± 10 K), where one would expect anomaly in resistivity due to critical point phenomena. To the best of our knowledge, these are the first such measurements on nanowires near the T$_C$. With the decrease in diameter, we observed a decrease in the transition temperature. The data analysis shows that the observed resistance data shows critical behavior down to the lowest diameter wire of 20 nm and the anomaly close to T$_C$ can be treated in the frame work of resistance behavior near a critical region. However, there is a suppression of the critical behavior of the resistivity including a decrease in the magnitude of $\alpha$ and the ratio $A_+ / A_- \rightarrow -1$ as one would expect in a quasi one-dimensional spin system.

# CHAPTER 5

# Low temperature electrical transport in ferromagnetic Ni nanowires

In this chapter we report the electrical transport properties of arrays of single crystalline nickel nanowires (diameters ranging from 55 nm to 13 nm) fabricated by electrodeposition into the cylindrical pores of anodic alumina membranes. X-ray and Transmission Electron Microscope based characterization ensured single-crystalline nature of the nanowires. We have measured the resistance of the Nickel nanowires in the temperature range 3 K to 300 K with the specific aim to probe the effect of size reduction on the temperature dependence of resistivity. As the lateral dimension decreases, deviations from the bulk resistivity are observed. Our work reveals intrinsic differences in the transport mechanisms taking place in these wires when the diameter of the wires is brought down from bulk to nano regime. The resistance data has been analyzed using the Bloch-Wilson function and the Debye temperature ($\theta_R$) was calculated from the fits. The Debye temperature showed a systematic decrease with decrease in diameter, which we note is a trend in other FCC metals like Ag and Cu. We observed an increase in the residual resistivity as the diameter is decreased due to surface scattering. We further observe a strong suppression in the spin wave contribution to the resistivity of the magnetic nanowires as the diameter is decreased. We discuss the likely causes for these changes.

This page is intentionally left blank

## 5.1 Introduction

Research on nanostructures has led to the exploration of novel physics and material properties at reduced physical dimensions. The influence of reduced physical dimensions is of both fundamental and technological interest. Nanoscale arrays of magnetic materials have large potential in wide range of applications including magnetic data storage, nanoelectromechanical systems (NEMS) sensors and actuators, nanoscale spintronic devices etc. Magnetic nanowires represent an important variety of magnetic nanostructures. The resistance of a magnetic nanowire becomes a subject matter of immense interest when applications such as electronic and spintronic devices are concerned. While a lot of work has been done on the magnetic properties of Nickel nanowires [1–3] the electrical transport properties remain vastly unexplored. In the regime where the diameter of the nanowires is few tens of nm and the carrier mean free path is limited by the wire dimensions, the resistivity depends not only on the material but also on its size [4–8]. Though the transport becomes of quantum nature when the diameter of nanowire gets closer to the molecular size, there is a considerable size range (few tens of nanometer in case of metals) where the Boltzmann transport theory is applicable [9], yet it is modified by finite size effects. This typically occurs in metallic nanowires with diameter < 100 nm. Interestingly, this is also the diameter regime where technical applications are envisaged.

In this chapter we present a detailed study of the electron transport in single crystalline nanowires of various diameters (55 nm, 35 nm, 18 nm, 13 nm). There are three essential points which this paper attempts to address, (a) what happens to the electron-phonon interaction and whether a single Debye temperature description can explain the data, (b) the effect of surface scattering on the resiudal resistivity and (c) what happens to the electron-magnon interaction as the diameter of the nanowire is reduced. To study the effect of surface scattering, it is important to have the nanowires to be single crystalline to avoid the grain boundary resistance. So here we made efforts to synthesize single crystalline nanowires for the present study by pulsed electrodeposition. In this work, we measured the resistance of oriented single crystalline nickel nanowires in the temperature range 3 K to 300 K. The wires are ferromagnetic with $T_C$ above room temperature. The resistance of a pure nickel bulk wire (~50 μm) is also measured as a reference. To the best of our knowledge, such a detailed investigation of resistivity measurements of Nickel nanowires with diameters well below 100 nm in the full temperature range of 3 K to 300 K and the subsequent quantitative analysis have not been done before. We note that developing a scheme by which temperature dependence of resistivities of nanowires with diameters 10 nm or above, can be estimated is a fruitful exercise because it gives a predictive test that resistivities can be estimated without actually measuring it in every sample. We also show that the experiment gives us a way to determine the Debye temperature ($\theta_R$) for the nanowires and gives us a way to predict what happens to the elastic modulus on size reduction. The results obtained by us also show that the magnetic contribution to resistivity is considerably suppressed on size reduction.

## 5.2 Electrical Resistivity of 3d ferromagnetic metals

For simple non-magnetic metals like Cu, Ag and Au, the temperature dependence of electrical resistivity is well described by the Bloch-Grüneisen (BG) function, $f\left(T/\theta_R\right)$, where $\rho_0$ is the residual resistivity arising out of the temperature-independent scattering of electrons from lattice defects and impurities. $\rho_L$ originates from the scattering of conduction electrons by lattice phonons [10].

$$\rho = \rho_0 + \rho_L \qquad (5.1)$$

$$\rho_L = \alpha_{el-ph} f_{BG}\left(\frac{T}{\theta_R}\right) \qquad (5.2)$$

$$f\left(\frac{T}{\theta_R}\right) = \left(\frac{T}{\theta_R}\right)^n \int_0^{\frac{\theta_R}{T}} \frac{x^n}{\left(e^x - 1\right)\left(1 - e^{-x}\right)} dx \qquad (5.3)$$

where $\theta_R$ is the Debye temperature as observed from resistivity. In most cases it closely matches with $\theta_D$ as obtained from other methods like heat capacity measurement. Generally, $n = 5$ for simple metals and alloys. $\alpha_{el-ph}$ is a constant $\propto \dfrac{\lambda_{tr}\omega_D}{\omega_p^2}$ where $\lambda_{tr}$ is the electron-phonon coupling constant, $\omega_D$ is the Debye frequency and $\omega_p$ is the plasma frequency [11] in the simple theory of electron-phonon interaction. It has been shown by our group before that in nonmagnetic metallic nanowires like Cu or Ag, the temperature dependent resistivity can be quantitatively described by the BG formula[9] even for wires with diameter as low as 15 nm. The size reduction however, leads to a change in the Debye temperature ($\theta_R$). In case of magnetic materials, in addition to electron-phonon interactions, other mechanisms of scattering exist which can give rise to additional contribution to the temperature dependence of $\rho$. Unlike simple metals, the electrical resistivity of 3d ferromagnetic transition metals like Ni, Co, Fe have a magnetic contribution $\rho_M$ as the electrons get scattered by magnons at low temperatures. It is of interest to check what happens to the magnetic contribution, $\rho_M$, when the size is reduced below 100 nm. The total resistivity ($\rho$) is the sum of the contributions (Matthiessen's rule) due to potential scattering ($\rho_0$), phonon scattering ($\rho_L$) and magnetic scattering ($\rho_M$).

$$\rho = \rho_0 + \rho_L + \rho_M \tag{5.4}$$

In case of a ferromagnetic metal the *s* and *d* bands overlap at the Fermi level. So for both lattice and magnetic scattering, conduction electrons might undergo *s-s* (intraband) as well as *s-d* (interband) transitions. Typically, for magnetic metals and alloys with large d-band density of states giving rise to electron-phonon scattering involving *s-d* transitions, the scattering of conduction electrons by phonons is described by the Bloch-Wilson formula with n=3 in equations 1-3[10]. The magnetic scattering arises from the exchange interaction between the conduction electrons and the more localized magnetic electrons (3d), commonly called *s-d* interaction. Using a spin wave description, Mannari [12] found the following expression for $\rho_M$:

$$\rho_M = BT^2; \quad B = \frac{3\pi^5 S\hbar}{16e^2 k_F}\left(\frac{\mu}{m}\right)^2 \frac{\left(k_B NJ(0)\right)^2}{E_F^4} \tag{5.5}$$

where S (= 1/2) is spin of the conduction electron, $\mu$ is the effective magnon mass, *m* is the electron mass. For Ni, the ratio $\mu/m \approx 38$, $E_F$ is the Fermi energy, $k_F$ is the Fermi wave vector, and *NJ(0)* is the strength of the *s-d* interaction in the long wavelength limit ($\approx 0.48$ eV for Ni), *N* being the number of spins. For Ni, the above relation gives $B \approx 1.1$ X $10^{-13}$ $\Omega$.mK$^{-2}$ which matches very well with that obtained experimentally [13] in bulk Ni (1.1 X $10^{-13}$ $\Omega$.mK$^{-2}$).

In the bulk 3d transition metals and alloys, $\rho_M \propto T^2$ (considering the *s-s* electron magnon scattering) behavior for *T* < 15 K has been reported [12, 13]. At temperatures *T* > 15 K, there are additional terms and the scattering process has complicated temperature dependence. In this chapter, we investigated the effect of size reduction on the magnetic term in the electrical resistivity in case of nickel nanowires. Our analysis shows how the resistivity ($\rho$) changes as we decrease the diameter of the nanowires where the electron mean free path becomes comparable to the diameter of the nanowires.

## 5.3 Experimental

Though, the synthesis, characterization and measurement techniques are well described in Chapter 2 and Chapter 3, for the sake of completeness, here we give a brief description. The Nickel nanowire arrays with diameters in the range of 55 nm to 13 nm were prepared using a pulsed potentiostatic electrodeposition technique using commercially available standard anodic alumina templates [14]. The pore diameters of the templates were checked using Scanning Electron Microscopy (SEM). The templates have a thickness of 50 μm-60 μm. A layer of 200 nm of silver is evaporated on to one surface of the nanoporous templates in a thermal evaporation chamber. The coated side of the template was used as the working electrode with the uncoated surface facing the electrolyte in a three electrode potentiostatic deposition bath. Saturated calomel electrode (SCE) and Platinum electrode were the reference and the counter electrodes respectively. The depositions were carried out with a 300g/l $NiSO_4.6H_2O$, 45g/l $NiCl_2.6H_2O$, 45 g/l $H_3BO_3$ solution as the electrolyte with the working electrode at a pulse potential of -1 V with respect to the SCE with 80% duty cycle and pulse period of 1 second. Electrodeposition continued till the Nickel filled the pores and came out to the side exposed to electrolyte as observed by an increase in the current. The extra layer of Nickel deposited on the template side facing electrolyte is removed by polishing. The wires thus prepared were characterized using X-Ray Diffraction (XRD), Transmission Electron Microscopy (TEM) and Scanning Electron Microscopy (SEM). Powder X-Ray Diffraction using CuKα radiation was carried out by retaining the wires inside the alumina templates. The magnetic characterization of the nanowires is presented in Chapter 2 shows that the nanowires are magnetic with characteristic hysteresis loops and coercivities. For Scanning Electron Microscopy, the template was partially etched out by keeping them immersed in 3M NaOH solution for about 10 min and then washing out the dissolved parts with distilled water. For Transmission Electron Microscopy, the templates were etched out completely by keeping them immersed in 6M NaOH solution for about 30 min. The nanowires in the templates were annealed at 600 K in $10^{-6}$ mbar vacuum. The resistance of the wires was measured in the temperature range 3 K to 300 K using pseudo four probe method[9] as described in details in chapter 3, in a pulsed-tube based closed cycle refrigerator[15] using a low frequency (174.73 Hz) AC current of 100 μA with a resolution of nearly 2 ppm.

## 5.4 Results

### 5.4.1 Structural characterization

The structural and crystallographic nature of the wires constitutes an important part in the analysis of the data. The wires used in this investigation are single crystalline and oriented in nature. This has been established by techniques such as x-ray diffraction (XRD) and high-resolution transmission electron microscopy.

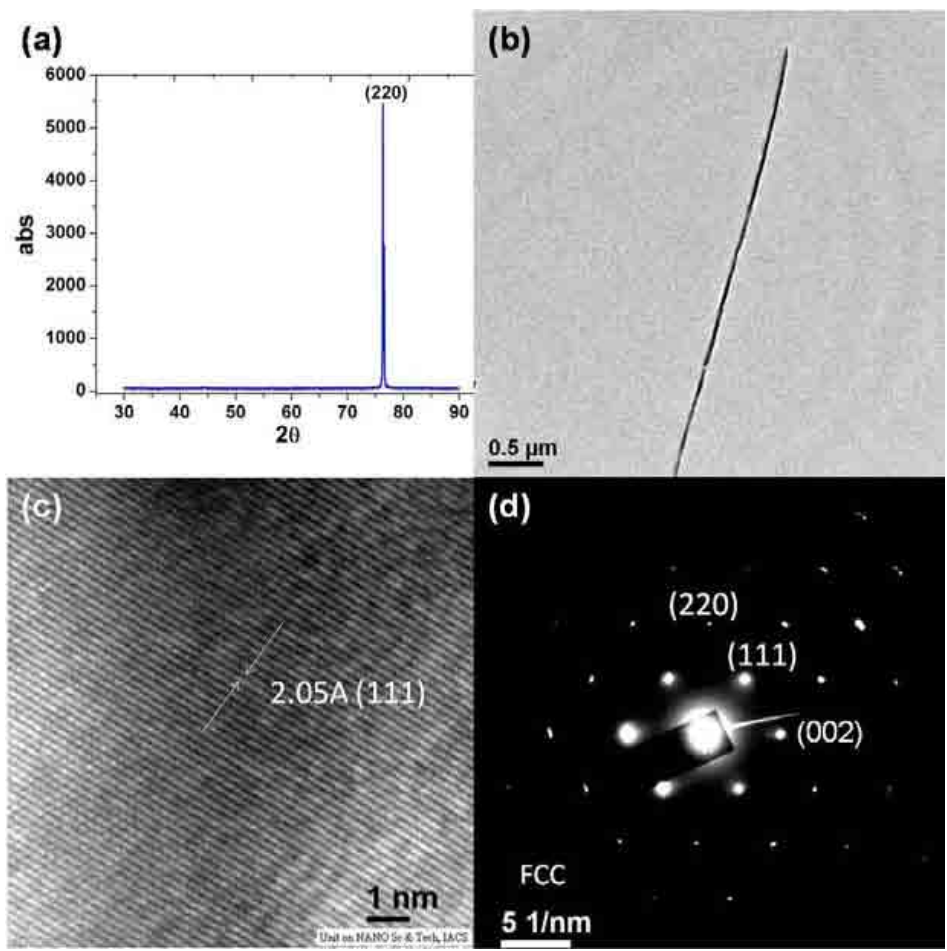

**Figure 5.1:** (a) XRD pattern of 55 nm diameter nanowire arrays, (b) TEM image of a 55 nm diameter nanowire, (c) HRTEM image of the lattice planes in a 55 nm diameter nanowire, (d) The electron diffraction pattern of a 55 nm diameter nanowire.

The structural data have been shown in Fig. 5.1. The data shown in Fig. 5.1 is for a 55 nm diameter wire and it is typical for other wires as well. X-ray diffraction pattern shown in Fig. 5.1(a) reveals that the nanowires grown have a preferential direction ((220)) of growth. All the wires show ((220)) growth direction as presented in the chapter 2 dedicated to growth and characterization. Fig. 5.1(b) shows a Transmission electron micrograph of a 55 nm diameter nanowire. A High Resolution Transmission Electron Micrograph (HRTEM) is shown in Fig. 5.1(c). The electron diffraction pattern shown in Fig. 5.1(d) confirms that the nanowires are single crystalline and the FCC nature of the crystal structure. The fact that the wires are single crystalline is very important because this ensures that the residual resistivity ($\rho_0$) comes predominantly from the surface scattering of electrons in absence of scattering at any grain boundary. As a check on the lattice parameter the value of lattice constant obtained from the XRD data is ~ 3.55 Å for different nanowires.

## 5.4.2 Electrical transport

The electrical resistance measurements were carried out by retaining the wires within the alumina membrane. The resistance of the nanowire arrays is measured by pseudo four probe method in which two electrical leads were attached to each of the two sides of the membrane containing Nanowires using silver epoxy. The method and the issues related to contact resistance and contact noise have been discussed in details in previous publications by our group [9, 16, and 17]. In the case of 50 µm diameter bulk nickel wire, the resistance is measured by simple four probe method.

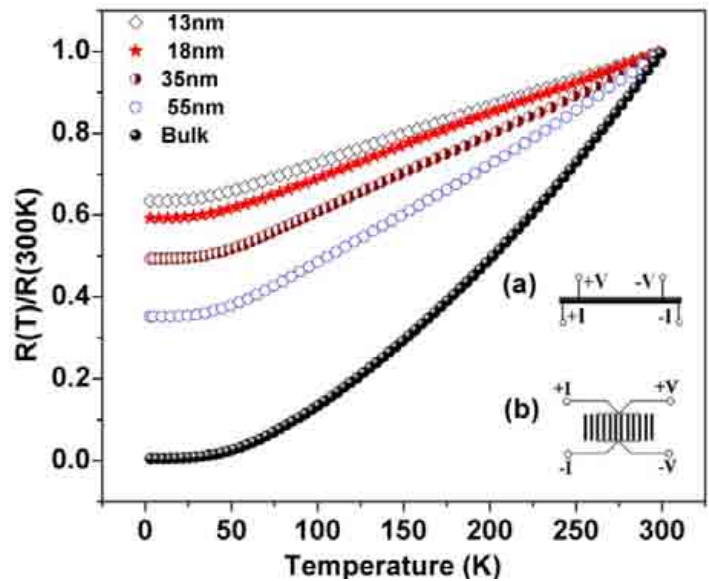

**Figure 5.2:** Normalized plot of resistance data of the nanowires as compared with the bulk wire; Schematic representations of (a) Four probe method and (b) Pseudo four probe configurations.

The resistance data of nanowires (normalized by resistance at 300 K) along with the bulk nickel wire is shown in Fig. 5.2. Schematic representations of the four probe and pseudo four probe methods are shown in the corresponding insets (Fig. 5.2(a) and 5.2(b)) respectively. A first look into the plot shows that the wires become more resistive as the diameter decreases. We could successfully put 4-probes on a single nanowire of diameter ~ 55 nm using Focused Ion Beam and the resistivity determined by the 4-probe method matches well with that determined by retaining them in the array in the template. The results of measurement on single nanowire is presented in the next chapter. Thus the contact resistance is not significant. We comment on this later on. In the following sub-section we analyze the resistance data.

### 5.4.3 Analysis of electrical resistance data for $T > 15$ K

The first part of our analysis is to investigate the temperature dependence of the resistance $R$. The electrical resistivity of a 3d ferromagnetic material can be described by Eq. (5.4). At very low temperatures ($T < 15$ K), the magnetic part varies with temperature as $\rho_M = BT^2$. Beyond 15 K, the magnetic contribution has complicated temperature dependence because of additional intraband s-d electron-magnon scattering. However, well below the Debye temperature and above 15 K, the main temperature dependent term that dominates the resistivity is the electron-phonon interaction term ($\rho_L$). We explore whether the temperature dependence of the resistance can be explained in the frame work of the Boltzmann transport theory using Eqs. (5.1) - (5.3). As discussed, $\rho_M$ for $T > 15$ K is generally much less than $\rho_L$. We thus fit the resistance data for 15 K $\leq$ T $\leq$ 100 K, using Eq. (5.6). The best fit is obtained with $n = 3$ (Bloch-Wilson formula).

$$R = R_0 + \alpha \left(\frac{T}{\theta_R}\right)^n \int_0^{\frac{\theta_R}{T}} \frac{x^n}{\left(e^x - 1\right)\left(1 - e^{-x}\right)} dx \qquad (5.6)$$

$\alpha$ is related to $\alpha_{el\text{-}ph}$ (Eqs. (5.1)- (5.3)) by the same geometric factor that relates $R$ to $\rho$. The analysis as in Eq. (5.6) using $R$ instead of $\rho$ allows us to explore the temperature dependence and determine $\theta_R$, without the uncertainty in the absolute value of $\rho$. The fit parameters $R_0$, $\alpha$ and $\theta_R$ were optimized to give a relative fit error (defined in % as $\frac{R_{measured} - R_{fit}}{R_{measured}} \times 100$) of less than 0.4

% over the entire range. The fitting plots for wires of various diameters are shown in Fig. 5.3.

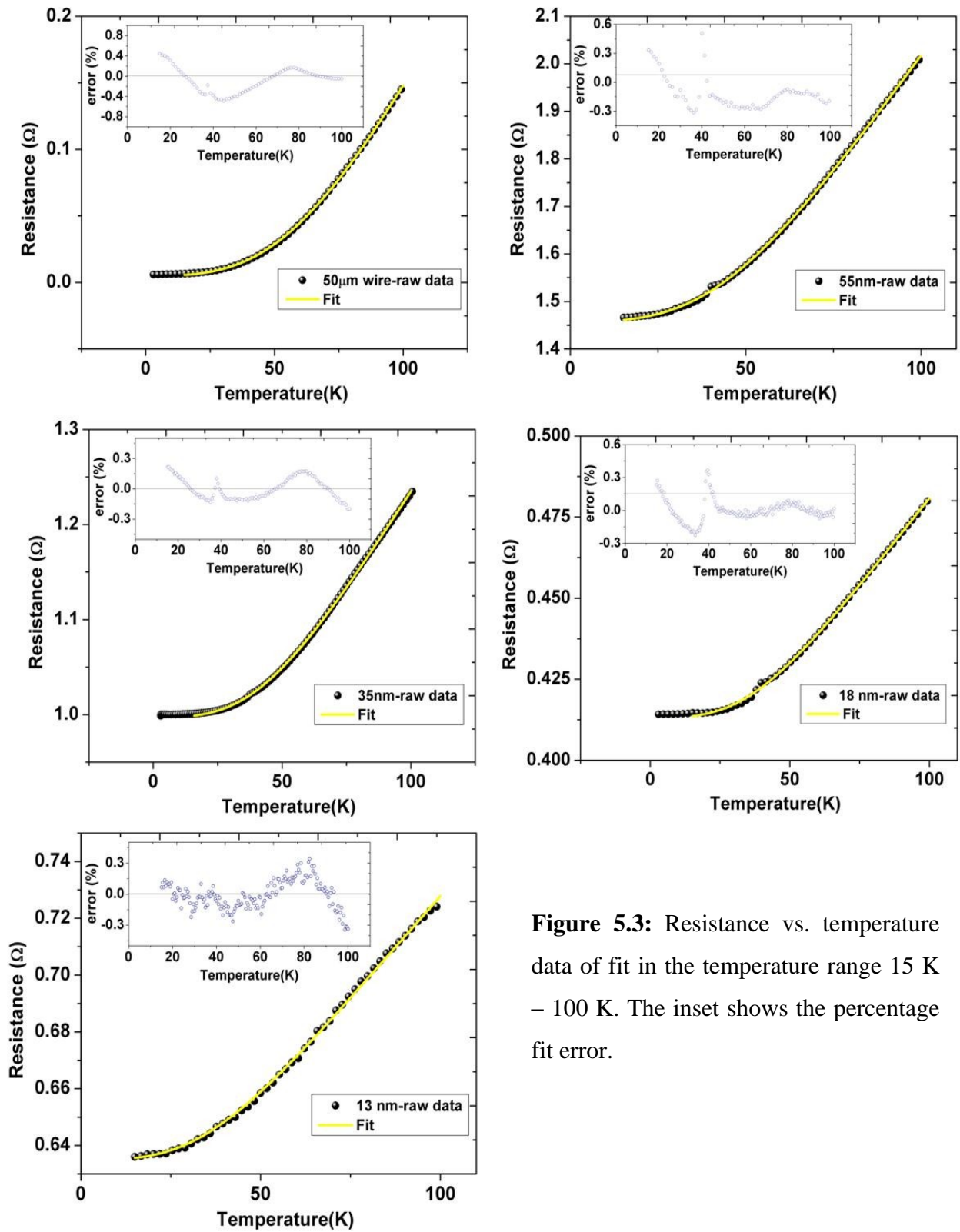

**Figure 5.3:** Resistance vs. temperature data of fit in the temperature range 15 K – 100 K. The inset shows the percentage fit error.

The Debye temperature of 471 K for the bulk wire of 50 μm diameter as estimated from the fitting matches with accepted value [10]. The values of Debye temperature obtained for wires of various diameters are tabulated in Table 5.1. We will discuss the size dependence of $\theta_R$ in the section on Discussion. We note that in the data we obtained small kinks in the data obtained for nanowires at temperature ~ 40 K. The reason for this is not very clear and we wish to do a high resolution transmission electron microscopy on our samples in future to understand if any structural transition is seen at the said temperature. However, the data taken on 55 nm diameter single nanowires (presented in Chapter-6) does not show such appreciable kink at 40 K. The reason behind it can be the considerable noise in case of the measurements performed using low current densities in case of single nanowires. Nevertheless our fitting results did not get affected by these kinks as we understood when we analyzed the removing/including the kink and the results remained the same in both the cases.

**Table 5.1** Summary of the analyzed resistance data.

| Wire Diameter (nm) | ρ4.2 K (Ωm) | ρ300 K (Ωm) | $\theta_R$ | B (ΩmK²) |
|---|---|---|---|---|
| 50000 | $0.02 \times 10^{-8}$ | $6.88 \times 10^{-8}$ | 471 | $(1.53\pm0.07) \times 10^{-13}$ |
| 55 | $3.85 \times 10^{-8}$ | $1.09 \times 10^{-7}$ | 351 | $(1.40\pm0.09) \times 10^{-13}$ |
| 35 | $6.54 \times 10^{-8}$ | $1.32 \times 10^{-7}$ | 342 | $(9.73\pm0.42) \times 10^{-14}$ |
| 18 | $1.68 \times 10^{-7}$ | $2.76 \times 10^{-7}$ | 297 | $(6.13\pm0.46) \times 10^{-14}$ |
| 13 | $1.79 \times 10^{-7}$ | $2.85 \times 10^{-7}$ | 284 | $(4.75\pm0.09) \times 10^{-14}$ |

To check whether a single parameter description of $\rho_L$ using $\theta_R$ is valid for the nanowires we used the following procedure. From Eqs. (5.1)-(5.3), we obtain the following relation:

$$\rho_L\left(T\right) = \rho - \rho_0 \tag{5.7}$$

$$\frac{\rho_L\left(T\right)}{\rho\left(\theta_R\right)} = \frac{\rho - \rho_0}{\rho\left(\theta_R\right)} = \frac{R - R_0}{R\left(\theta_R\right)} \tag{5.8}$$

This shows that if the BG equation is valid, then $\dfrac{\rho - \rho_0}{\rho(\theta_R)}$ and hence, $\dfrac{R - R_0}{R(\theta_R)}$ is a unique function of $\dfrac{T}{\theta_R}$. $R(\theta_R)$ and $\rho(\theta_R)$ are the values of phonon contribution to the resistance and the resistivity respectively evaluated at Debye temperature $\theta_R$. If indeed the ratio in Eq. (5.8) is a unique function of $\dfrac{T}{\theta_R}$, then for all the nanowires the resistance curves can be scaled into a single curve provided all of them have the same electron-phonon coupling constant $\alpha_{el\text{-}ph}$. In Fig. 5.4, we plot the data for all the nanowires $\dfrac{\rho - \rho_0}{\rho(\theta_R)}$ vs. $\dfrac{T}{\theta_R}$. It can be seen that the data for all the wires collapse into a single curve. This ensures that a single parameter description using $\theta_R$ is valid. This also confirms that $\alpha_{el\text{-}ph}$ is independent of the wire diameter. We note that size independence of $\alpha_{el\text{-}ph}$ has also been seen in two other FCC metal nanowires (Cu and Ag) investigated by our group [9]. This is an important observation and it allows us to estimate the absolute value of resistivity of nanowires as explained in the next section.

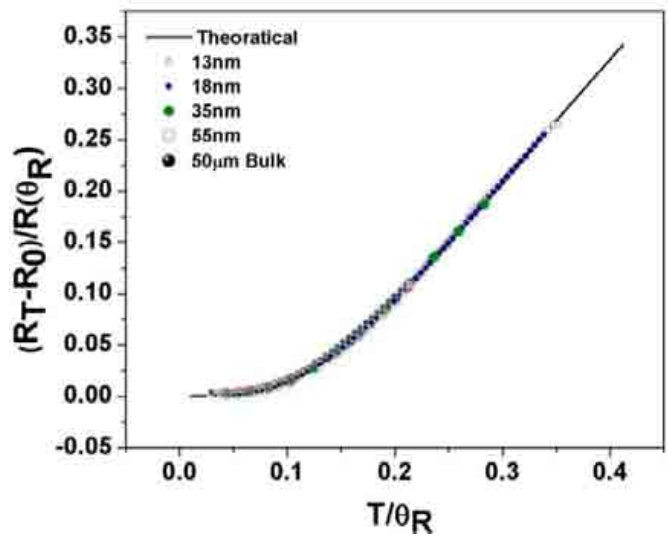

**Figure 5.4:** Plot of $R_D = \dfrac{R - R_0}{R(\theta_R)}$ as a function of $\dfrac{T}{\theta_R}$. $R(\theta_R)$ is the phonon contribution to the resistance at $T = \theta_R$.

## 5.4.4 Absolute value of resistivity

Determination of the absolute value of the resistivity ($\rho$) of the nanowire from the observed resistance $R$ of the array is not a straight forward issue because of the uncertainty in the determination of the number of wires in the array. A method has been used by our group [9] to determine the value of $\rho$ that does not need counting the number of wires. We use this method here, and more importantly also validate the method by measurement of the resistivity of a single nanowire of a given diameter. The method is a general method and uses the fact that the temperature dependence of the resistivity at $T < \theta_R$ is determined by the electron-phonon contribution only. For nanowires of non-magnetic materials (where electron-phonon interaction is the only cause of temperature dependence) this can be successfully employed over the whole range of T. For magnetic nanowires this can also be used to a temperature up to which the contribution by magnetic scattering compared to the electron-phonon scattering is negligible. We extend the method to the case of magnetic nanowires and restrict our analysis to a moderate temperature of 100 K (to avoid any considerable contribution due to spin wave scattering of both s-s and s-d electron-magnon interactions which increase as the temperature is increased) and preferred exact calculations instead of approximations which have been done earlier [9].

In the temperature of interest (15 K to 100 K), the predominant temperature dependent part of the resistivity in the magnetic nanowires arises from the electron-phonon interaction part (because Eq. (5.6) fits very well up to 100 K). This can be further understood that in nanowires (later from Fig. 5.9) where we observe a suppression in spin wave resistivity as the diameter of the wire is decreased, the temperature dependence predominantly comes from the electron-phonon interaction. As a result we can use for $T < 100$ K,

$$\frac{d\rho}{dT} = \frac{d\rho_L}{dT} \tag{5.9}$$

$$\frac{d\rho}{dT} = \alpha_{el-ph} f^1 \tag{5.10}$$

where $f^1 = \frac{df}{dT}$. If $\rho_B$ and $\rho_N$ are the resistivities of the bulk and nanowire respectively at any given temperature, then

$$\frac{\frac{d\rho_N}{dT}}{\frac{d\rho_B}{dT}} = \frac{f_N^1}{f_B^1} \tag{5.11}$$

$$\frac{\rho_N (TCR)_N}{\rho_B (TCR)_B} = \frac{f_N^1}{f_B^1} \tag{5.12}$$

$$\rho_N = \frac{f_N^1 (TCR)_B}{f_B^1 (TCR)_N} \rho_B \tag{5.13}$$

where $TCR = \frac{1}{\rho}\frac{d\rho}{dT} = \frac{1}{R}\frac{dR}{dT}$ is the temperature coefficient of resistivity and the subscripts *B* and

*N* refer to the bulk and nanowire respectively. Both these quantities are directly obtained from measurements. The $f_N^1$ and $f_B^1$ can be directly calculated since we know the values of the respective $\theta_R$. Thus all the quantities on the right hand side of Eq. (5.13) are known. In our case the resistivity of bulk Nickel wire was $8.6 \times 10^{-9}$ $\Omega$m at 100 K. Using this value, the resistivity of nanowires were calculated in the temperature range of interest using Eq. (5.13). Fig. 5.5 shows the resistivities of the nanowires evaluated. We name this method of estimation of resistivity a **"slope method".** Though the resistivity of the nanowires is of the same order as reported earlier [18], our samples are less resistive because of better crystallinity.

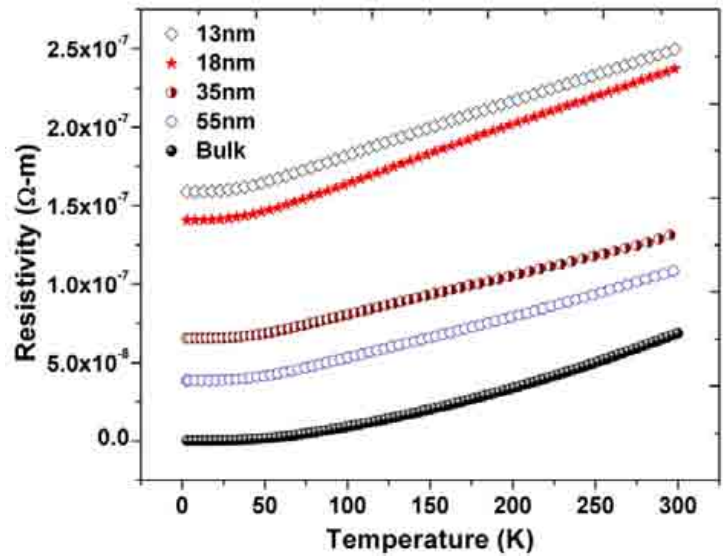

**Figure 5.5:** Resistivities of the nanowires as compared with the bulk wire.

## 5.4.5 Resistivity at 4.2 K and surface scattering

The determination of the absolute value of the resistivity, and in particular, the absolute value at 4.2 K ($\rho_{4.2\ K}$) allows us to investigate the role of surface scattering in determination of the residual resistivity. To understand the effect of surface scattering, we have fit $\rho_{4.2\ K}$ with the surface scattering model [19] for wires of diameter $d \ll l$ given as

$$\frac{\rho(d)_{4.2K}}{\rho_0} = \frac{1-p}{1+p}\frac{l}{d} \qquad (5.14)$$

where $\rho_{4.2\ K}$ and $\rho_0$ are the resistivity of the nanowire of diameter $d$ and that of the bulk metal respectively at 4.2 K and $l$ is the electron mean free path of the bulk wire. $p$ is the specularity coefficient, which is the fraction of electrons getting elastically scattered from the wire boundary. The dependence of $\rho(d)_{4.2\ K}$ on $d$ can be seen in Fig. 5.6. From the fit to Eq. (5.14) (shown in Fig. 5.6) we obtained $p = 0.05$. This $p$ has been assumed to be more or less the same for all the wires. The fit of the data to Eq. (5.14), validates this assumption. The value of $p$ obtained by us is much less in comparison to that in noble metals.

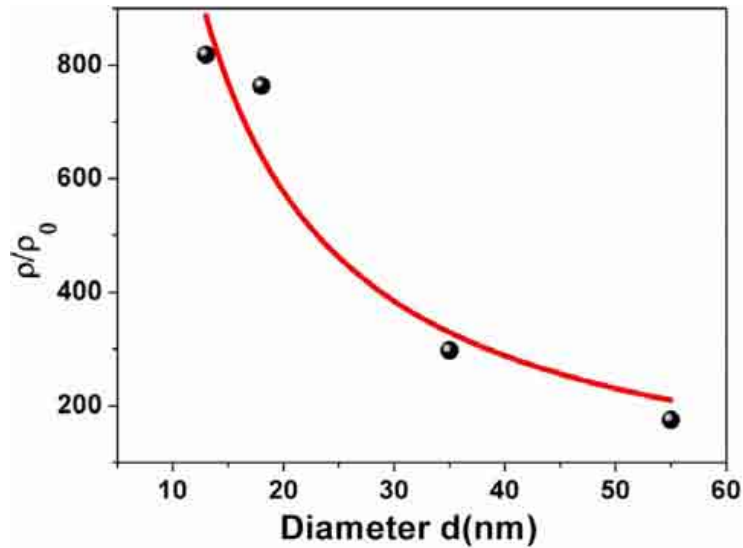

**Figure 5.6:** Fit of Eq.(14) to the resistivity data of the nanowires at $T$ = 4.2 K.

Determination of the specular coefficient $p$ allows us to find the mean free path of electrons ($l_d$) in a nanowire of diameter $d$ at 4.2 K. The following relation [20], valid for $d << l$, was used.

$$l_d = \frac{1+p}{1-p}\ d \qquad (5.15)$$

From Eq. (5.15), with $p = 0.05$, the electron mean free path in nanowires at 4.2 K comes out to be 1.1 times the diameter of the nanowires. This is an interesting result because the mean free path of the electrons in the nanowires is as large as the diameter of the wire. Since the wire diameter puts a limit to electron mean free path, the surface scattering is dominant. The low value of $p$ implies that the surface scattering is diffused in nature. The above observation shows that the electrons do not suffer appreciable scattering within the volume of the nanowire since the nanowires are single crystalline.

## 5.4.6 Magnetic contribution to the resistivity

At very low temperatures ($T < 15$ K) the magnetic part varies with temperature as $\rho_M = BT^2$ in 3d ferromagnets as given in Eq. (5.5). The resistivity data for temperatures $T < 15$ K is analyzed using Eq. (5.4) to estimate the contribution due to magnetic scattering arising out of spin fluctuations.

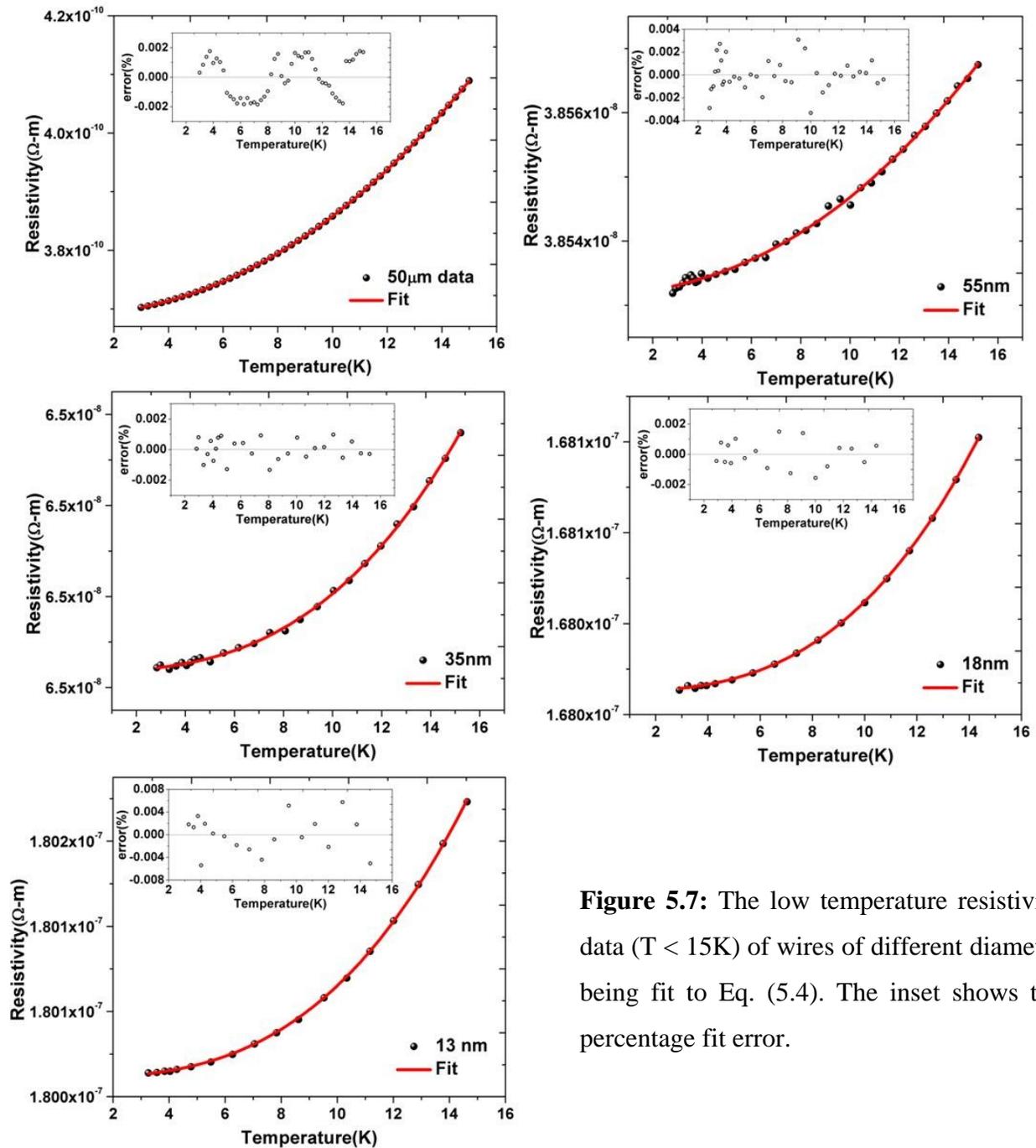

**Figure 5.7:** The low temperature resistivity data (T < 15K) of wires of different diameter being fit to Eq. (5.4). The inset shows the percentage fit error.

For the case of bulk nickel wire (50 μm), we found that the resistivity $\rho$ indeed has a quadratic dependence on $T$ and $B = 1.53 \times 10^{-13}$ Ω.m K$^{-2}$ which is in excellent agreement with previous theoretical and experimental reports [11, 12]. Fitting results of Eq. (5.4) are shown in Fig. 5.7 for wires of various diameters. The inset shows the fit percentage error, which is less than 0.001%. In Fig. 5.8 we show $\rho_M$ ($T < 15$ K), the magnetic contribution to the resistivity, (obtained after removing the residual and the lattice part of resistivity) of the nanowires below $T < 15$ K. Since below the mentioned temperature the magnetic contribution to the resistivity is mainly due to *s-s* electron-magnon kind which follows Eq. (5.5), we plot $\rho_M$ ($T < 15$ K) as a function of $T^2$. For all the nanowires the $T^2$ dependence persists, and there is a systematic decrease of $B$ as the diameter is reduced (Fig. 5.10(b)). $B$ is considerably lower for lowest diameter wires used in our study compared to that in the bulk. The values of $B$ are tabulated in Table 5.1.

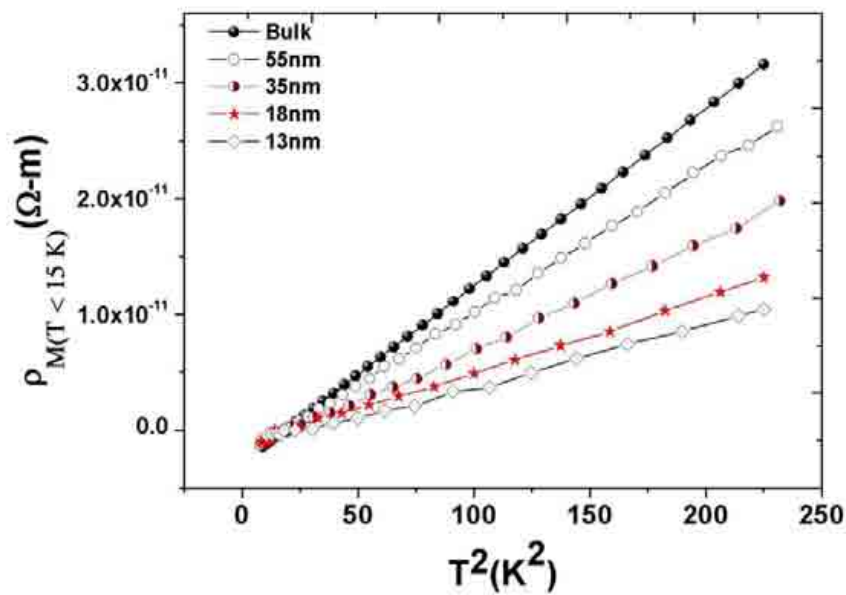

**Figure 5.8:** Plot of magnetic part of resistivity, $\rho_M$ ($T < 15$ K) versus $T^2$ for $T < 15$ K.

For temperatures $T > 15$ K and above, the magnetic contribution to the resistivity, $\rho_M$ ($T > 15$ K) is expected to show a complicated behavior due to both intra band (*s-s*) and interband (*s-d*) kind of electron-magnon scattering[21]. To assess the contribution of the magnetic scattering in Fig. 5.9 we show the $\rho_M$ ($T > 15$ K), which is obtained from the observed resistivity after removing the residual resistivity and the lattice part of resistivity. This is plot as a function of $T$.

In this case also we see a suppression of magnetic contribution over the complete temperature range as the wire diameter is lowered. In the bulk wire there is sufficient contribution of magnetic scattering for $T > 100$ K. In nanowires it is much less compared to the temperature dependent lattice part of the resistivity ($\rho_L$).

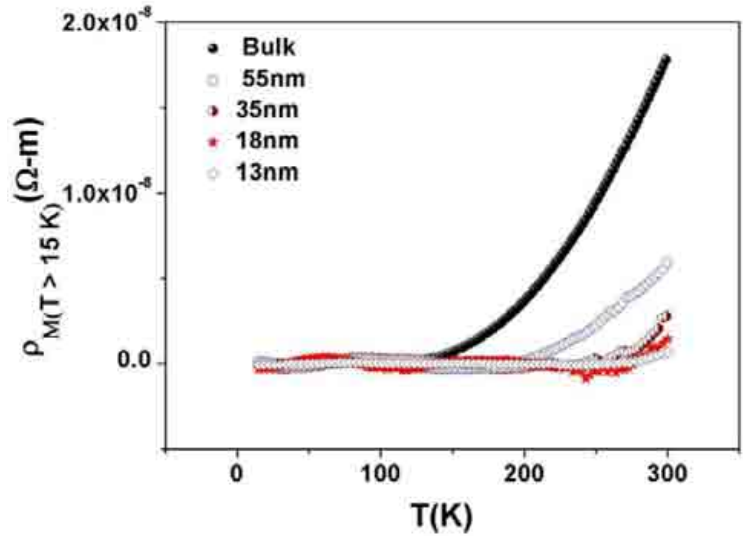

**Figure 5.9:** Plot of magnetic part of resistivity, $\rho_M$ ($T > 15$ K) versus $T$ for $T > 15$ K.

## 5.5 Discussion

The Debye temperature ($\theta_D$) of a solid as determined from the heat capacity is determined mainly by the sound velocity and, hence, the elastic modulus. An enhancement of the modulus will enhance the sound velocity and hence $\theta_D$. Similar consideration applies to the Debye temperature determined from the resistivity ($\theta_R$). This work establishes that proper analysis of the resistivity data gives a viable way to evaluate the Debye temperature. In Fig. 10(a) we collect the data on Ni determined in this investigation as well as the data obtained by our group on two other FCC metals, Cu and Ag. The data are plotted as the ratio $\left( \dfrac{\theta_R(d)}{\theta_R(Bulk)} \right)$ where the symbols in brackets refer to the diameter $d$ and the bulk solid respectively. It can be seen that in all the three solids the ratio decreases as $d$ is reduced (For the two Ag nanowires of smallest diameter, the wires were not FCC but had presence of HCP phase as well.). The results show that the reduction in $\theta_R(d)$ would indicate that the elastic modulus decreases on reduction of $d$. The fact that in all the three FCC solids similar behavior are observed establish that the diameter dependence is robust. Previous experimental studies, those which addressed the issue of size dependence of modulus or the Debye temperature were done using nano-indentation or by looking at force-displacement relation using contact mode Atomic Force Microscopy (AFM). In terms of dependence of elastic modulus on size in these materials experimental studies offer no convergence. For instance in hydrothermally synthesized Ag nanowires [22] (crystallinity not checked) the elastic modulus increased on size reduction as determined by AFM. For very thin films of Ni [23] and Cr[24], the modulus as determined by AFM were found to decrease on size reduction. Determination of Young's modulus in Ag[25] and Au[26] nanowires showed no clear trend as the error bars were large. In many of these experiments the nanowires were polycrystalline or their state of crystallinity were not determined. Interestingly in a very early [27] determination of heat capacity of lead nanoparticles there was observation of enhancement of the $T^3$ term and hence lowering of $\theta_D$. We note that in ultrathin Au films grown epitaxially (thickness 2 nm -70 nm)[28], the Debye temperature ($\theta_R$) was evaluated from the $\rho$ vs. T curve using BG theory and $\theta_R$ decreases as diameter ($d$) is reduced. In the present investigation the use of resistivity to find investigate the Debye temperature of metallic nanowires is, thus, a fresh approach and it allows higher precision. The nanowires used here also were well characterized and were single crystalline in nature.

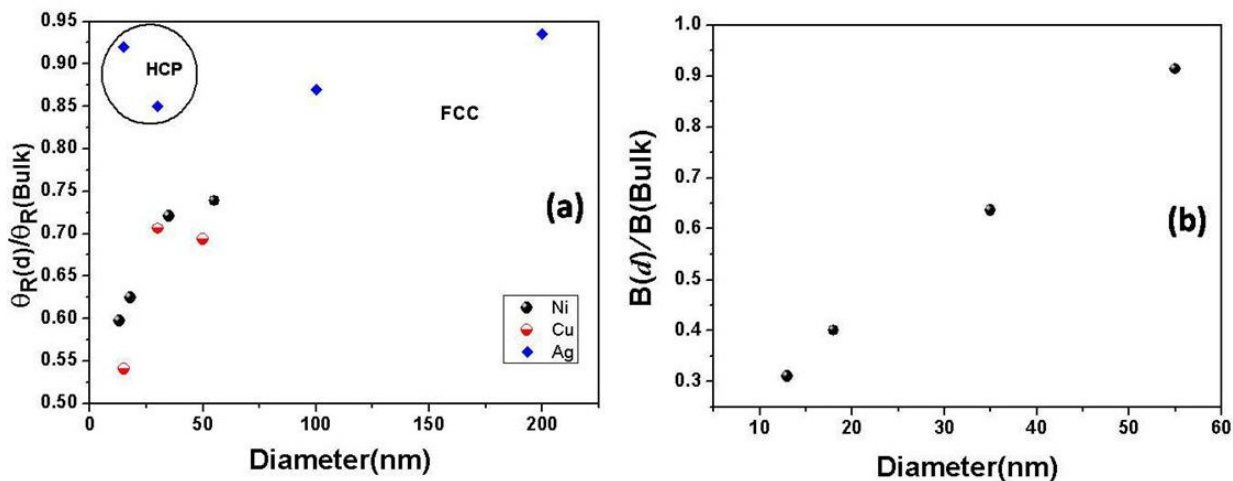

**Figure 5.10:** **(a)** Normalized value of Debye temperature $\left( \dfrac{\theta_R(d)}{\theta_R(Bulk)} \right)$ of Ni, Cu, Ag nanowires as a function of the wire diameter **(b)** Normalized value of magnetic resistivity constant $B$ $\left( \dfrac{B(d)}{B(Bulk)} \right)$ as a function of the wire diameter $d$.

The theoretical reasons for lowering of modulus of the elasticity has been attributed to different types of surface contributions [29, 30]. For example, using bond-order-length-strength correlation mechanism it was shown that the Young's modulus as a function of size will reduce or increase as the size is reduced depending on the temperature as well as the change in bond energy in the surface [29]. Using the parameters for metal nanowires, it can be shown that the weakening of surface energy and the enhanced contribution of surface bonds on smaller diameter sample are expected to reduce the elasticity modulus on size reduction at low temperatures. Similar results were also obtained from cohesive energy calculations [30]. These theoretical predictions agree very well with our observation of reduction of $\theta_R$ in FCC metal nanowires.

The observation of decrease in $B$ by a large factor when the size is reduced can arise mainly from two factors: (a) A reduction in the $s$-$d$ interaction $NJ(0)$ and (b) a reduction of the magnon effective mass $\mu$. The last alternative seems to be less likely because the spring stiffness constant $D \propto \mu^{-1}$. This implies that size reduction leads to enhancement of $D$. However, in a ferromagnet, $T_C \sim D$. Thus, an enhancement of $D$ on size reduction looks unlikely because experimentally we find that size reduction actually leads to a reduction in $T_C$. It thus appears that our experiment suggests that $NJ(0)$ is reduced on size reduction. We suggest this as a likely mechanism. In nanowires, the contribution of the spins at the surface becomes increasingly important as the size is reduced. These spins sitting in a disordered environment near or at the surface may not support long wavelength propagating spinwaves that are needed for the temperature dependent resistivity of the type that gives $\rho_M$. Instead, they may be overdamped spinwaves which will not make any contribution to the spin wave scattering. This will reduce $N$ effectively and will reduce $NJ(0)$. Since $NJ(0)$ appears as a quadratic term, a small decrease can lead to a relatively larger effect in $B$ (The damped surface spinwaves, however, can lead to a static temperature independent scattering. The fact that we see very large diffused electron scattering from the surface leading to low specularity factor $p$ may suggest that this indeed is the case).

## 5.6 Conclusion

In conclusion, we have synthesized Nickel nanowires within anodic alumina membranes with pore diameter varying from 55 nm to 13 nm. Resistance measurements were carried out in the temperature range 3 K to 300 K. The resistance data in the temperature range 15 K-100 K has been analyzed using Bloch-Wilson function. The values of Debye temperatures obtained for nanowires decrease with a decrease in the wire diameter which we consistently find for other FCC metal nanowires. However, we find the electron-phonon coupling constant is nearly unchanged on size reduction down to 13 nm. The experiment allowed us to determine the absolute value of the residual resistivity (at 4.2 K). Using the Dingle and Sondheimer theories, we analyzed the data and found that the surface scattering plays a major role in the high resistivity values in case of nanowires. The electron mean free path is limited mainly by the diffused surface scattering and the highly crystalline nature of the wire ensures that there are negligible scattering of electrons by defects within the bulk of the wires. The data for temperature $T < 15$ K, can be analyzed including the magnetic scattering term and observed a large change as one goes down to the nanoscale. Even for temperature $T > 15$ K, we observed a decrease in the magnetic contribution. Thus, over the whole temperature range, we observed a suppression in the magnetic contribution to the resistivity as we lower the wire diameter.

# CHAPTER 6

# Temperature dependent electrical resistivity of a single strand of ferromagnetic nanowire (single crystalline)

We have measured the electrical resistivity of a <u>single strand</u> of a ferromagnetic Ni nanowire of diameter 55 nm using a 4-probe method in the temperature range 3K-300 K. The wire used is chemically pure and is a high quality oriented single crystalline sample. The four probes were attached using FIB assisted Pt deposition. Precise evaluation of the temperature dependent resistivity ($\rho$) allowed us to identify quantitatively the electron-phonon contribution (characterized by a Debye temperature ($\theta_R$) as well as the spin-wave contribution which is significantly suppressed upon size reduction. The temperature independent residual resistivity was determined predominantly by surface scattering. The differences as well as similarities between the results of single nanowire measurement and measurements made on arrays of nanowires are discussed.

This page is intentionally left blank

## 6.1 Introduction

The resistivity ($\rho$) of magnetic nanowires is of immense interest both from scientific and technological points of view. The reduction in size can lead to both qualitative and quantitative changes in comparison to electrical transport in wires of much larger dimensions. While a lot of work has been reported on the magnetism in the ferromagnetic nanowires, the electrical transport (in particular, the temperature dependent part) is largely unexplored. In particular, there has been lack of extensive data on high quality single crystalline ferromagnetic nanowires over a large temperature range that allows a quantitative evaluation of the resistivity data and the contributions from different sources of electron scattering. In the last chapter we have presented the measurements done on arrays of nanowires of single crystalline nickel nanowires of various diameters by a pseudo four probe method. The pseudo four method of resistance measurement involves an additional contribution of contact resistance. Though this is not much a problem because of low contact resistance when compared with the actual resistance of the sample, it is equally important to verify the physics in a single nanowire instead of ensemble of nanowires. This is to address an important question- how does the observed results in arrays tally with the results obtained in single nanowire measurements. The electrical and thermal measurements on single nickel nanowire with lateral dimension 100 nm X 180 nm, have been reported [1] earlier. However these measurements were carried out on a polycrystalline wire with much higher resistivity which makes the relative contribution of the temperature dependent terms much weaker and the evaluation of magnetic contribution in particular becomes difficult. We present, in this chapter, a concise study of electrical transport over the temperature range 3 K to 300 K in an oriented ((220)), single crystalline cylindrical nickel nanowire of much smaller diameter (55 nm). The single crystallinity is important because we can avoid the grain boundary contribution to scattering which often leads to high residual resistivity and thus masks other important phenomena at low temperatures. The reduction of grain boundary scattering leads to lowering of residual resistivity so that we can observe the temperature dependent term that arises from the spin wave contribution typically below 15K. If the residual resistivity is not lowered one may not see this term. The results and the analysis presented below allow us to clearly separate out all the contributions to the resistivity quantitatively. The results show that in such single crystalline nanowires the electrons can reach a mean-free path of the order of 1.1 times the diameter. The size reduction leads to a reduction of the Debye temperature ($\theta_R$) by nearly 30% and substantial reduction of the magnetic contribution to electrical resistivity, which has never been reported before.

## 6.2 Synthesis and Characterization

The nanowires used in this experiment were prepared by pulsed potentiostatic electrodeposition of Ni inside commercially[2] obtained nanoporous anodic alumina templates of thickness ~ 56 μm with average pore diameter of ~ 55 nm . The deposition was carried out in a bath containing a 300 g/l $NiSO_4.6H_2O$, 45 g/l $NiCl_2.6H_2O$, 45 g/l $H_3BO_3$ electrolyte with the working electrode (a 200 nm silver layer evaporated on one side of the template) at a pulse potential of -1 V with respect to the reference electrode (Saturated Calomel), with 80% duty cycle and a pulse period of 1 second. The nanowires formed inside the cylindrical pores of the templates were characterized by X-ray Diffraction XRD) and Transmission Electron Microscopy (TEM) and found to be single crystalline FCC with a preferential growth direction along (220) direction. The representative TEM data and the XRD data are shown in Fig. 6.1 (a) and 6.1(b) respectively. The wires have average diameter of ≈ 55 nm as measured from the TEM data.

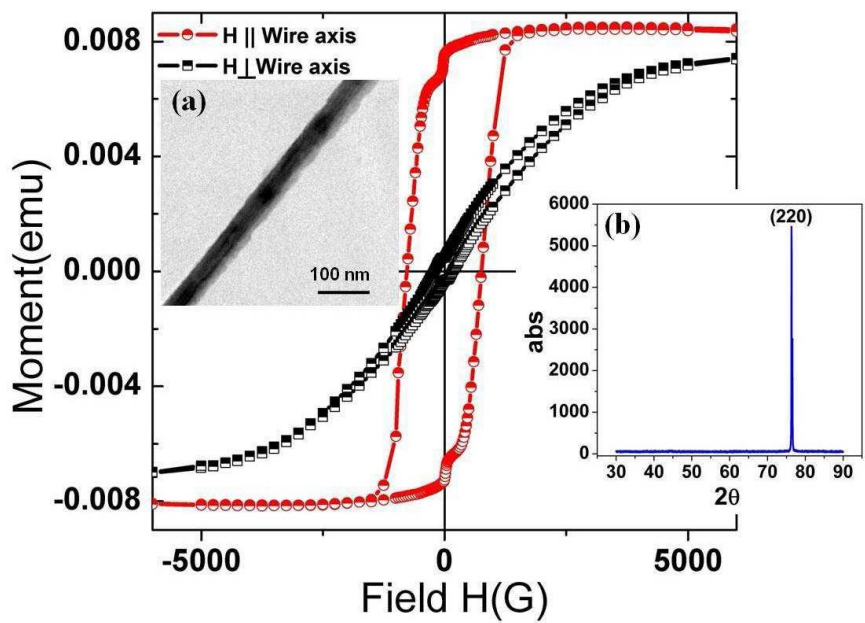

**Figure 6.1:** M-H curves of the nanowire arrays with measuring field (H) parallel and perpendicular to the wire axis. Inset (a) TEM of 55 nm oriented Ni nanowire. Inset (b) XRD of the sample.

The high resolution TEM data show absence of grain boundary over the length of the wire. The wires so grown are ferromagnetic, as established by the M-H curves. The typical M-H curves at 300 K shown in Fig. 6.1 were taken at H parallel and perpendicular to the wire long axis by retaining the Ni wires in the alumina template. The M-H curve reveals the highly anisotropic magnetic nature of the array of wire. It establishes that the wires are strongly ferromagnetic and have their spins aligned to the axis of the wire. Fig. 6.2 shows a HRTEM image of a 55nm diameter nanowire lateral edge. The lattice fringes can be clearly seen in the image.

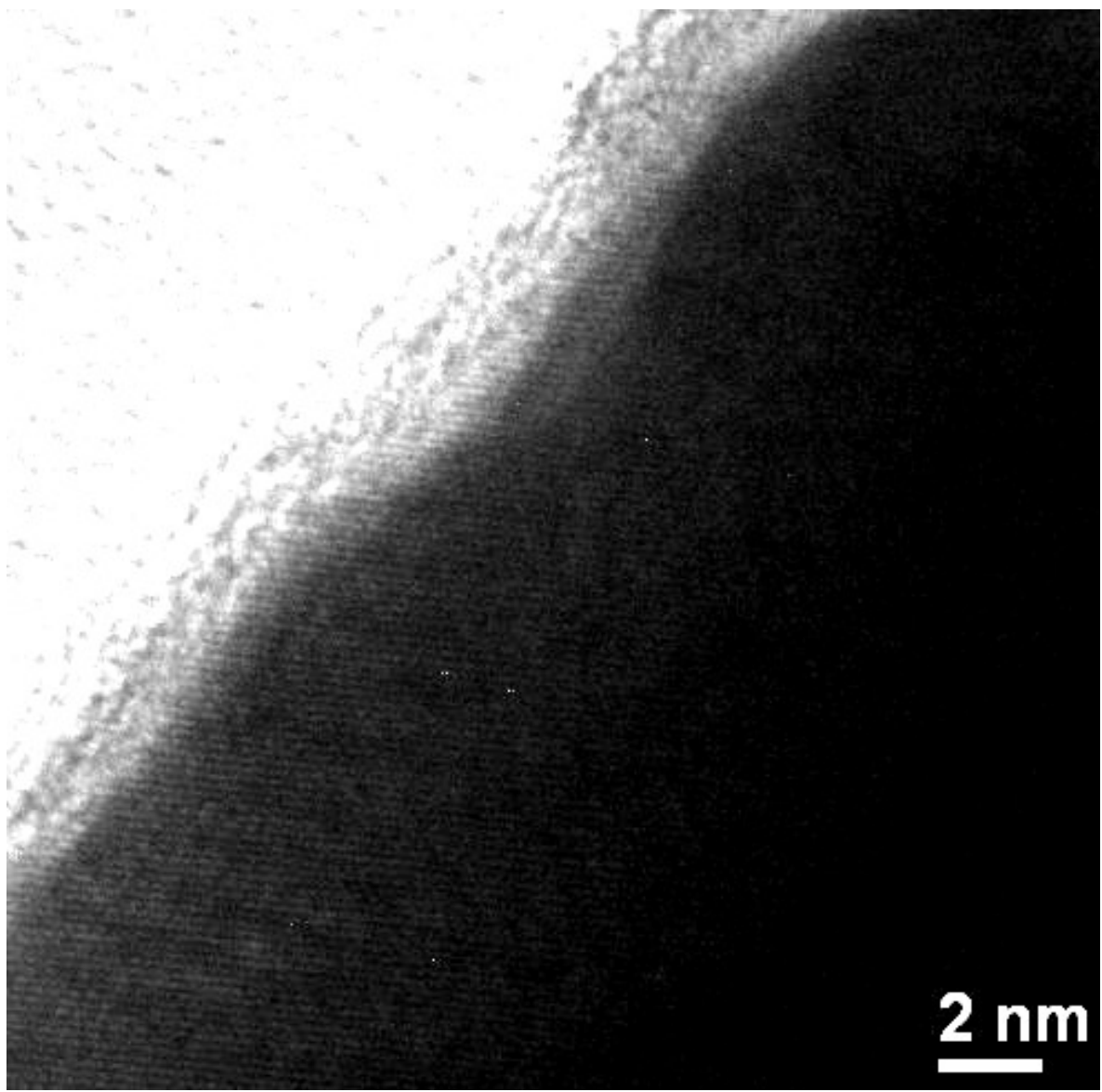

**Figure 6.2:** HRTEM image of a 55 nm diameter nickel nanowire.

## 6.3 Electrical transport

The sample preparation and contact making methods consisting of various steps are clearly described in chapter 3 of this thesis. Here, we give a brief overview once again for the sake of completeness. For electrical measurements on a single nanowire, the wires were removed from the template by dissolving the latter in a 6M NaOH solution and subsequently washing with millipore water several times. One to two drops of the suspension containing the Ni nanowires were sprayed in the middle portion of a silicon substrate (with 300 nm oxide layer) containing gold contact pads of thickness 500 nm which we made by UV lithography. A relatively long nanowire was chosen under the electron microscope and the probes (~ 750 nm wide and 300 nm thick) were attached to the nanowire connecting them to the bigger gold contact pads by focused ion beam (FIB) assisted platinum deposition.

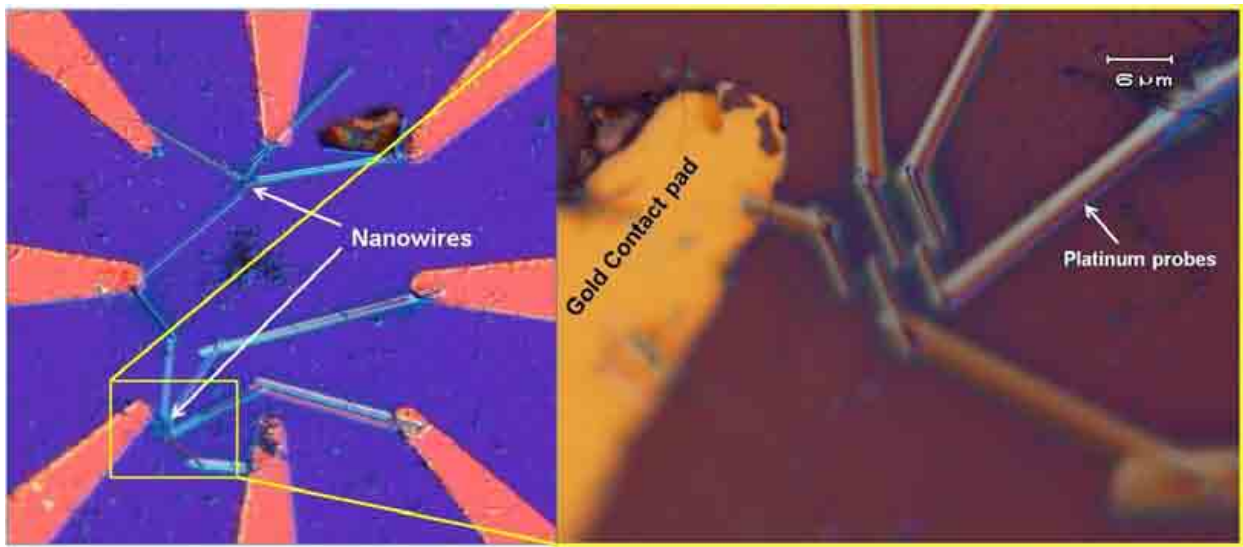

**Figure 6.3:** Optical microscope images of the device with gold contact pads and platinum contacts to the nanowire at different magnification.

Fig. 6.3 shows the optical microscope images of two single nanowires (One with four platinum probes and other with five contact probes) along with the macroscopic gold contact pads prepared by photolithography as explained in details in chapter 3. Fig. 6.4 shows the SEM image of a nickel nanowire of diameter 55 nm with 5 probes on it (the same region shown in optical microscope). 4-probes were used for the measurement of the resistance of the single strand of the nanowire. The details are explained in chapter 2 section 3.2.1.

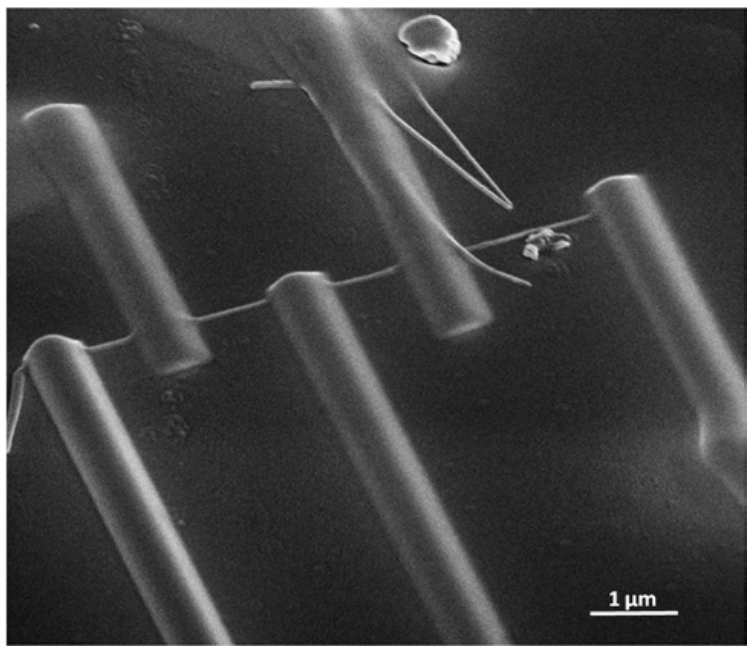

**Figure 6.4:** Scanning Electron Microscope image of the nanowire connected to 5 Pt probes made using FIB assisted platinum deposition.

To avoid damage due to electro-migration we used an AC signal with a low current amplitude of $10^{-6}$A (current density $\approx 4$ X $10^{8}$ A.m$^{-2}$) with a frequency of 174.73 Hz. The resistance (typically 20-30 Ohms) was measured using a phase sensitive detection scheme with a resolution nearly 2ppm.

## 6.4 Results and analysis

The normalized electrical resistivity of the single Ni nanowire measured from 3 K-300 K as compared to a 50 μm thick nickel wire which is the "bulk" reference is shown in Fig. 6.5. It should be noted that this is the first report of electrical measurement on a single strand of an oriented single crystalline nickel nanowire of this dimension. It is evident from Fig. 6.5 that the residual resistivity ratio (RRR) of the single nanowire $\approx 2.3$ is much less than that of the bulk wire ($\approx 312$). The residual resistivity ($4.45 \times 10^{-8}$ Ω.m) of the nanowire (even though lower in diameter) is much lesser than reported earlier ($17.6 \times 10^{-8}\,\Omega.m$) [1] because of its better crystallinity.

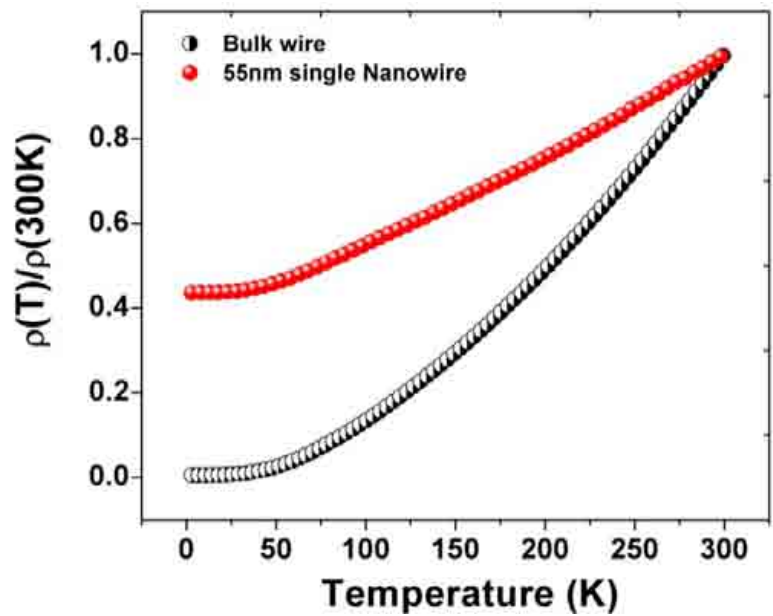

**Figure 6.5:** The normalized electrical resistivity of the single Ni nanowire measured from 3 K-300 K as compared to a 50 μm thick nickel wire.

The resistivity as a function of temperature of the single nanowire is shown in Fig. 6.6. The data is fitted to Eq. (5.1) (refer to chapter 5). We carried out the same kind of analysis as we did in case of arrays of nickel nanowire in the last chapter. For 15 K < $T$ < 100 K, the temperature dependence arises mainly from the lattice contribution $\rho_L$ when compared with $\rho_M$. Nevertheless at high temperatures ($T$ > 150 K), $\rho_M$ becomes significant again. It can be seen from Fig. 6.6, the deviation of the Bloch Wilson fit from the data arises due the magnetic scattering which eventually gives rise to the critical contribution near $T_C$. In the temperature range 15 K- 100 K we could use the BG formula (Eq. (5.3) with n=3) to fit the behavior of $\rho(T)$ neglecting $\rho_M$. We obtained reasonably good fit (Fig. 6.6) with an error of less than 0.3% shown as an inset in Fig. 6.6. The data thus can be described by a single parameter, namely the Debye temperature ($\theta_R$), which as obtained from our analysis is 345 K which is close to the value of 351 K obtained from analysis of resistivity data taken in an for arrays of nanowires of the same diameter. This is less than that of the value 471 K for our reference bulk wire which matches with those obtained from specific heat measurements (475 K) [3].

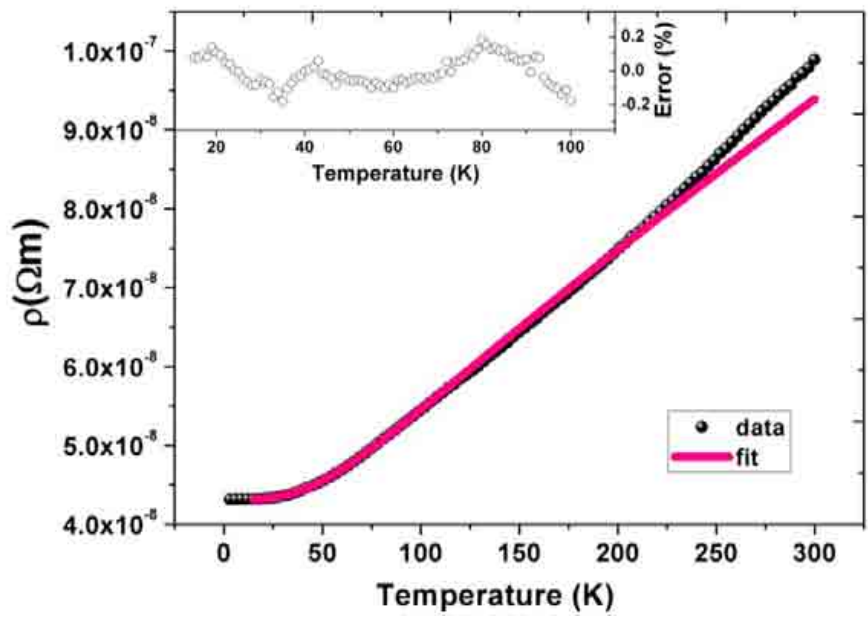

**Figure 6.6:** Electrical resistivity of the single Ni nanowire along with the fit to the Bloch-Wilson relation up to 100 K. The inset shows the fit error (%).

The data for temperature $T < 15$ K was fitted with Eq. (5.5) with the magnetic contribution $\rho_M = BT^2$.

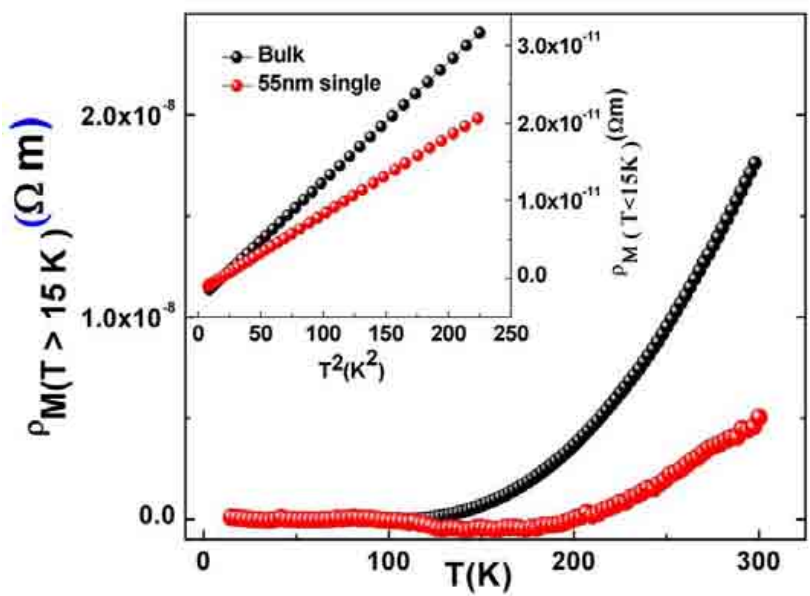

**Figure 6.7:** Magnetic part of resistivity $\rho_M = \rho(T) - \rho_0 - \rho_L$ for $T > 15$ K. The inset shows the magnetic part of resistivity with quadratic temperature dependence for $T < 15$ K.

The constant $B$ for the single nanowire was found to be 1.01 X $10^{-13}$ $\Omega$.mK$^{-2}$. Similar analysis for the reference bulk wire data gave $B = 1.529$ X $10^{-13}$ $\Omega$.mK$^{-2}$ which is close to the values reported in bulk nickel [4, 5]. The data on the nanowire thus shows a significant reduction in the magnetic contribution at low T as shown in the inset of Fig. 6.7. The magnetic contribution ($\rho_M$) for $T > 15$ K is evaluated by subtracting out the extrapolated $\rho_L$ and $\rho_0$ from the total measured $\rho$ as shown in Fig. 6.7. The reference bulk data are also shown for comparison. $\rho_M$ has a negligible value in the temperature range 15 K -150 K. For $T > 150$ K also there is a significant suppression of the magnetic scattering in comparison to that seen in the reference bulk wire. We can clearly see an effective decrease in the magnetic contribution to the resistivity as one goes from bulk to nanowires. This suppression of the magnetic contribution $\rho_M$ in the nanowires is a new observation that has not been reported earlier. The physical reason for the reduction of $B$ has been discussed in previous chapter.

To evaluate the effect of surface scattering, we have used the surface scattering model [6] given for wires of diameter $d << l$, $l$ being the electron mean free path in the bulk sample as described in Chapter 5. Using Eq. (5.14), we obtained $p = 0.018$ (as compared to $p = 0.05$ in case of measurement done on array of nanowires), the specularity coefficient, which is the fraction of electrons getting elastically scattered from the wire boundary ($p = 1$ for completely specular surface and for diffused scattering $p \rightarrow 0$). With $p = 0.018$, we estimated the mean free path at 4.2 K given by [7] as we did in case of arrays in Chapter 5 using Eq. (5.15). The mean free path $l_{NW} \approx 1.037d$ implies that the mean free path is determined predominantly by surface scattering and the electrons do not suffer significant scattering within the volume of the nanowire because of high purity and fewer defects.

## 6.5 Discussion

The most significant thing to be observed here is how the measurement on single nanowire compares with that obtained from measurement on arrays. In case of a single nanowire, the measurement is carried out on an insulating substrate where the wire is relatively free to volume change under temperature change. However in case of the measurement done on arrays of nanowires, there is always a strain arising out of unequal volume changes of the wire material and the matrix in which the wires are embedded. However in this whole exercise of data analysis we found the results obtained in case of single nanowire to be very close to the results obtained in case of arrays of nanowires of the same diameter. In Fig. 6.8 we show the resistivity of single nanowire and array of nanowires compared with bulk wire. The inset shows the normalized plot of resistivities.

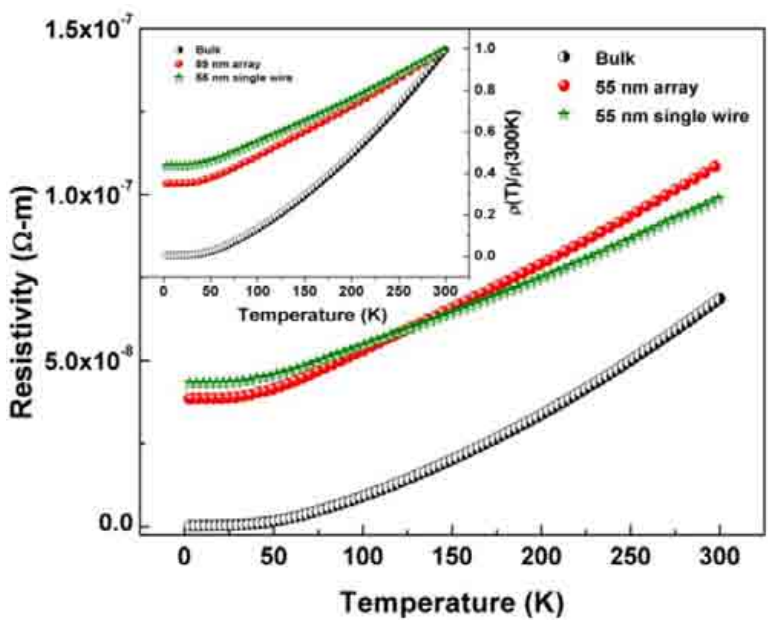

**Figure 6.8:** Resistivity of single nanowire and array of nanowires compared with bulk wire. The inset shows the normalized plot of resistivities.

One can see a difference in the slope of the resistivity vs. temperature data of single nanowire and array of nanowires with the temperature coefficient of resistance less for single nanowire. However, the absolute values of the resistivities are very close and differ by less 15% throughout the whole temperature range of measurement. The different configuration of the wires can bring about these small differences. Moreover, in case of arrays of nanowires, the aspect ratio of nanowires is nearly 60 times more than that in case of single nanowire. The strain imposed by the alumina matrix can also be attributed to this difference.

Another important attempt we did in this work is to understand the effectiveness of the **slope method** in the determination of resistivity from raw resistance data without the use of sample dimensions (as employed in case of arrays of nanowires in last chapter). We used the **slope method (Eq. (5.13))** described in the chapter 5, which uses the Debye temperature and temperature coefficient of resistance to estimate the resistivity of single nanowire raw resistance data. Surprisingly the resistivity obtained from this indirect way of estimation is found to be 3.2% higher in the single nanowire case. The result shown in Fig. 6.9 proves that the slope method is indeed effective in estimation of resistivity when the actual dimension is not known.

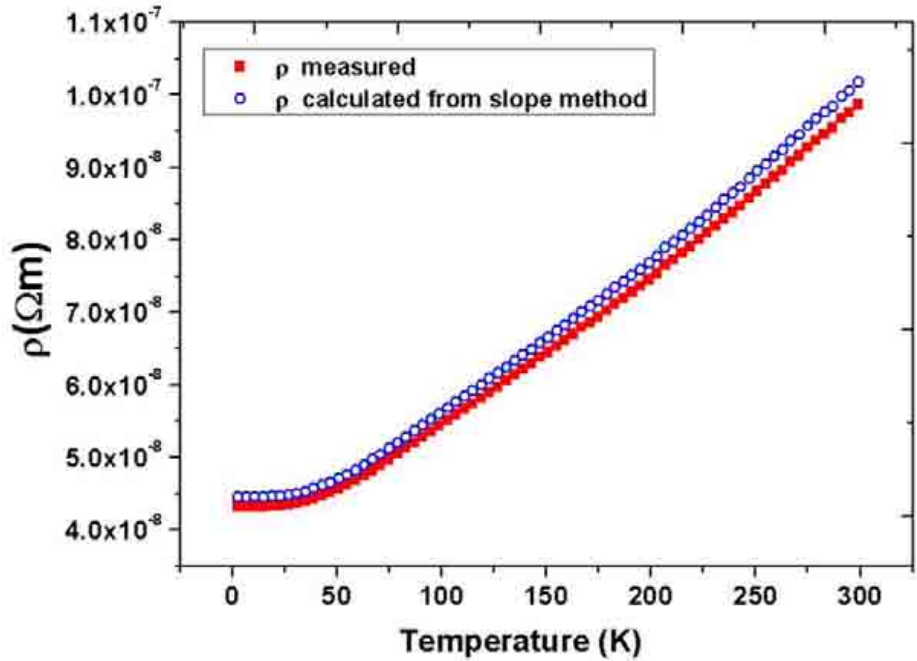

**Figure 6.9:** A comparison of the resistivity measured from the dimensions of the wire to the resistivity obtained from the indirect slope method.

## 6.6 Conclusion

In conclusion, we have reported the electrical resistivity in a single crystalline and oriented Ni nanowire. The resistivity in the temperature range 3 K - 300 K was measured by a four-probe method with FIB-deposited Pt contacts. The single crystalline nature of the wire ensures that the temperature independent residual resistivity is determined mainly by the surface scattering. We find that the decrease in diameter significantly decreases the Debye temperature ($\theta_R$). The magnetic part of the resistivity is remarkably suppressed in the case of the nanowire as seen in case of measurement on array of nanowires of same diameter. The single crystalline Ni nanowires with the temperature dependent resistivity being almost linear for $T > 100$ K and predominantly determined by a single parameter $\theta_R$, might be excellent temperature sensors with nanometric dimensions and thus, with very rapid response time.

# CHAPTER 7

## Synthesis of copper nanotubes and their electrical transport properties

*In this chapter, we present a novel, versatile and general approach for preparing metal nanotube (MNT) arrays. The method uses a template like anodic alumina and has full control on length, diameter and wall thickness of the nanotube. The method is general enough and in principle, can be applied to prepare single metal nanotubes of all metals which can be deposited by electrodeposition technique. This method in principle can also be used to deposit nanotubes of many semiconductors which can be prepared by electrodeposition. As a generic example, we present the synthesis of single crystalline copper nanotube arrays by electrodepositing copper into the pores of porous anodic alumina template. In the later part of this chapter we discuss the electrical transport in these nanotube arrays.*

## 7.1 Introduction

The discovery of carbon nanotubes[1] in 1991 initiated the interest on tubular nanostructures because of their immense fundamental importance as well as their potential applications in the nanoscale devices, sensors, catalysis, thermal materials, structural composites, field emission, biomedicine and energy storage/conversion[2-7]. Porous membranes such as anodic alumina, polycarbonate membranes, block copolymers etc. have opened up the possibilities of the synthesis of arrays of ordered nanowires of a number of materials. Use of electrodeposition to synthesize ordered arrays of nanowires in such templates is well studied. As mentioned in the chapter 1, there have been very few reports of synthesis of metal nanotube arrays. **There are very few reports of the synthesis of arrays of metal nanotubes[8-15] using nanoporous templates**. Most of these earlier works in this area involve chemical modification of the pore surface of porous templates to enhance the deposition of the metal on the pores. Such chemical modifications often add impurities to the nanotubes[12, 13]. These methods are often specific for a particular kind of metal to be deposited and cannot be used for a variety of materials. Often it is also specific to an application. Another method [16, 17], where the nanotubes of metals are synthesised by electrodeposition inside the pores of a partially coated template where the working electrode is a thin deposited noble metal. However, there are problems associated with time of coating and often the tubes are of diameter > 200 nm. The processes are time consuming because of the complexities involved[12, 13]. Thus a general and efficient method for the synthesis of metal nanotubes (with controlled length, diameter and wall thickness) still remains a challenge particularly in templates which allow synthesis of large ordered arrays. In this chapter we describe an electrodeposition based method that in a single step can make well ordered arrays of metal nanotubes.

## 7.2 Principle

The basic principle of our method is the programmed motion of ions inside the pores of nanoporous templates. The method of synthesis of metal nanotube arrays described here exploits the basic principle of electrodeposition in a rotating electric field. As shown in Fig. 7.1, the electrodeposition is carried out in the presence of a lateral rotating electric field which is applied in addition to the longitudinal d.c electrodeposition current. The applied rotating field as our simulation shows (described later on) makes the ions graze the surface of the pores in helical paths and thus makes the deposition selectively occuring in the region near the wall of the nanopores.

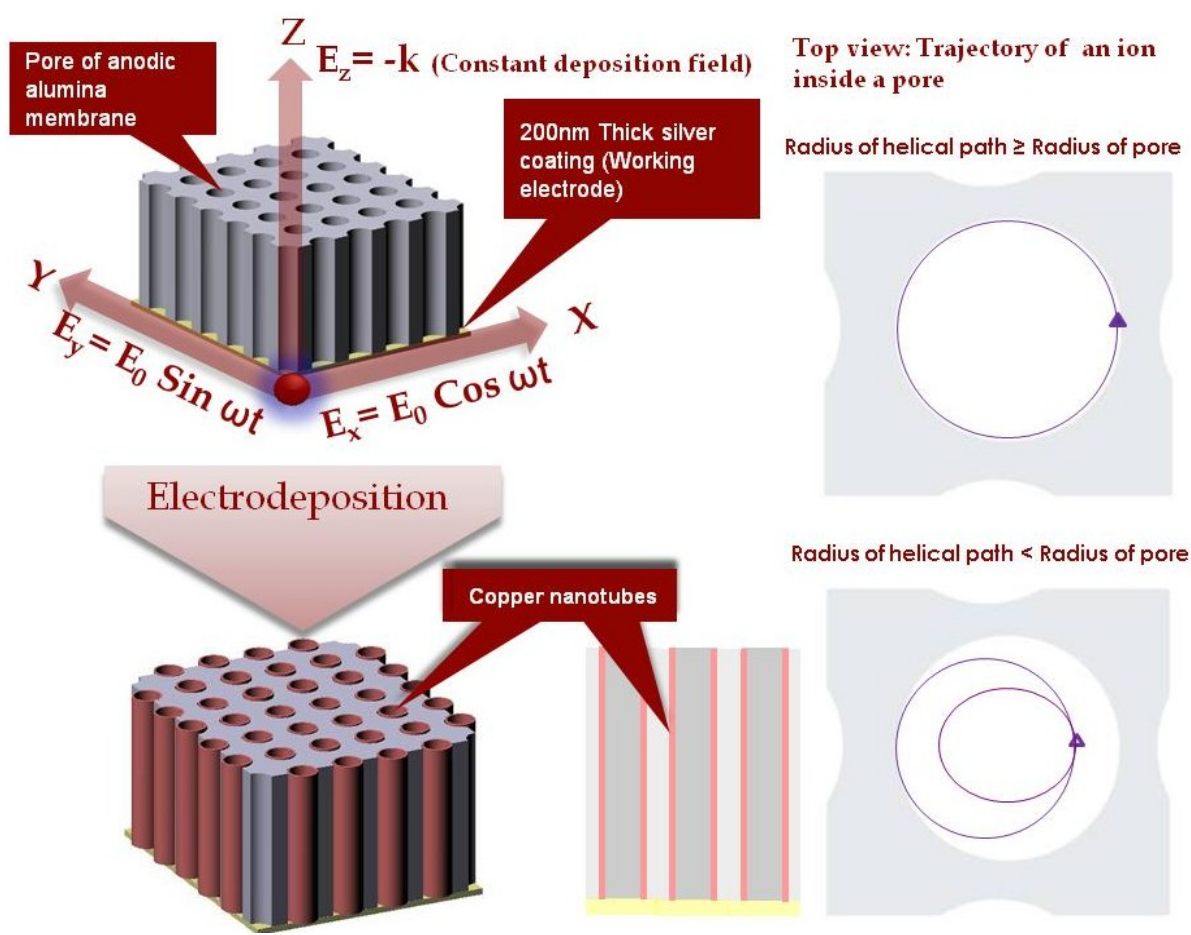

**Figure 7.1:** (a) Detailed principle of synthesis of metal nanotubes. (b) The trajectory of an ion with the radius of its helix ≥ pore radius. (c) The trajectory of an ion with the radius of its helix < pore radius.

It is important to understand the idea through a simple model and simulation, which are presented in the next section before the discussing the experimental results in the same way the problem was approached.

## 7.3 Modelling & Simulation

The mechanism of the formation of nanotubes can be understood in the following way. Taking μ, the mobility of ions, the velocity V of an ion in an electric field E is proportional to the velocity, with μ, being the proportionality constant[18]

$$\vec{V} = \mu \vec{E} \qquad (7.1)$$

We first consider the effect of the rotating field. With two sinusoidal electric fields acting along the x-axis and y-axis respectively as shown in Fig. 7.1, the two velocity components are

$$V_x = \frac{dx}{dt} = \mu E_0 \cos \omega t \qquad (7.2)$$

&

$$V_y = \frac{dy}{dt} = \mu E_0 \sin \omega t \qquad (7.3)$$

These equations essentially imply

$$X = \frac{\mu E_0}{\omega} \sin \omega t + X_0 \qquad (7.4)$$

&

$$Y = -\frac{\mu E_0}{\omega} \cos \omega t + Y_0 \qquad (7.5)$$

So the motion of the ion due to the superposition of the electric fields follows a circular orbit with a radius R

$$R = \frac{\mu E_0}{\omega} \qquad (7.6)$$

When the process of electrodeposition is started, an additional electric field $E_Z$ acts along the z-axis making the trajectory of each ion a helix. The radius of such a helix is given by R which can easily be tailored by the field amplitude $E_0$ and frequency $\omega$. Thus each ion gets deposited following such a trajectory. Within the pores of the template the motion of the ions gets constrained by the pore walls if the radius R is more than or equal to the pore radius $R_0$. This way ions, irrespective of their initial positions, upon reaching the pore walls will move staying close to the surface wall of the pores. For a given pore radius $R_0$ we can define a critical field $E_C = \omega R_0 / \mu$ where $R = R_0$. When $E_0 \geq E_C$ the deposition will occur preferentially at the pore wall.

## 7.3.1 Correction to the Field

The above simple picture of the formation of nanotube with a rotating field of constant amplitude ($E_0$) at all points along the radius of the tube, however, gets modified in the real situation. The field seen by the ions is not the same as the applied electric field because of the much larger dielctric constant of the electrolyte compared to the surrounding alumina. If $E_0$ is the amplitude of the applied field, the amplitude of the field deep inside the electrolyte is $E_{Elec} = E_0 \left( 2 K_{Alumina} / (K_{Alumina} + K_{Elec}) \right)$ (refer to appendix-I), where K denotes the dielectric constant of the medium concerned. However, this field reduction will not occur abruptly at the alumina − electrolyte interface. The applied electric field decays exponentially to $E_{Elec}$ with a characterisitc Debye screening length [19] $\lambda = \sqrt{\dfrac{\varepsilon_r \varepsilon_0 kT}{e^2 C_0}}$ , where $\varepsilon_0$ is the permittivity of free space, $\varepsilon_r$ is the dielectric constant, $k$ is the Boltzmann's constant, $T$ is the temperature, $e$ is the elementary charge, and $C_0$ is the ionic concentration. Using the relevant numbers we find that the screening length $\lambda$ is comparable to the radius of the pore and thus the field is not screened abruptly at the interface. This makes the electric field dependent on the distance from the tube wall and thus a function of the position. As a result one cannot use the simple relation as given in Eq. (7.6) to find the radius of the tube formed.

To test the above mechanism of formation of the nanotubes, we made a computer simulation of the experiment using the above model and the field amplitude profile given by the relation $E(r) = E_{Elec} + (E_0 - E_{Elec})e^{-\frac{(R_0 - r)}{\lambda}}$ (refer to appendix) instead of simply $E_0$. r is the distance from the centre along the radius and $\lambda$ is the characteristic screening length.

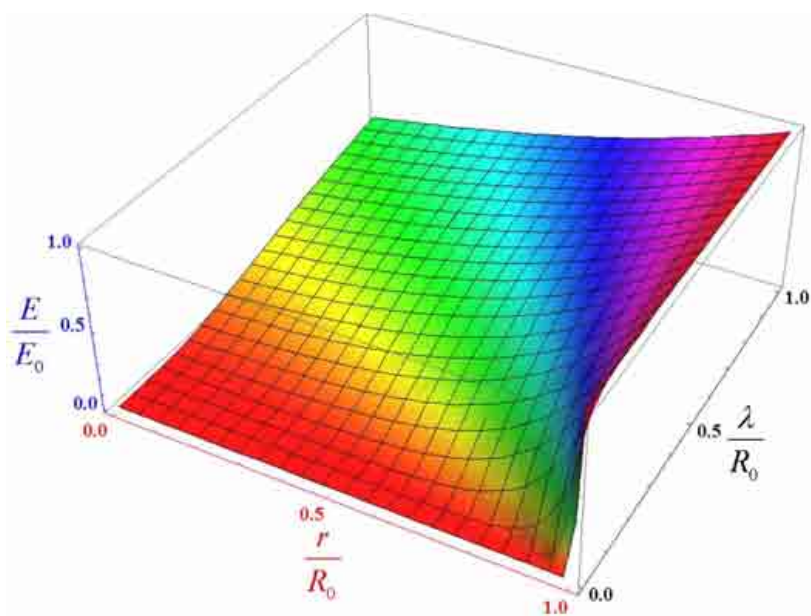

**Figure 7.2:** A general field profile showing the variation of electric field with distance from the centre (r) of the pore of radius $R_0$ and Debye screening length $\lambda$ in the regime ($E_0 >> E_{Elec}$).

The results from the simulation show that the thickness of the nanotube $\Delta R/R_0$ (expressed as fraction of pore radius $R_0$) has a simple dependence on the parameter $\lambda/R_0$. For a given applied field and as one would expect $\Delta R/R_0$ increases as $\lambda/R_0$ becomes smaller. For a given $\lambda/R_0$ the wall thickness $\Delta R/R_0$ also depends on $E_0$. Our computer simulations reveal that for $\lambda/R_0 \geq 0.47$, with $E_0 \geq E_C$, any ion, irrespective of its initial position, manages to reach the pore walls and traverses a helix grazing the surface of the pore (Fig. 7.1(b)) before getting deposited when it comes in contact with the atoms of the working electrode or the growing deposition front.

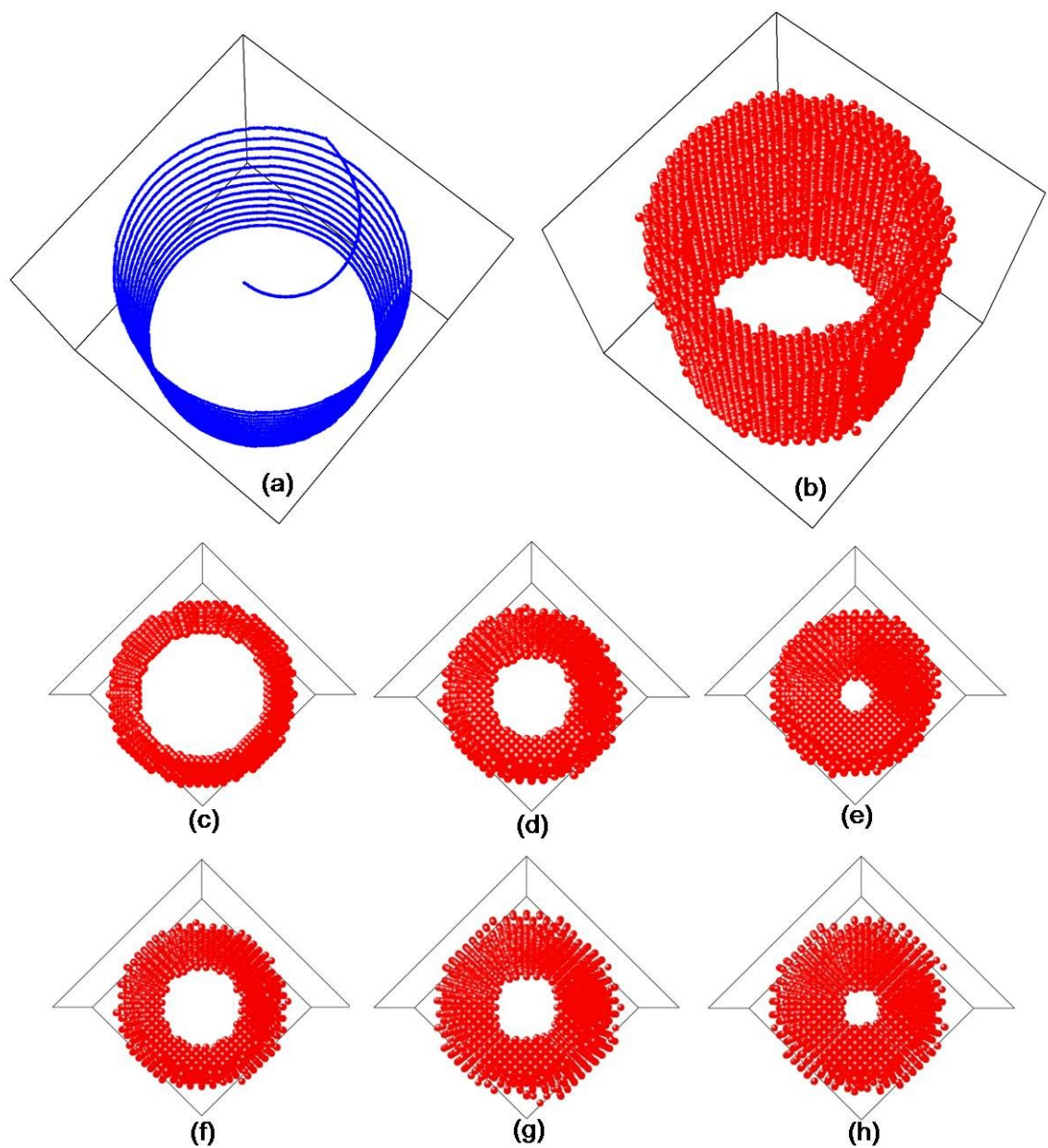

**Figure 7.3:** (a) Typical trajectory of a single ion. (b) Nanotube formed by electrodepositing 50000 atoms in a pore having 10nm diameter. Frequency of the rotating electric field is 20Hz. (c), (d), (e) correspond to the tubes formed for descending lateral field amplitudes $1.13E_c$, $E_C$, $0.86E_c$ for the same $\lambda/R_0 = 0.47$. (f), (g), (h) represent the tubes formed for 0.47, 0.4, 0.33 of $\lambda/R_0$ respectively at $E_0 = E_C$.

A typical trajectory of a single ion for such a case is shown in Fig. 7.3(a). For other values of $\lambda/R_0$ and $E_0$ (unless $E_0$ is too high such that $E_{Elec} \sim E_C$), the trajectory becomes helical with fluctuating radii (Fig.7.1(c)). Formation of a nanotube with diameter ~10nm, with 50000 atoms (whose initial positions and velocities are randomized) is shown in Fig. 7.3(b). The nanotube so formed in the simulation has a wall thickness ~2nm. This is a typical example of a metal nanotube formation. In the Fig. 7.3(c), 7.3(d), 7.3(e), we show development of tubes of different thicknesses for the same $\lambda_0/R_0=0.47$ but for different $E_0$. Our simulations also reveal that for $E_0=E_C$, the tube formation initiated even with values starting from $\lambda/R_0 \sim 0.2$ and for fields $E_0 > E_C$, tube formation can occur even with $\lambda/R_0 \leq 0.2$. It can be understood further from the plot shown in Fig. 7.4 where we show the fractional thickness as a function of fractional Debye screening length. For a fixed field ($E_0 = E_C$) the development of tubes for different $\lambda/R_0$ is shown in Fig.7. 3(f), 7.3(g), 7.3(h).

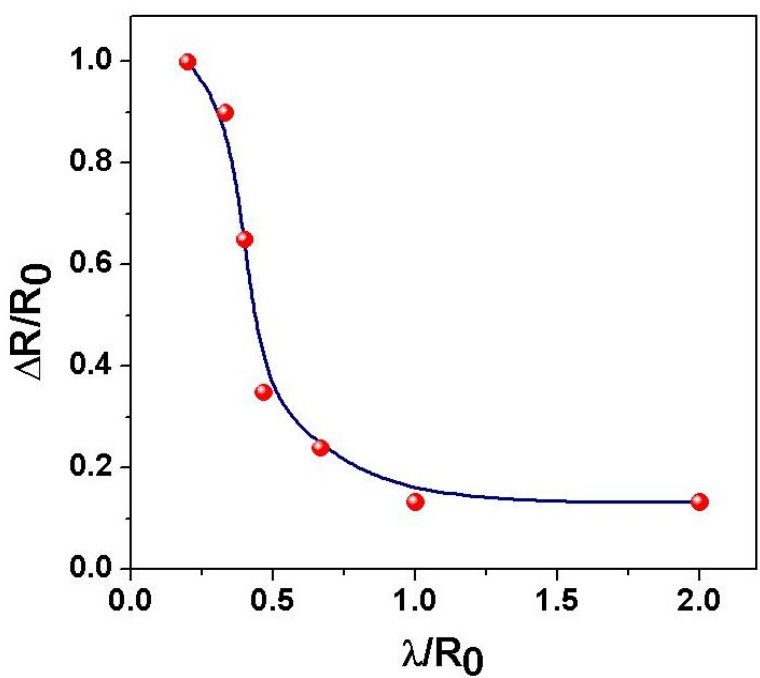

**Figure 7.4:** The evolution of thickness at $E=E_C$ as a function of Debye screening length ($\lambda$) expressed in terms of radius of the pore ($R_0$).

## 7.4 Experimental procedure for synthesis of copper nanotube arrays

The single crystalline copper nanotube arrays are prepared by electrodeposition in nanoporous anodic alumina membranes placed in the plane of a rotating electric field. Two sinusoidal electric fields of the same amplitude with a phase difference of $\pi/2$ (as shown in Fig. 7.1) constitute the rotating electric field. The principle is well illustrated in Fig. 7.1. Anodic alumina were used to prepare the copper nanotube arrays. The practical implementation of the scheme is shown in Fig. 7.5(a).

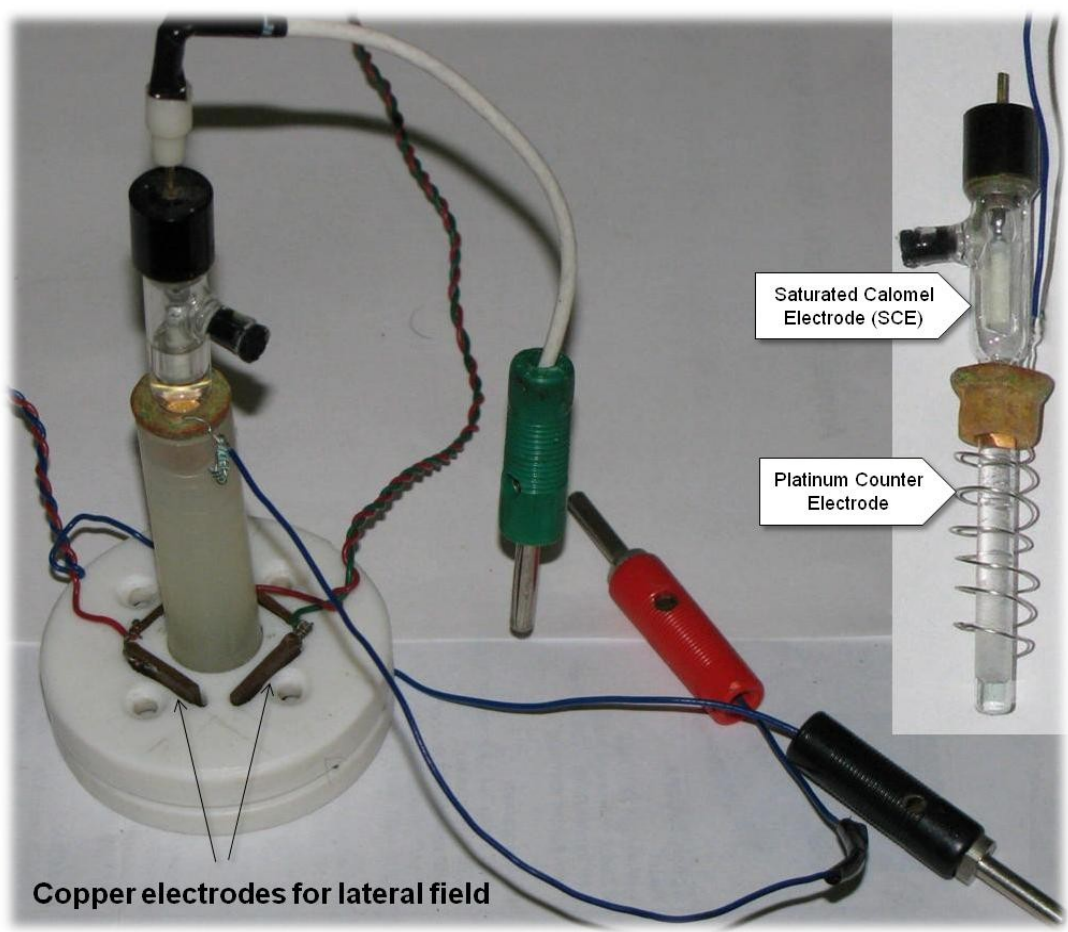

Figure 7.5(a): Experimental setup used for the synthesis of nanotube arrays.

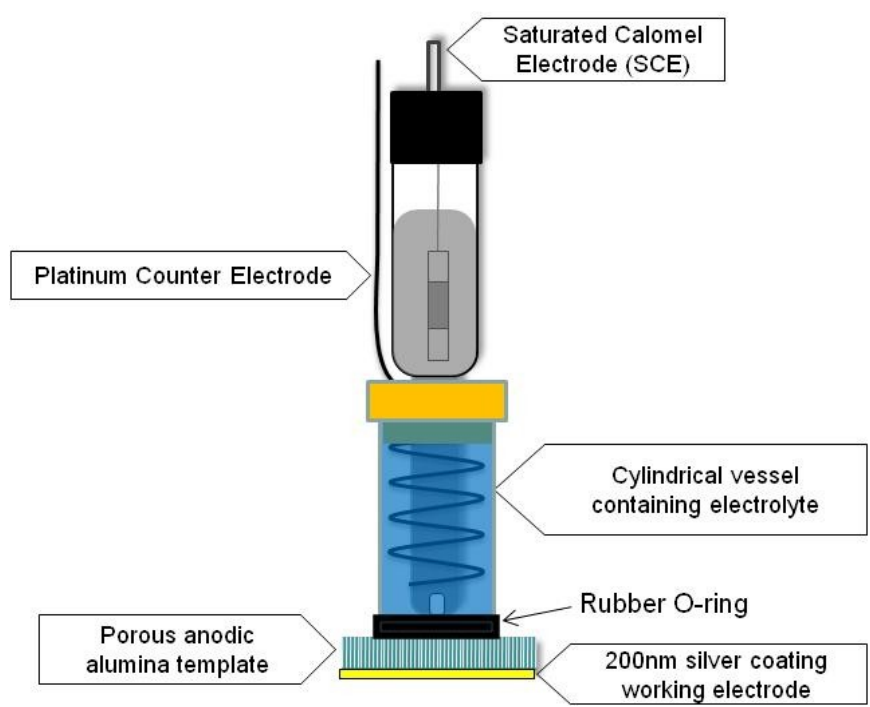

Figure 7.5(b): The vertical electrodeposition scheme used for the synthesis of nanotube arrays.

The anodic alumina membranes specified as 200 nm are observed to have most of the pores having diameters ranging between 200-240 nm. A 200nm thick layer of silver was sputtered on to one of the surfaces of the membrane. This silver layer acts as the working electrode in the three electrode potentiostatic electrodeposition as shown in Fig. 7.5(b). A Saturated Calomel Electrode (SCE) was used as the reference electrode. A platinum wire was used as the counter electrode. A 1M $CuSO_4.5H_2O$ (99.995% purity, procured from Sigma Aldrich) is slowly injected in the millipore water taken in the electrodeposition cell through a nozzle at the top of the cell during electrodeposition. The deposition is stopped when an abrupt rise in current is observed as indication of growth of tube over the complete lengths of the pores. The electrodeposition was carried out at a potential of -0.3 Volt with respect to the SCE. The membrane was placed in the plane of a rotating electric field during electrodeposition. The lateral rotating electric field was created by two pairs of parallel copper electrodes perpendicular to each other and separated by Teflon spacers as shown in Fig. 7.5(a). If the two pairs of electrodes are placed along the x and y axis, then the deposition is carried out along the z axis with the template placed in the middle of the four electrodes making the sides of a square. A sinusoidally varying voltage from a signal generator was applied to one of the pair of parallel copper electrodes. A similar signal was phase shifted by $\pi/2$ and was applied to the other pair of copper electrodes. The frequency used in the experiment was 10 Hz. With average direct current densities (Faradic current that does the actual deposition) of 6mA/cm$^2$, the time of deposition of the tubes in 200 nm pore diameter anodic alumina membranes (membrane area = 1cm$^2$, membrane thickness = 60 μm) is 15-20 min. This shows that the method is relatively faster than other reported methods [12, 13].

For imaging by scanning electron microscope (SEM), the template containing the nanotubes is etched partially with 3M NaOH solution to dissolve most of the aluminium oxide layer. The remaining template after washing several times with Millipore water was used for the imaging. SEM images were taken with the Quanta 200 FEG SEM (FEI Co.) For Transmission Electron Microscopy (TEM), the template was completely etched with 6M NaOH solution and washed 10 times with Millipore water before spraying on a carbon coated copper grid. Images were obtained by a JOEL, JEM-2010 having $LaB_6$ electron gun for operation between 80-200kV. Structural characterization was done by PANALYTICAL X-ray powder diffractometer with CuKα radiation (λ=1.5418Å) with the nanotubes remaining embedded in the template.

## 7.5 Experimental Results

This method gives very high quality MNT arrays as established by various structural chracterization tools. Fig. 7.6(a) shows a typical array of copper nanotubes fabricated by the method described, as imaged by a **Scanning Electron Microscope (SEM)**. Fig. 7.6(d) shows a closer view of the tubes protruding out from the partially etched anodic alumina template. The tubes shown in Fig. 7.6, were fabricated in porous alumina templates with pore diameters specified as 200 nm. The wall thickness of these tubes as observed (upon further zooming into the image) is ~ 20 nm. The Energy-dispersive spectrometry (EDS) of the tips of the tubes revealed that these tubes are composed of the element copper.

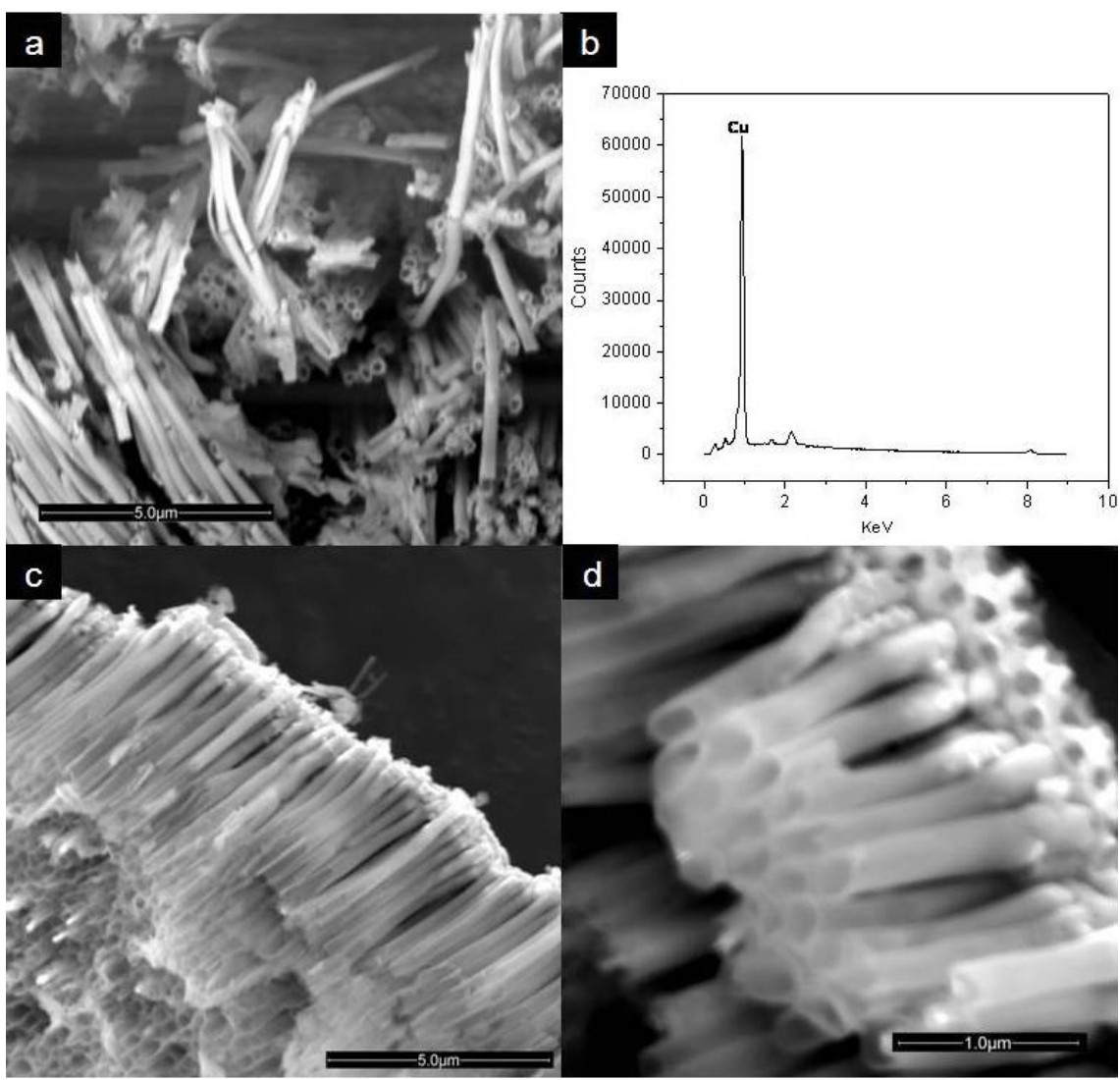

**Figure 7.6:** (a) Copper nanotubes after the removal of alumina templates. (b) EDS Spectrum of the copper nanotubes. (c) SEM image of a large array of copper nanotubes. (d) Side view of the nanotubes (The wall thickness is clearly visible).

The Transmission Electron Microscope (TEM) image (Fig. 7.7(a)) shows clearly that these tubular structures have constant wall thickness throughout their length. Fig. 7.7(b) shows a 160nm diameter tube with a thickness of approximately 15nm. The selective area diffraction pattern of a single nanotube is shown in Fig. 7.7(c). The diffraction data is indexed into the (220) plane. The TEM data shows that these tubes are single crystalline in nature.

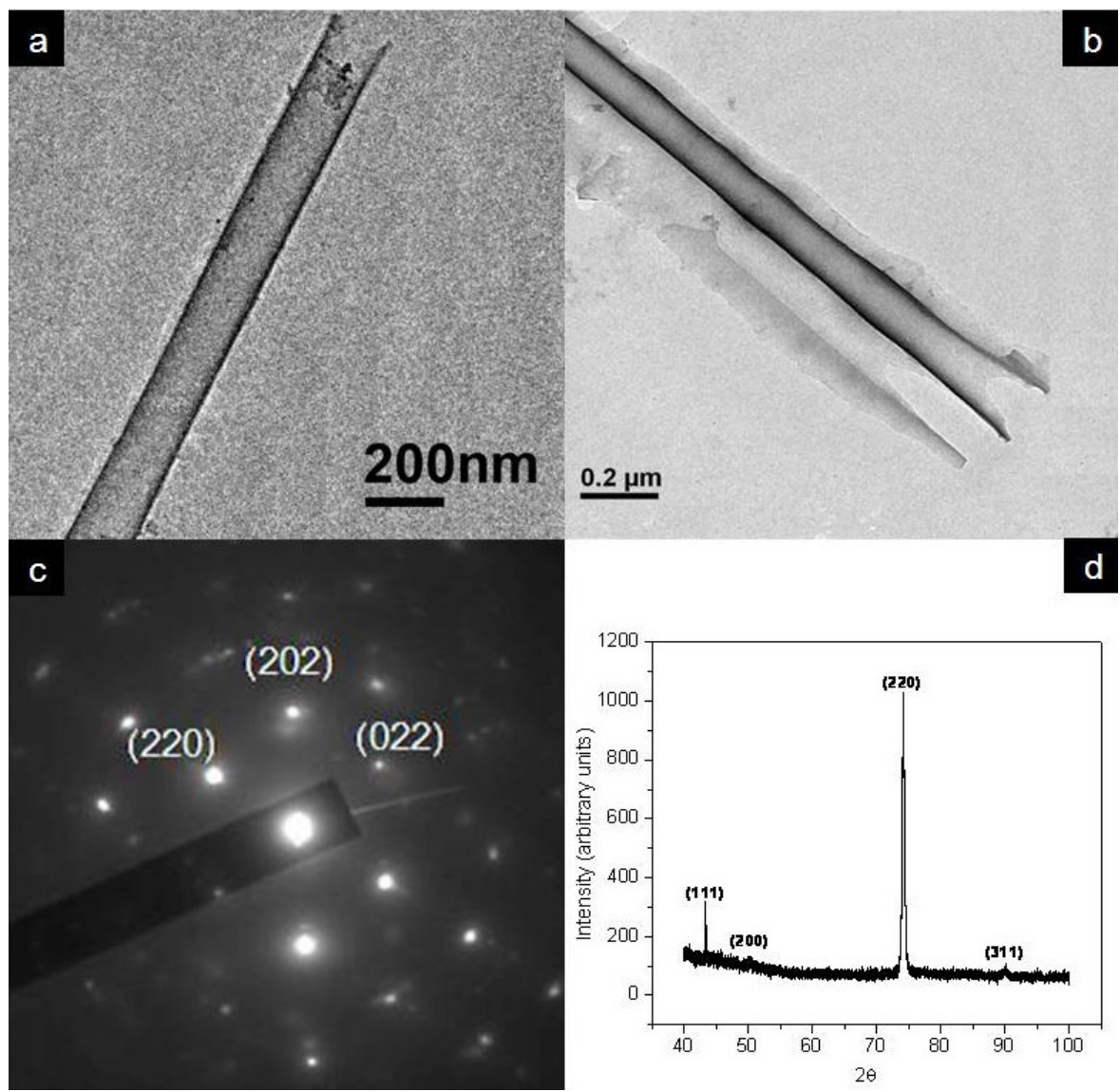

**Figure 7.7:** (a), (b) TEM images of copper nanotubes after being separated from the alumina template. The tube in (b) is partially broken to show the wall of the tube. c) Electron diffraction pattern of a tube. (d) XRD pattern of the copper nanotubes.

The single-crystalline nature of the MNT arrays have been further estbalsihed by X-Ray diffraction (XRD). A typical XRD pattern of the as synthesized samples is shown in Fig.7.7(d). The XRD data were taken by retaining the MNTs within the template. The reflections are indexed to the face-centered cubic (fcc) structure (Space group: Fm3m). No other peaks were obtained except those of copper which indicates the purity of the nanotubes formed. The XRD pattern (Fig. 7.7(d)) shows that the (220) reflection is the only prominent peak in comparison to the other reflections, indicating the fact that the array of tubes have a preferential growth direction along the (220) plane. The XRD data are in conformity with the TEM diffraction data presented before.

## 7.6 Variation of tube thickness with lateral field amplitude ($E_0$)

The electrodeposition technique that is widely used to make metal nanowires in templates uses the deposition field which is essentially longitudinal so that they are along the axis of the pore. Our innovation is based on controlling the motion of ions during electrodeposition and restricting the ions to the walls of the pores. This is achieved by a rotating electric field which is always perpendicular to the electrodeposition field and thus the extra field forces the ions to graze along the surface of the walls. The rotating electric field is produced by perpendicular superposition of two sinusoidal electric fields (of identical amplitude and frequency) to each other, differing by a phase of $\pi/2$. This is in accordance with the Lissajous figures where two sinusoidal signals with identical amplitude and frequency give rise to a circle when superposed perpendicularly with a phase difference of $\pi/2$. Since the pores to be filled have circular cross section, we used a phase difference of $\pi/2$ between the lateral two sinusoidal fields. The electric fields acting on the ions are shown in Fig. 1.

The frequency of rotating electric field determines the number of revolutions an ion makes grazing the wall surface of the pore before getting deposited. This is very important in the formation of a tube with uniform wall thickness. In our experiment, for the actual growth we chose a frequency of around 10Hz in such a way that the ions make enough revolutions along the pore walls through a column of 60μm pore depth before getting deposited. The lateral rotating electric field is produced by two pairs of copper plates, each pair being perpendicular to the other. With these specifications and taking the standard mobility of copper ions [18] as $5.56 \times 10^{-6}\,\mathrm{m^2 s^{-1} V^{-1}}$ we chose the voltage amplitude ($V_0$) of the rotating electric field ($E_0 = V_0/d$; $d$ being the distance between the electrode pairs) as 3V ($E_0 > E_C$ corresponding to 100nm) to get a helical path of ion perfectly grazing the wall of a pore of radius 100nm.

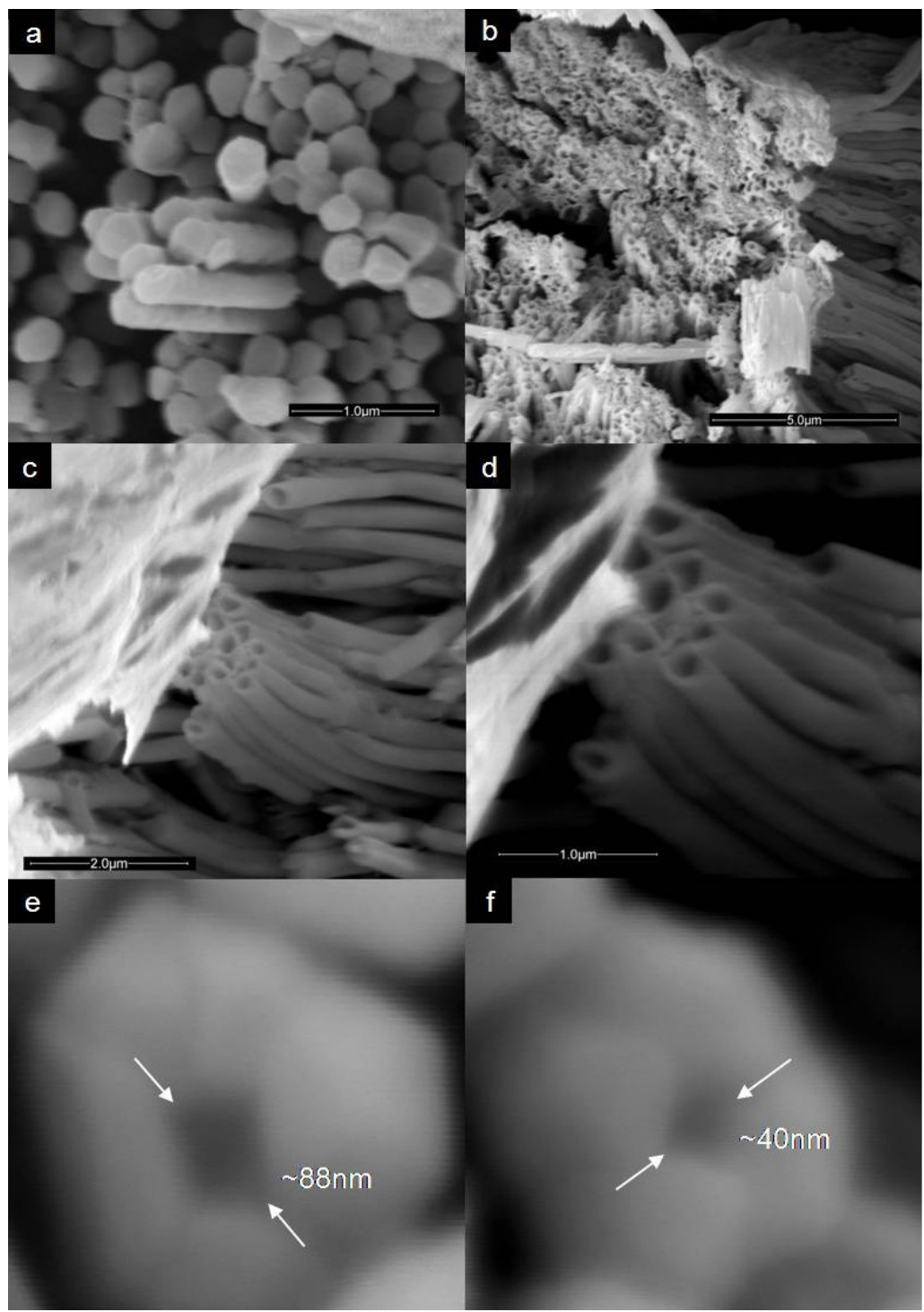

**Figure 7.8** (a) SEM image of Copper nanowires formed with the voltage amplitude of lateral rotating electric field zero. (b) A large array of copper nanotubes formed with the voltage amplitude 3V. c) Side view of the copper nanotubes (Voltage amplitude 3V). (d) Closer view of copper nanotubes (Voltage amplitude 3V). (e) SEM images of single copper nanotube of 230nm diameter (Voltage amplitude 2V). (f) SEM images of single copper nanotube having 230nm diameter (Voltage amplitude 1.5V).

From the simulation results, we infer that for a radius of 100nm, a screening length $\lambda \geq$ 20 nm will result in the formation of a tube at $E_0 = E_C$. For $E_0 > E_C$, the tube formation can take place even with $\lambda < 20$ nm. This is the typical value of the screening length for electrolytes with ionic concentrations closer to millimoles. Due to the depletion of ions due to deposition and inhomogeneity (refer to experimental), the electrolyte inside pores is of approximately millimolar concentration and this initiates the formation of the tubular structures.

Fig. 7.8(a) shows the result of electrodeposition in the porous membrane in the absence of a lateral electric field. In the absence of a rotating field one obtains nanowires, as expected. The effect of the lateral rotating electric field of 3V (voltage amplitude) is shown in Fig. 7.8(b), which shows a large array of MNT with wall thickness ~ 15-20 nm. A decrease in the lateral electric field is seen to increase the wall thickness, as evident from Fig. 7.8(e) ($V_0 = 2V$) and Fig. 7.8(f) ($V_0 = 1.5V$). This will correspond to the tube formation for the simulation shown in Fig. 7.3(e) ($E < E_c$). The resulting tubes have wall thicknesses of 70nm and 95nm respectively with outer tube diameters being 230 nm.

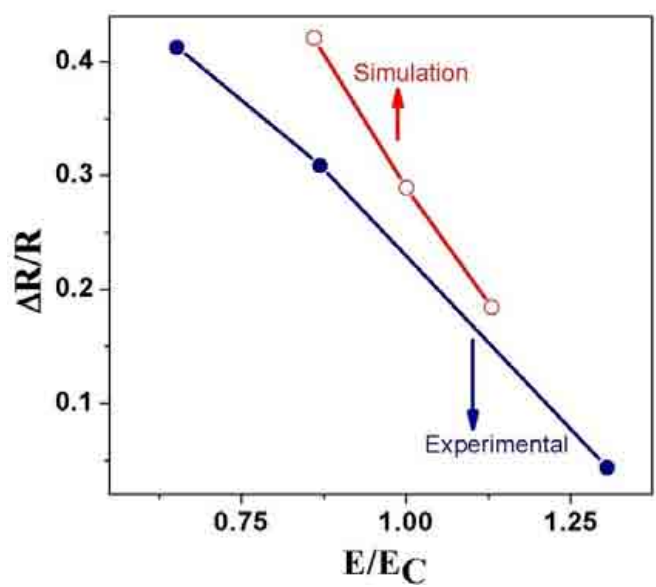

**Figure 7.9:** Comparision between experiment and simulation results.

From the simulation we expect the wall thickness to be $\approx$ 85nm and 110nm, respectively for $V_0 = 2V$ and 1.5V. Thus our simulation results agree reasonably well with the experimental results. The validity can be further clearly undertsood from Fig. 7.9 where we plot the fractional thickness of the nanotubes as a function of electric field expressed in terms of fraction of the critical field $E_C$. This establishes the fact that the basic mechanism proposed for the formation of the tube is correct. The slight deviation can be attributed due to various factors like time, field lines modification etc. The thickness of deposition also depends on the time of deposition. After the initiation of the tube wall growth, the modified electrodeposition field lines which will mostly terminate on the top edge of the tube wall periphery, will favour the growth of the tube further. The exact formulation of the tube thickness on the time and modified field lines are issues being considered for further study.

The principal benefit of this approach is that it is general, and it does not need any chemical modification or partial coating of the pores for synthesizing the nanotubes. The method also does not alter the chemistry of the standard electrodeposition that is used for a given material. It only changes the external electric field configuration. Thus this can be applied to the synthesis of any metal/compound nanotubes which can be electrodeposited. The simulation of the method gives a physical insight into how metal nanotubes can be formed inside porous templates. To our knowledge this is the first method which is not only simple and fast but also based on designing the shape of an electrodeposited metal by contolling the ionic dynamics inside an electrolyte. The MNT have a constant wall thickness throughout the length as seen in the TEM images. Thus they can act as a hollow nanoelectrode and give options of filling them with other materials like semiconductors and high dielectric constant materials for such applications in fields like nanoelectronics, solar cells, supercapacitors etc.

## 7.7 Electrical Transport Measurements

An important issue in formation of metallic nanotubes is its chemical purity in addition to its structural integrity. To test that we measured the resistance of the nanotubes from 4.2K to 300K to understand the temperature dependence of resistance in case of these nanotubes. In the previous chapters we studied the size dependence of electrical resistivity and related phenomena. In case of nanotubes we have an enhancement of surface to volume ratio, an important factor often responsible for the finite size effects and novel properties in case of nano materials. The growth of Cu nanotube and the ability to measure its electrical resistivity gives uas an oppurtunity to study the effect of surface to volume ratio on physical propeorties (using electrical measurements) that have not been possible before. This is thus a novel aspect of the thesis. It is also important to understand how does the resistivity of a nanotube array various from the nanowire array of the same metal. Debye temperature $(\Theta_R)$ as we discussed in Chapter 5 is found to be dependent upon diameter of the nanowires because of enhanced surface to volume ratio in case of nanowires. Also the residual resistivity is found to increase with enhancement in surface scattering. These effects make nanotubes interesting systems from the standpoint of electrical measurements because they have much larger surface to volume ratio but with same or comparable diameter as well as aspect ratio.

Electrical measurement of the nanotube arrays has been done by a pseudo four probe method[20]. The normalised resistance data is shown in Fig. 7.10 as compared with that of a nanowire array of the same diameter (average diameter ~ 230 nm) and bulk copper. The electrical resistance increases as a function of temperature, typical of metallic behaviour, with residual resistivity ratio ($R_{300K}/R_{4.2K}$) of 3.5.

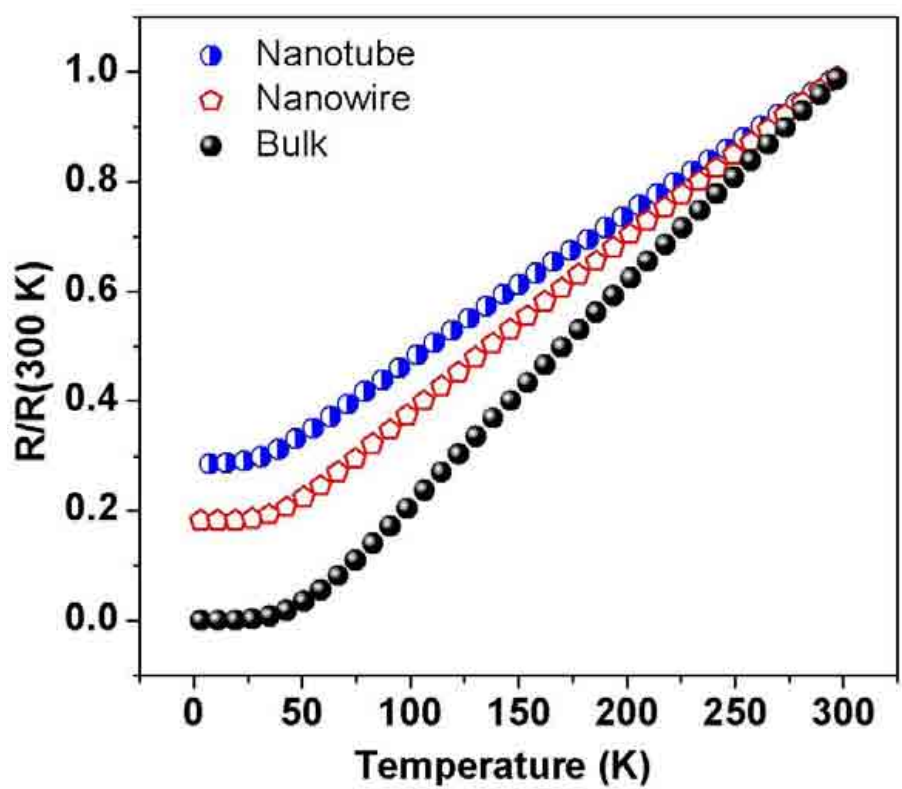

**Figure 7.10:** Normalized resistance of the nanotube as compared with nanowire of the same diameter and bulk copper.

This residual resistivity ratio is typical of copper films and is indicative of a reasonable chemical purity and absence of significant structural defects. The metallic behaviour makes these single crystalline copper nanotube arrays highly promising materials for nanoelectronic applications. A more careful analysis of the data using the Bloch-Gruneisen formula (n=5) as described in chapter 6 (Eqn. 6.13), reveals that the Debye temperature ($\Theta_R$) estimated from the resistance data in case of nanotubes is 221K as compared to 265K and 318K in case copper nanowire of similar diameter (Fig. 7.8(a) and bulk copper. A typical fit for the nanotube arrays is shown in Fig. 11 with the error (%) less than 0.3% is shown in the inset.

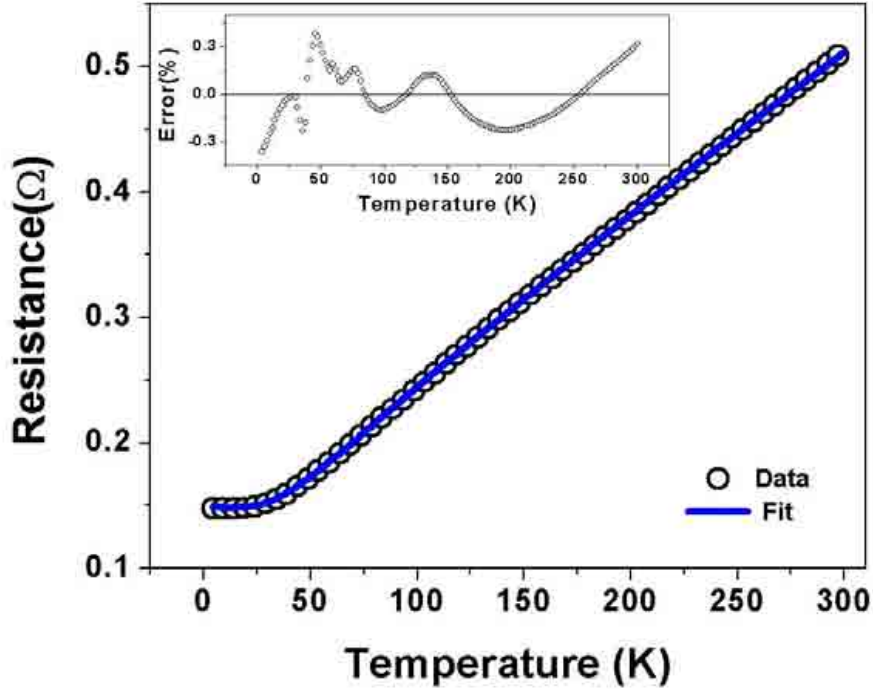

**Figure 7.11:** Fitting of the resistance data using Eq. 6.13 (chapter 6).

The obtained value the Debye temperature is close to the temperatures reported for nanowires of similar diameter[20] as the tube thickness with similar FCC structure. We tried to estimate the resistivity of the nanotubes using the method described in Chapter 6, but in this case it is more similar to the calculations reported by Aveek Bid *et al.* [20]. The resistivity of a non magnetic metal consists of two main contribution, the residual resistivity arising from impurities and imperfection of the crystal lattice and a temperature dependent lattice part arisiong out of the elctrons scattered by the lattice phonons at any given temperature.

$$\rho = \rho_0 + \rho_{el-ph}(T) \qquad (7.7)$$

$$\frac{d\rho}{dT} = \frac{d\rho_{el-ph}(T)}{dT} \qquad (7.8)$$

The resistivity of a simple metal shows a linear trend as it approaches the Debye temperature ($\Theta_R$), above which it becomes linear. Thus it easily follows from Equation. 5.2, that

$$\frac{d\rho_{el-ph}(T)}{dT} \propto \frac{1}{\theta_R} \qquad (7.9)$$

Thus

$$\frac{d\rho(T)}{dT} \propto \frac{1}{\theta_R}$$
$$\Rightarrow \frac{1}{\rho}\frac{d\rho(T)}{dT} \propto \frac{1}{\rho\theta_R} \qquad (7.10)$$

$$\frac{1}{\rho}\frac{d\rho(T)}{dT} = \frac{1}{R}\frac{dR(T)}{dT} = \beta, \quad \text{is the temperature coefficient of resistivity. If we attach suffix}$$

'n' and 'b' representing the nano and bulk sample to the above equation, then we can arrive at a relation.

$$\beta_b \propto \frac{1}{\rho_b\,\theta_{R_b}} \quad \& \quad \beta_n \propto \frac{1}{\rho_n\,\theta_{R_n}} \tag{7.11}$$

$$\Rightarrow \rho_n = \frac{\beta_b\,\theta_{R_b}}{\beta_n\,\theta_{R_n}}\,\rho_b \tag{7.12}$$

Thus knowing $\rho_b$, one can easily calculate the resistivity of the nanosamples.

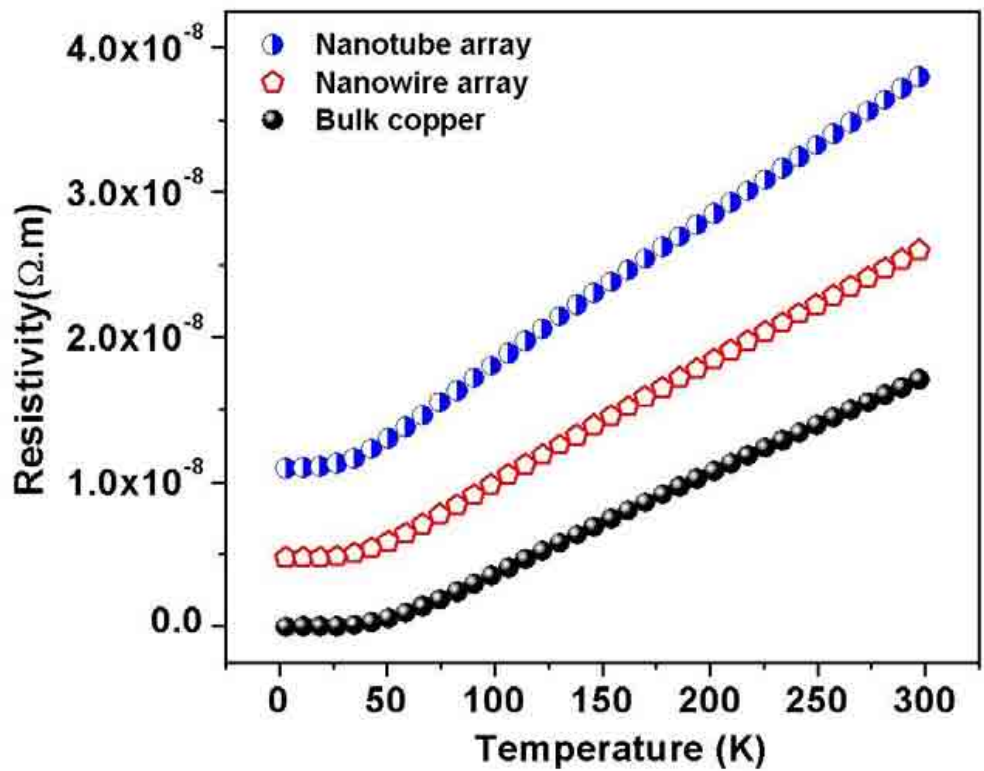

**Figure 7.12:** Resistivities of nanotube & nanowire arrays estimated using Eqn.7.12

The resistivities estimated using Eqn. 7.12 are plotted in Fig. 7.12. The electron mean free paths determined using the Drude's free electron model is found to be ~ 60 nm at 4.2 K in case of nanotube with thickness 15-20 nm as compared to a mean free path of ~140 nm in case of the nanowires of the same diameter.

As stated earlier, the Debye temperature obtained for the nanotube is close to the reported [20] values for nanowire of same diameter as the thickness. This particular observation establsihes the importance of the surface to volume ratio as the factor that affects the elastic modulus (and hence the Debye temperature) and not the diameter of the wire. This can be elaborated in the following way. The Debye temperature depends on surface to volume ratio (S/V) which is ~2/t (Where, t is the thickness of the tube having inner and outer radii as r and R respectively and with high aspect ratio; see Fig. 7.13(a) . One can clearly see from Fig. 13(b), a dependence as when the $\Theta_R$ of the nanotube (thickness~20nm) is plotted along with that of the available nanowire data [20] of the same crystal structure.

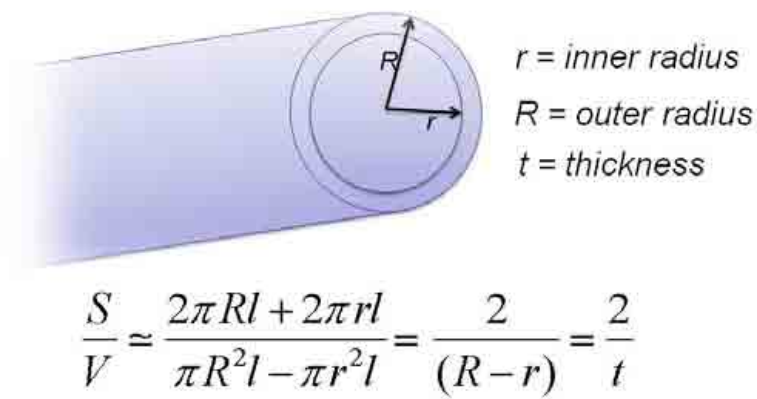

$$\frac{S}{V} \simeq \frac{2\pi Rl + 2\pi rl}{\pi R^2 l - \pi r^2 l} = \frac{2}{(R-r)} = \frac{2}{t}$$

**Figure 7.13(a):** Surface/Volume ratio of a nanotube with high aspect ratio.

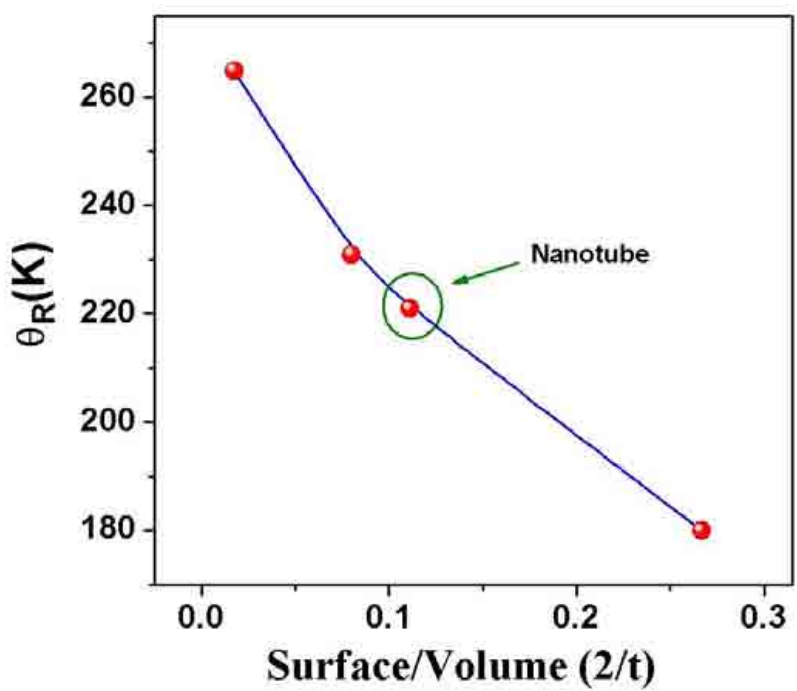

**Figure 7.13(b):** Surface/Volume ratio of a nanotube of high aspect ratio as compared to the case of nanowires.

## 7.8 Conclusion

In summary, ordered arrays of single crystalline copper nanotubes have been prepared by a novel potentiostatic electrodeposition technique in nanoporous templates in the presence of a lateral rotating electric field. The wall thickness of the metal nanotubes so obtained are in the range of 15-20 nm and can be controlled by changing the amplitude of the rotating field. The X-ray diffraction results show that the copper nanotubes grown have a preferential direction of growth. The electron diffraction results show that the tubes are single crystalline in nature. The value is more close to that of nanowire with diameter nearly equal to the thickness of the tube. The synthesis method is a simple innovation that controls the ion dynamics during electrodeposition. We believe that this is a general method for growth of other metal nanotube arrays and can be applied to many materials which can be grown by electrodeposition technique, including compound material nanotube arrays. The study of electrical resistance as a function of temperature shows that these tubes have a metallic behaviour. The Debye temperature as estimated from resistance data is found to decrease from the nanowire to nanotube of the same diameter for the FCC structure and is found to depend on the surface to volume ratio. The increase in resistivity from the nanowire to tube array can be understood to be arising from enhanced surface scattering from the enhanced surface area.

# Summary and conclusions of the thesis

## Experimental contributions made in the thesis

❖ The synthesis of high quality single crystalline metal nanowires and nanotubes is perfected in this thesis using both DC electrodeposition and pulsed electrodeposition . The nanowires in the diameter range between 13nm to 200 nm have been grown inside the nanoporous templates of anodic alumina templates. The wires grown were extensively characterized for the structure and microstructure using tools like XRD,TEM and HETEM.

❖ Low temperature and high temperature resistance measurement setups have been prepared to study the effect of diameter on the electrical resistivity in these nanowires and nanotubesof different crystallographic nature. The necessary software for measurement automation was developed.

## Physics contributions made in the thesis

❖ The nanowires are of metals, it is unlikely that they show any quantum effect, but one would expect classical size effects to show up in these nanowires. The thesis made an extensive measurement of electrical transport properties followed by quantitative evaluation in magnetic nanowires that has not been done before.

❖ High temperature (above 300K) resistivity measurements have never been done in nanowires. While in non-magnetic nanowires, one would expect only phonon contributions to dominate, in magnetic nanowires one would expect scattering from spins to dominate and this show up as the critical behaviour near $T_C$. We reported in this thesis, the first investigation of the high temperature resistance in nickel nanowires to understand the effect of finite size on the critical phenomena near ferromagnetic to paramagnetic phase transition. A decrease in $T_C$ with the decrease in diameter following the scaling relation of Curie temperature shift with a characteristic shift exponent is observed. A precise measurement of the resistance near $T_C$ followed by detailed quantitative analysis of the resistance anomaly revealed a systematic change in the critical exponent of specific heat $\alpha$ as extracted. Systematic change in the values of ratios of the amplitudes of the leading term and corrections to the scaling indicating that the size reduction makes the system approach a quasi one dimensional case.

❖ The low temperature resistance measurements were done on single crystalline oriented (220) nickel nanowires of various diameters ranging from 55 nm to 13 nm synthesized pulsed electrodeposition in anodic alumina templates with a specific aim to understand the effect of surface scattering on the resistivity and to establish whether one can have a quantitative estimation of the same. The detailed analysis on the residual resistivity using Dingle and Sondheimer theories shows that the enhancement in the residual resistivity with the decrease in size is due to limitation imposed electron mean free path by the finite wire diameter. A systematic analysis of the resistivity variation with T was done using Bloch-Gruneissen theory to obtain the Debye temperature. The Debye temperature as estimated from the low temperature data is found to decrease with the decrease in diameter indicating the enhanced surface to volume ratio leads to softening of the modulus of elasticity which in turn reduces the Debye temperature. The magnetic contribution to the resistivity, observed below 15K was estimated and was found to be suppressed with the decrease in wire diameter over the whole temperature range. Four probe based resistivity measurements done on single nanowire of 55 nm and subsequent analysis yielded similar results.

❖ A new method of synthesis of arrays of metal nanotubes arrays based on electrodeposition in anodic alumina templates in presence of a rotating electric field is invented. As a generic example, copper nanotube arrays have synthesised and well characterized by XRD, TEM, SEM. The dependence of tube thickness on the voltage amplitude of the rotating electric field is studied by varying the later. Systematic modelling and simulation results are found to be in reasonable agreement with the experimental results.

❖ The thesis reports the first measurement of resistivity of metal nanotubes over the complete temperature range from 3K to 300K. The copper nanotubes are found to be metallic in nature as seen from the resistance measurements. A detailed analysis of the resistivity shows an enhancement in nanotube resistivity in comparison to nanowire of the same diameter synthesised under similar condition but in absence of rotating electric field. A closer analysis involving surface to volume ration revealed that the Debye temperature as estimated from the resistance data using Bloch-Grüneisen formula is found to be dependent only on the thickness of the nanotube, matching the value close to that of nanowires of diameter equal to the thickness of the tube.

## Scope for further work

❖ In the nanowire fabrication front, it will be challenge to device the electrochemistry of making nanowires of metals with low magnetic Curie temperature like Gadolinium for ease in measurements near the critical temperatures and understand similar physics as studied in case of nickel nanowires in this thesis.

❖ It is also a challenging to prepare alloy nanowires of specific stoichiometry. Particularly novel binary alloys like NiTi (Nitionol) shape memory alloys can have immense applications potential in NEMS based devices.

❖ It is interesting to study the effect of magnetic ion concentration doped in noble metal nanowires synthesised by electrodeposition to study the effects like Kondo effect, RKKY interaction etc. in such systems.

❖ In addition devices like spin valves nanowires made by successive electrodeposition can be interesting problems.

❖ In the case of nanotube arrays synthesis, there are a number of materials which can be deposited by electrodeposition and can be tested to make their arrays of nanotubes. In addition coaxial nanotubes, core-shell nanotubes of various metals and semiconductors can be explored and their electrical properties can be studied. Possible application of such systems can be explored.

❖ The Ni nanowires can also be excellent systems to study the basic physics of current induced magnetization reversal, an important building block of Spintronics.

# Appendix-I

## Gauss-Hermite Quadrature

**Gauss–Hermite quadrature[1]** is an extension of Gaussian quadrature ( integration in numerical analysis) method for approximating the value of integrals of the following form

$$\int_{-\infty}^{+\infty} e^{-x^2} f(x)\, dx.$$

In **Gauss–Hermite quadrature**

$$\int_{-\infty}^{+\infty} e^{-x^2} f(x)\, dx \approx \sum_{i=1}^{n} w_i f(x_i)$$

where $x_i$ is the $i$-th root of Hermite polynomial $H_n(x)$

$$H_n(x) = (-1)^n e^{x^2} \frac{d^n}{dx^n} e^{-x^2}$$

and the weight $w_i$ is given by

$$w_i = \frac{2^{n-1} n! \sqrt{\pi}}{n^2 [H_{n-1}(x_i)]^2}.$$

# Appendix-II

## Electric field inside a cylindrical cavity

The electric field inside a cylindrical cavity of radius $a$ with dielectric constant K$_1$ kept in a medium of dielectric constant K$_2$ can be calculated solving Laplace equations.

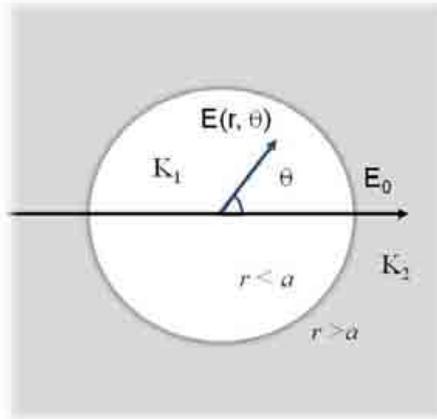

To calculate the field inside the cylinder, we have to solve the Laplace equation $\left(\nabla^2 \phi = 0\right)$ in cylindrical polar coordinates. $\Phi$ is a function of $r$ and $\theta$ only.

Thus,

$$\phi = \sum_{n=1}^{\infty}\left[A_n \cos n\theta + B_n \sin n\theta\right]r^n + \sum_{n=1}^{\infty}\left[C_n \cos n\theta + D_n \sin n\theta\right]r^{-n} + A_0 + C_0 \ln r$$

With the following boundary conditions

   i.   $\Phi_1$ and $\Phi_2$ satisfy Laplace's equation i.e. $\nabla^2 \phi_1 = 0$ and $\nabla^2 \phi_2 = 0$.

   ii.   Inside the sphere i.e. for $r < a$, the potential $\Phi_1$ is finite.

   iii.   As $r \to \infty$, the potential $\Phi_2$ must tend to $-E_0 z = -E_0 r \cos\theta$.

   iv.   At $r = a$ , $\Phi_1 = \Phi_2$.

   v.   At $r = a$ , $\varepsilon_1 \dfrac{\partial \phi_1}{\partial r} = \varepsilon_2 \dfrac{\partial \phi_2}{\partial r}$.

Using (ii), we get $C_n = D_n = C_0 = 0$. Thus, $\phi_1 = A_0 + \sum_{n=1}^{\infty}\left[A_n \cos n\theta + B_n \sin n\theta\right]r^n$ .

Using (iii), we get, $A_0^{'} = 0, C_0^{'} = 0, A_1^{'} = -E_0$.

Thus, $\phi_2 = -E_0 r \cos\theta + \sum_{n=1}^{\infty} \left[ C_n^{'} \cos n\theta + D_n^{'} \sin n\theta \right] r^{-n}$.

At $r = a$ , $\Phi_1 = \Phi_2$, therefore,

$$A_0 + \sum_{n=1}^{\infty} \left[ A_n \cos n\theta + B_n \sin n\theta \right] a^n = -E_0 a \cos\theta + \sum_{n=1}^{\infty} \left[ C_n^{'} \cos n\theta + D_n^{'} \sin n\theta \right] a^{-n}$$

.

Thus, $A_n a^n = C_n^{'} a^{-n}$, for $n \neq 1$ and $B_n a^n = D_n^{'} a^{-n}$.

Thus, $A_n = 0 = C_n$, for $n \neq 1$ and $B_n = D_n = 0$; $A_0 = 0$.

$$A_1 a = -E_0 a + \frac{C_1^{'}}{a} \ldots\ldots\ldots\ldots\ldots\ldots\ldots\ldots\ldots(1)$$

$\phi_1 = A_1 r \cos\theta$

$\phi_2 = -E_0 r \cos\theta + C_1 r^{-1} \cos\theta$

$\because \varepsilon_1 \frac{\partial \phi_1}{\partial r} = \varepsilon_2 \frac{\partial \phi_2}{\partial r}, \therefore E_1 A_1 \cos\theta = E_2 \left\{ -E_0 \cos\theta - \frac{C_1}{a^2} \cos\theta \right\}$

$\Rightarrow \varepsilon_1 A_1 = -\varepsilon_2 E_0 - \varepsilon_2 \frac{C_1}{a^2}$

$$\text{Thus, } \frac{\varepsilon_1}{\varepsilon_2} A_1 = -E_0 - \frac{C_1}{a^2} \ldots\ldots\ldots\ldots\ldots\ldots\ldots(2)$$

From (1) and (2), we get,

$A_1 \left( 1 + \frac{\varepsilon_1}{\varepsilon_2} \right) = -2E_0$

$\Rightarrow A_1 = \frac{-2E_0}{\left( 1 + \dfrac{\varepsilon_1}{\varepsilon_2} \right)} = \frac{-2E_0}{\left( 1 + \dfrac{\kappa_1}{\kappa_2} \right)} = \frac{-2\kappa_2 E_0}{\kappa_1 + \kappa_2}$ .

$\therefore \phi_{inside} = A_1 r \cos\theta = \frac{-2\kappa_2}{\kappa_1 + \kappa_2} E_0 r \cos\theta$

Hence, the field inside the cavity is given by

$$E_{inside} = -\frac{\partial \varphi_{inside}}{\partial z} = \frac{2\kappa_2}{\kappa_1 + \kappa_2} E_0$$